%% file: viikinkoski_2017.tex
\documentclass{aa}
\usepackage{amsmath,url,longtable}
\usepackage{amssymb}
\usepackage{paralist}
\usepackage{natbib}
\bibpunct{(}{)}{;}{a}{}{,}
\usepackage{lmodern}
\usepackage{siunitx}
\usepackage[maxfloats=50]{morefloats}
\usepackage{graphicx,mathtools,subcaption,caption}
\usepackage{float}
\usepackage[varg]{txfonts}
\usepackage[section]{placeins}

\usepackage{ulem}

\newcommand{\rem}[1]{\textcolor{red}{\sout{#1}}}

\renewcommand{\rem}[1]{}
\usepackage{pstricks}

\begin{document}

\title{Adaptive optics and lightcurve data of asteroids: twenty shape models and information content analysis}
\titlerunning{Shape models of twenty asteroids}
\date{Received x-x-2017 / Accepted x-x-2017}
   
\author{M.~Viikinkoski\inst{1}
\and J.~Hanu{\v s}\inst{2}
\and M.~Kaasalainen\inst{1}
\and F.~Marchis\inst{3}
\and J.~\v{D}urech\inst{2}
}
\institute{Department of Mathematics, Tampere University of Technology, PO Box 553, 33101 Tampere, Finland
  \and
    Astronomical Institute, Faculty of Mathematics and Physics, Charles University, V Hole\v{s}ovi\v{c}k\'ach 2, 18000 Prague, Czech Republic
  \and
    SETI Institute, Carl Sagan Center, 189 Bernado Avenue, Mountain View CA 94043, USA
}

\abstract{We present shape models and volume estimates of twenty asteroids based on relative photometry and adaptive optics images. We discuss error estimation and the effects of myopic deconvolution on shape solutions.  For further analysis of the information capacities of data sources, we also present and discuss ambiguity and uniqueness results for the reconstruction of nonconvex shapes from photometry.}

\maketitle

\section{Introduction}
In our previous paper \citep{Hanus2017b}, we derived 3D shape models for about 40 asteroids by combining disk-integrated photometry, adaptive optics (AO) images, and stellar occultation timings by the All-Data Asteroid modelling \citep[ADAM\footnote{\url{https://github.com/matvii/ADAM}},][]{VK15} procedure. Occultation chords are useful for the 3D shape construction and especially the size estimate since they provide information on the silhouette of the asteroid independently of the scattering law \citep{Durech2011}. 

There are several asteroids with resolved AO images obtained by the Near InfraRed Camera (Nirc2) mounted on the W.M.~Keck II telescope and optical disk-integrated photometry that can be used for the 3D shape model determination, but lacking observed occultation timings. In this paper, we determine shape models and spin states for twenty such asteroids. Because the size estimates are based exclusively on the AO data of varying quality, special attention must be paid when assessing their uncertainties. For instance, erroneous or badly resolved images may inflate the size of the shape model and hide details present in other images. Since the obtaining of additional data is seldom an option, effects of bad quality data must be identified and mitigated. Statistical resampling methods facilitate the detection of outlier images and the estimation of the uncertainties of reconstruction.    
 
Another interesting topic is the effect of myopic deconvolution on constructed shape models. Since the level of detail of full shape models is necessarily lower than the apparent resolution of AO images, it is natural to question the necessity of myopic deconvolution as a preprocessing step when data are used for shape modelling. Furthermore, while deconvolution is useful for the visual inspection of individual images, processed images are potentially biased by prior assumptions and contaminated by artifacts. To investigate the usefulness of myopic deconvolution, we construct shape models using both unprocessed and deconvoluted images, and compare their differences.

This paper is organized as follows. In Sect.~\ref{sec:methods} we describe the data reduction steps and outline the ADAM shape optimization procedure. We discuss our results and comment on some of the individual asteroids in Sect.~\ref{sec:results}. We present a procedure for uncertainty and information content estimation, in particular noting that the post-processing of AO images, while useful for the visual investigation of an image, is not necessarily needed or even desirable in 3D asteroid modelling. 
In Sect.~\ref{sec:nonc}, we further build on the information content analysis by revisiting the case of photometry only, with analytical results that are very scarce for nonconvex shapes. We sum up our work in Sect.~\ref{sec:conclusion}.

\section{Methods}
\label{sec:methods}
\subsection{Data reduction}
All the adaptive optics images used in this paper, together with the calibration data, were downloaded from the Keck Observatory Archive (KOA). The AO images were acquired with the Near Infrared Camera (NIRC2) mounted at the W.M Keck II telescope.
While the diffraction-limited angular resolution of the telescope is 45~mas, the actual achievable detail varies with the seeing conditions. 

Raw data were processed using the typical data reduction pipeline of dark frame and flat-field correction followed by the bad pixel removal. Finally, to obtain data with sufficient signal-to-noise ratio, several short-exposure frames were shift-added together.

The compensation of atmospheric turbulence offered by the AO instrument is only partial, and the point-spread function (PSF) is partially unknown. In order to mitigate the blurring effect caused by the PSF, AO images are usually post-processed with a myopic deconvolution algorithm.
 
Deconvolution is an ill-posed problem in the sense that high-frequency information is irrevocably lost during the imaging process. To acquire an estimate of the original data, prior assumptions adapted to the structure of the observed object must be used. Typically, an $L_2\!-\!L_1$ regularization \citep[e.g., total variation, see][]{St06}, which smooths out small gradients in the image and preserves large gradients, is preferred.
 
In this paper, we apply the AIDA \citep{Ho07} deconvolution algorithm. The AIDA algorithm uses the Bayesian framework and iterative optimization to produce an estimate of the true image and PSF, given the initial image and the partially known PSF obtained from the instrument.

In addition to the AO images (see Table~\ref{table0}), we use the optical lightcurves available in the Database of Asteroid Models from Inversion Techniques \citep[DAMIT\footnote{\url{http://http://astro.troja.mff.cuni.cz/projects/asteroids3D}},][]{Durech2010}. While recovering nonconvex features based on lightcurves alone is seldom possible, as we discuss below and in the corresponding references, photometry is crucial for complementing the AO data and stabilizing the shape optimization process. Moreover, prior knowledge of the sidereal rotation period and the spin axis orientation are useful inputs for the ADAM modelling. These are available in DAMIT as well. 

\subsection{Shape modelling}
An intuitively direct approach to asteroid shape reconstruction is contour matching \citep{Ca10,Ka11}, in which the asteroid boundary contour is extracted from the deconvolved image and compared with the corresponding plane-of-sky model boundary. This facilitates model fitting independently of the scattering law, but the contour identification in the AO images is problematic due to smearing and other imaging artifacts. Besides, the boundary of a non-convex shape observed at high phase angles may break into disconnected contours. 

In this paper we use the shape reconstruction algorithm ADAM \citep{VK15} that has been successfully applied to many asteroids \citep{Vi15, Ha17, Hanus2017b, Ma17}. The method used in ADAM circumvents the boundary identification problem by minimizing the difference between the projected model and the image in the Fourier domain. In this way, contour extraction is not required since the boundary is determined automatically during the optimization, based on all the available data. Moreover, ADAM can use both deconvolved and unprocessed images, making the comparison between reconstructed models possible.

ADAM facilitates the simultaneous use of various data types with different weightings. In this case, we combine disk-integrated photometry with the adaptive optics images. Utilizing the Levenberg-Marquardt optimization algorithm \citep{Press}, ADAM minimizes the objective function
\begin{equation}\label{eq:fit}
\chi^2:=\lambda_{ao}\chi^2_{ao}+\lambda_{lc}\chi^2_{lc}+\sum_i\lambda_i\gamma_i^2,
\end{equation}
where
\[
\chi^2_{ao}=\sum_{i,j}\left\Vert V_i(u_{ij},v_{ij})-e^{2\pi\imath\left(o^x_i u_{ij}+o^y_i v_{ij}\right)+s_i}\,\mathcal{F}\!P_i(u_{ij},v_{ij})\mathcal{F}\!M(u_{ij},v_{ij})\right\Vert^2,\]
and $\mathcal{F}\!M(u_{ij},v_{ij})$ is the Fourier transform of the plane-projected  model $M$ evaluated at the $j$th frequency point $(u_{ij},v_{ij})$ of the $i$th image, $V_i$ is the Fourier transform of the $i$th AO image, and $\mathcal{F}\!P_i$ is the Fourier transform of the PSF corresponding to the AO image. The offset $(o^x,o^y)$ within the projection plane and and the scale $s_i$ are free parameters determined during the optimization. The term $\chi^2_{lc}$ is a square norm measuring the model fit \citep{KT01a} to the lightcurves. The last term corresponds to regularization functions $\gamma_i$ and their weights $\lambda_i$ \citep{VK15}. 

Shapes are represented by triangular meshes, where the vertex locations are given by the parametrization. ADAM uses two different shape supports: octantoids \citep{KV12}, which utilize spherical harmonics, and $\sqrt{3}$-subdivision surfaces \citep{Ko02}.

\section{Results and discussion}
\label{sec:results}
We use a shape modelling approach similar to that in our previous study \citep{Hanus2017b}: For each asteroid, we choose initial weight terms in Eq.~\eqref{eq:fit}. Fixing the image data weight $\lambda_{AO}$, we decrease regularization weights $\lambda_i$ and the lightcurve weight $\lambda_{LC}$ until the image data fit $\chi^2_{AO}$ ceases to improve \citep{KV12}. As the final step, the model fit to AO images is assessed visually. This procedure is repeated for both deconvolved and raw AO images using two shape supports. The usage of different shape representations allows us to separate the effects of parametrization from the actual features supported by the data. If the available data constrain the shape sufficiently, both octantoid and subdivision parametrizations should converge to similar shapes \citep{VK15}.   

Given the shape models, we compute the volume-equivalent diameter. To gauge the stability of the derived  model, we use the jackknife resampling method \citep{Efr81}. We produce $n$ additional shape models by leaving out one of the $n$ available AO images. 
For asteroids with only one observation, the jackknife method cannot obviously be used (28~Belona, 42~Isis, and 250~Bettina). In these cases the uncertainty estimates are based on the variability observed with different shape supports.

In Table~\ref{table0}, we have listed the mean pixel size in kilometers (adjusted to the asteroid distance), diameters of the reconstructed models based on both deconvolved and raw AO images, and the variability intervals obtained by resampling. The estimated spin parameters, together with their uncertainties, are listed in Table~\ref{tab:results}. These uncertainty estimates result from combined variations within all the shape supports. 

It is apparent that the resampling method seems to detect potential outlier images (e.g. the fuzzy images of 121~Hermione), which typically result in large variations in diameters. However, it is also evident that not all shape uncertainty can be revealed by resampling. For instance, sampling from a set of observations with similar observation geometries can lead to practically same shapes even if half of the asteroid were unobserved (see also the discussion on sampling in shape space in \cite{VK15}). In general, the uncertainty in the shape models is also affected by the image resolution; it is unreasonable to expect the accuracy of the diameter estimate to exceed the pixel quantization error (i.e. half of the pixel size).   
   
We have derived shape models and estimated their sizes for twenty asteroids. The disk-resolved data allowed us to remove the pole ambiguity (i.e., reject one of the two pole solutions that usually fit the optical lightcurves equally well) for four asteroids -- (7)~Iris, (48)~Doris, (65)~Cybele, and (283)~Emma. Mass estimates for most asteroids studied here are available in the literature, so we were able to combine those values with our volume determinations, which gave us bulk density estimates. Unfortunately, some mass estimates are affected by large uncertainties (72~Feronia, 354~Eleonora, or 1036~Ganymed). Derived sizes, adopted masses, asteroid taxonomy \citep{Tholen1989, Tholen1989b, Bus2002, DeMeo2009} and bulk densities are listed in Table~\ref{tab:densities}.


\paragraph{7 Iris} 
Iris seems to be the third largest S-type asteroid (after 15~Eunomia and 3~Juno). Its size of $D=216\pm7$~km is consistent with previous estimates based on stellar occultations \citep[$D=210\pm30$~km,][]{Durech2011} or radar data \citep[$D=223\pm37$~km,][]{Ostro2010}. We provide the most accurate size estimate so far, which allowed us to derive a reasonably constrained bulk density of $\rho=(2.4\pm0.5)$ gcm$^{-3}$. This value is typical for S-type asteroids. We note that some of images manifest severe ringing artifacts caused by the deconvolution algorithm.

A non-convex shape model based on range-Doppler radar images was reconstructed in \cite{Ostro2010}. Due to  limited radar coverage, a significant portion of the model  was based dynamical constraints rather than observations. While the resolution of AO images is insufficient for resolving non-convex details present in the  radar model, the dimensions and the large-scale features are consistent between the models.

\paragraph{15 Eunomia}
Eunomia is the largest S-type asteroid. Its large size together with the location in the inner main belt made Eunomia a great target for imaging. Indeed, all the AO images of Eunomia are nicely resolved. The surface is rather smooth without any obvious features. All this resulted into one of the best shape models we have ever derived. Our spin solution is similar to the one of \citet{Kaasalainen2002b} based on optical lightcurves only. Our size ($D=275\pm5$~km) is slightly larger than most previous estimates based mostly on thermal models ($\sim250$~km). The bulk density of $\rho=(2.9\pm0.2)$ gcm$^{-3}$ corresponds to typical values within the S-types.

\paragraph{23 Thalia}
We have two images with almost identical observing geometries. Based on visual estimation of the fit, the second pole from Table~\ref{tab:results} seems more likely.

\paragraph{24 Themis}
Our spin state solution of Themis is consistent with the previous determinations of \citet{Higley2008} and \citet{Hanus2016a}. We present the first non-radiometric size of Themis of $D=215\pm15$~km, which is somewhat larger than previous estimates. The low bulk density of $\rho=(1.1\pm0.4)$ gcm$^{-3}$ is in an agreement with the C/B taxonomic classification. The quality of AO images is not sufficient for differentiating between poles. However, the first pole from Table~\ref{tab:results} seems marginally better.

\paragraph{65 Cybele}
Asteroid Cybele is a member of the so called Cybele group of minor bodies, which orbit outside the main belt at semi-major axis of $\sim3.5$ au. Our spin state and shape model solution is in agreement with the previous convex shape model determination of \citet{Franco2015}. However, the AO data allowed us to remove the pole ambiguity leaving us with a unique solution. Our size of $D=296\pm25$~km is significantly larger than the estimate of \citet{Carry2012b} based on interpretation of various literature values. The only non-radiometric solution available so far published by \citet{Muller2004b} is based on a thermophysical modelling of thermal infrared data and reports $D=273\pm11$~km. This value, though, based only on a ellipsoidal shape and rather preliminary pole solution, is smaller than our estimate, but still agrees within the large uncertainties. The bulk density of $\rho=(1.0\pm0.3)$ gcm$^{-3}$ is low, but still reasonable, for a C-complex object. The only reliable bulk density of another P-type asteroid is available for asteroid (87)~Sylvia \citep[$\rho=1.34\pm0.21$ and $1.39\pm0.08$ gcm$^{-3}$,][]{Be2014,Hanus2017b}. Both values are rather consistent. We note that the mass estimate of Cybele is based on astrometric observations only, so it might not be fully reliable. 

The diameter of the octantoid model constructed from raw images is approximately ten percent larger than the others (cf. Table ~\ref{table0}), which is typically indicative of discrepancies in the data. By discarding the most likely outlier image, we obtain somewhat smaller diameter estimate $D=284\pm 25$~km, where the uncertainty estimate is again affected by the octantoid model.  

\paragraph{121 Hermione}
A C-type asteroid Hermione is actually a binary system with a $\sim12$km large moon \citep{Merline2002}. Our rotation state results are consistent with previous solutions of \citet{Descamps2009, KV12, Hanus2016a}. We drop images that are badly smeared or are inconsistent with other observations. Of available 13 images, we use nine for shape determination. Some of the images suggest non-convex features, but they are not visible in all of the images. By further reducing the data set, a non-convex shape model similar to \citet{Descamps2009} can be constructed.

\paragraph{511 Davida}
The Keck disk-resolved images of Davida were already used for the rotation state and ellipsoidal shape model determination by \citet{Conrad2007}. Our shape modelling that includes also a comprehensive optical lightcurve dataset is consistent with the one of \citet{Conrad2007}, apart from our size ($D=311\pm5$ km) is larger by about seven percent (c.f., $D=289\pm21$ km). The resulting bulk density of $\rho=(2.1\pm0.4)$ gcm$^{-3}$ is rather high considering Davida is a C-type asteroid. On the other hand, such high bulk density is similar to those of another large C-type asteroids (1)~Ceres and (10)~Hygiea \citep{Park2016, Hanus2017b}. This could suggest that Davida's interior is not homogeneous in composition.

\paragraph{1036 Ganymed}
We derived a first shape model of a NEA based on ground-based disk-resolved images. Ganymed is the largest NEA, but also the smallest minor body that was spatially resolved by 10m class telescopes equipped with adaptive optics systems. On the other hand, the low mass and a non-existence of a known moon resulted into a poor mass estimate, therefore, our bulk density is essentially meaningless. Our size of $D=38\pm3$~km is similar to the size of $D=36\pm3$~km derived by an analysis of the WISE thermal infrared data in the means of a thermophysical model performed by \citet{Hanus2015a}.

\section{Uniqueness and nonuniqueness in photometry}
\label{sec:nonc}

From the discussion and results above, we see that it is important to have a general understanding of the information content or capacity of a given type of data set and source. This concerns especially the uniqueness and the resolution level of the solution. For instance, the use of post-processed AO data does not necessarily extract more information from the raw AO data when making a physical model. Since photometry is the most important data source for targets without much disk-resolved data, we consider in this section the information limits of photometry.

Information content analysis, including uniqueness proofs of the reconstructions of bodies based on various projectionlike data sources as well as the weighting of those sources (in, e.g., the simultaneous use of AO and photometry), has been presented in a number of works \citep[see, e.g.,][and references therein]{KL06,Ka11,VK14,NK17}.

We revisit here the reconstruction of nonconvex shapes from photometry only. The numerical results discussed, e.g., in \cite{KT01a,DK03,KD06}  give a good practical overview of the inverse problem. The problem, however, is more a mathematical than a computational one, so any nonuniqueness or uniqueness proofs are invaluable for understanding the information capacity of the data. We present here fundamental results that require a somewhat special setup, but they give insight to the general problem, and are among the very rare proofs that can be given about the problem in the first place.

In the following, the term {\it illuminated projection area} denotes the total area of the projections of the visible and illuminated (hereafter VI) parts of a body in the viewing direction (from which the illumination direction can differ). {\it Brightness data} is the generalization of this, where the surface elements contributing to the illuminated projection area are each weighted by a scattering function depending on the local viewing and illumination conditions (the scattering of illuminated projection areas is called geometric). Brightness data are also called generalized projections \citep{KL06}. In two dimensions, the body is a planar curve and the projection area is the sum of the widths of the VI parts of the curve seen from the viewing direction. 

{\it Tangent-covered bodies} or TCBs are bodies for which each point on the surface has at least one tangent that does not intersect any other part of the body (but can be tangent to them). The {\it tangent hull} of a body is the set of surface points for which the above criterion is true, augmented by their tangents to form a closed, connected surface of a TCB. By a {\it concavity}, we mean a part of the surface of a body that is not part of its tangent hull. As discussed in \cite{Ka11}, tangent-covered bodies (and the tangent hull of a body, also called its profile hull) are reconstructable from their disk-resolved silhouette or profile curves. TCBs are thus the set of all convex bodies and all nonconvex bodies without concavities. By definition, bodies with concavities are not TCBs, a TCB is identical to its tangent hull, in three dimensions convex bodies are a subset of TCBs, and in two dimensions the sets of convex bodies and TCBs are identical.

{\bf Two-dimensional nonconvex bodies cannot be uniquely determined from their brightness data.} By definition, the tangent hull of a 2D body is its convex hull, so any nonconvexities of a 2D body are concavities. Each concavity is covered by a line that is part of the convex hull, and all parts of a line have the same visibility and illumination. Thus the effect of the concavity on brightness data can be replaced by smaller concavities, covered by the same line, that are isomorphic to the original concavity whose length along the line equals their combined lengths. It is also possible to have concavities of different shapes along the same line that together produce a shadow effect that can be attributed to one concavity of still another shape (see below). Therefore the brightness data of any nonconvex 2D body can be reproduced by infinitely many other nonconvex versions. We call this {\it scale ambiguity}.

{\bf Concavities of three-dimensional bodies cannot be uniquely determined from the brightness data of the bodies.} For simplicity, we assume here any concavity to be contained in a plane that is part of the convex hull of the body. This is just to avoid lengthy discussions of special shadowing conditions that are not material to the argument. Then the 3D case is a direct generalization of the scale ambiguity of the 2D case: the effect of the concavity on brightness data can be replaced by smaller isomorphic (or possibly other) concavities in the plane whose combined surface area equals that of the concavity. We assume that the plane is suitably larger than the part of it occupied by the concavity so that the smaller concavities can be arranged within the plane. Note that this arrangement is also nonunique: even if one uses a size constraint for the concavities, their locations in the plane cannot be deduced from brightness data.

An interesting corollary of this nonuniqueness is that even a large-scale concavity is actually indistinguishable from a locally rugged surface: in other words, a concavity within a plane can be replaced by the same plane with scattering properties caused by small-scale roughness \citep[cf. the discussion in][]{Ka04}.

{\bf A subset of tangent-covered bodies can be determined from their illuminated projection areas at least with a simple scale constraint.} Here we show that the set of bodies essentially reconstructable from their illuminated projection areas is larger than the set of convex surfaces. Essentially means here the use of a natural constraint. The uniqueness proof for the brightness data of convex bodies is discussed in \cite{KL06} and references therein.

Proving anything about the integrals over a nonconvex body dependent on directions is notoriously difficult for the simple reason that usually such integrals are not analytically calculable, requiring numerical ray-tracing or the finding of the roots of equations containing high-order functions. Therefore we must resort to a number of special assumptions, considering first the case in two dimensions. We can construct a proof there since in 2D the location of the shadow boundary point is essentially the same as its projection area (length in 2D). In 3D, the shadow boundary is resolved in 2D \citep[see][]{Ka11}, and the area requires the computation of an integral (as would brightness data in 2D). 

Our reconstruction consists of three parts: {\it 1.} the uniqueness of a concavity up to scaling in 2D, {\it 2.} the separation of the uniqueness of this shape and that of the rest of the 2D curve, and {\it 3.} the transformation of the 2D curve into a 3D TCB, and the scale constraints.

{\it 1.} For simplicity, we consider an isolated concave section $\cal S$ of a curve $\cal C$ in the $xy$-plane, with end points on the $x$-axis, one at the origin and the other at $(L,0)$, and the rest of $\cal S$ is below the $x$-axis. We are interested in the shadow caused by the end point at $(0,0)$.  Let the illumination direction be $\phi$ and the viewing direction $\theta$, with $\phi-\theta=\alpha$. When $\alpha\ne 0$, $\cal S$ can be uniquely reconstructed by finding the point $(x_s,y_s)$ on $\cal S$ separating the shadow and the illuminated part for successive values of $\phi$ and $\theta$. We require $\cal S$ to be starlike w.r.t. the origin: this enables simple shape parametrization and yields only one shadow and one illuminated section on the concavity. Further, we require VI parts of $\cal S$ to cover the whole of the line between $(x_s, y_s)$ and $(L,0)$ projected in $\theta$, and that there are $\theta$ covering the whole of the motion of $(x_s,y_s)$ from $(0,0)$ to $(L,0)$ as $\theta$ increases. A simple sufficient but by no means necessary arrangement for this is to let $\cal S$ be an inverted convex curve with the angles between the $y$-axis and the tangents of $\cal S$ both $\alpha/2$ at $(0,0)$ and $(L,0)$. 

\begin{figure}
\centering
\includegraphics[clip=true,trim=0 80 0 0,scale=0.3]{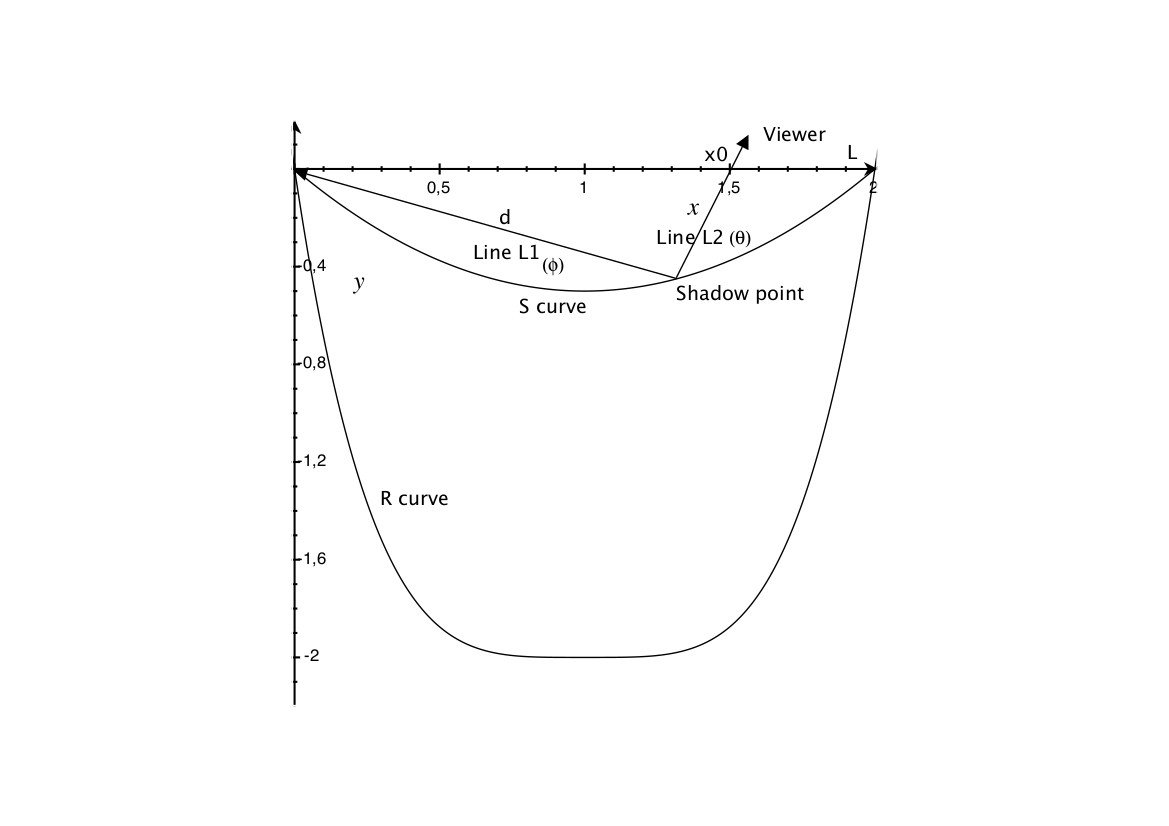}
\caption{Sketch of the geometry of the nonconvex shape in the $xy$-plane.}
\label{noncfig}
\end{figure}

The shadow point $(x_s,y_s)$ is the intersection point of two lines $L_1$ and $L_2$. $L_1$ is the shadow line through the origin in $\phi$, and $L_2$ is the line in $\theta$ such that its distance from the corresponding line through $(L,0)$ is the observed illuminated projection length $l$. Thus (see Fig.\ \ref{noncfig})
\begin{equation}
x_s=-d\cos\phi=x_0-s\cos\theta,\quad
y_s=-d\sin\phi=-s\sin\theta,
\end{equation}
where $d$ and $s$ are the length parameters of the lines $L_1$ and $L_2$, respectively, and $L_2$ passes through $x_0$ for which
\begin{equation}
l=(L-x_0)\sin\theta.
\end{equation}
Combining these, one obtains the unique solution for $d$, and thus $(x_s,y_s)$ at given $\theta$, $\alpha$, $L$, and observed $l$. Parametrizing $l$ with the usual polar angle $\varphi$, obtained directly from $\varphi=\phi-\pi$, we have
\begin{equation}
d(\varphi)\sin\alpha=L \sin\theta-l(\theta), \quad \theta=\varphi-\alpha+\pi, \quad 0\le\theta\le\pi-\alpha,
\end{equation}
so
\begin{equation}
[x(\varphi), y(\varphi)]_{\cal S}=[d(\varphi)\cos\varphi,d(\varphi)\sin\varphi],
\end{equation}
and $d(\varphi)=0$ for some values of $\theta$ when $l(\theta)=L\sin\theta$; $d(\varphi)>0$ for some interval of $\varphi$ ending at $\varphi=2\pi$ when $d=L$ and $l=0$.

This result shows formally the scale ambiguity: $d$ is unique only if there is one concavity with $d$, since otherwise one may define
\begin{equation}
\sum_i d_i\sin\alpha=\sum_i L_i \sin\theta-l(\theta), \quad\sum_i L_i=L,\quad\sum_i d_i=d,
\end{equation}
so scaled-down arrangements of adjacent concavities with similar or various $d_i(\varphi)$ will produce the same $l(\theta)$.

{\it 2.} Recall that any convex curve can be defined by its curvature function $C(\psi)$, where the direction angle of the outward surface normal is given by $0\le\psi<2\pi$. The length element is $ds=C(\psi)\,d\psi$, so the points $[x(\psi),y(\psi)]$  on the curve are given by
\begin{equation}
x(\psi)=-\int_0^\psi C(\psi')\sin\psi'\, d\psi',\quad
y(\psi)=\int_0^\psi C(\psi')\cos\psi'\, d\psi'.
\end{equation}
The observed projection length of the VI part of the curve at $\theta$ is
\begin{equation}
l(\theta)=\int_{-\pi/2+\alpha}^{\pi/2}\,C(\psi+\theta)\cos\psi\,d\psi.
\end{equation}
Expanding $ C(\psi)=\sum_n a_n\cos n\psi+b_n\sin n\psi,\, n\ge 0, \, a_1=b_1=0, $ and similarly
$ l(\theta)=\sum_n c_n\cos n\theta+d_n\sin n\theta $, we have \citep{OC84,Ka16}

\begin{equation}
c_n=a_n I^c_n(\alpha)+b_n I^s_n(\alpha), \quad 
d_n=b_n I^c_n(\alpha)-a_n I^s_n(\alpha),
\end{equation}
where
\begin{equation}
I^c_n(\alpha)=\int_{-\pi/2+\alpha}^{\pi/2}\cos n\psi\cos\psi\, d\psi, \,
I^s_n(\alpha)=\int_{-\pi/2+\alpha}^{\pi/2}\sin n\psi\cos\psi\, d\psi,
\end{equation}
so the solution of the inverse problem is
\begin{equation}
a_n=\frac{I^c_n c_n-I^s_n d_n}{(I^c_n)^2+(I^s_n)^2}, \quad
b_n=\frac{I^s_n c_n+I^c_n d_n}{(I^c_n)^2+(I^s_n)^2}.
\end{equation}
The denominator $(I^c_n)^2+(I^s_n)^2$ is nonzero at $0<\alpha<\pi$, so the convex curve is uniquely obtained from the $l(\theta)$-data at any $\alpha$.

When the integral defining $l(\theta)$ mixes parts of $\cal S$ and other parts of $\cal C$, there is no simple way to parametrize $\cal C$ so as to allow analytical computations of $l(\theta)$. Thus we separate the observations of $\cal S$ from the rest of the curve $\cal C$ denoted by ${\cal R}={\cal C}\setminus \cal S$. We assume that $\cal R$ is convex and without linear sections. Then, denoting the angle interior to $\cal C$ between the $x$-axis and the tangent of $\cal R$ at the origin and $(L,0)$ by, respectively, $\gamma_0$ and $\gamma_L$, we must have $\gamma_0\le\alpha, \gamma_L\le\alpha$. 
With these assumptions, either parts of $\cal S$ or parts of $\cal R$ are VI at any given $\theta$. The data from $\cal S$ cannot be mimicked by a convex curve since they are like those of a line for some range of $\theta$.

Given a full range of observations $l(\theta)$ of $\cal C$, one can proceed as follows. First find the values of $\theta$ at which $l(\theta)$ vanishes at the given $\alpha$. Assuming $\cal R$ not to contain straight lines ending at sharp corners elsewhere, there are two intervals (or points) in $\theta$ with $l(\theta)=0$, one ending at $\theta=0$ and one starting at $\theta=\alpha$ (the lengths of the intervals are $\alpha-\gamma_0$ and $\alpha-\gamma_L$). Between these, $l(\theta)$ is sinusoidal for an interval of $\theta$ starting at the end of one interval of $l=0$: $l=L\sin\theta$. This reveals the existence of the corners at the origin and $(L,0)$. It also fixes the $xy$-frame in our convention, and gives the size of $\cal S$: $L=l/\sin\theta$ from the sinusoidal interval. After the shadow effect appears at some $\theta$, the shape profile of $\cal S$ is uniquely obtained from $l(\theta)$-data at $0\le\theta\le \pi-\alpha$. Observations of $l(\theta)$ in the range $\pi-\alpha\le\theta\le 2\pi$ are sufficient to determine $C(\psi)$ of $\cal R$ (or any convex curve) in the interval $3\pi/2-\alpha\le\psi\le 3\pi/2+\alpha$ since $C(\psi)$ in this interval does not affect $l(\theta)$ at other values of $\theta$. For this range of $\theta$, these data are exactly the same as those from $\cal C$ with $\cal S$ a straight line from the origin to $(L,0)$ would be.

This concludes the determination of $\cal C$ up to the scale ambiguity on $\cal S$: one obtains $\cal R$ and the shape $\cal S$, but there is no way of telling from $l(\theta)$ whether the data are due to $\cal S$ or
smaller concavities on the line section $L_0$ between the origin and $(L,0)$. 

{\it 3.} Next we define a class of 3D bodies by extending the 2D curve $\cal C$ to apply to some interval $z_0\le z\le z_1$, and covering the cylindrical shape of height $z_1-z_0$ with planes at $z_0$ and $z_1$. This surface is a TCB, and we call the shape class cylindrical TCBs. Concavities in 2D are changed into saddle surfaces rather than concavities in 3D, with much richer information possibilities. Viewing directions along the $z$-axis then reveal the true area $A$ between the real concavity curve and $L_0$: if $A_P$ is the observed projected area of the body and $A_C$ the area contained inside the curve formed of $\cal R$ and $L_0$, $A=A_C-A_P$. This information strongly constrains the possible scale ambiguities. Now we must have $\sum_i A_i=A$ in addition to $\sum_i L_i=L$ for each concavity $i$, and the combined area of many concavities is smaller than the area of one due to the quadratic scaling of the areas $A_i$ w.r.t. the lengths $L_i$. Also, viewing and illumination directions other than those along the $z$-axis or in the $xy$-plane offer additional information. In these other observing geometries, scale ambiguity is at least partly removed for the saddle-surface versions of the concavities, but it is difficult to show the exact properties analytically. 

If $A$ equals the area $A_S$ by $S$ (on the whole of $L_0$) or is close to it, the assumption of one concavity is well justified. Indeed, if $A=A_S$, only one occurrence of the concavity shape $\cal S$ is possible if only this shape is considered. A simple way to enforce the assumption is to attribute almost all area $A$ to one scaled concavity $\cal  S$ and fill the remaining length on $L_0$ by negligibly small-scale roughness.

The above construction applies also to other arrangements than that of $\cal S$ and $\cal R$. For example, the profiles of the concave sides of a "nonconvex Reuleaux triangle" are solvable at least for $\alpha\ge \pi/6$. Viewing directions tilted from the $z$-axis can be used to isolate the area information for each concave section of the surface by placing them in the opposite azimuthal direction.

{\bf Not all tangent-covered bodies can be uniquely determined from their brightness data.} Even the use of all observing geometries and size constraints cannot resolve the case where sections on $L_0$ are without concavities (just straight lines). Then it is impossible to say where along the line the sections and concavities are. Another example of TCB ambiguity:
consider the surface of a wedge-shaped 3D tangent-covered body (or a convex body with a wedgelike part), with two intersecting planes forming the wedge. A new tangent-covered surface can be formed by making a tangent-covered hole that is contained between the two planes. The hole need not be cylindrical as long as it is tangent-covered so that the new body is still a TCB. If the hole is suitably smaller than the wedge, all brightness data of the body can be reproduced by having a collection of smaller isomorphic holes within the wedge arranged such that their combined area equals that of the original concavity. This is possible because the wedge shape allows an infinite number of isomorphic holes to be created (two parallel planes instead of a wedge would not allow this). A corollary of this is that a hole can be replaced by arbitrarily small-scale perforation.

Even when formally removable, ambiguities lead to instabilities near limit conditions. For example,
if the straight line on which two scale-ambiguous concavities are adjacent is bent between the concavities, the above uniqueness results are obtained for both separately. However, this requires $\alpha$ to be at least as large as $\pi-\beta$ where $\beta$ is the bending angle. A small $\beta$ is a typical depiction for many nearby nonconvex features on asteroid surfaces, so reconstructions are unstable even at high $\alpha$. 

{\bf The reconstruction of a nonconvex body from its brightness data is fundamentally nonunique.}
Scale and location ambiguities are inherent to the inverse problem. Nevertheless, the reconstructable class of bodies from brightness data is, in a certain sense and with constraints, larger than the set of convex shapes especially due to the projection area information from TCBs. This corroborates the numerical success in simulations such as those in \cite{DK03}. However, the reconstruction of nonconvex bodies from photometry has neither the fundamental uniqueness properties of the convex case nor the Minkowski stability that pertains to the global shape and applies to both data and model errors. These aspects are illustrated by, e.g., the case of the asteroid Eros. Eros, with its sizable nonconvex TCB-like feature, can be roughly approximated by the simple cylindrical model when viewed from the direction of its rotation axis. One might thus expect its photometry to yield a unique nonconvex solution. Even so, as discussed in \cite{KD06}, the convex model fits the data as well as a nonconvex one and, above all, better than the real shape with usual scattering models. There are various nonconvex shapes that fit the data equally well. This underlines the ambiguous properties and the instabilities of the photometry of nonconvex bodies and the need for large $\alpha$.

A convex surface actually represents merely a nonconvex case in which regularization suppressing local nonconvex features has been given infinite weight. This, however, does generally not deteriorate the fit as discussed in \cite{DK03}; so far, the only asteroid requiring a nonconvex shape to explain its photometry is Eger \citep{DV12}. All others can be explained down to the noise level by convex shapes, which means that, from the point of view of regularization theory, there is no optimal regularization weight so it is best to use full weight to avoid the inevitable instabilities and nonuniqueness at lower weights. Statistically, no result between the two extreme weights can be shown to be the best one, so the safest result is a convex shape because of its strong uniqueness and stability properties. From the Bayesian point of view, the problem is the lack of proper statistics to cover the systematic data and model errors (dominating over the data noise) and the difficulty of finding shape sampling covering the whole of the shape space \citep[single or a few shape supports cannot do this properly in the sense of Markov chain Monte Carlo; see][]{VK15}.

\section{Conclusions}
\label{sec:conclusion}
We have determined the spins and shape models of about twenty asteroids and estimated their diameters (see Tables~\ref{tab:results}~and~\ref{tab:densities}). Derived bulk densities are usually consistent within the asteroid's taxonomic classifications -- $\sim$1--2~gcm$^{-3}$ for C-complex members and $\sim$2--3~gcm$^{-3}$ for S-complex asteroids. The notable exception is slightly larger bulk density of the C-type asteroid (511)~Davida, which is similar to bulk densities of the largest C-type asteroids (1)~Ceres and (10)~Hygiea. This might suggest at least some degree of differentiation.

However, one should keep in mind that a model is always a sample of a set of probable solutions based on an amalgamation of data, prior assumptions and subjective judgment. Provided the data coverage and quality is sufficient, it is possible to construct a plausible model that could have produced the observed data. Nevertheless, one should be careful when drawing conclusions based on model features. For instance, based on only one image, one can seldom conclusively deduce whether an apparent local detail is due to imaging artifacts or actual surface features (e.g., shadowing effects from craters).   
   
We have used the jackknife resampling method to generate a set of plausible models, but the number of samples is strictly limited to the number of data images. The obvious next step would the utilization of the  bootstrap method, which is tantamount to setting the image data weights randomly. This would allow a more complete sampling of the solution space, but would require manual intervention, since not all the models produced by the bootstrap method are physically plausible. Besides, this is still not sufficient for sampling the solutions since the available shape representations are constrained. 

Based on the sample of asteroids in this paper, the shape models constructed from deconvolved and raw AO images are visually almost indistinguishable. There is no apparent reason to prefer deconvolved data in shape reconstruction: the deconvolution process does not uncover any additional information. Rather, it may introduce spurious details. Moreover, the sharp cutoffs caused by deconvolution determine the position of the boundary contour beforehand, while it would be more natural to relegate that choice to the shape optimization algorithm, which sees all the data. Discrepancies in the diameters could be attributed to inherent uncertainties present in the data, and should be taken into account when estimating the size uncertainty. 

With the advent of the SPHERE adaptive optics system, inconsistencies caused by the myopic deconvolution in shape modelling are likely to disappear. Indeed, results obtained in \cite{Vi15,Ha17,Ma17} already demonstrate that the quality of SPHERE images easily surpasses that of older Keck data as a result of greatly improved Strehl ratio. However, data from the older AO instruments continue to be useful in the foreseeable future, since they often provide complementary information due to a long range of observation epochs. 

Tangent-covered bodies offer the least ambiguous information for reconstruction from lightcurves only. We have shown that at least a subclass of TCBs can be thus reconstructed (with suitable constraints and accepting a number of ambiguities). However, we also showed that not all TCBs are reconstructable from photometry, and unlike convex bodies, nonconvex shapes have fundamental ambiguities in this context also when $\alpha>0$. This is important for understanding the inherent ambiguity and instability of nonconvex solutions from photometry, and why the combination of photometry and even one AO image is already considerably more informative than lightcurves only. This applies both to size estimation and to the extraction of local surface features. On the other hand, at least a few AO images at various geometries are much more conclusive than one. Indeed, as shown in \cite{Ka11}, boundary curves of images (that contain the bulk of information) allow the unique reconstruction of a class of bodies more extensive than that of TCBs (for $\alpha>0$; for $\alpha=0$, this class is exactly TCBs). In addition to ruling out wrong shape possibilities and making the result more robust against imaging errors, a set of images allows the actual resolution level of the full 3D model to approach the potential of the AO resolution.

\begin{acknowledgements}
This research was supported by the Academy of Finland Centre of Excellence in Inverse Problems. JD was supported by the grant 15-04816S of the Czech Science Foundation.

This research has made use of the Keck Observatory Archive (KOA), which is operated by the W. M. Keck Observatory and the NASA Exoplanet Science Institute (NExScI), under contract with the National Aeronautics and Space Administration.
\end{acknowledgements}

\bibliography{diss}   
\bibliographystyle{aa}

\input{tabs/tab1.tex}

\input{tabs/tab2.tex}
\begin{appendix}

\section{Online tables and figures}
In this appendix we list the disk-resolved AO observations used for shape modeling (Table~\ref{tab:ao}) and the diameters with their uncertainties estimated using the resampling method (Table~\ref{table0}). Figures~\ref{fig7}--\ref{fig1036} show the deconvolved AO images and the  model projections. Images are numbered from left to right, corresponding to the observation dates listed in Table~\ref{tab:ao}.
\input{tabs/tab0.tex}
\input{figures1.tex}
\input{tabs/tab4.tex}

\end{appendix}

\end{document}

%% file: tabs/tab1.tex
\begin{table*}[htb]
\centering
\caption{\label{tab:results}Rotation state parameters $\lambda_{\mathrm{a}}$, $\beta_{\mathrm{a}}$, $P_{\mathrm{a}}$ with a reference to the corresponding publication that we used as initial inputs for the modeling with ADAM, rotation state parameters $\lambda$, $\beta$, $P$ derived by the ADAM algorithm, and the number of available light curves $N_{\mathrm{lc}}$ and disk-resolved images $N_{\mathrm{ao}}$.}
\resizebox*{!}{0.4\textheight}{
\begin{tabular}{r l r r r r r r r r r}
\hline 
\multicolumn{2}{c} {Asteroid} & \multicolumn{1}{c} {$\lambda_{\mathrm{a}}$} & \multicolumn{1}{c} {$\beta_{\mathrm{a}}$} & \multicolumn{1}{c} {$P_{\mathrm{a}}$} & \multicolumn{1}{c} {Reference} & \multicolumn{1}{c} {$\lambda$} & \multicolumn{1}{c} {$\beta$} & \multicolumn{1}{c} {$P$} & \multicolumn{1}{c} {$N_{\mathrm{lc}}$} & \multicolumn{1}{c} {$N_{\mathrm{ao}}$} \\
\multicolumn{2}{l} {} & [deg] & [deg] & \multicolumn{1}{c} {[hours]} &  & [deg] & [deg] & \multicolumn{1}{c} {[hours]} &  & \\
\hline\hline


  7 &         Iris &   16 &  15   & 7.138843 &  \citet{Kaasalainen2002b} &   18$\pm$4 &  19$\pm$4 & 7.138843 &  39 & 14 \\
  7 &         Iris &  196 &   2   & 7.138843 &  \citet{Kaasalainen2002b} &    \multicolumn{3}{c} {Rejected} &  39 & 14 \\
 12 &     Victoria &  174 & $-$17 & 8.66034 &        \citet{Hanus2016a} &  177$\pm$4 & $-$34$\pm$4 & 8.66034 &  53 &  8 \\
 14 &        Irene &   95 & $-$11 & 15.02986 &         \citet{Hanus2011} &   91$\pm$7 & $-$14$\pm$4 & 15.02988 &  29 &  3 \\
 15 &      Eunomia &    3 & $-$67 & 6.082753 &  \citet{Kaasalainen2002b} &    0$\pm$5 & $-$67$\pm$2 & 6.082752 &  48 &  8 \\
 23 &       Thalia &  159 & $-$45 & 12.31241 &        \citet{Torppa2003} &  158$\pm$2 & $-$45$\pm$4 & 12.31241 &  50 &  2 \\
 23 &       Thalia &  343 & $-$69 & 12.31241 &        \citet{Torppa2003} &  341$\pm$3 & $-$73$\pm$6 & 12.31241208 &  50 &  2 \\
 24 &       Themis &  137 &  59   & 8.374187 &        \citet{Hanus2016a} &  138$\pm$4 &  69$\pm$5 & 8.374187 &  46 &  4 \\
 24 &       Themis &  331 &  52   & 8.374187 &        \citet{Hanus2016a} &  328$\pm$5 &  70$\pm$2 & 8.374187 &  46 &  4 \\
 28 &      Bellona &  102 &  $-$8 & 15.70785 &        \citet{Durech2011} &  99$\pm$4 &  $-$16$\pm$6 & 15.70785 &  23 &  1 \\
 40 &     Harmonia &   22 &  31   & 8.908483 &         \citet{Hanus2011} &   22$\pm$2 &  38$\pm$5 & 8.908485 &  23 &  3 \\
 42 &         Isis &  106 &  40   & 13.58364 &         \citet{Hanus2011} &  108$\pm$1 &  47$\pm$3 & 13.58364 &  31 &  1 \\
 48 &        Doris &  108 &  47   & 11.8901 &        \citet{Hanus2016a} &    \multicolumn{3}{c} {Rejected} &  31 &  2 \\
 48 &        Doris &  297 &  61   & 11.8901 &        \citet{Hanus2016a} &  296$\pm$3 &  57$\pm$5 & 11.89010 &  31 &  2 \\
 56 &       Melete &  103 & $-$27 & 18.14817 &        \citet{Hanus2016a} &  103$\pm$1 & $-$26$\pm$2 & 18.14817 &  34 &  2 \\
 56 &       Melete &  282 &  $-$5 & 18.14817 &        \citet{Hanus2016a} &  283$\pm$1 &  $-$3$\pm$3 & 18.14817 &  34 &  2 \\
 65 &       Cybele &  208 &  $-$7 & 6.081434 &        \citet{Franco2015} &  208$\pm$1 &  $-$3$\pm$3 & 6.081435 &  59 &  7 \\
 65 &       Cybele &   27 & $-$14 & 6.081434 &        \citet{Franco2015} &    \multicolumn{3}{c} {Rejected} &  59 &  7 \\
 72 &      Feronia &  102 & $-$55 & 8.09068 &        \citet{Hanus2013a} &  100$\pm$6 & $-$51$\pm$6 & 8.09068 &  20 &  2 \\
121 &     Hermione &    4 &  13   & 5.550878 &      \citet{Descamps2009} &  362$\pm$2 &   13$\pm$1 & 5.550877 &  48 & 13 \\
146 &       Lucina &  305 & $-$41 & 18.5539 &        \citet{Durech2009} &  304$\pm$6 & $-$41$\pm$2 & 18.55387 &  22 &  2 \\
250 &      Bettina &  100 &  17   & 5.054420 &        \citet{Torppa2003} &  101$\pm$2 &  9$\pm$2 & 5.054414 &  22 &  1 \\
283 &         Emma &  251 &  22   & 6.895222 &   \citet{Michalowski2006} &  256$\pm$3 &  23$\pm$3 & 6.89523 &  29 &  5 \\
283 &         Emma &   85 &  37   & 6.895221 &   \citet{Michalowski2006} &    \multicolumn{3}{c} {Rejected} &  29 &  5 \\
354 &     Eleonora &  144 &  54   & 4.277186 &         \citet{Hanus2011} &  161$\pm$8 &  41$\pm$7 & 4.277186 &  64 &  3 \\
511 &       Davida &  297 &  26   & 5.129363 &        \citet{Torppa2003} &  299$\pm$1 &  24$\pm$1 & 5.129363 &  58 & 10 \\
1036 &      Ganymed &  190 & $-$78 & 10.31284 &       \citet{Hanus2015a} &  198$\pm$10 & $-$79$\pm$1 & 10.31304 & 177 &  5 \\
\hline
\end{tabular}
}%
\end{table*}

%% file: tabs/tab2.tex
\begin{table*}[htb]
\centering
\caption{\label{tab:densities}Bulk density estimates based on our volume estimated by the ADAM shape modeling from combined optical light curves and disk-resolved images.}
\resizebox*{!}{0.4\textheight}{
\begin{tabular}{r l c c c c c c c c }
\hline
 \multicolumn{2}{c} {Asteroid} & $D_{\mathrm{a}}$ & Reference & $D$ & $M$ & Reference & T1 & T2 & $\rho$ \\
 \multicolumn{2}{c} {} & [km] &  & [km] & [10$^{18}$ kg] &  &  &  &  [gcm$^{-3}$] \\ \hline\hline

%
  7 &         Iris &  203$\pm$24 &        \citet{Hanus2013b} &   216$\pm$7 &  12.9$\pm$2.1 &        \citet{Carry2012b} &  S &  S & 2.4$\pm$0.5 \\
  7 &         Iris &  203$\pm$24 &        \citet{Hanus2013b} &      \multicolumn{6}{c} {Rejected} \\
 12 &     Victoria &  124$\pm$8 &        \citet{Carry2012b} &   115$\pm$3 &  2.5$\pm$0.5 &        \citet{Carry2012b} &  S &  L & 3.1$\pm$0.6 \\
 14 &        Irene &  149$\pm$17 &        \citet{Hanus2013b} &   155$\pm$6 &  2.8$\pm$1.0 &        \citet{Fienga2014} &  S &  S & 1.4$\pm$0.5 \\
 15 &      Eunomia &  254$\pm$27 &        \citet{Hanus2013b} &   275$\pm$5 &  31.4$\pm$1.8 &        \citet{Carry2012b} &  S &  S & 2.9$\pm$0.2 \\
 23 &       Thalia &  107$\pm$12 &        \citet{Hanus2013b} &   120$\pm$8 &  2.0$\pm$0.1 &        \citet{Carry2012b} &  S &  S & 2.2$\pm$0.4 \\
 23 &       Thalia &  107$\pm$12 &        \citet{Hanus2013b} &   120$\pm$8 &  2.0$\pm$0.1 &        \citet{Carry2012b} &  S &  S & 2.2$\pm$0.4 \\
 24 &       Themis &  184$\pm$11 &        \citet{Carry2012b} &   215$\pm$15 &  5.9$\pm$1.9 &        \citet{Carry2012b} &  C &  B & 1.1$\pm$0.4 \\
 24 &       Themis &  184$\pm$11 &        \citet{Carry2012b} &   215$\pm$15 &  5.9$\pm$1.9 &        \citet{Carry2012b} &  C &  B & 1.1$\pm$0.4 \\
 28 &      Bellona &  121$\pm$11 &        \citet{Hanus2013b} &   135$\pm$7 &  2.6$\pm$0.1 &        \citet{Carry2012b} &  S &  S & 2.0$\pm$0.3 \\
 40 &     Harmonia &  123$\pm$12 &        \citet{Hanus2013b} &   113$\pm$7 &    &                     &  S &  S &   \\
 42 &         Isis &   97$\pm$10 &        \citet{Hanus2013b} &   102$\pm$4 &  1.6$\pm$0.5 &        \citet{Carry2012b} &  S &  L & 2.8$\pm$1.0 \\
 48 &        Doris &  212$\pm$11 &        \citet{Carry2012b} &      \multicolumn{6}{c} {Rejected} \\
 48 &        Doris &  212$\pm$11 &        \citet{Carry2012b} &   223$\pm$23 &  6.1$\pm$3.0 &        \citet{Carry2012b} & CG & Ch & 1.1$\pm$0.6 \\
 56 &       Melete &  114$\pm$8 &        \citet{Carry2012b} &   119$\pm$5 &  4.6$\pm$1.0 &        \citet{Carry2012b} &  P & Xk & 5.2$\pm$1.3 \\
 56 &       Melete &  114$\pm$8 &        \citet{Carry2012b} &   119$\pm$5 &  4.6$\pm$1.0 &        \citet{Carry2012b} &  P & Xk & 5.2$\pm$1.3 \\
 65 &       Cybele &  248$\pm$18 &        \citet{Carry2012b} &   296$\pm$25 &  13.6$\pm$3.1 &        \citet{Carry2012b} &  P & Xc & 1.0$\pm$0.3 \\
 65 &       Cybele &  248$\pm$18 &        \citet{Carry2012b} &      \multicolumn{6}{c} {Rejected} \\
 72 &      Feronia &   74$\pm$6 &        \citet{Hanus2013b} &    84$\pm$10 &  3.3$\pm$8.5 &        \citet{Carry2012b} & TDG &  $-$ & $-$ \\
121 &     Hermione &  187$\pm$6 &      \citet{Descamps2009} &   196$\pm$15 &  5.0$\pm$0.3 &        \citet{Carry2012b} &  C & Ch & 1.26$\pm$0.30 \\
146 &       Lucina &  119$\pm$11 &        \citet{Hanus2013b} &   131$\pm$15 &   &                   &  C & Ch &   \\
250 &      Bettina &  119$\pm$11 &        \citet{Hanus2013b} &   109$\pm$5 &    &                  &  M & Xk &   \\
283 &         Emma &  133$\pm$10 &        \citet{Carry2012b} &   142$\pm$14 &  1.4$\pm$0.0 &      \citet{Marchis2008a} &  X &  C & 0.9$\pm$0.3 \\
283 &         Emma &  133$\pm$10 &        \citet{Carry2012b} &      \multicolumn{6}{c} {Rejected} \\
354 &     Eleonora &  154$\pm$6 &        \citet{Carry2012b} &   169$\pm$30 &  7.2$\pm$2.6 &        \citet{Carry2012b} &  S & Sl & 2.8$\pm$1.8 \\
511 &       Davida &  289$\pm$21 &        \citet{Conrad2007} &   311$\pm$5 &  33.7$\pm$5.7 &        \citet{Fienga2014} &  C &  C & 2.1$\pm$0.4 \\
1036 &      Ganymed &  36$\pm$3 &         \citet{Hanus2015a} &    38$\pm$3 &  0.15$\pm$0.12 &             Fienga, privat com. &  S &  S & 5.2$\pm$4.4 \\
\hline
\end{tabular}
}%
\tablefoot{
The table gives previous size (surface- or volume-equivalent diameter) estimate $D_\mathrm{a}$ and its reference, volume-equivalent diameter $D$ of the shape solution derived here by ADAM, adopted mass $M$ and its reference, the Tholen \citep[T1,][]{Tholen1989, Tholen1989b} and SMASS II \citep[T2,][]{Bus2002} taxonomic classes, and our density determination $\rho$.}
\end{table*}
\twocolumn

%% file: tabs/tab0.tex
\begin{table}[htb]
 \caption{\label{table0}Models diameters and uncertainty estimates.} 
\resizebox{\columnwidth}{!}{
    \centering
    \begin{tabular}{|l l|c|c|c|c|c|c|c|}         
      \hline                     
      \multicolumn{2}{|c|} {Asteroid}  &  $\overline{\Delta p}$ & $D_{rw}$ &$\Delta D_{rw}$ &$D_{dc}$ &$\Delta D_{dc}$\\
      \hline
      \hline
      7 &Iris 	&		 8.3 & 220 220 &217--223 &213 212 &209--217 \\
       12& Victoria &		 6.9 & 116 117 &115--118 & 113 115 & 113--116\\
     14 &Irene  &		 11.9 &158 159 & 151--161 &151 154 &148--158\\
     15& Eunomia &		 10.7 & 276 278 & 274--279 & 271 275 &270--275 \\
      23& Thalia & 		 15.6 & 114 115 &113--118 & 126 124 & 120--130 \\
     24 &Themis & 		 16.2 & 215 225 & 200--243 &207 211 &199--229\\
      28& Bellona & 		 11.7 & 129 133 &-- & 134 143 & --\\
       40& Harmonia & 		 10 & 113 112 & 105--120 & 113 115 & 114--116\\
     42& Isis &			 6.6 & 102 106 & -- & 98 100& --\\
     48& Doris &			 18.7 &210 214 & 196--234 &238 229 & 218--241\\
     56& Melete &		10 & 114 116 &-- &122 124 & --\\
      65& Cybele (7 im.)&		21.3 & 297 314 &288--323 & 285 288 &281--303\\
      65& Cybele (6 im.)& 21.5 &278 305 &265--314 &276 278 & 268--294\\
     72&	Feronia &		 8 & 85 86 &81--94& 83 83& 80--95\\
     121& Hermione	(13 im.)	&17.8 & 199 200 &197--204& 193 192 &190--223\\
     121 &Hermione (9 im.) &	 	18.3 & 193 193 &189--200 & 193 193 &188--196\\
         146 &Lucina & 		 13.1 & 133 133 &128--159& 129 129 & 127--144\\
     250 &Bettina & 		 15.8 & 109 111 &-- & 116 118& --\\
          283& Emma &  		 12.3 &148 147 & 142--156 & 136 138 & 134--148\\
     354& Eleonora &  		 16.1 &162 164 & 151--203& 172 177&165--214\\
     511& Davida &   		14.9 & 312 313 &309--315 & 308 309 & 306--315 \\
     1036 &Ganymed & 		 3.3 & 39 39 & 38--41 &37 37&36--37 \\
      \hline
    \end{tabular}}
    \tablefoot{The mean pixel size in the AO images is $\overline{\Delta p}$ (in km), $D_{rw}$ and $D_{dc}$ are the model diameters based  respectively on raw and deconvolved images with two different parametrizations. Columns $\Delta D_{rw,dc}$ show the diameter variability estimated with the jackknife method.} 
\end{table}  

%% file: figures1.tex
\begin{figure}[htbp]
     \begin{subfigure}[b]{0.16\linewidth}
      \includegraphics[clip=true,trim=65 65 65 65,scale=0.39]{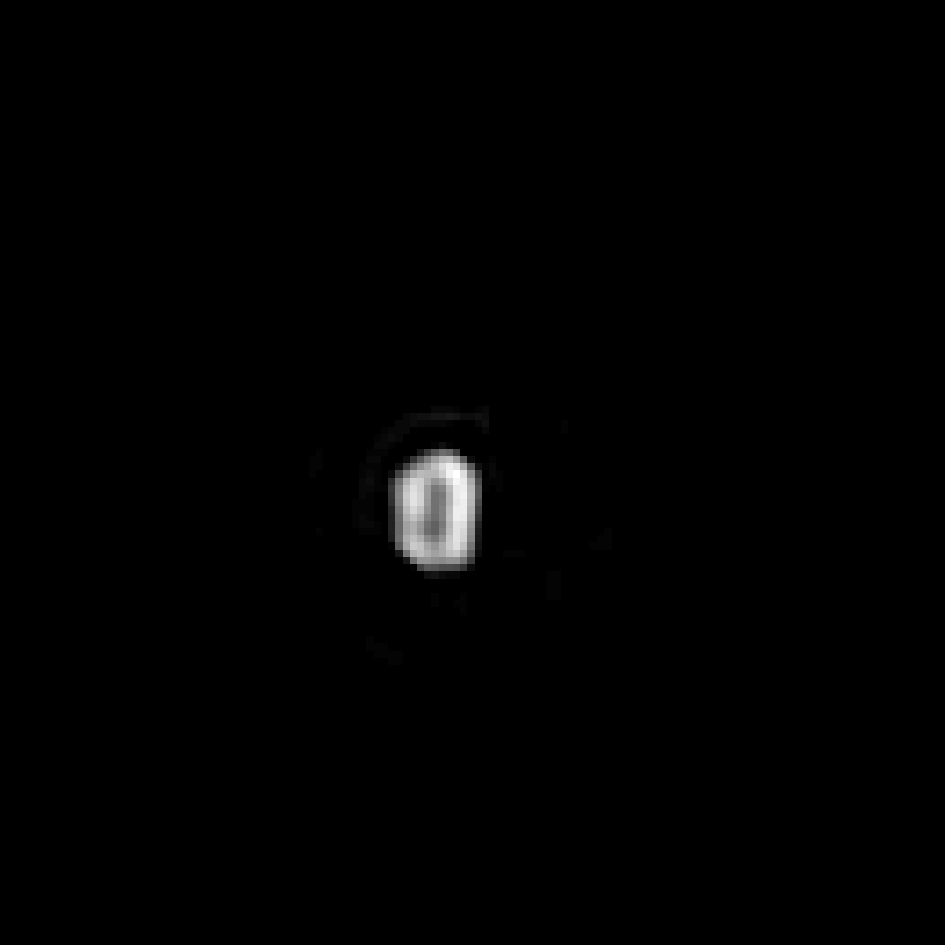} 
    \end{subfigure}%
     \begin{subfigure}[b]{0.16\linewidth}
     \includegraphics[clip=true,trim=70 75 60 55,scale=0.39]{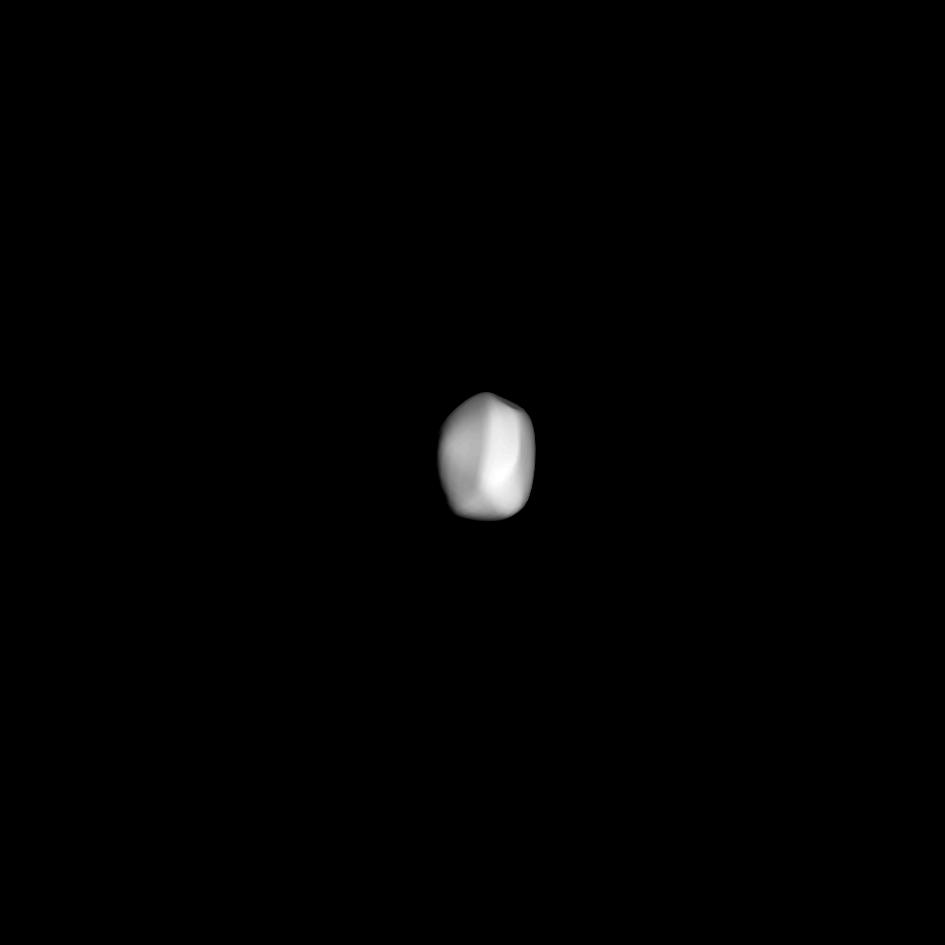}
     \end{subfigure}%
      \begin{subfigure}[b]{0.16\linewidth}
      \includegraphics[clip=true,trim=65 65 65 65,scale=0.39]{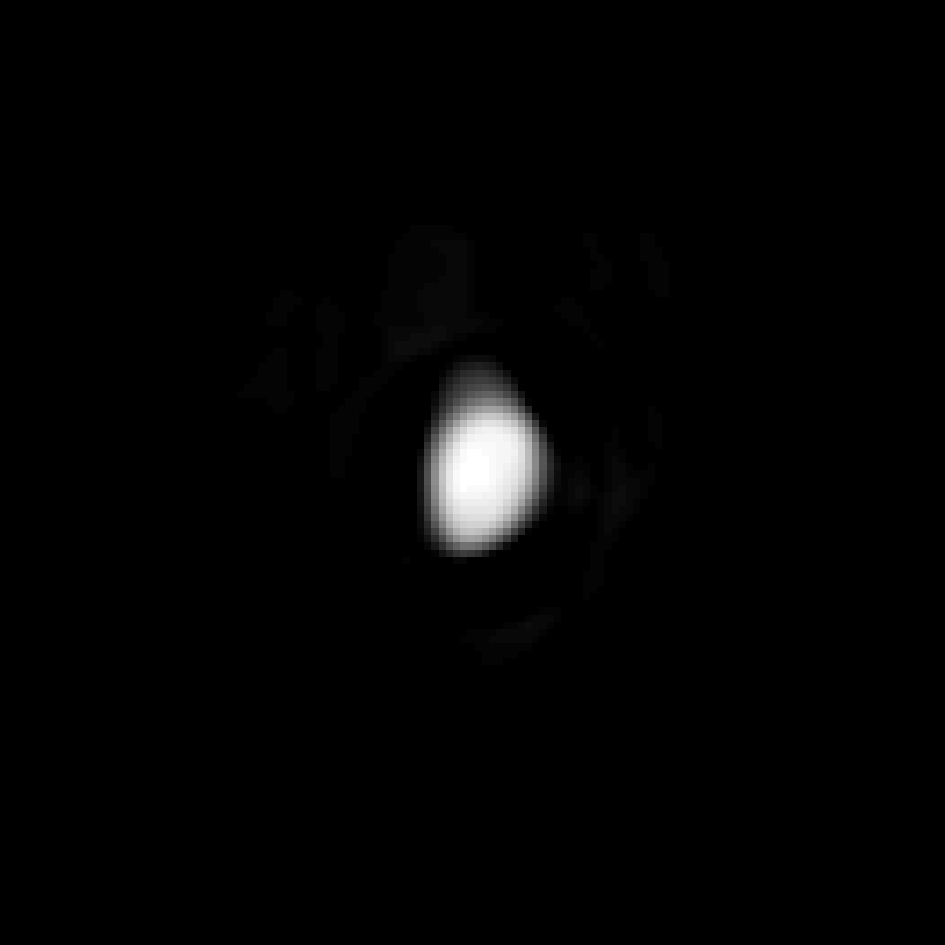}
    \end{subfigure}%
     \begin{subfigure}[b]{0.16\linewidth}
      \includegraphics[clip=true,trim=70 75 60 55,scale=0.39]{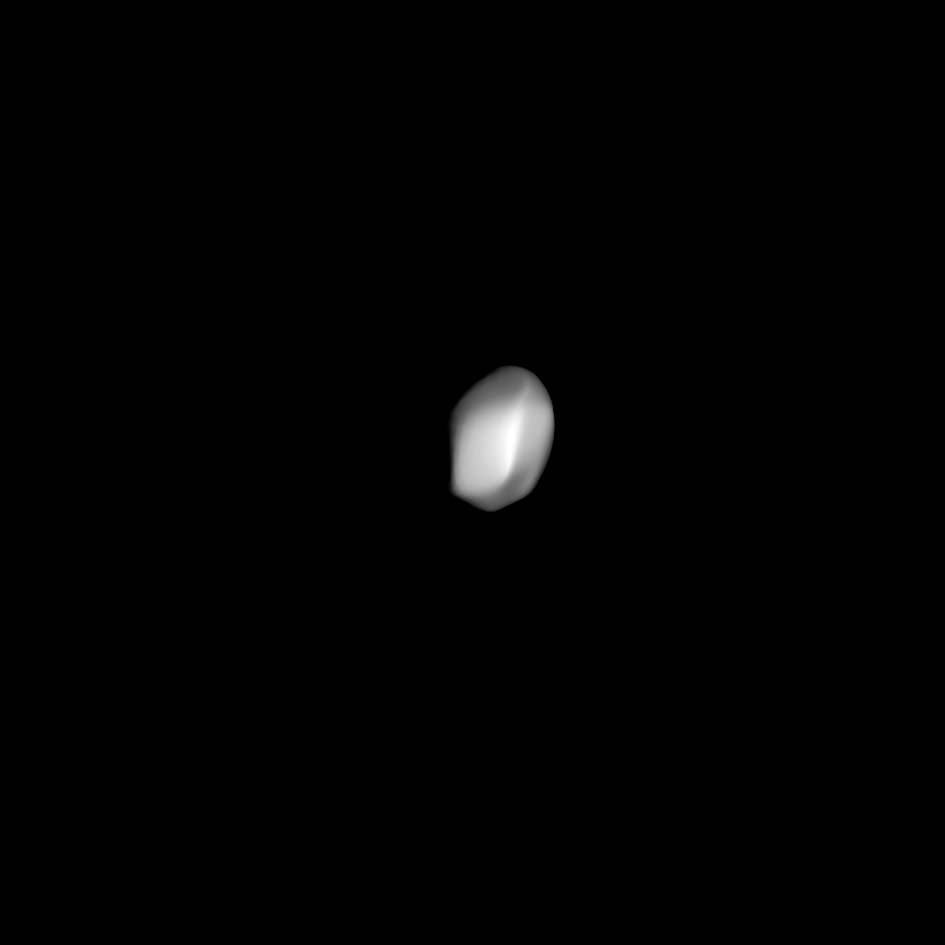}
    \end{subfigure}%
    \begin{subfigure}[b]{0.16\linewidth}
      \includegraphics[clip=true,trim=65 65 65 65,scale=0.39]{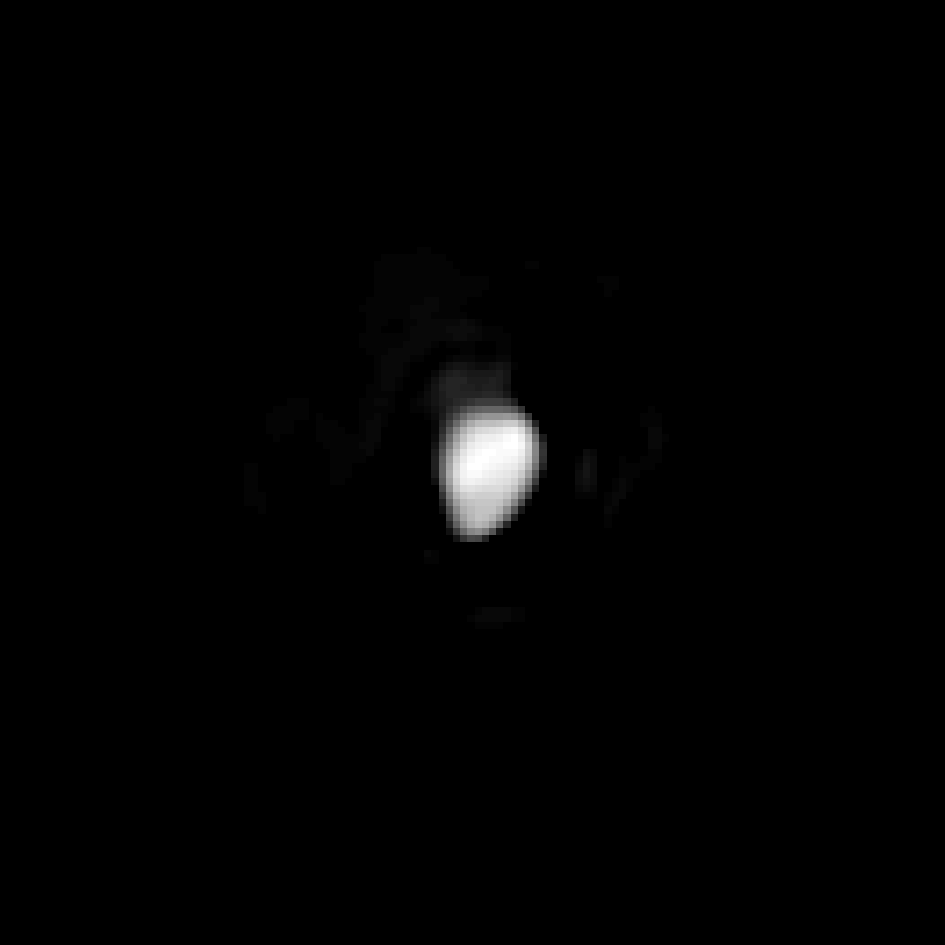} 
    \end{subfigure}%
     \begin{subfigure}[b]{0.16\linewidth}
     \includegraphics[clip=true,trim=70 75 60 55,scale=0.39]{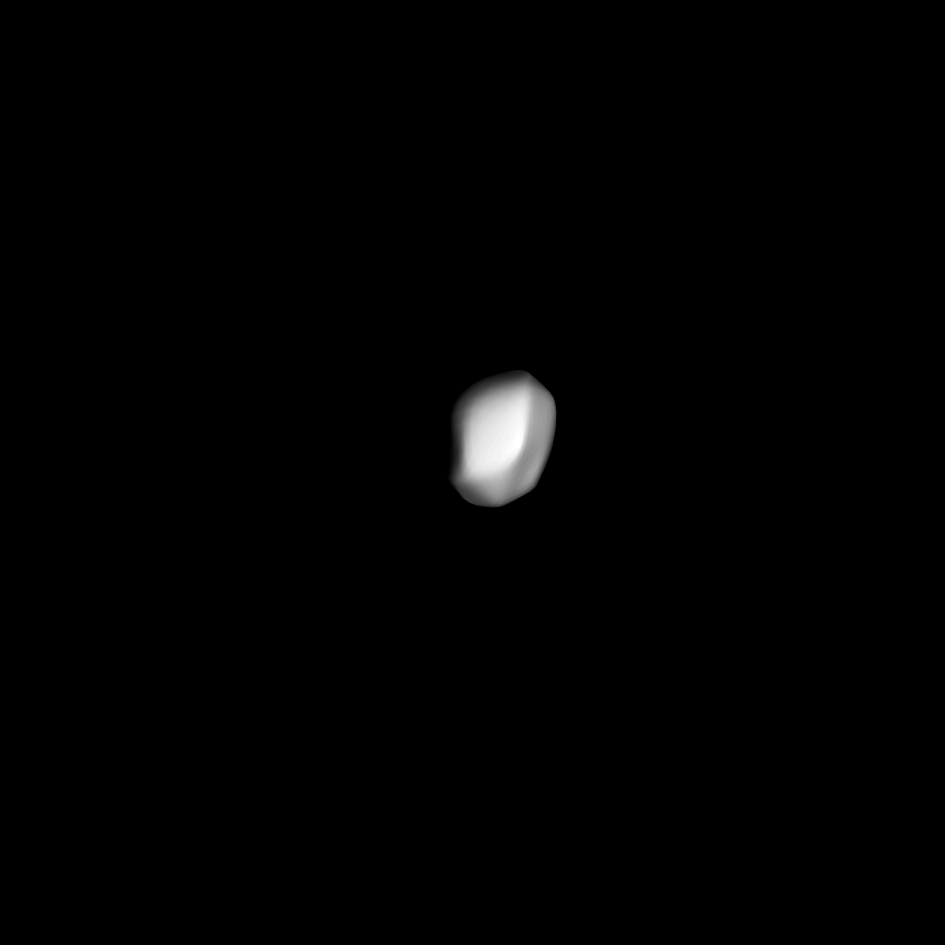}
     \end{subfigure}%
     
      \begin{subfigure}[b]{0.16\linewidth}
      \includegraphics[clip=true,trim=65 65 65 65,scale=0.39]{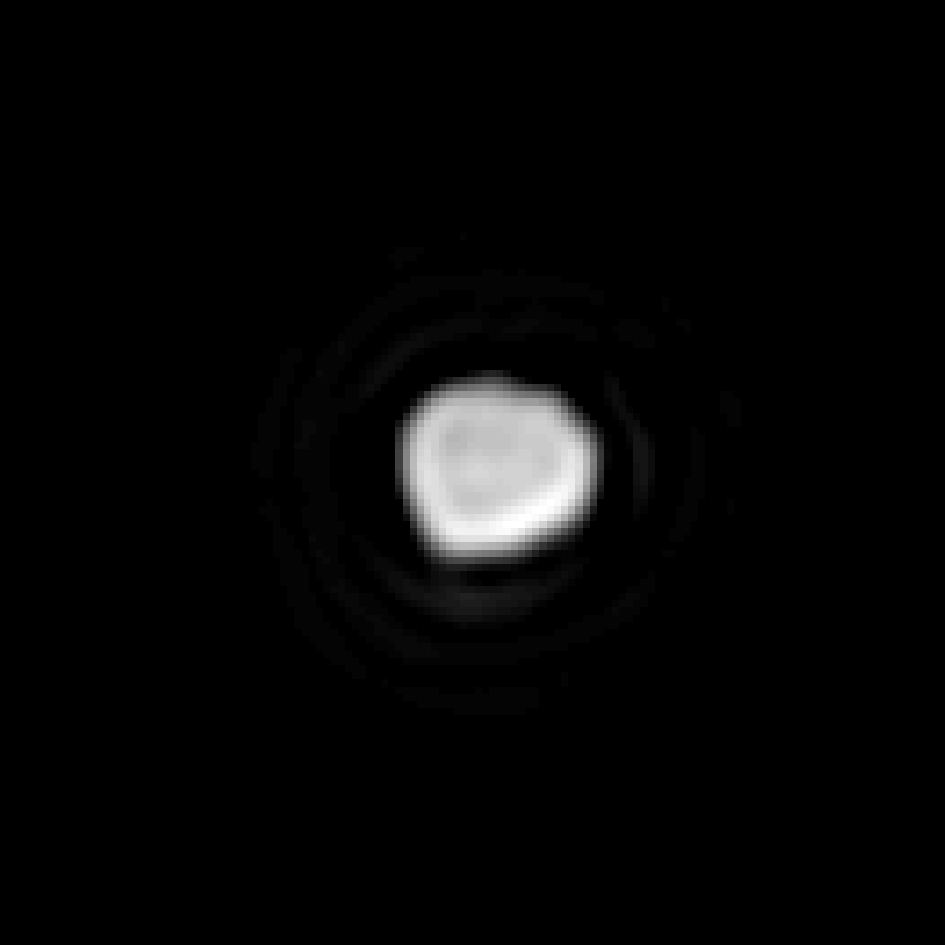} 
    \end{subfigure}%
     \begin{subfigure}[b]{0.16\linewidth}
     \includegraphics[clip=true,trim=70 75 60 55,scale=0.39]{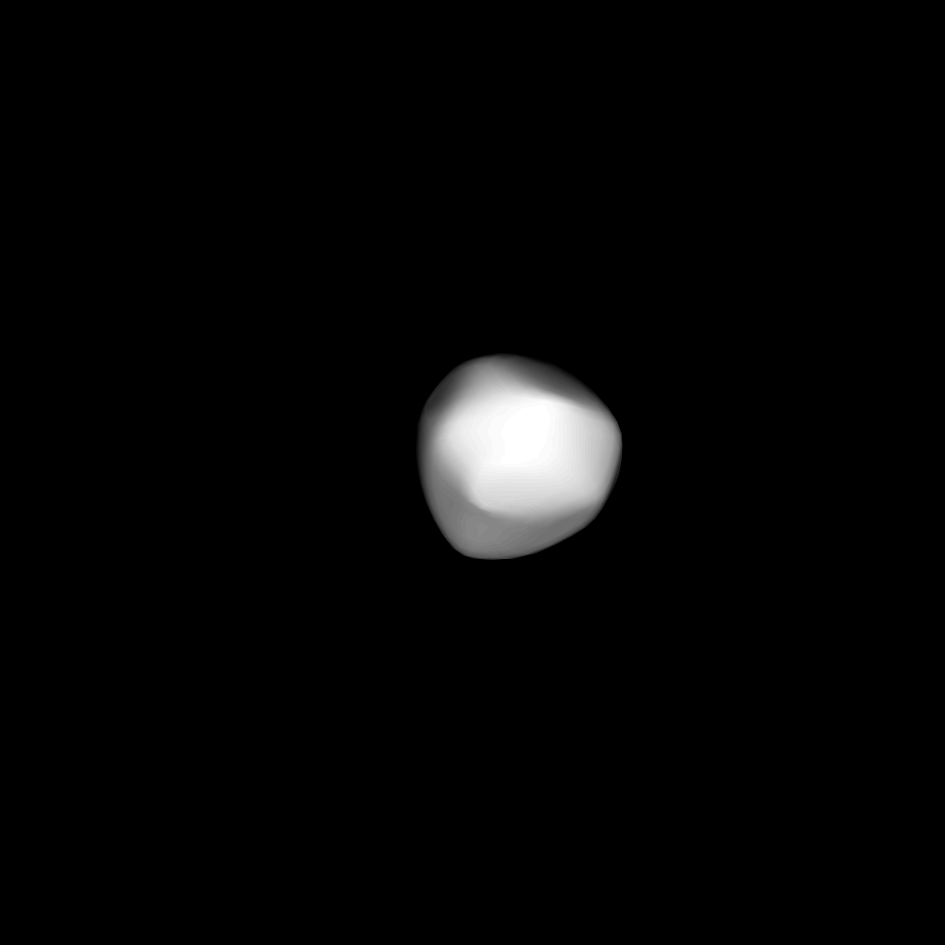}
     \end{subfigure}%
     \begin{subfigure}[b]{0.16\linewidth}
      \includegraphics[clip=true,trim=65 65 65 65,scale=0.39]{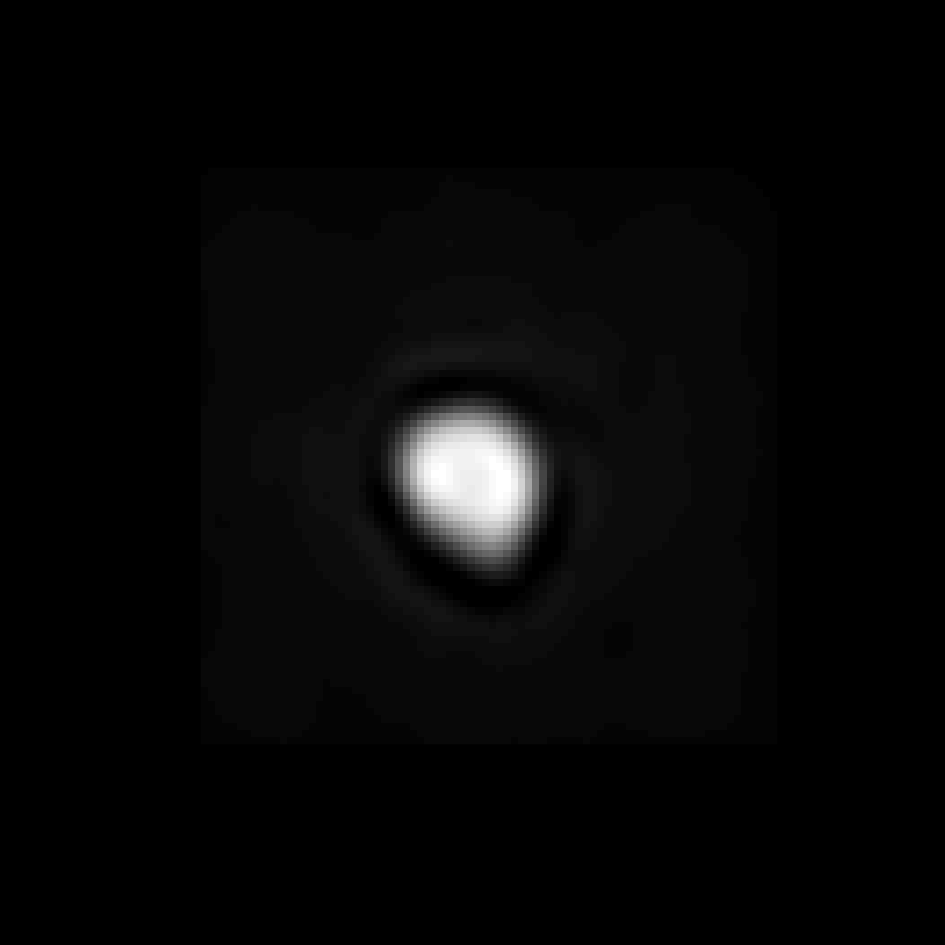} 
    \end{subfigure}%
     \begin{subfigure}[b]{0.16\linewidth}
     \includegraphics[clip=true,trim=70 75 60 55,scale=0.39]{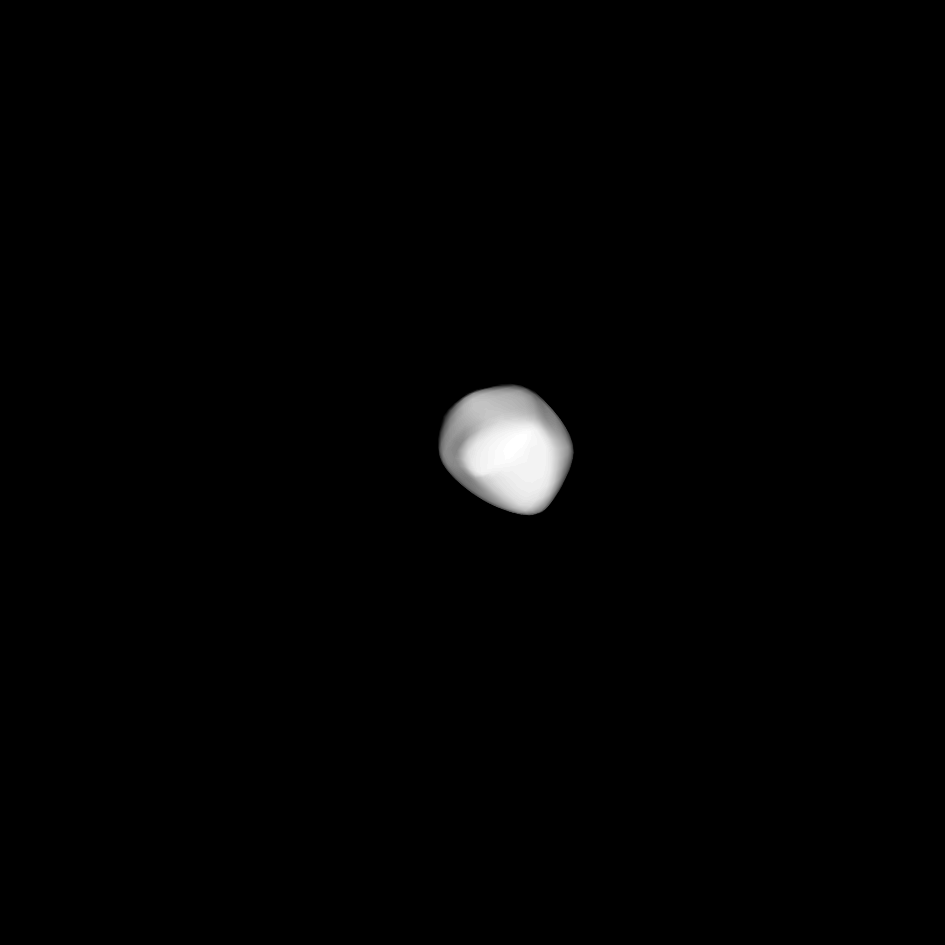}
     \end{subfigure}%
     \begin{subfigure}[b]{0.16\linewidth}
      \includegraphics[clip=true,trim=65 65 65 65,scale=0.39]{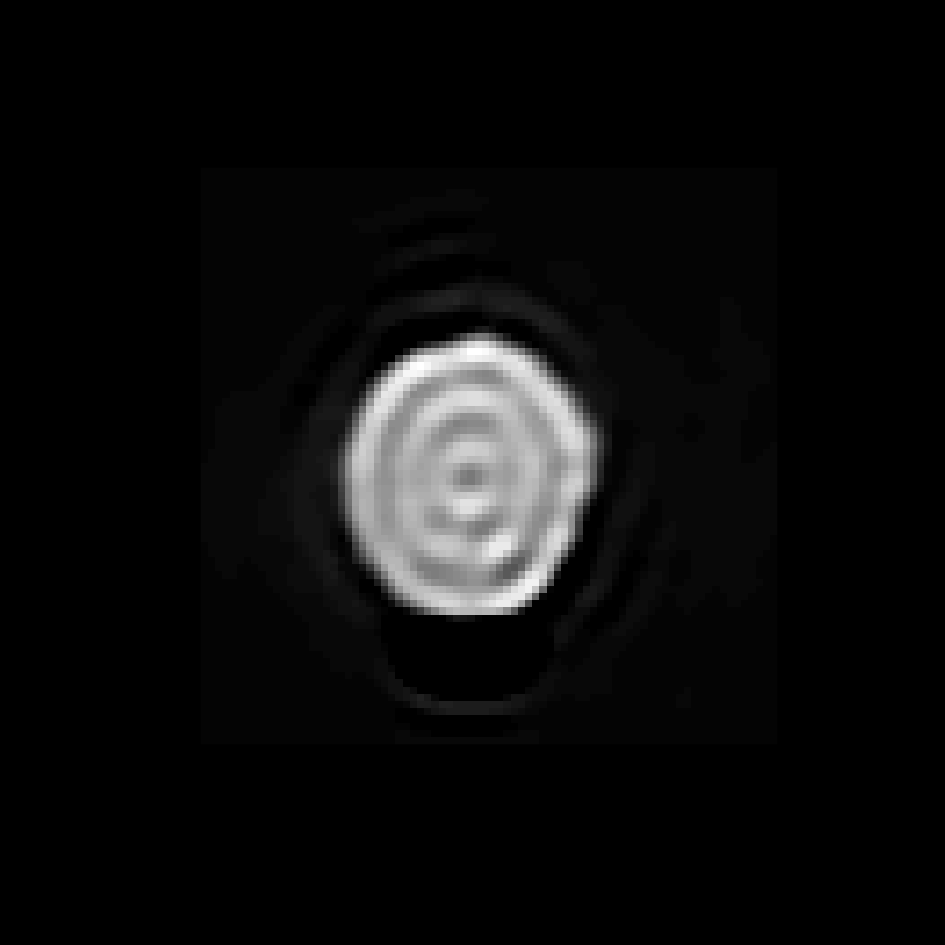} 
    \end{subfigure}%
     \begin{subfigure}[b]{0.16\linewidth}
     \includegraphics[clip=true,trim=70 75 60 55,scale=0.39]{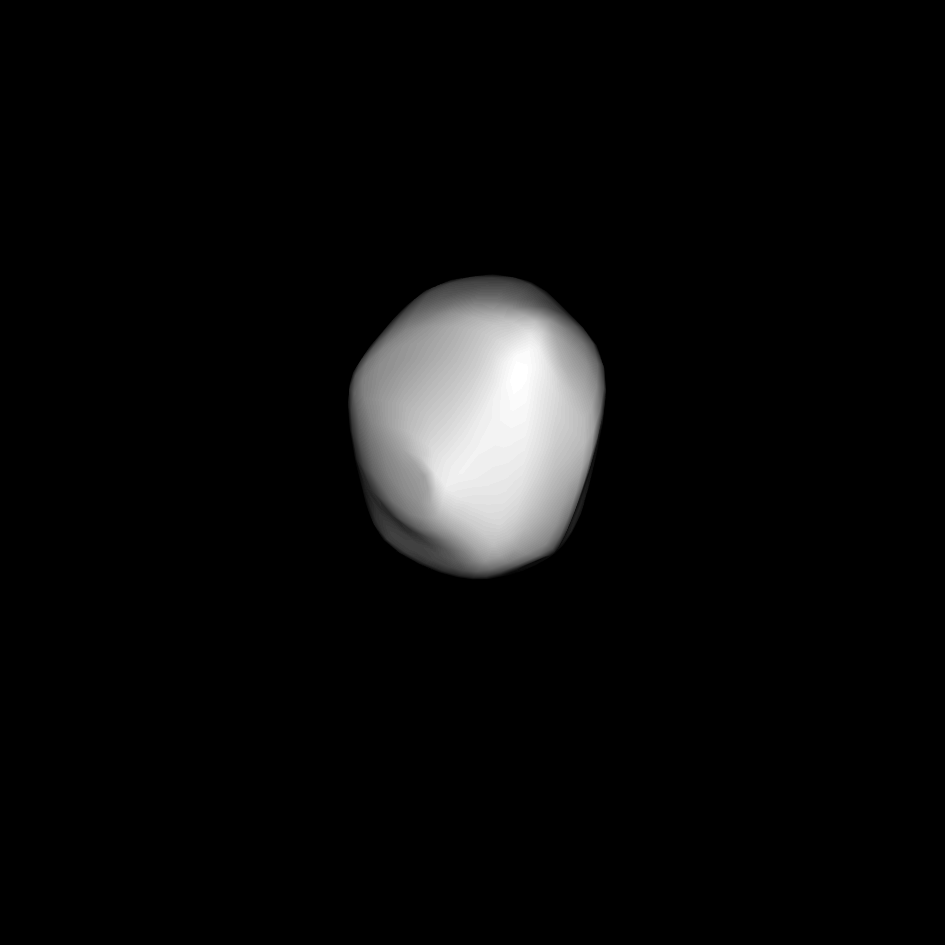}
     \end{subfigure}%
     
     \begin{subfigure}[b]{0.16\linewidth}
     \includegraphics[clip=true,trim=65 65 65 65,scale=0.39]{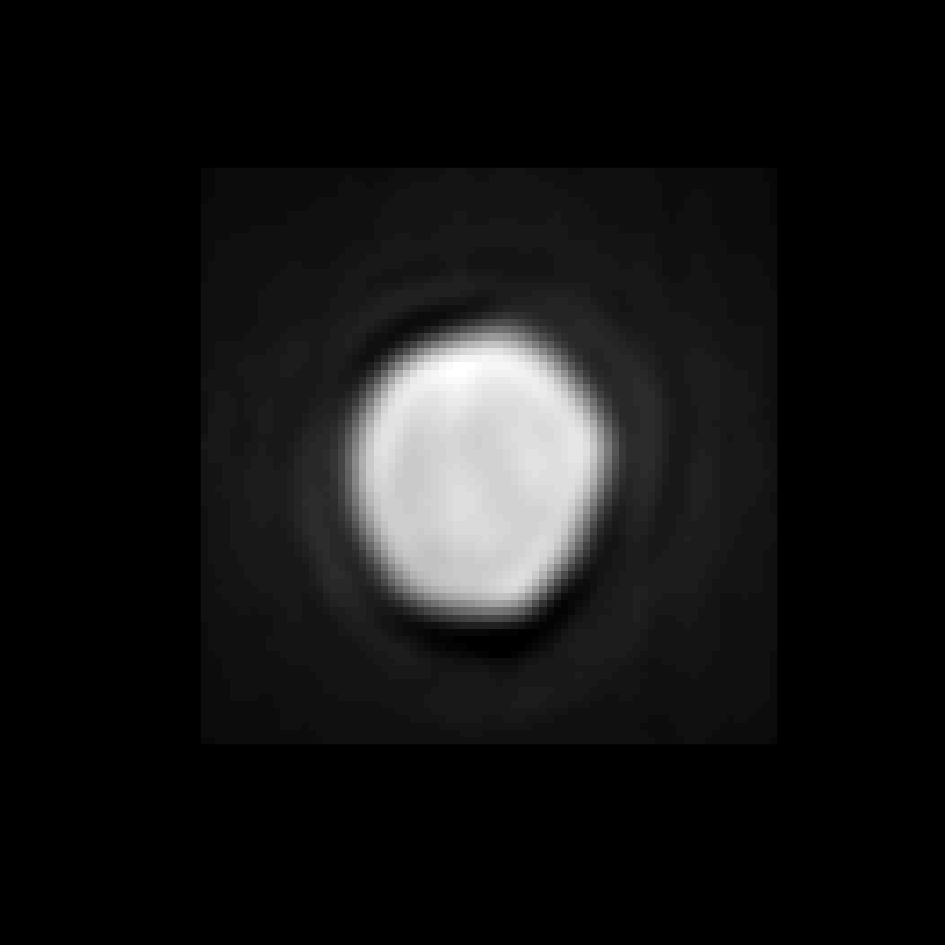}
     \end{subfigure}%
     \begin{subfigure}[b]{0.16\linewidth}
     \includegraphics[clip=true,trim=70 75 60 55,scale=0.39]{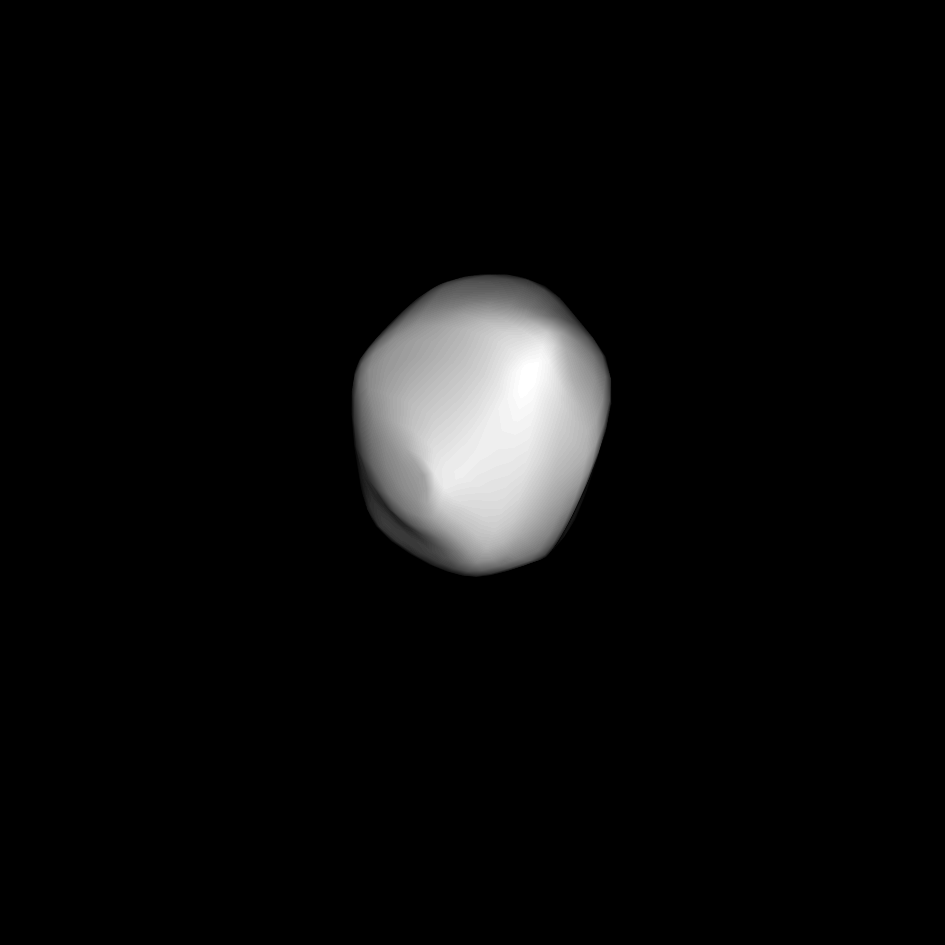}
     \end{subfigure}%
     \begin{subfigure}[b]{0.16\linewidth}
     \includegraphics[clip=true,trim=65 65 65 65,scale=0.39]{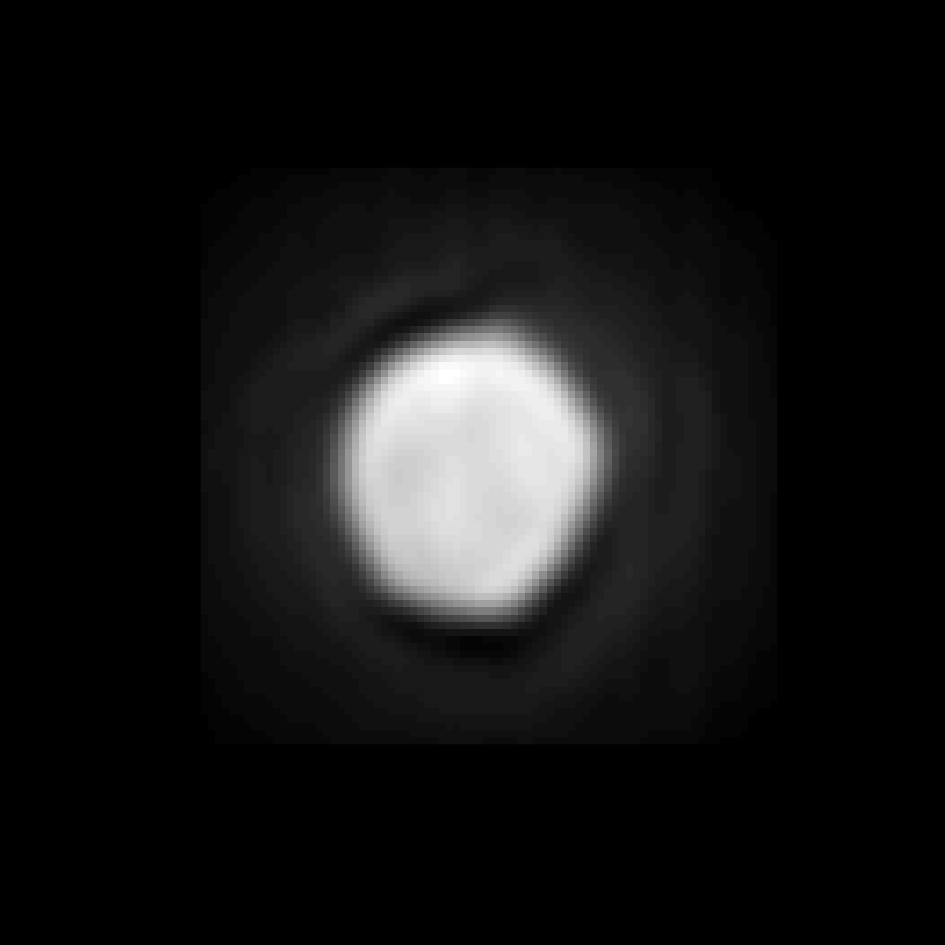}
     \end{subfigure}%
     \begin{subfigure}[b]{0.16\linewidth}
     \includegraphics[clip=true,trim=70 75 60 55,scale=0.39]{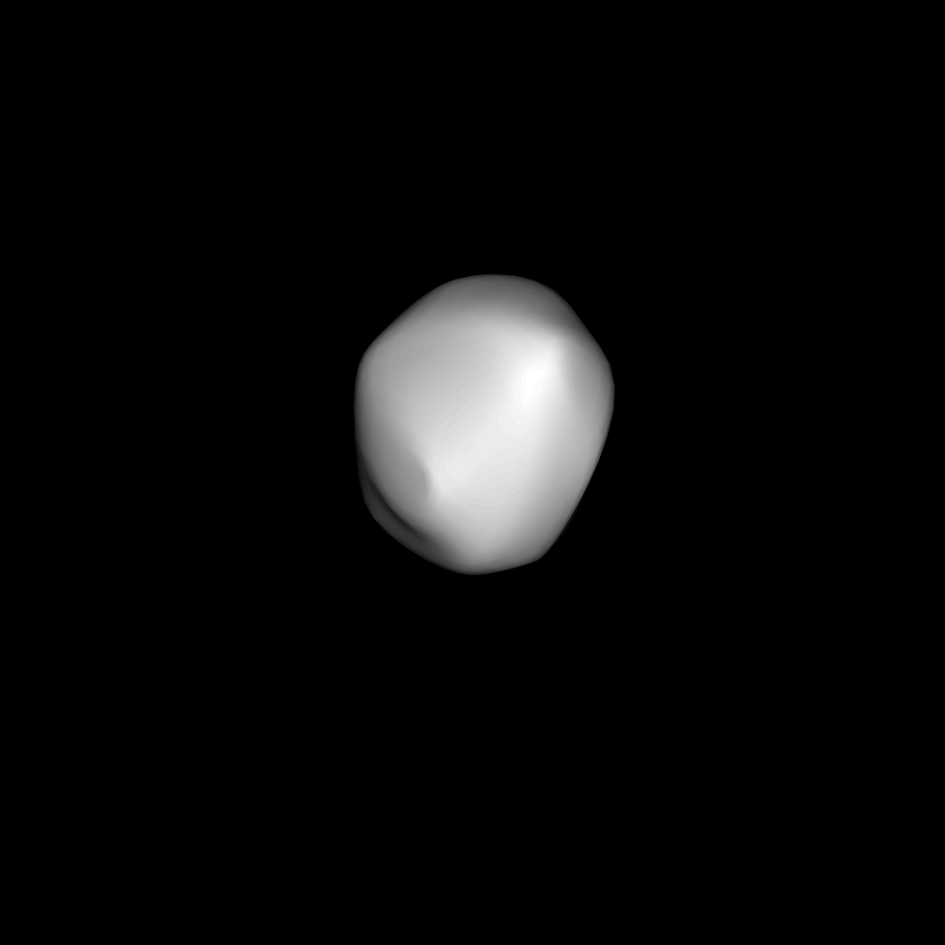}
     \end{subfigure}%
     \begin{subfigure}[b]{0.16\linewidth}
     \includegraphics[clip=true,trim=65 65 65 65,scale=0.39]{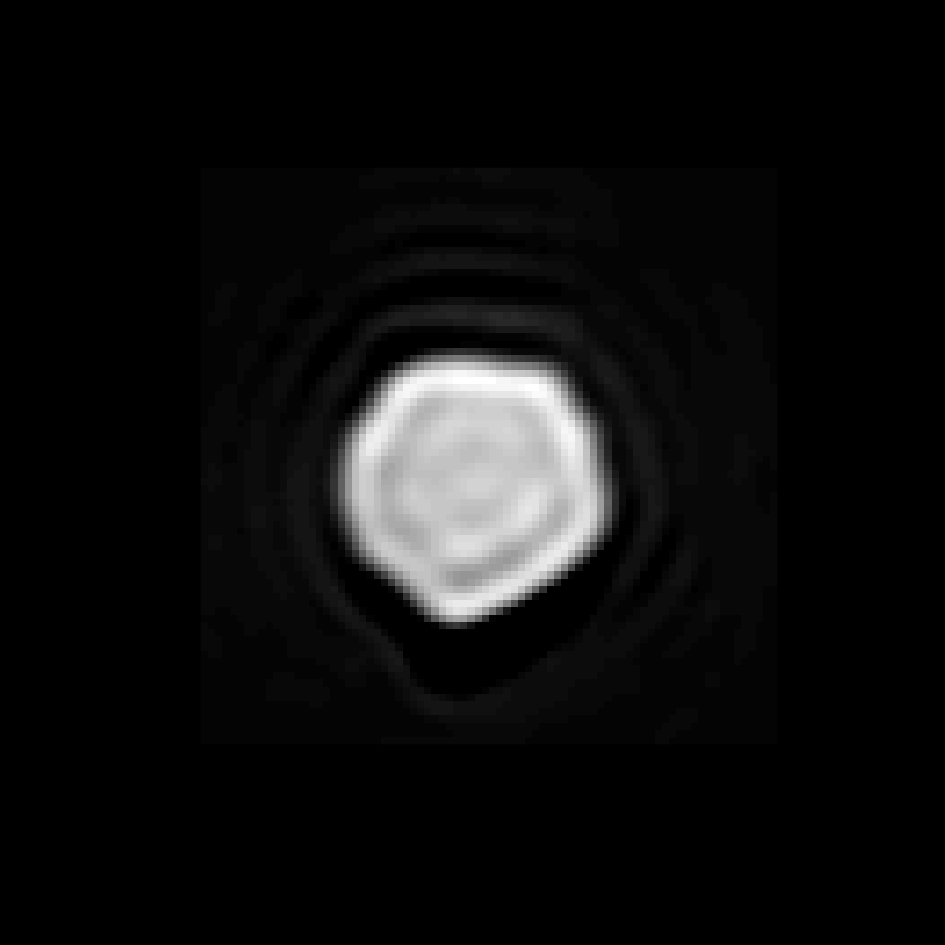}
     \end{subfigure}%
     \begin{subfigure}[b]{0.16\linewidth}
     \includegraphics[clip=true,trim=70 75 60 55,scale=0.39]{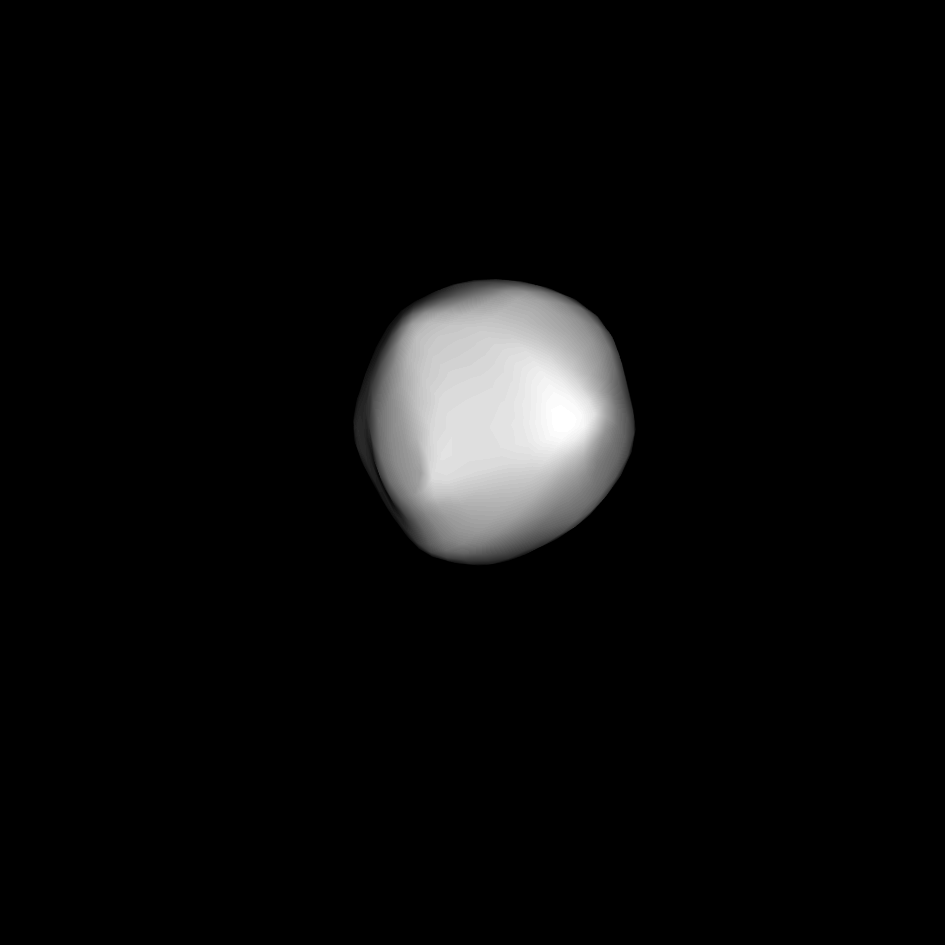}
     \end{subfigure}%
     
     \begin{subfigure}[b]{0.16\linewidth}
     \includegraphics[clip=true,trim=65 65 65 65,scale=0.39]{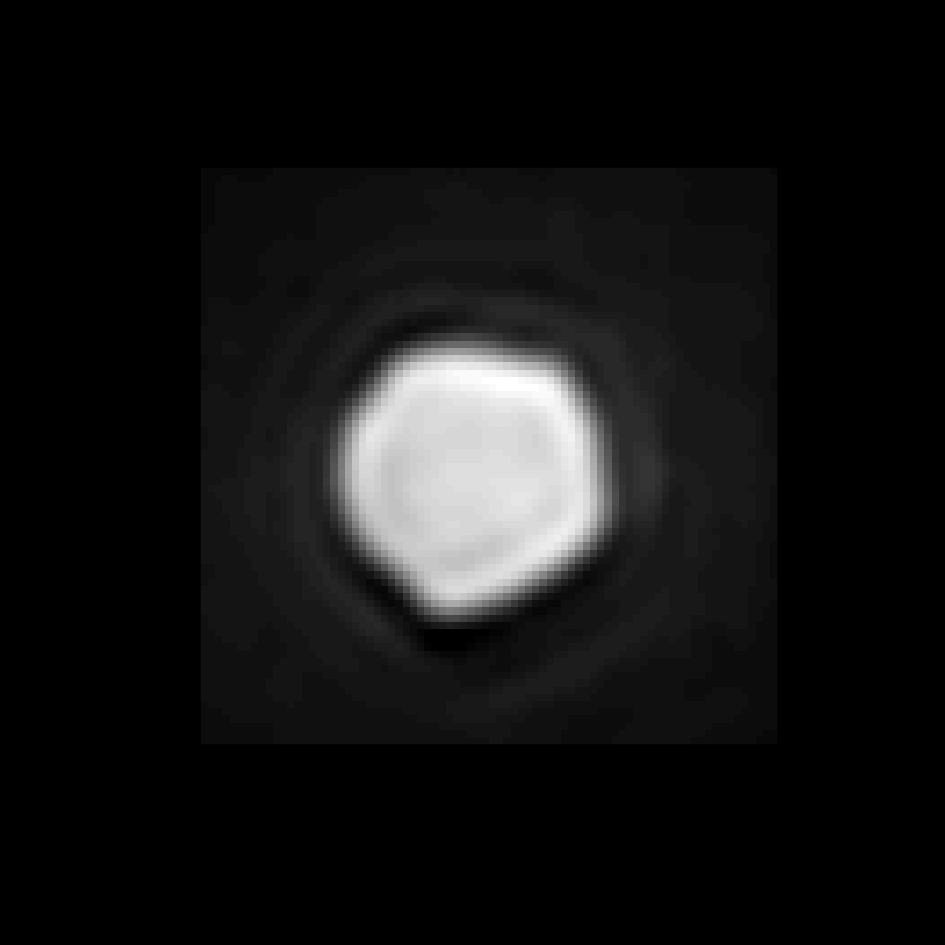}
     \end{subfigure}%
     \begin{subfigure}[b]{0.16\linewidth}
     \includegraphics[clip=true,trim=70 75 60 55,scale=0.39]{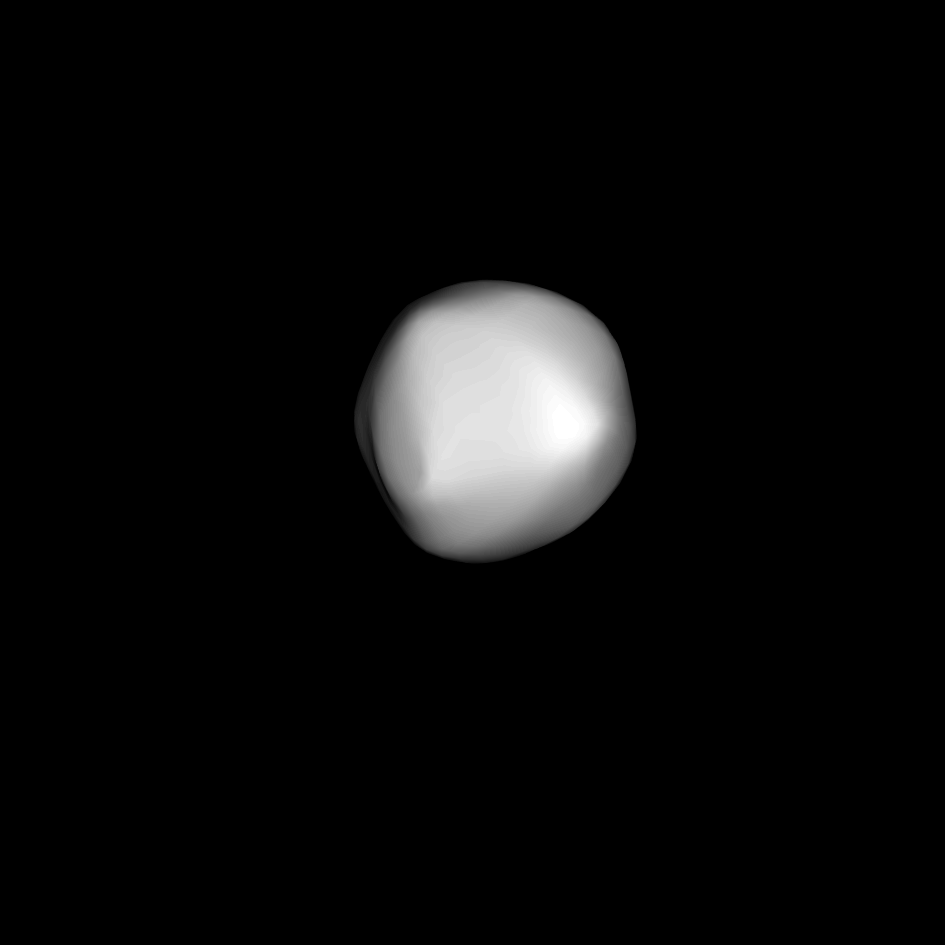}
     \end{subfigure}%
     \begin{subfigure}[b]{0.16\linewidth}
     \includegraphics[clip=true,trim=65 65 65 65,scale=0.39]{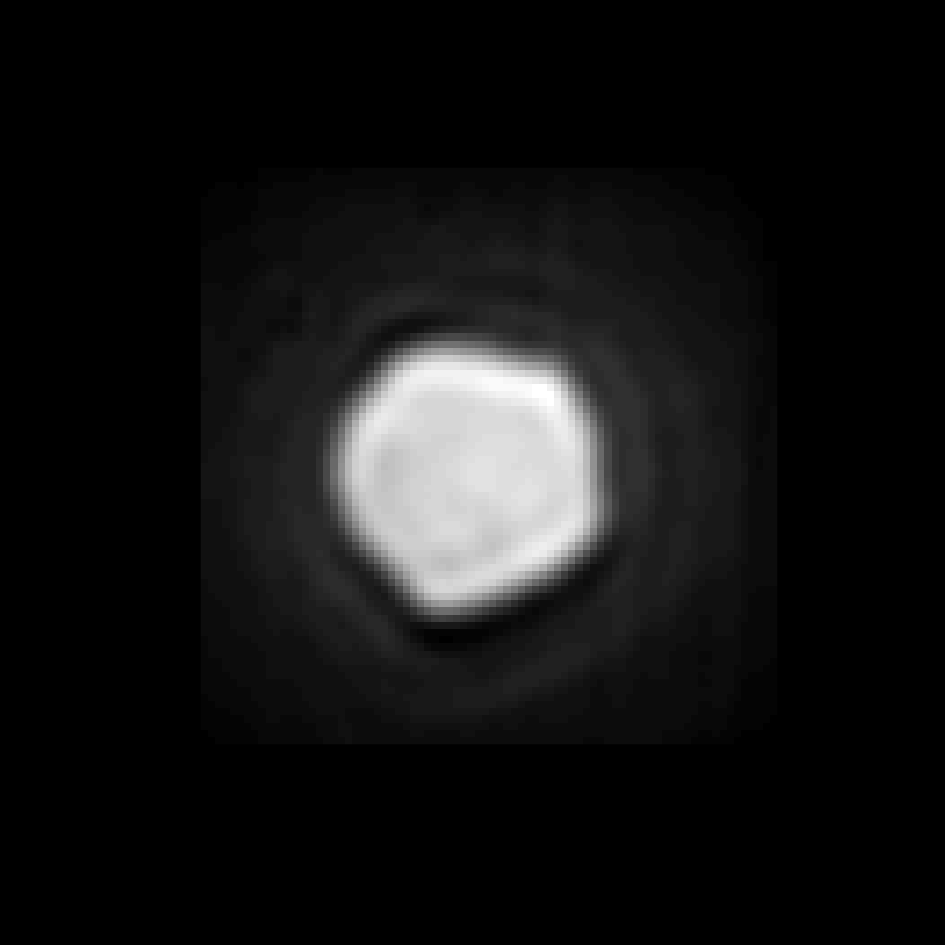}
     \end{subfigure}%
     \begin{subfigure}[b]{0.16\linewidth}
     \includegraphics[clip=true,trim=70 75 60 55,scale=0.39]{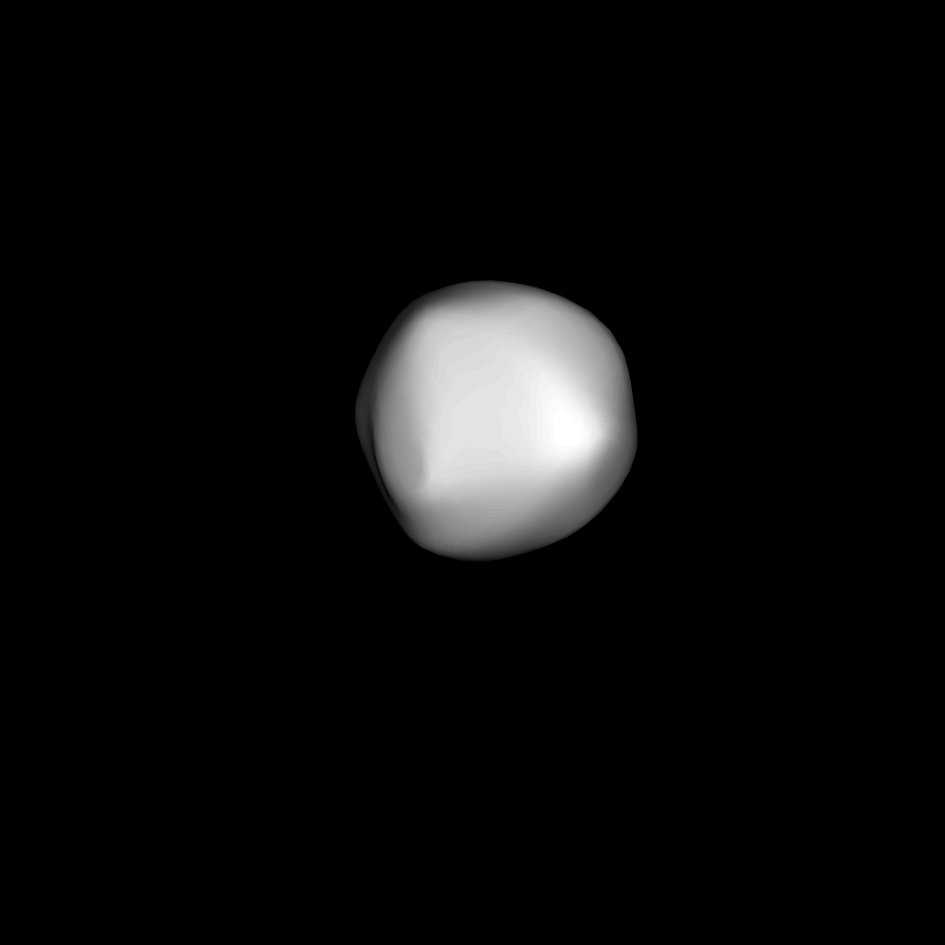}
     \end{subfigure}%
     \begin{subfigure}[b]{0.16\linewidth}
     \includegraphics[clip=true,trim=65 65 65 65,scale=0.39]{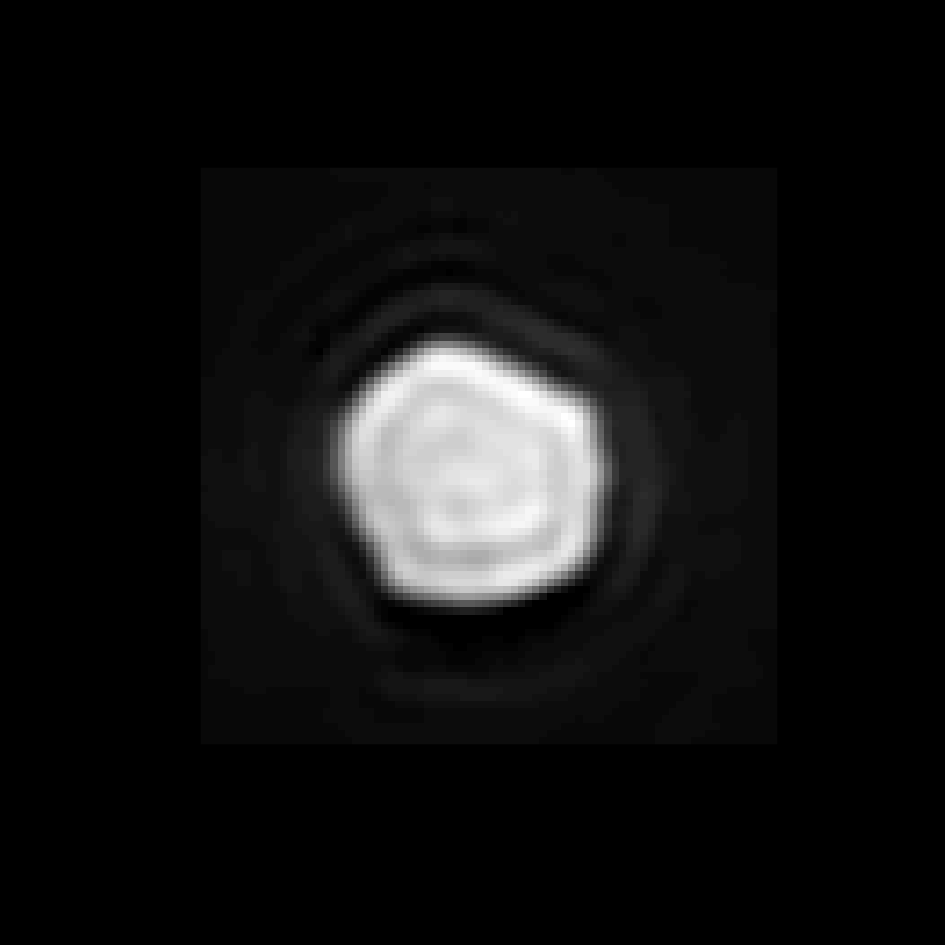}
     \end{subfigure}%
     \begin{subfigure}[b]{0.16\linewidth}
     \includegraphics[clip=true,trim=70 75 60 55,scale=0.39]{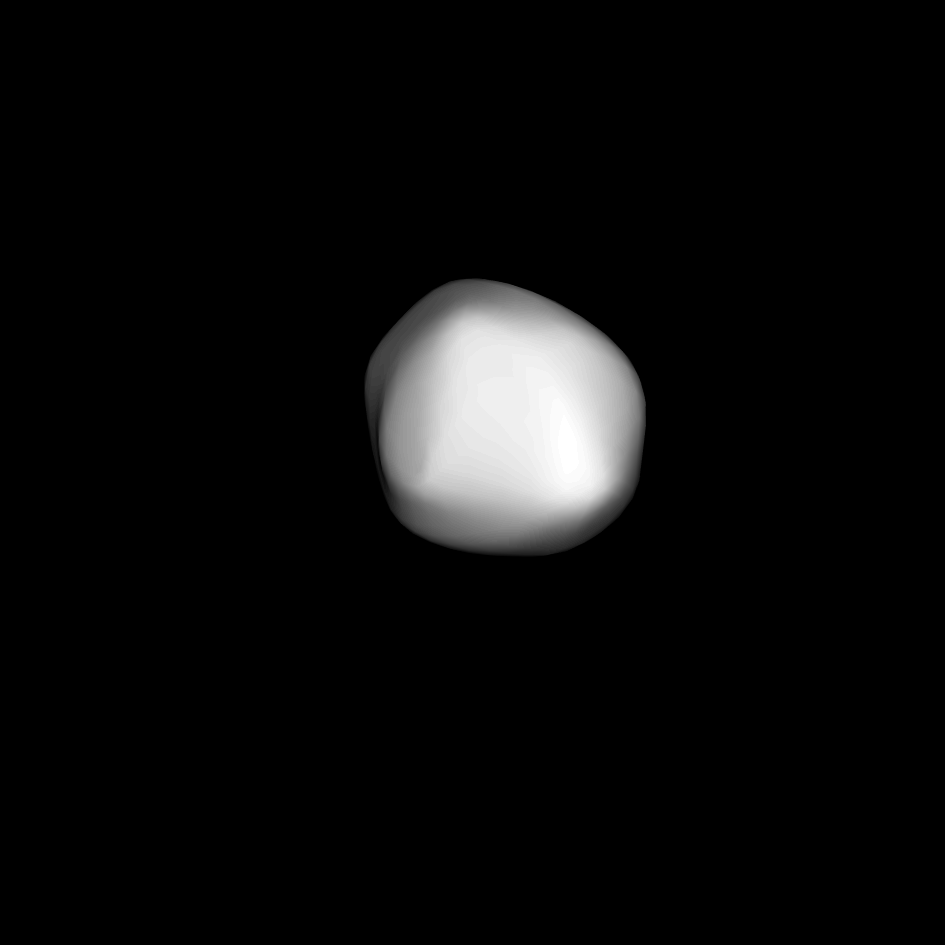}
     \end{subfigure}%
     
     \begin{subfigure}[b]{0.16\linewidth}
     \includegraphics[clip=true,trim=65 65 65 65,scale=0.39]{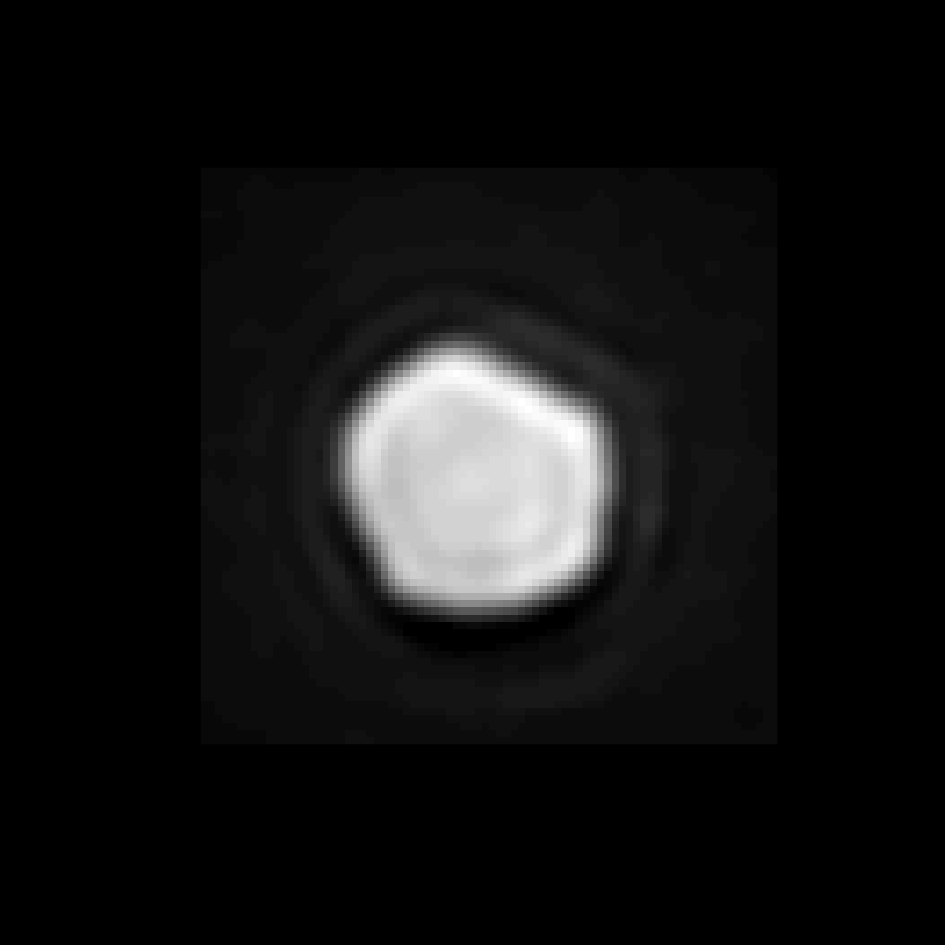}
     \end{subfigure}%
     \begin{subfigure}[b]{0.16\linewidth}
     \includegraphics[clip=true,trim=70 75 60 55,scale=0.39]{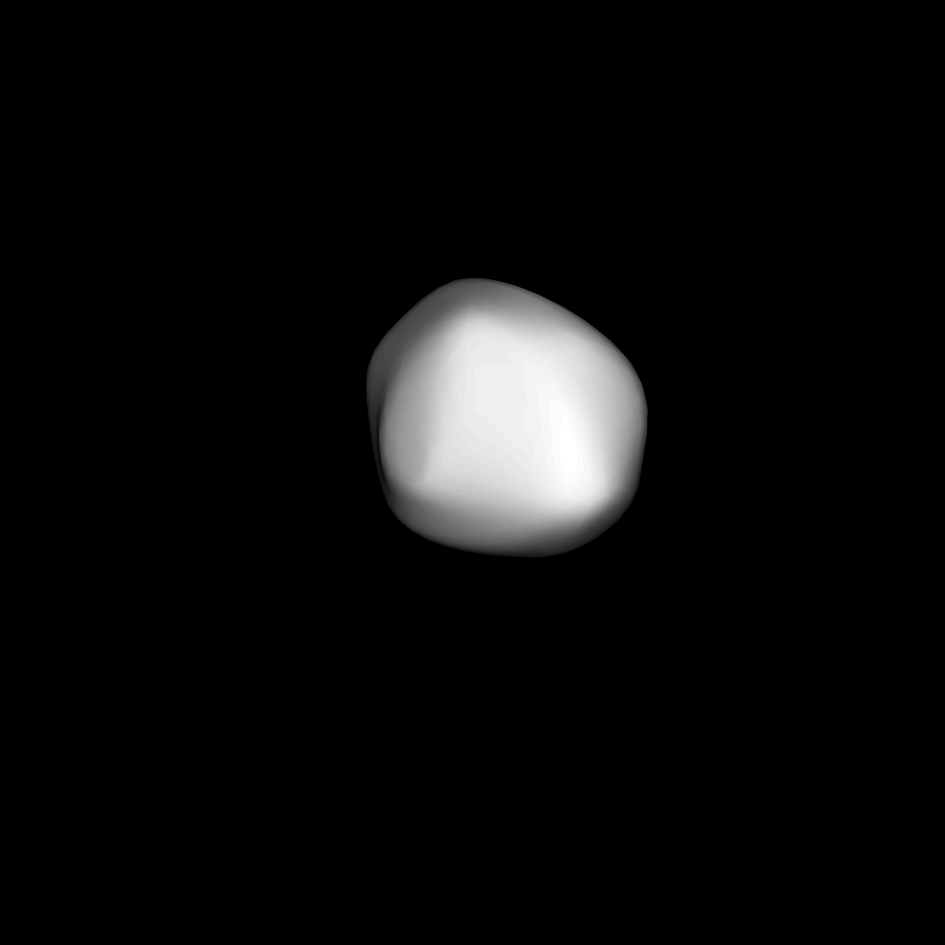}
     \end{subfigure}%
     \begin{subfigure}[b]{0.16\linewidth}
     \includegraphics[clip=true,trim=65 65 65 65,scale=0.39]{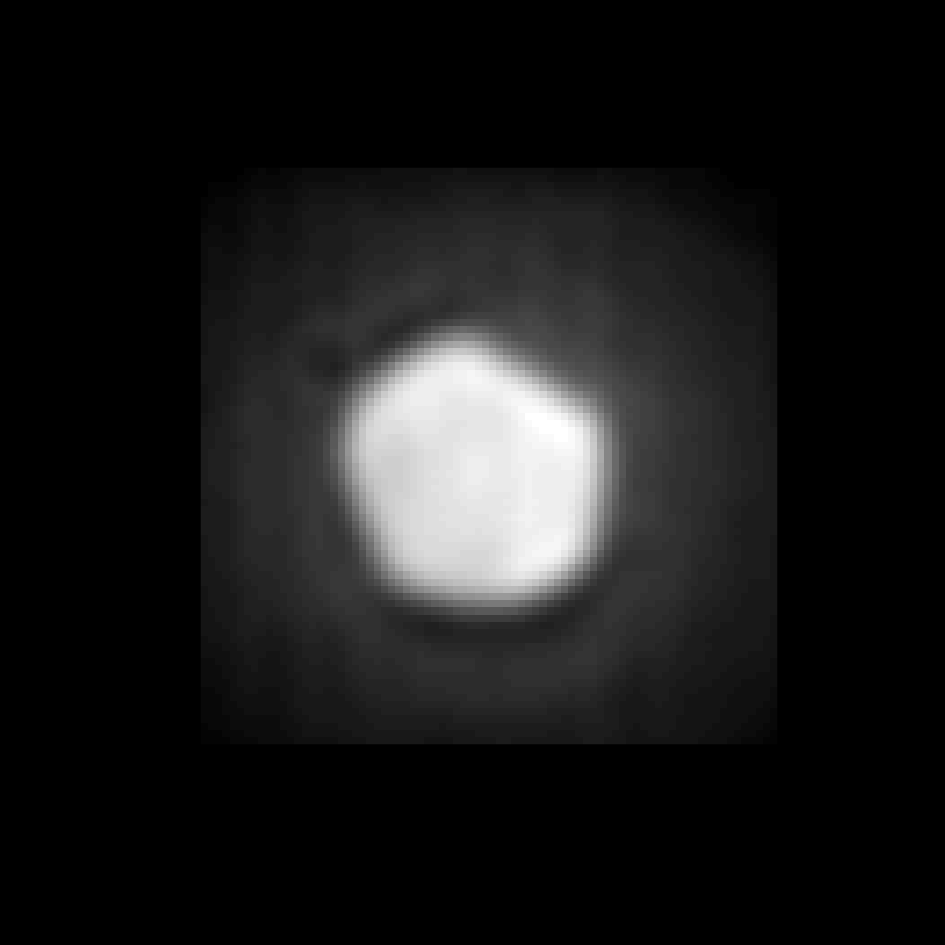}
     \end{subfigure}%
     \begin{subfigure}[b]{0.16\linewidth}
     \includegraphics[clip=true,trim=70 75 60 55,scale=0.39]{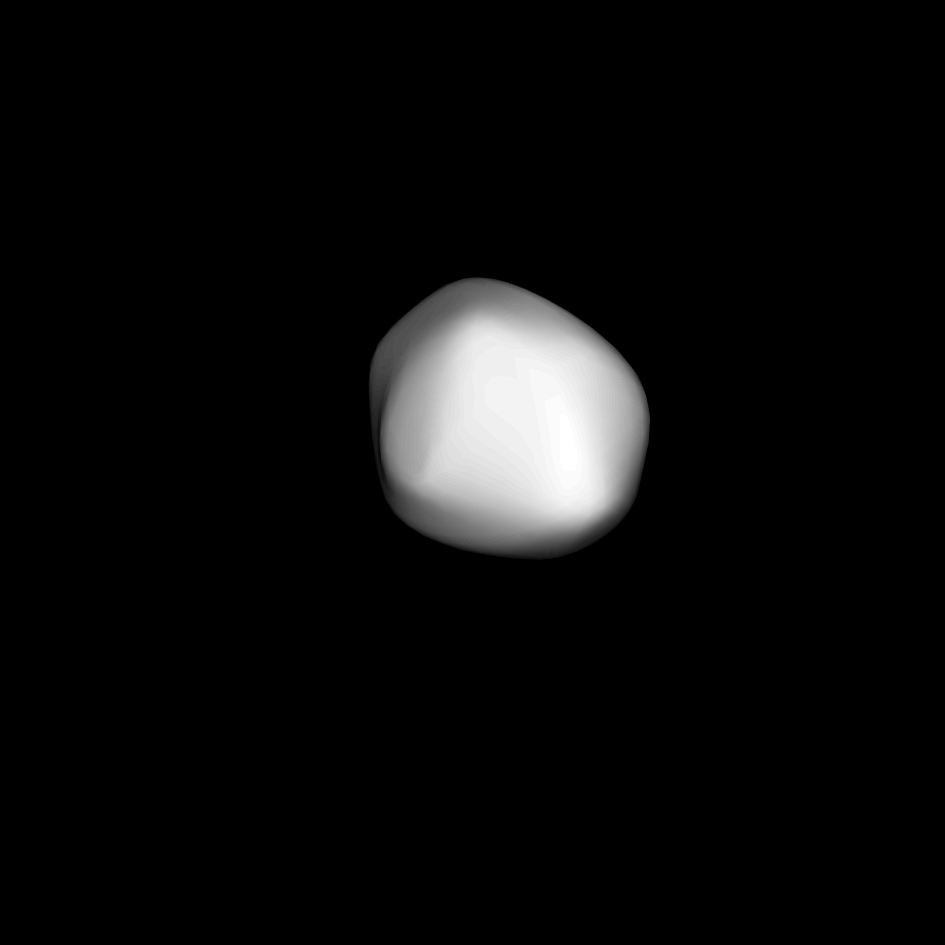}
     \end{subfigure}%
     \caption{\label{fig7}7 Iris}
    \end{figure}

\begin{figure}[t]
     \begin{subfigure}[b]{0.16\linewidth}
      \includegraphics[clip=true,trim=85 85 85 85,scale=0.66]{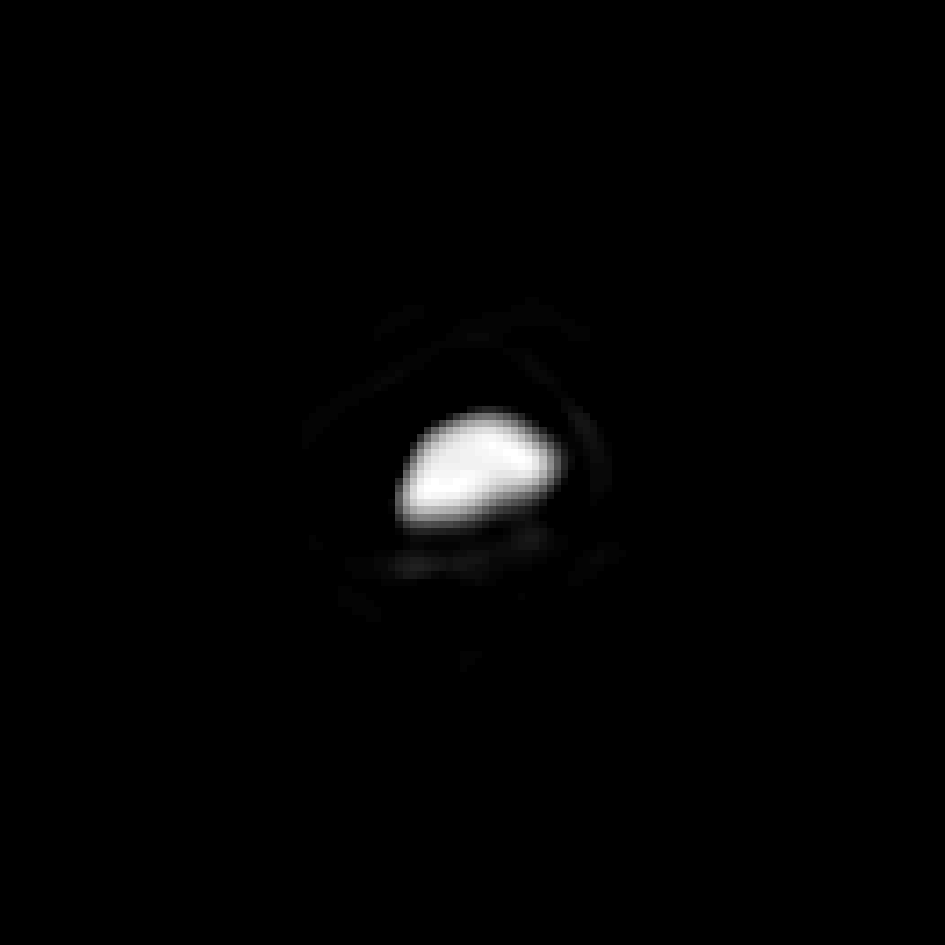} 
    \end{subfigure}%
     \begin{subfigure}[b]{0.16\linewidth}
     \includegraphics[clip=true,trim=90 90 80 80,scale=0.66]{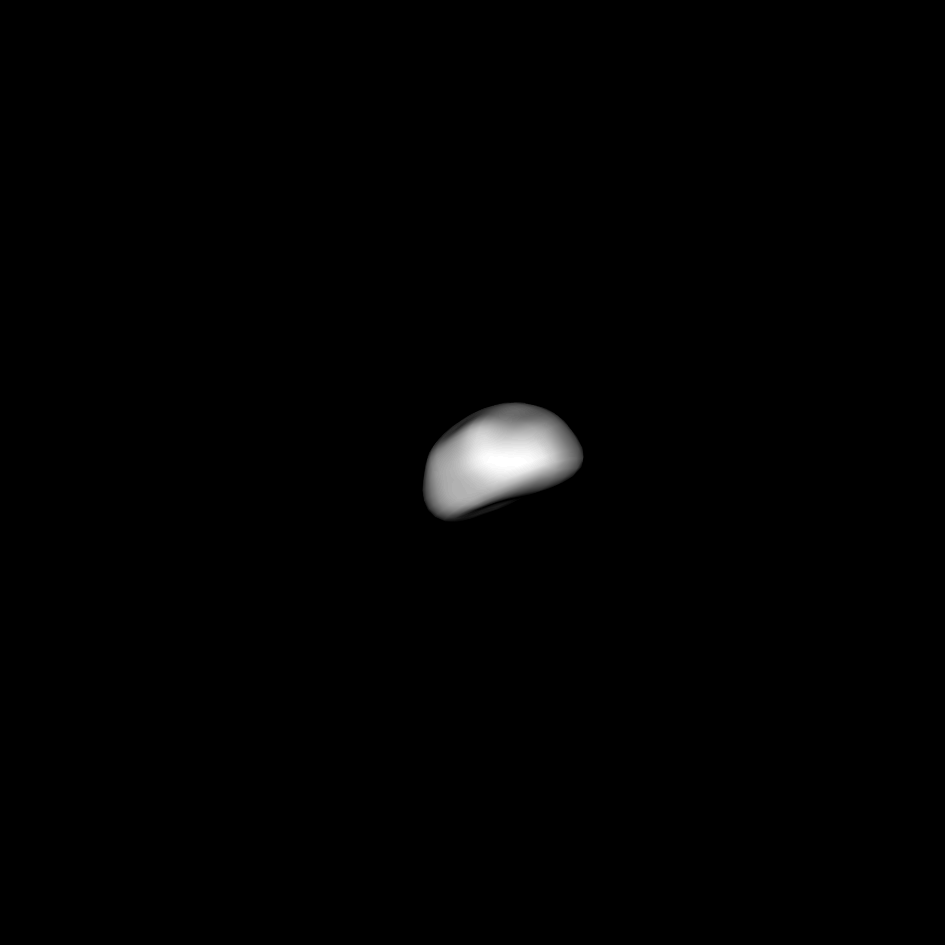}
     \end{subfigure}%
      \begin{subfigure}[b]{0.16\linewidth}
      \includegraphics[clip=true,trim=85 85 85 85,scale=0.66]{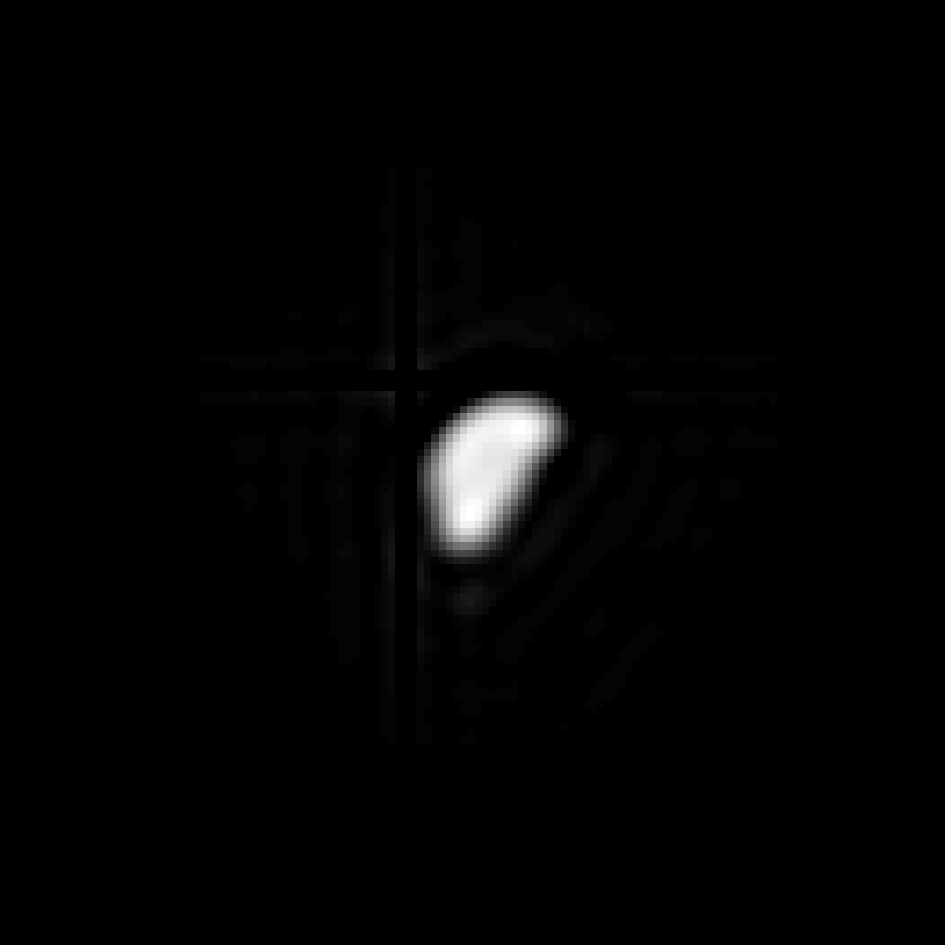}
    \end{subfigure}%
     \begin{subfigure}[b]{0.16\linewidth}
      \includegraphics[clip=true,trim=90 90 80 80,scale=0.66]{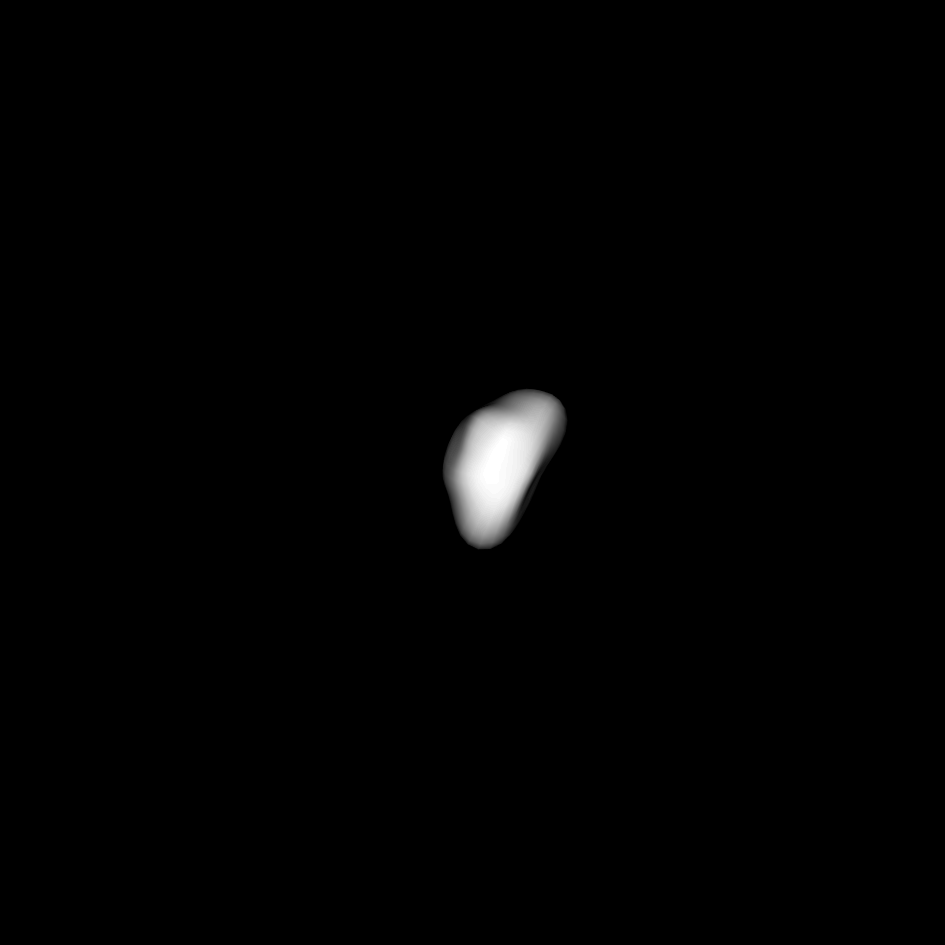}
    \end{subfigure}%
    \begin{subfigure}[b]{0.16\linewidth}
      \includegraphics[clip=true,trim=85 85 85 85,scale=0.66]{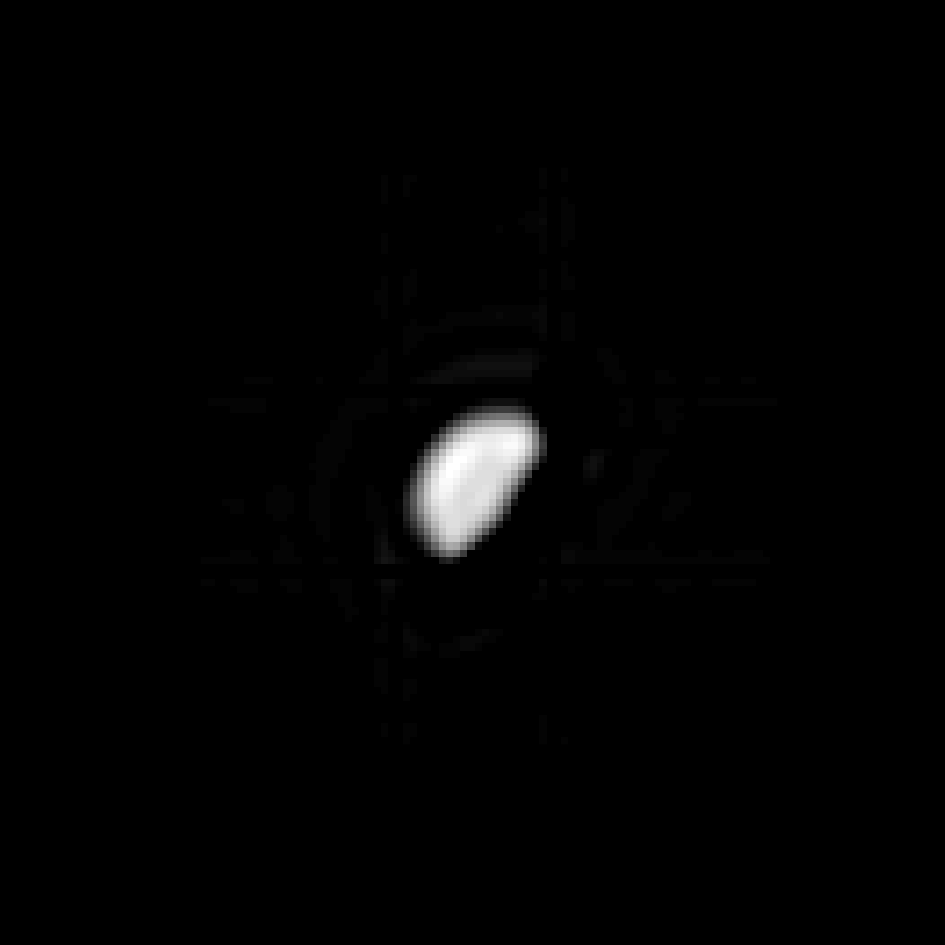} 
    \end{subfigure}%
     \begin{subfigure}[b]{0.16\linewidth}
     \includegraphics[clip=true,trim=90 90 80 80,scale=0.66]{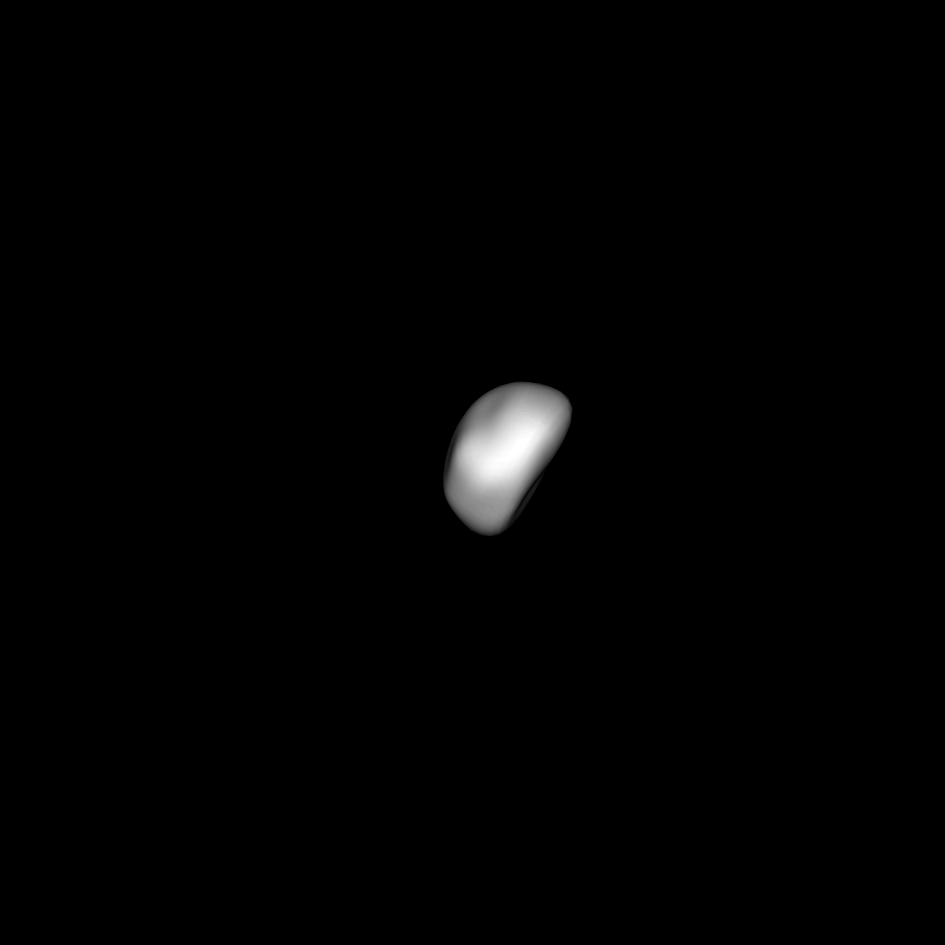}
     \end{subfigure}%
     
      \begin{subfigure}[b]{0.16\linewidth}
      \includegraphics[clip=true,trim=85 85 85 85,scale=0.66]{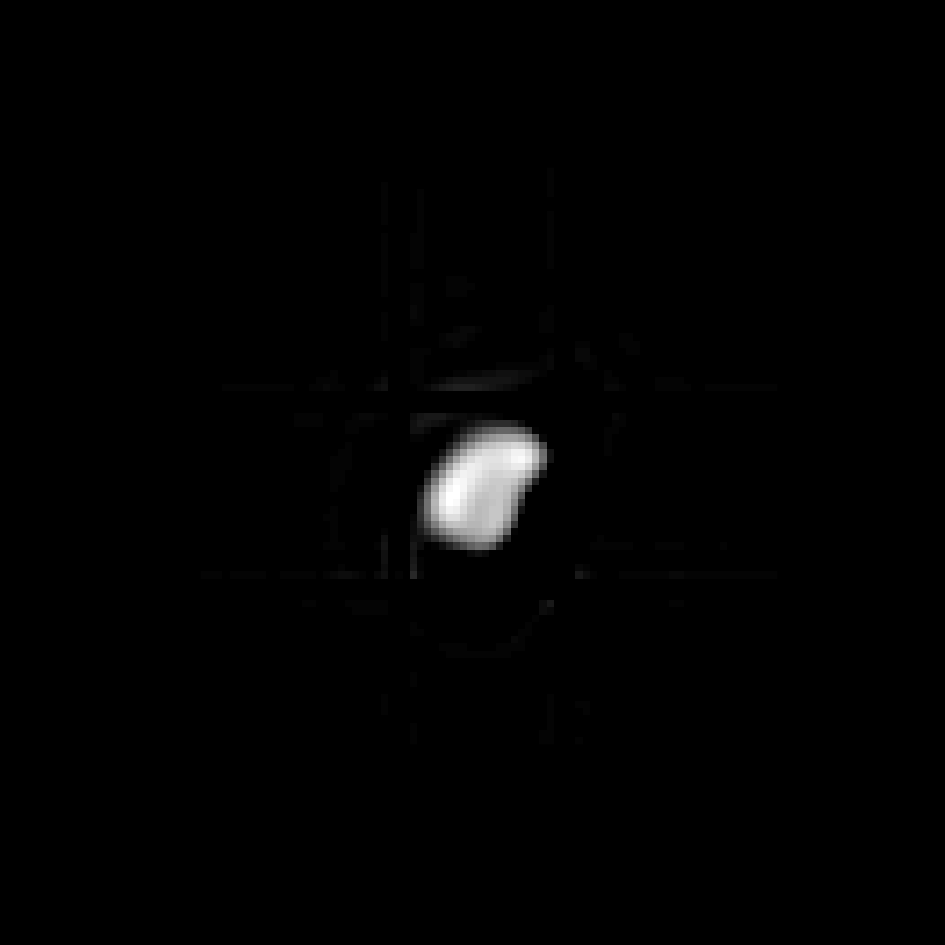} 
    \end{subfigure}%
     \begin{subfigure}[b]{0.16\linewidth}
     \includegraphics[clip=true,trim=90 90 80 80,scale=0.66]{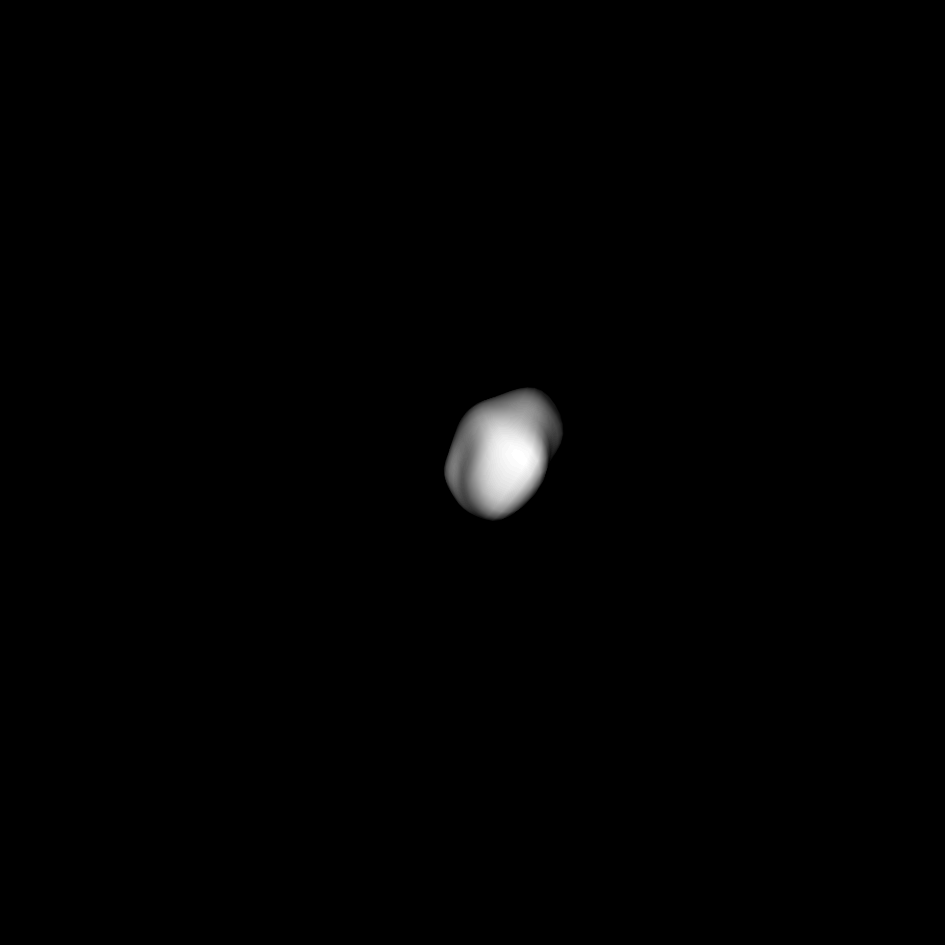}
     \end{subfigure}%
     \begin{subfigure}[b]{0.16\linewidth}
      \includegraphics[clip=true,trim=85 85 85 85,scale=0.66]{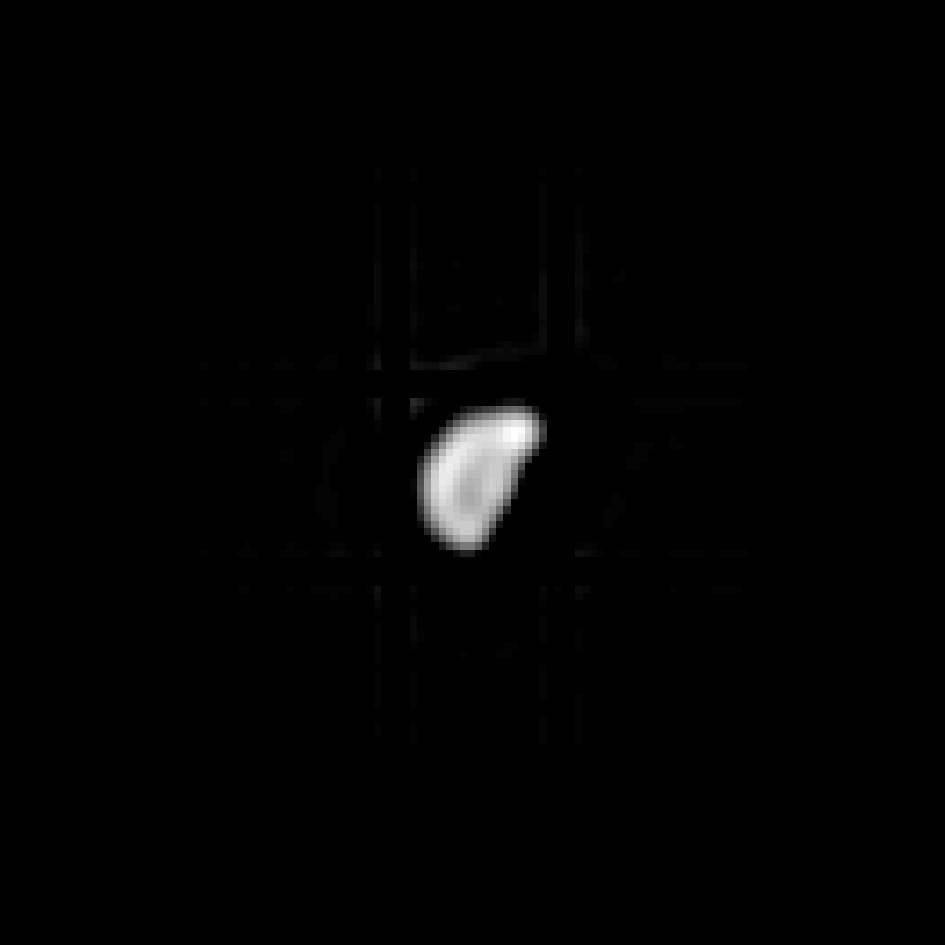} 
    \end{subfigure}%
     \begin{subfigure}[b]{0.16\linewidth}
     \includegraphics[clip=true,trim=90 90 80 80,scale=0.66]{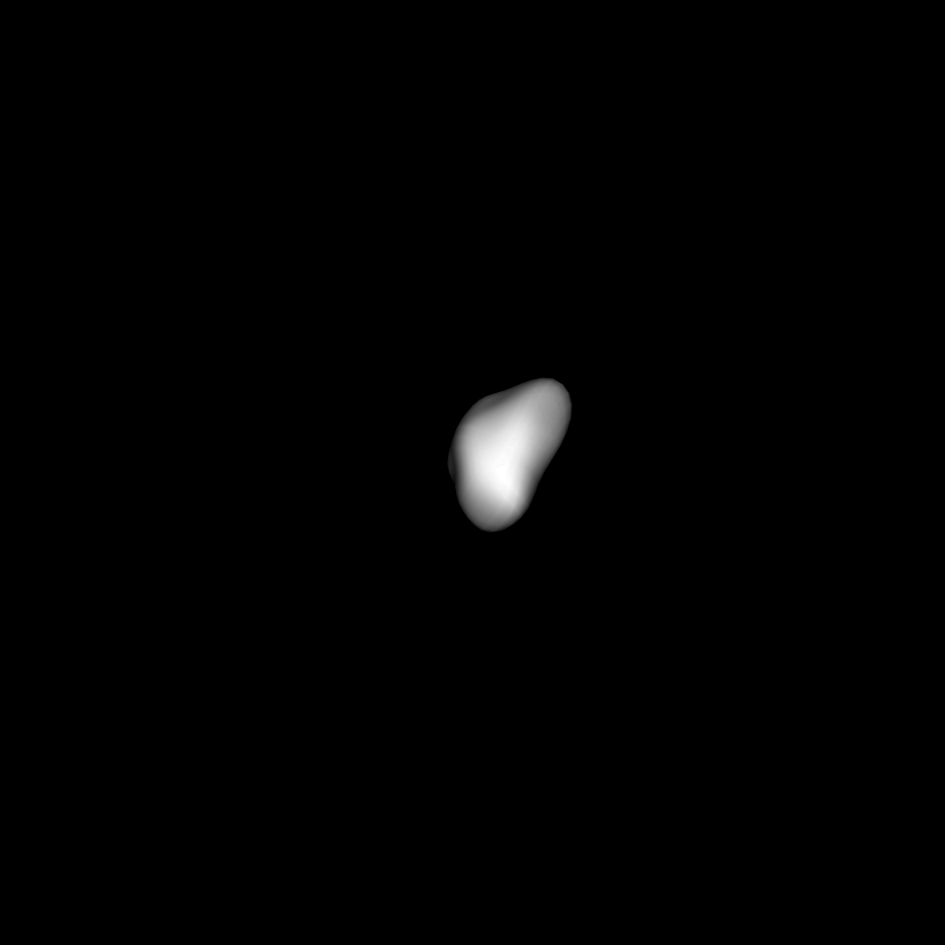}
     \end{subfigure}%
     \begin{subfigure}[b]{0.16\linewidth}
      \includegraphics[clip=true,trim=85 85 85 85,scale=0.66]{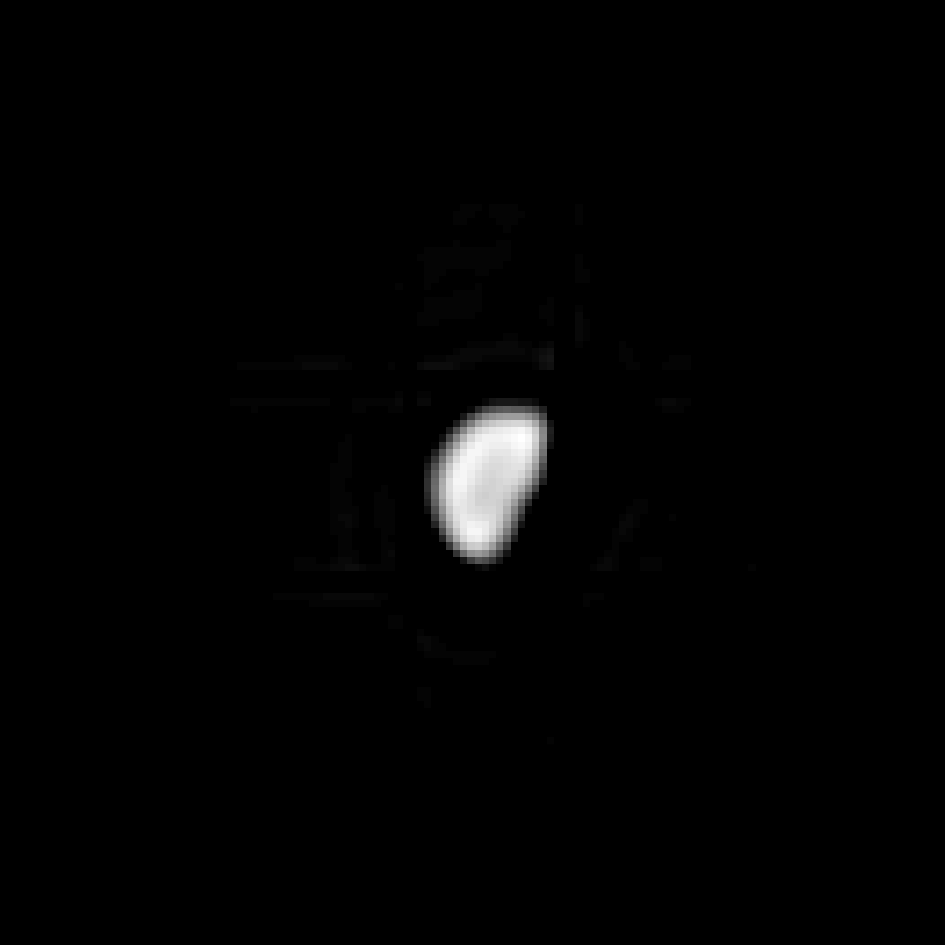} 
    \end{subfigure}%
     \begin{subfigure}[b]{0.16\linewidth}
     \includegraphics[clip=true,trim=90 90 80 80,scale=0.66]{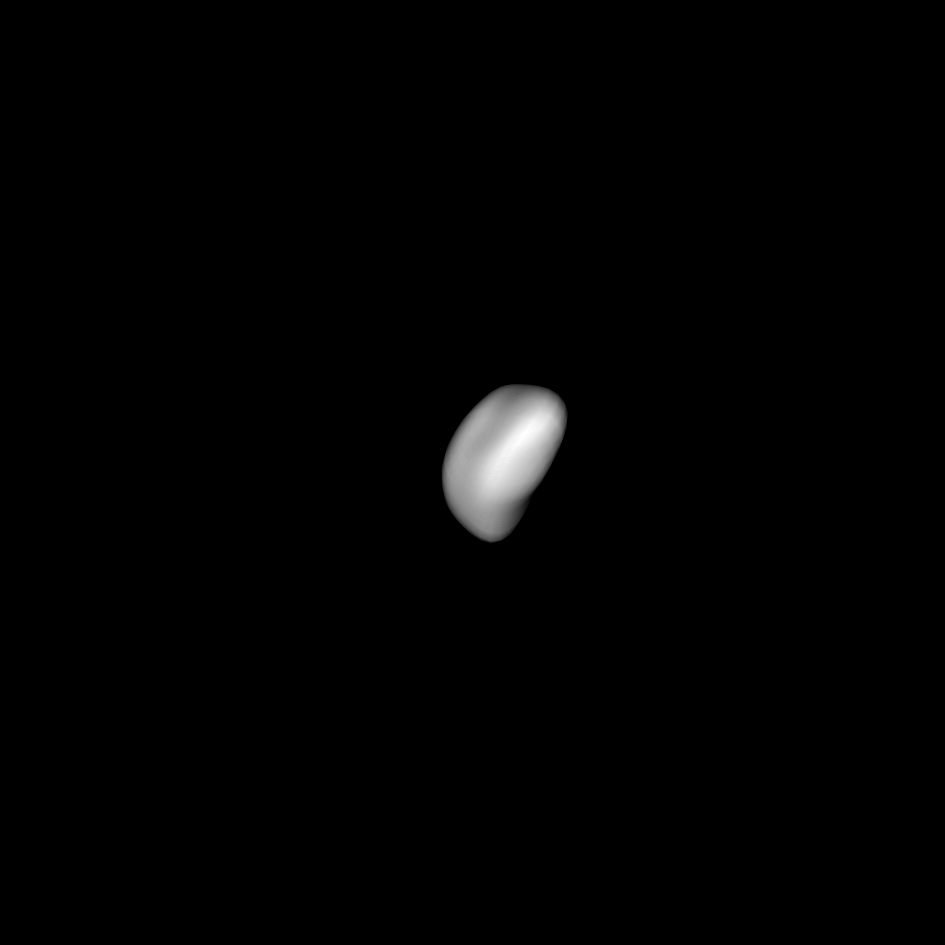}
     \end{subfigure}%
     
     \begin{subfigure}[b]{0.16\linewidth}
     \includegraphics[clip=true,trim=85 85 85 85,scale=0.66]{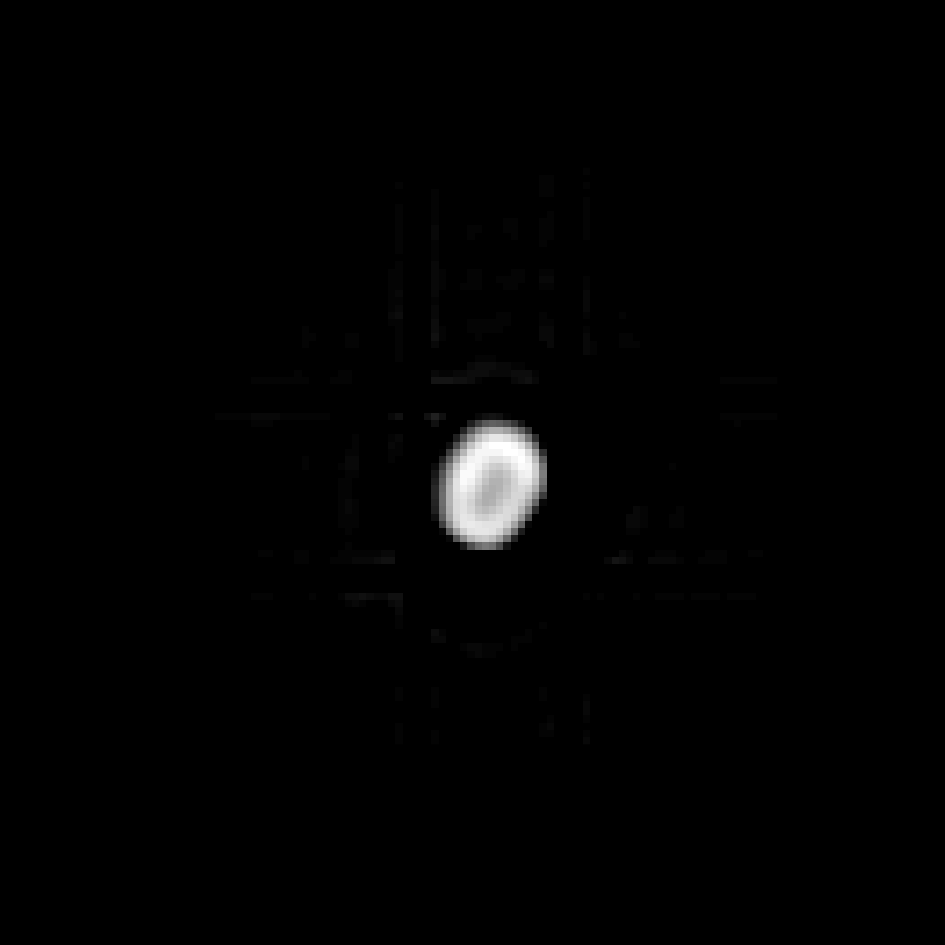}
     \end{subfigure}%
     \begin{subfigure}[b]{0.16\linewidth}
     \includegraphics[clip=true,trim=90 90 80 80,scale=0.66]{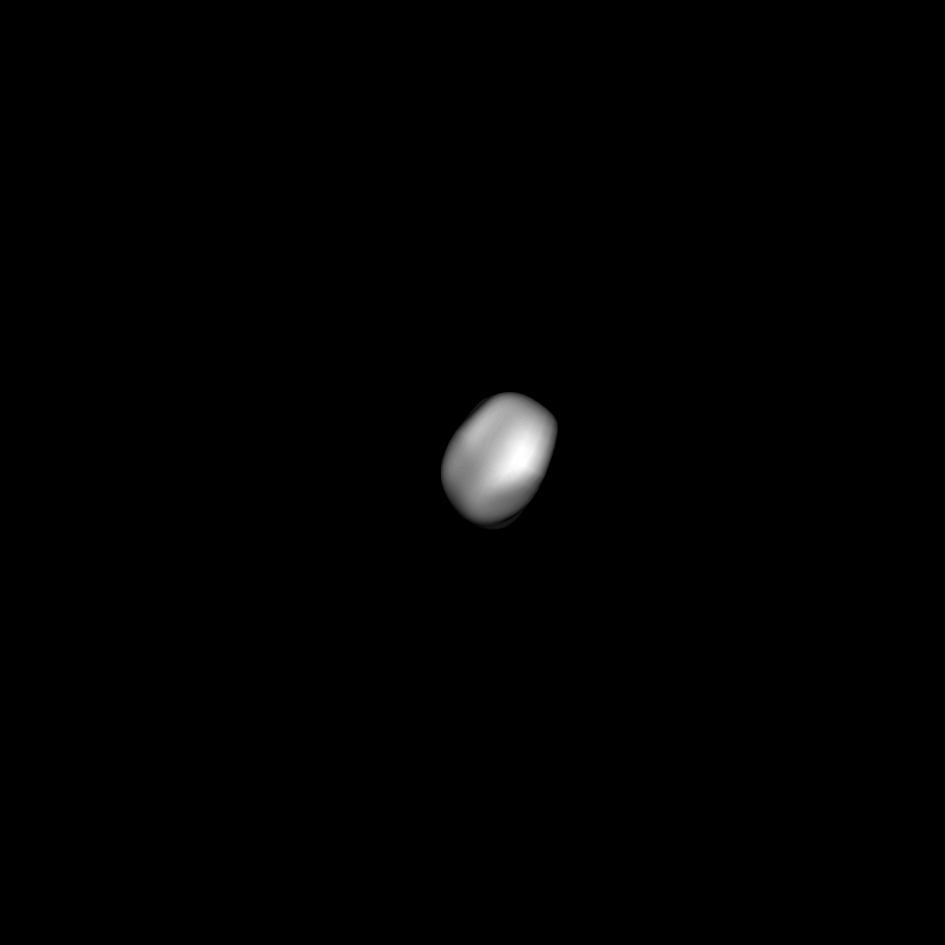}
     \end{subfigure}%
     \begin{subfigure}[b]{0.16\linewidth}
     \includegraphics[clip=true,trim=85 85 85 85,scale=0.66]{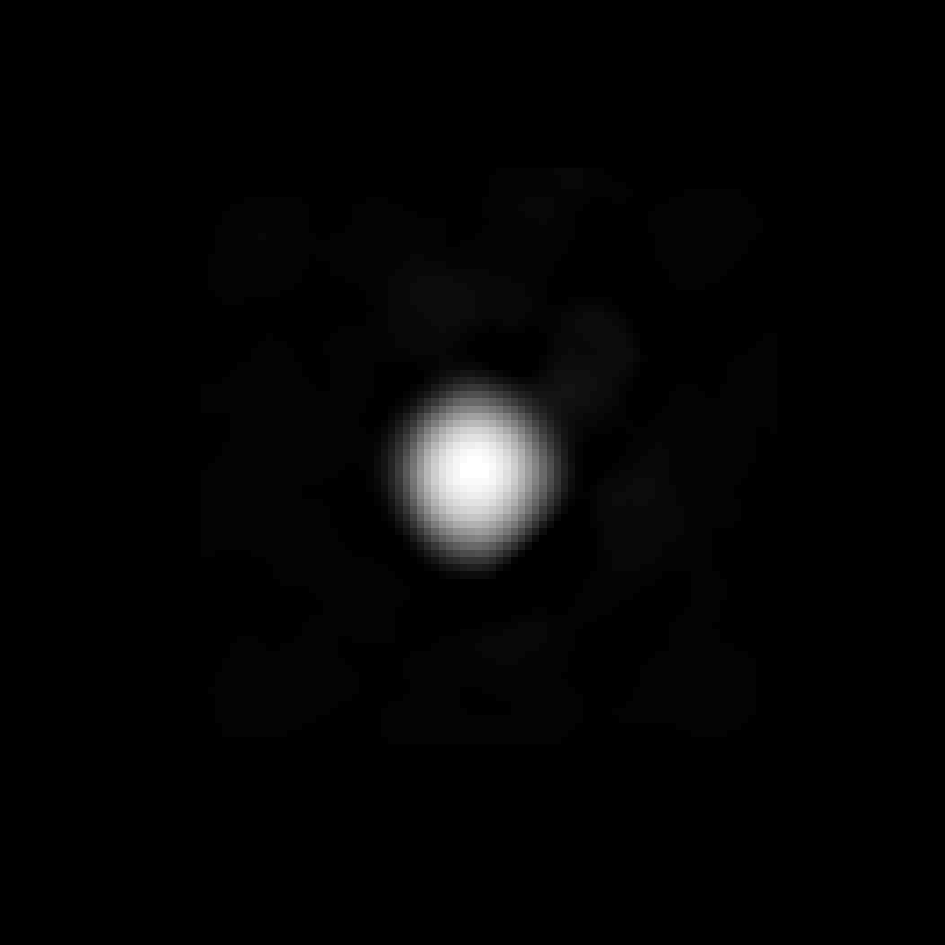}
     \end{subfigure}%
     \begin{subfigure}[b]{0.16\linewidth}
     \includegraphics[clip=true,trim=90 90 80 80,scale=0.66]{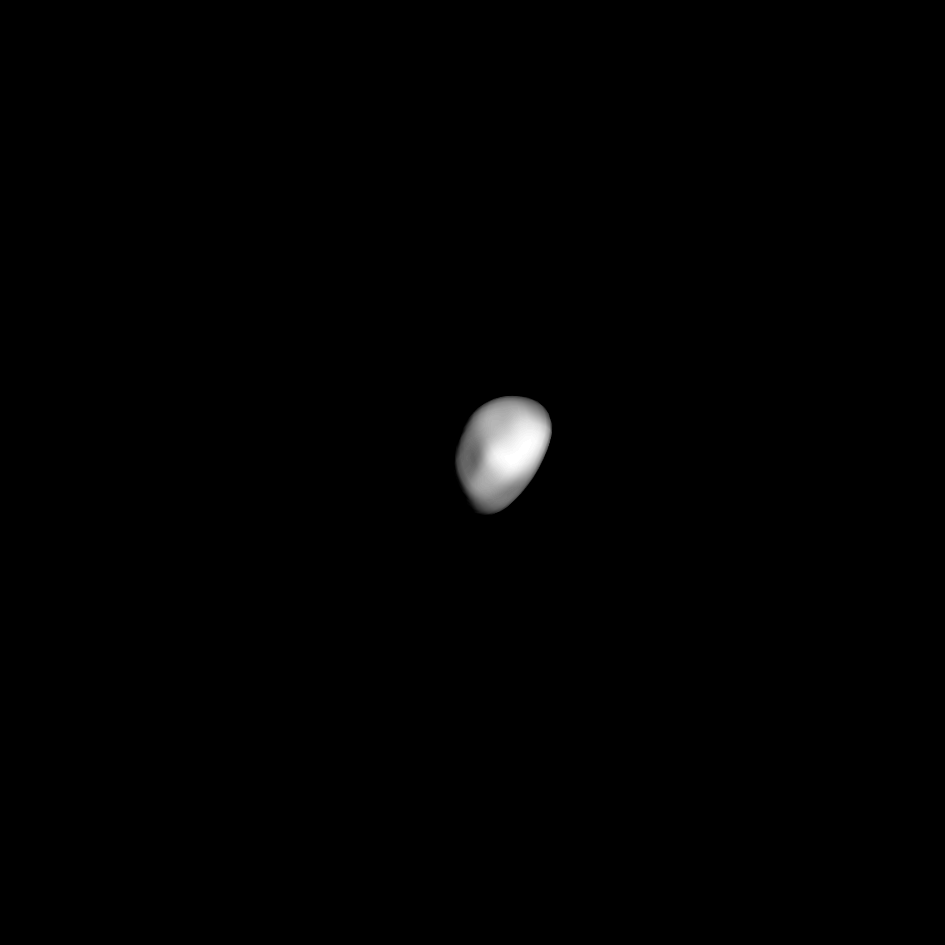}
     \end{subfigure}%
     \caption{\label{fig12}12 Victoria}
    \end{figure}
\begin{figure}[t]
     \begin{subfigure}[b]{0.16\linewidth}
      \includegraphics[clip=true,trim=90 85 80 85,scale=0.66]{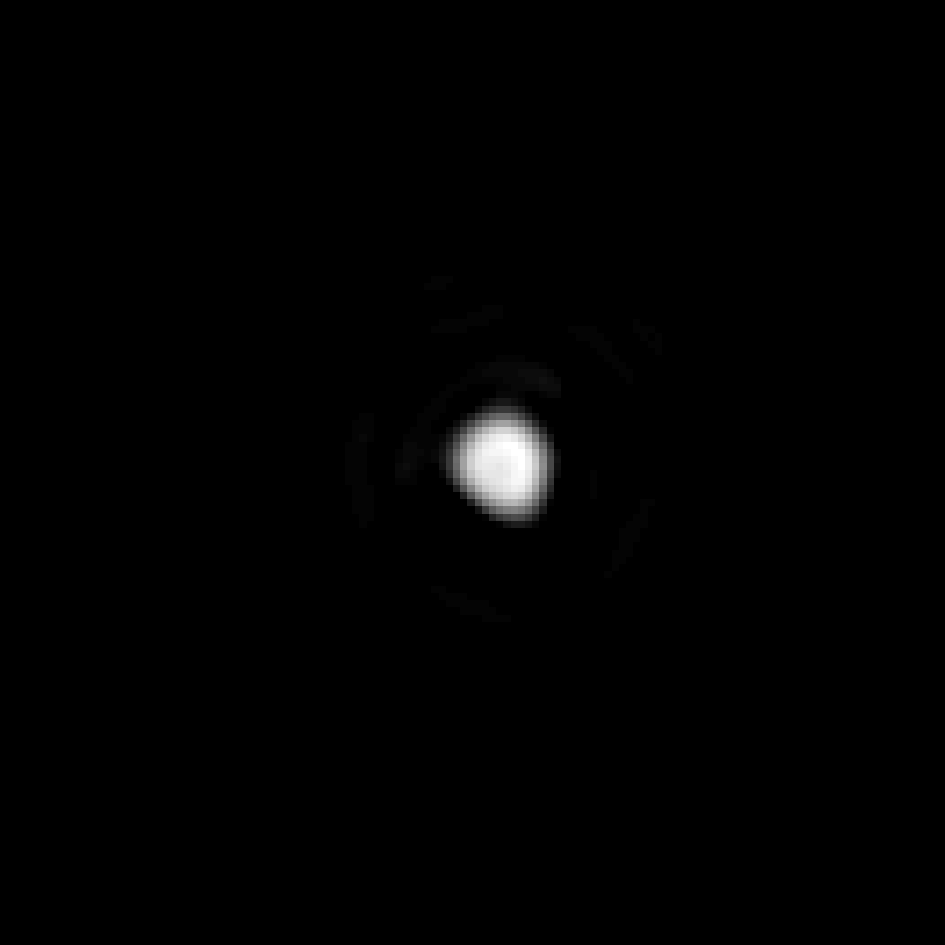} 
    \end{subfigure}%
     \begin{subfigure}[b]{0.16\linewidth}
     \includegraphics[clip=true,trim=90 90 80 80,scale=0.66]{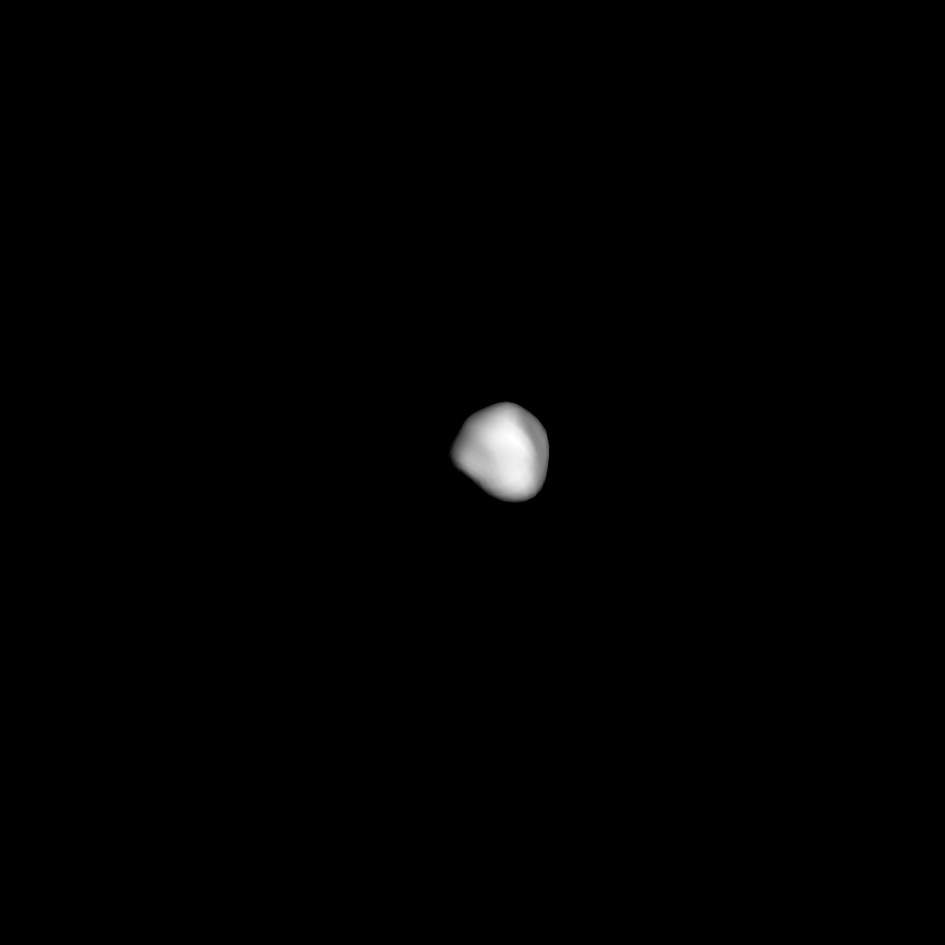}
     \end{subfigure}%
      \begin{subfigure}[b]{0.16\linewidth}
      \includegraphics[clip=true,trim=90 85 80 85,scale=0.66]{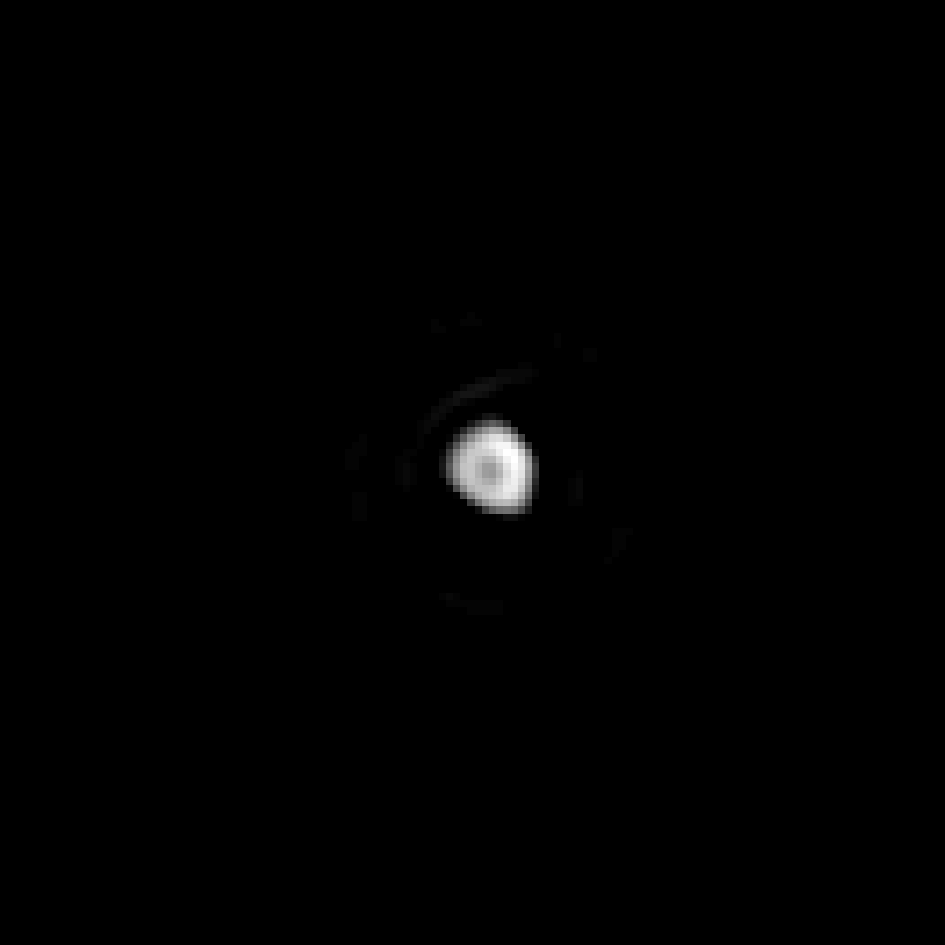}
    \end{subfigure}%
     \begin{subfigure}[b]{0.16\linewidth}
      \includegraphics[clip=true,trim=90 90 80 80,scale=0.66]{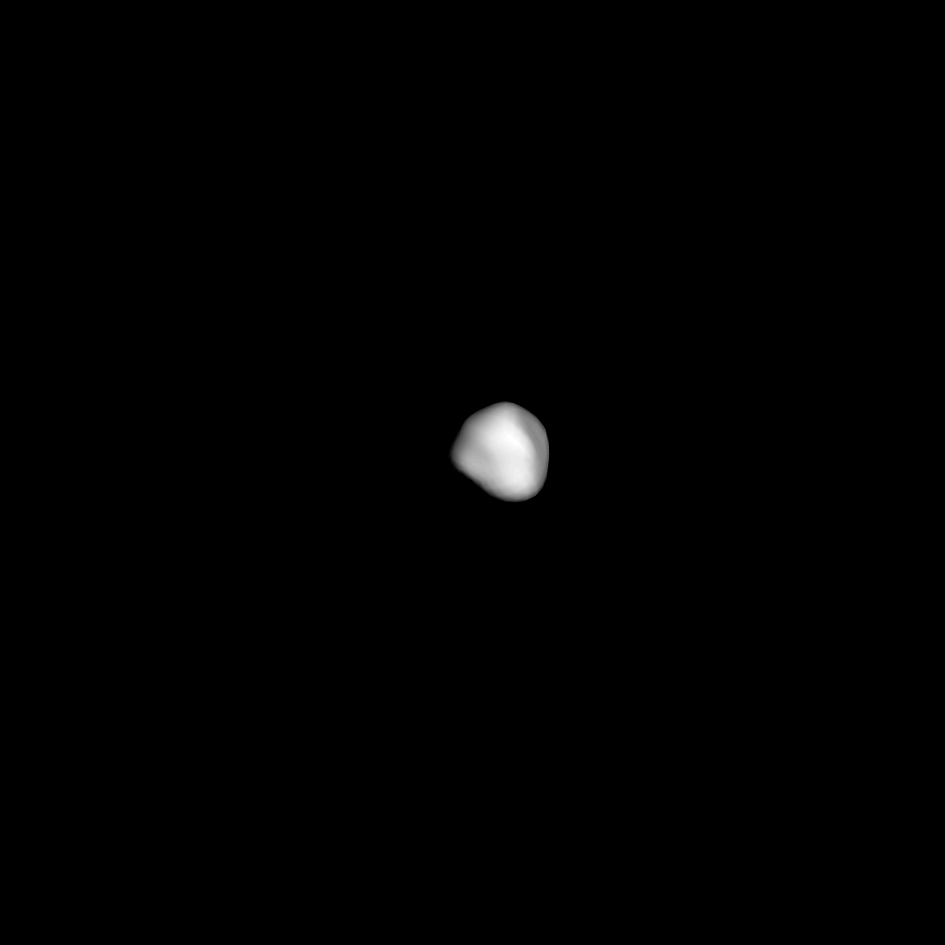}
    \end{subfigure}%
    \begin{subfigure}[b]{0.16\linewidth}
      \includegraphics[clip=true,trim=90 95 80 75,scale=0.66]{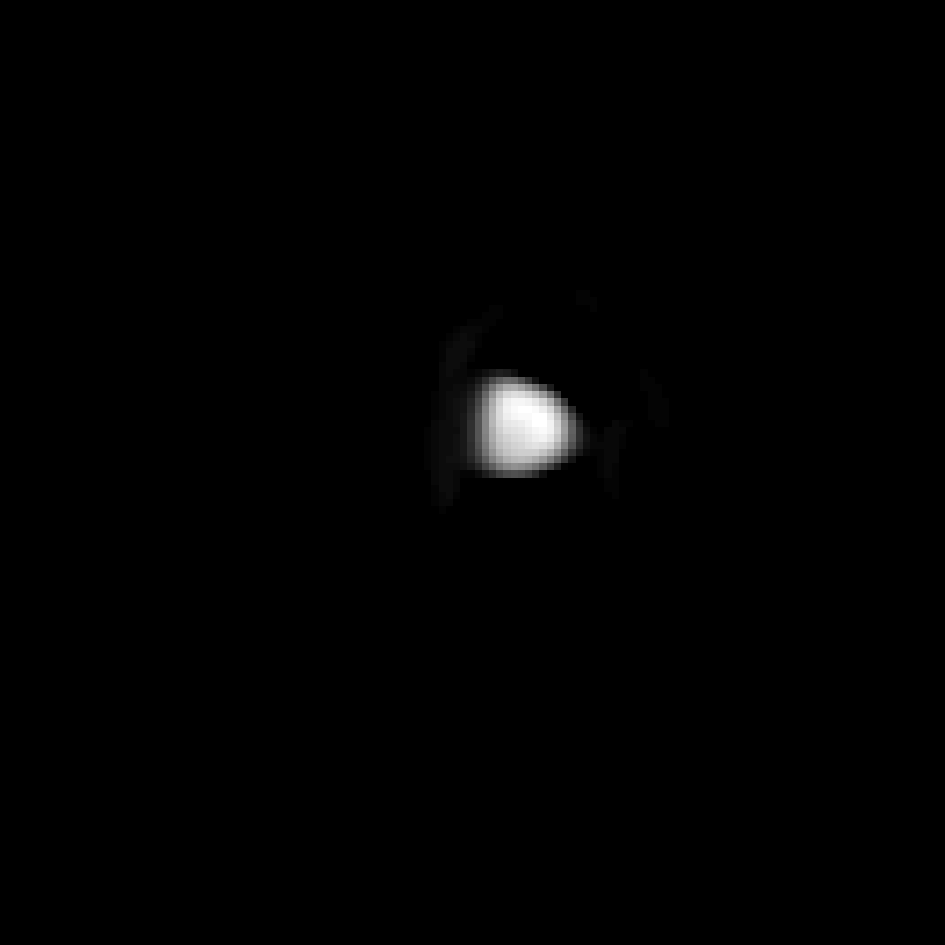} 
    \end{subfigure}%
     \begin{subfigure}[b]{0.16\linewidth}
     \includegraphics[clip=true,trim=90 90 80 80,scale=0.66]{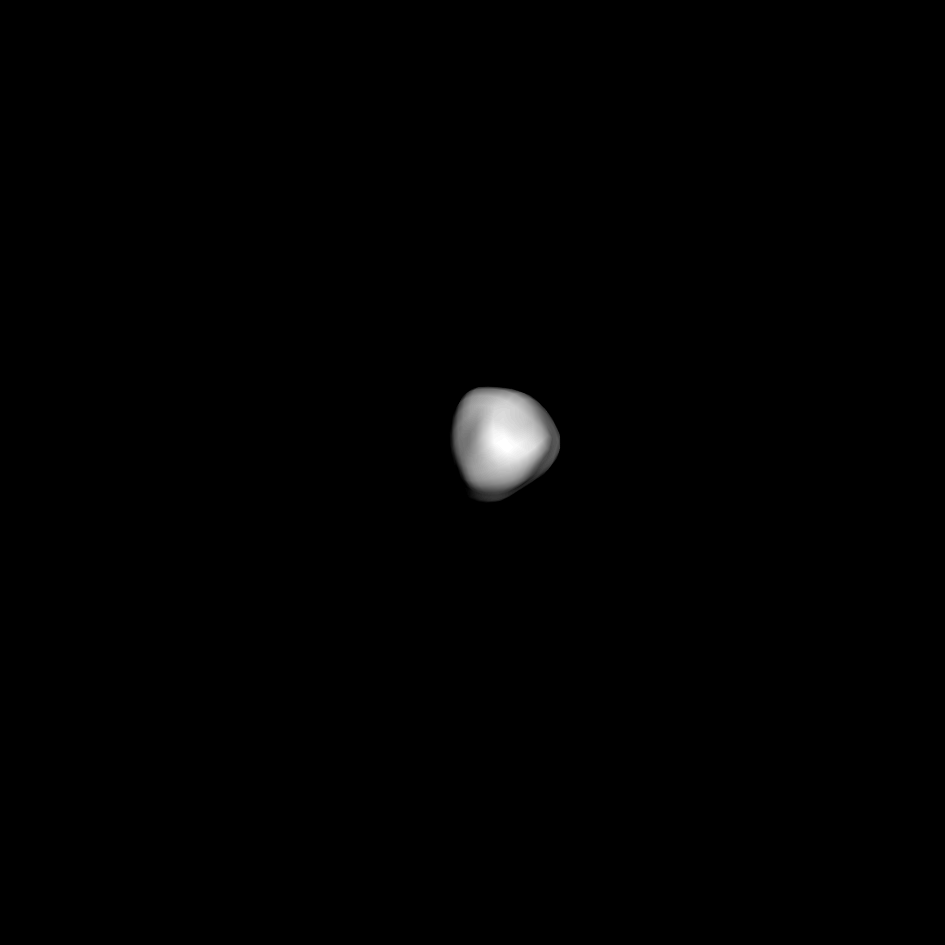}
     \end{subfigure}%
     \caption{\label{fig14}14 Irene}
\end{figure}
\begin{figure}[t]
     \begin{subfigure}[b]{0.16\linewidth}
      \includegraphics[clip=true,trim=65 65 65 65,scale=0.39]{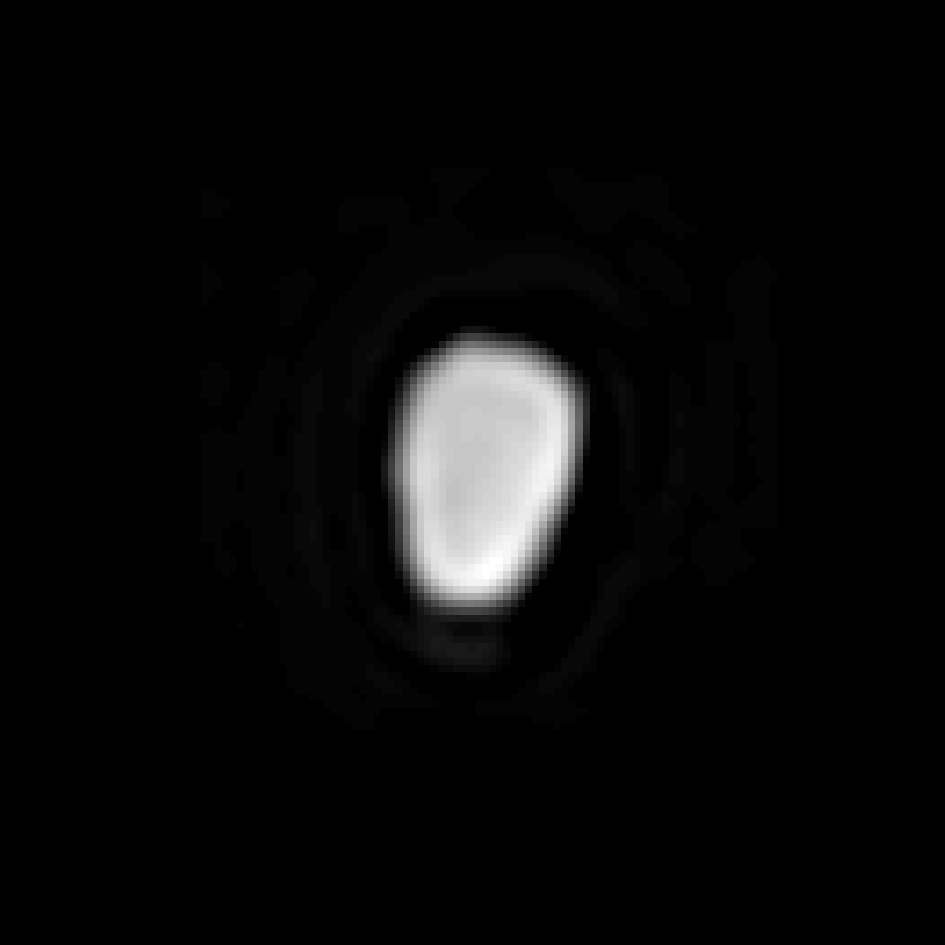} 
    \end{subfigure}%
     \begin{subfigure}[b]{0.16\linewidth}
     \includegraphics[clip=true,trim=65 65 65 65,scale=0.39]{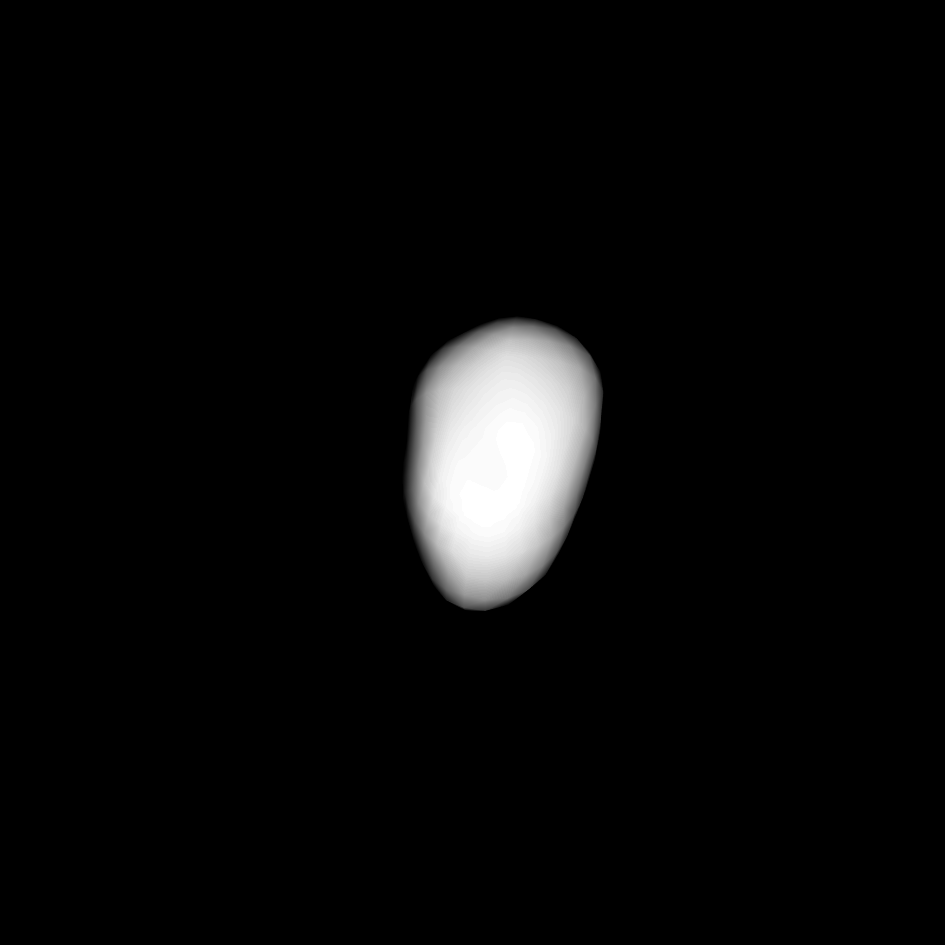}
     \end{subfigure}%
      \begin{subfigure}[b]{0.16\linewidth}
      \includegraphics[clip=true,trim=65 65 65 65,scale=0.39]{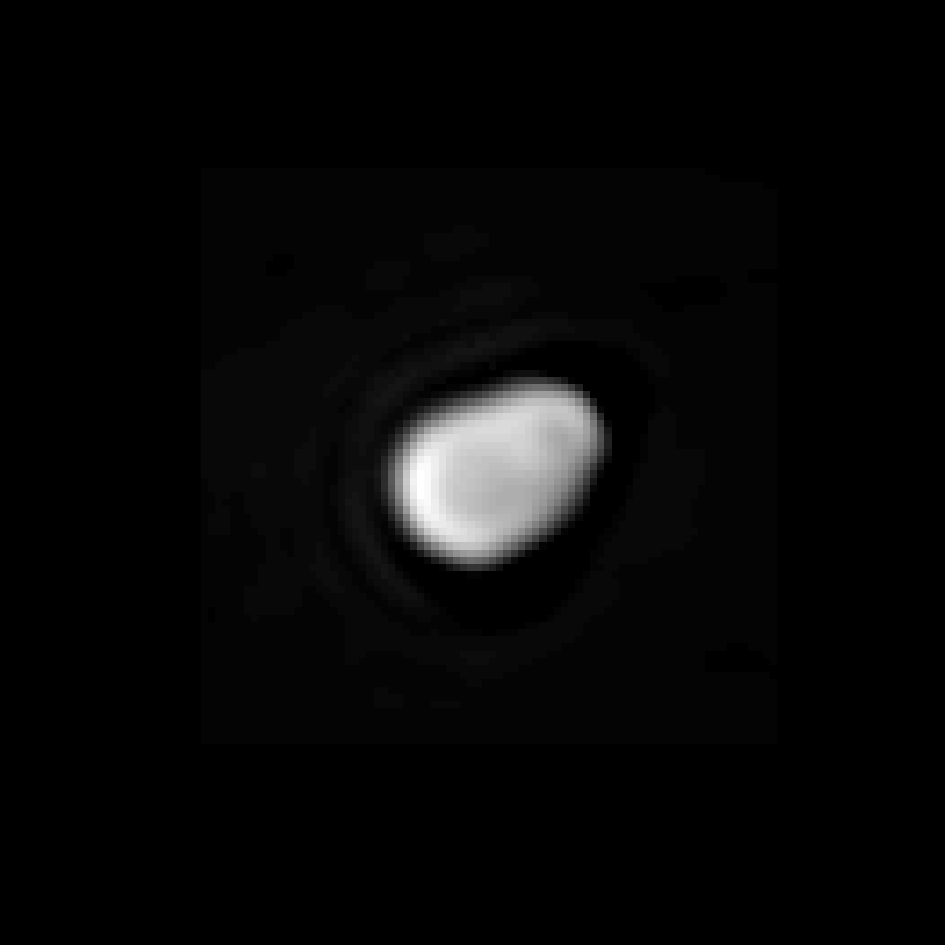}
    \end{subfigure}%
     \begin{subfigure}[b]{0.16\linewidth}
      \includegraphics[clip=true,trim=65 65 65 65,scale=0.39]{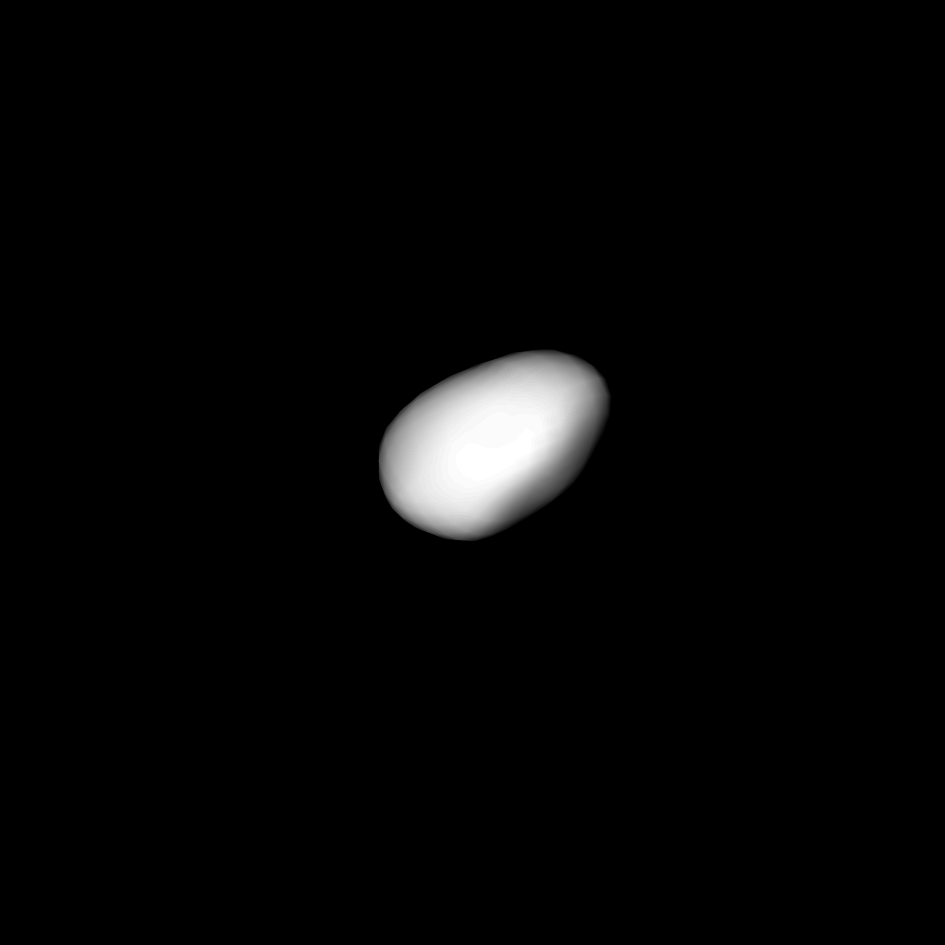}
    \end{subfigure}%
    \begin{subfigure}[b]{0.16\linewidth}
      \includegraphics[clip=true,trim=65 65 65 65,scale=0.39]{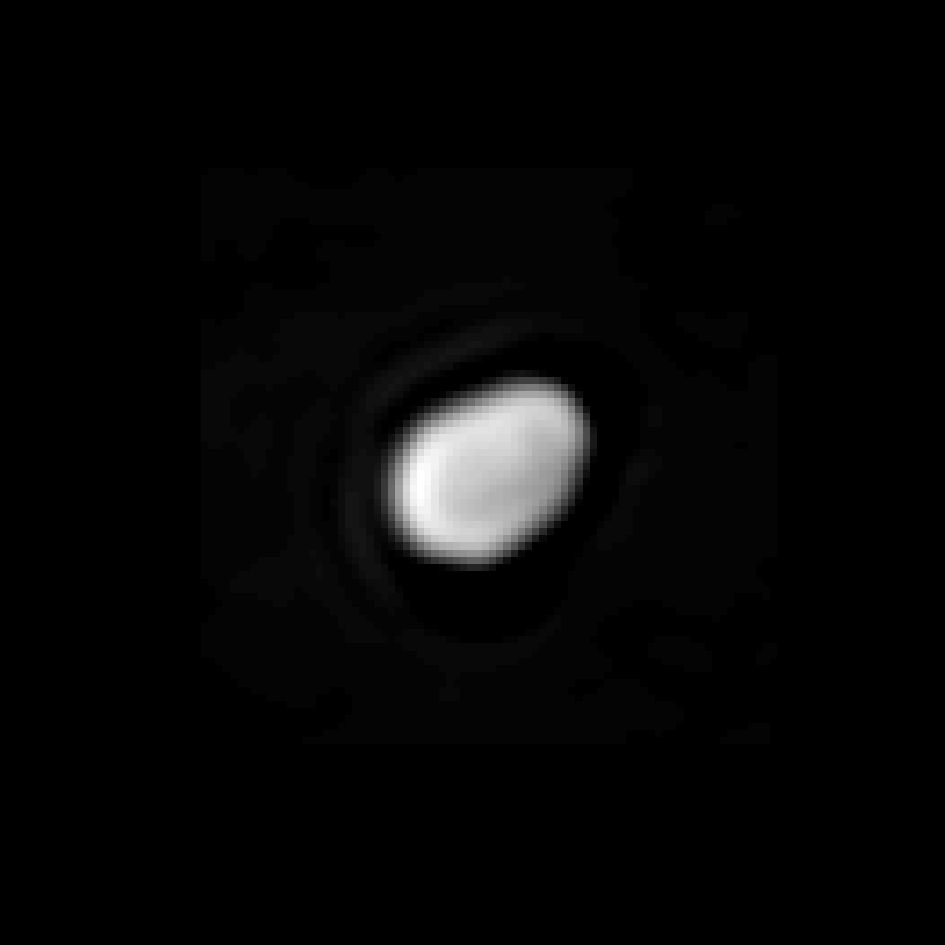} 
    \end{subfigure}%
     \begin{subfigure}[b]{0.16\linewidth}
     \includegraphics[clip=true,trim=65 65 65 65,scale=0.39]{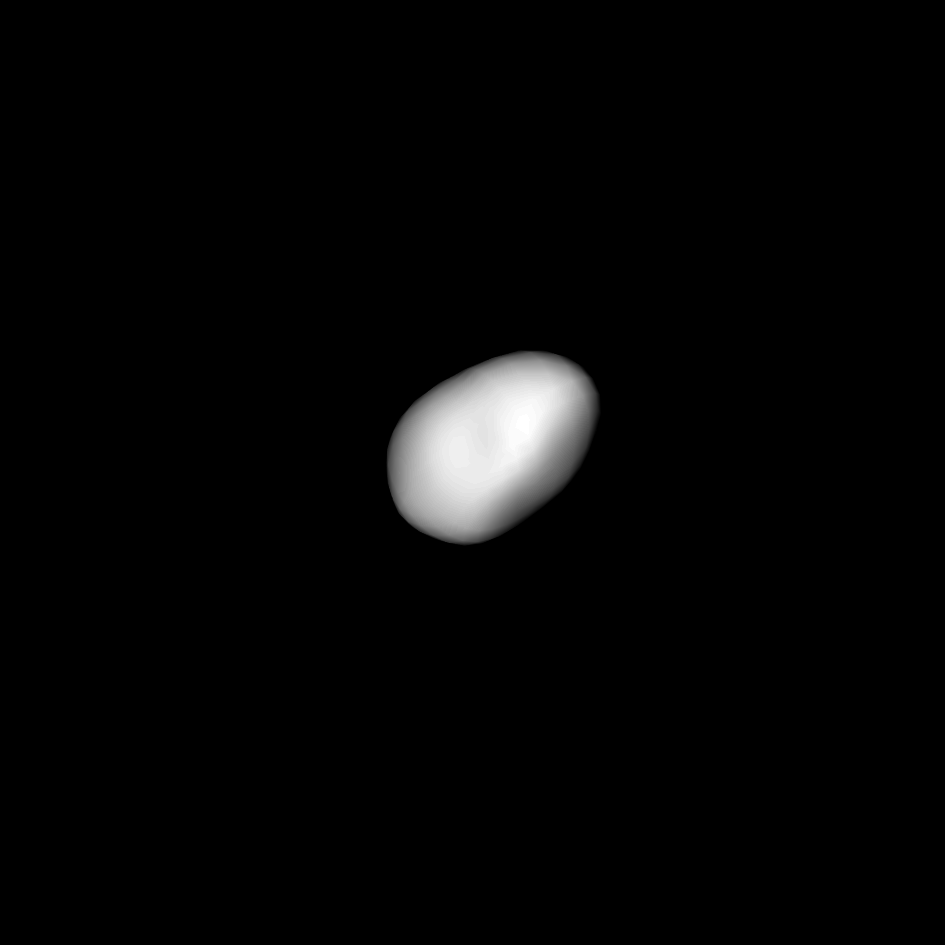}
     \end{subfigure}%
     
      \begin{subfigure}[b]{0.16\linewidth}
      \includegraphics[clip=true,trim=65 65 65 65,scale=0.39]{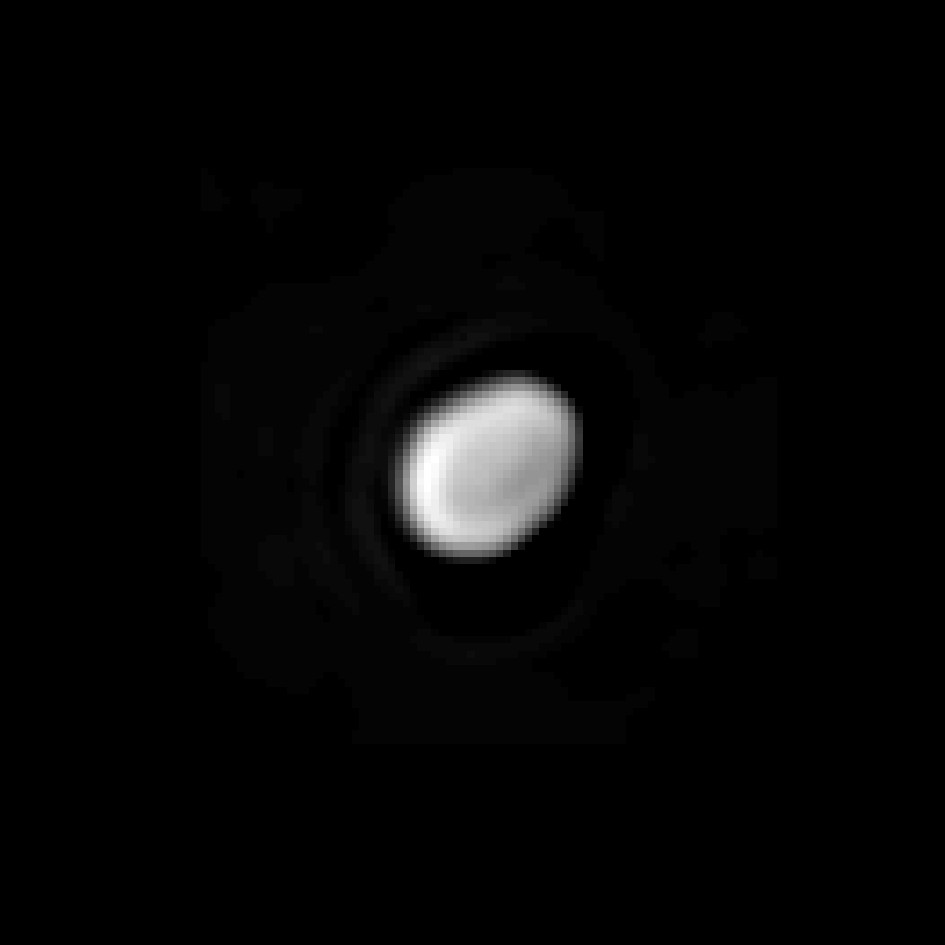} 
    \end{subfigure}%
     \begin{subfigure}[b]{0.16\linewidth}
     \includegraphics[clip=true,trim=65 65 65 65,scale=0.39]{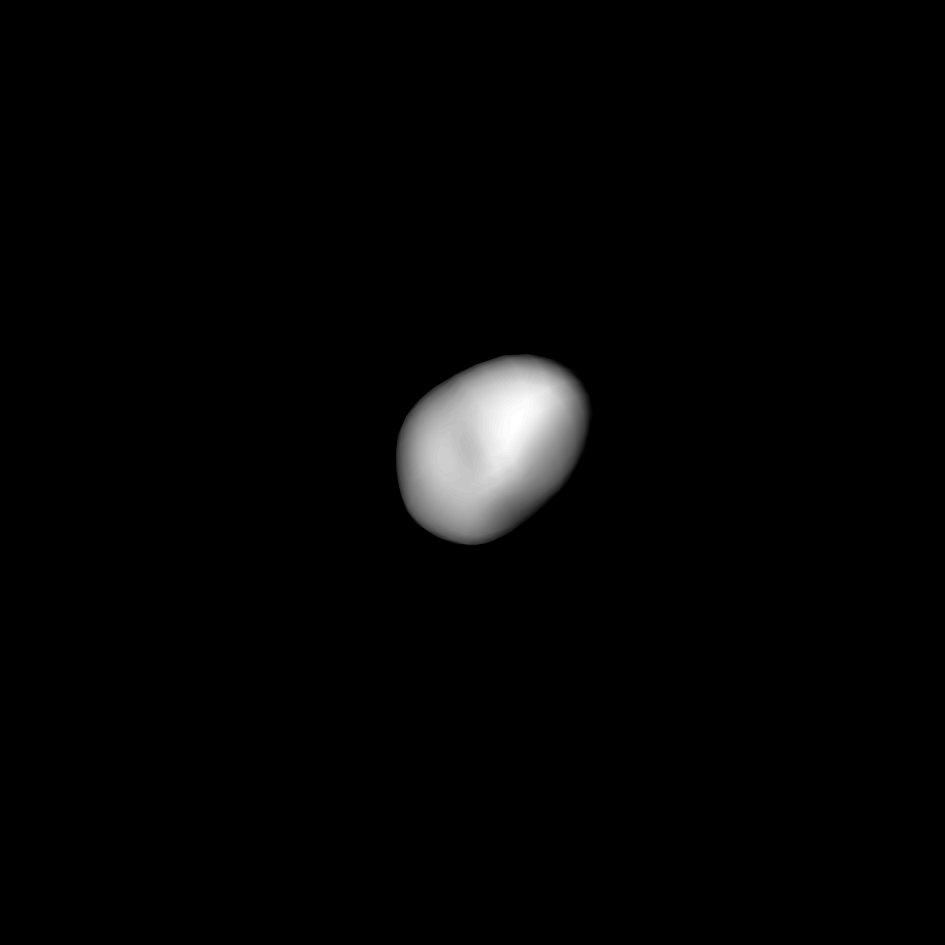}
     \end{subfigure}%
     \begin{subfigure}[b]{0.16\linewidth}
      \includegraphics[clip=true,trim=65 65 65 65,scale=0.39]{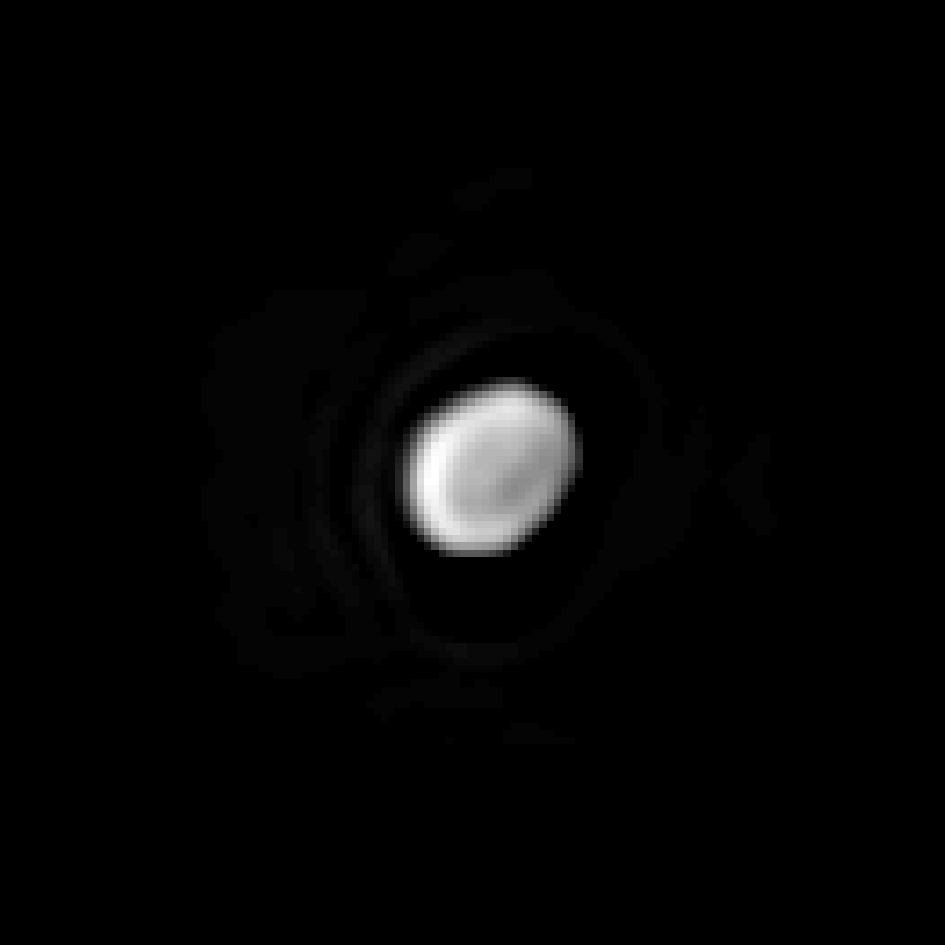} 
    \end{subfigure}%
     \begin{subfigure}[b]{0.16\linewidth}
     \includegraphics[clip=true,trim=65 65 65 65,scale=0.39]{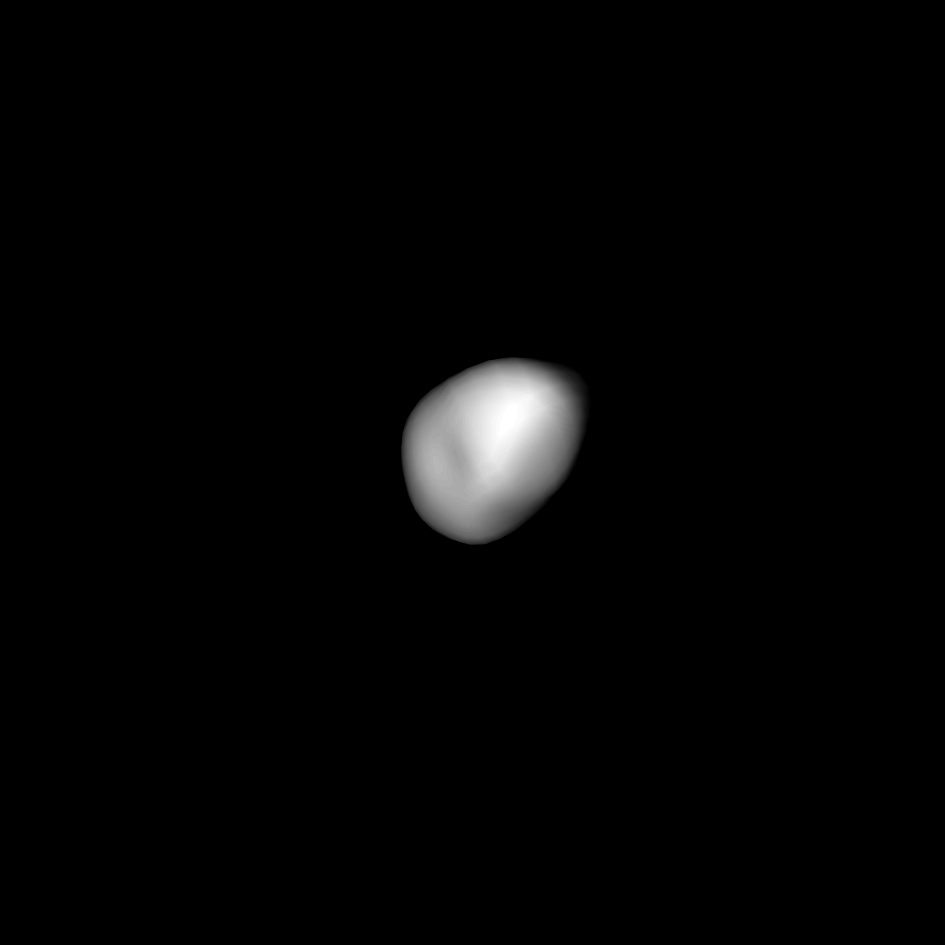}
     \end{subfigure}%
     \begin{subfigure}[b]{0.16\linewidth}
      \includegraphics[clip=true,trim=65 65 65 65,scale=0.39]{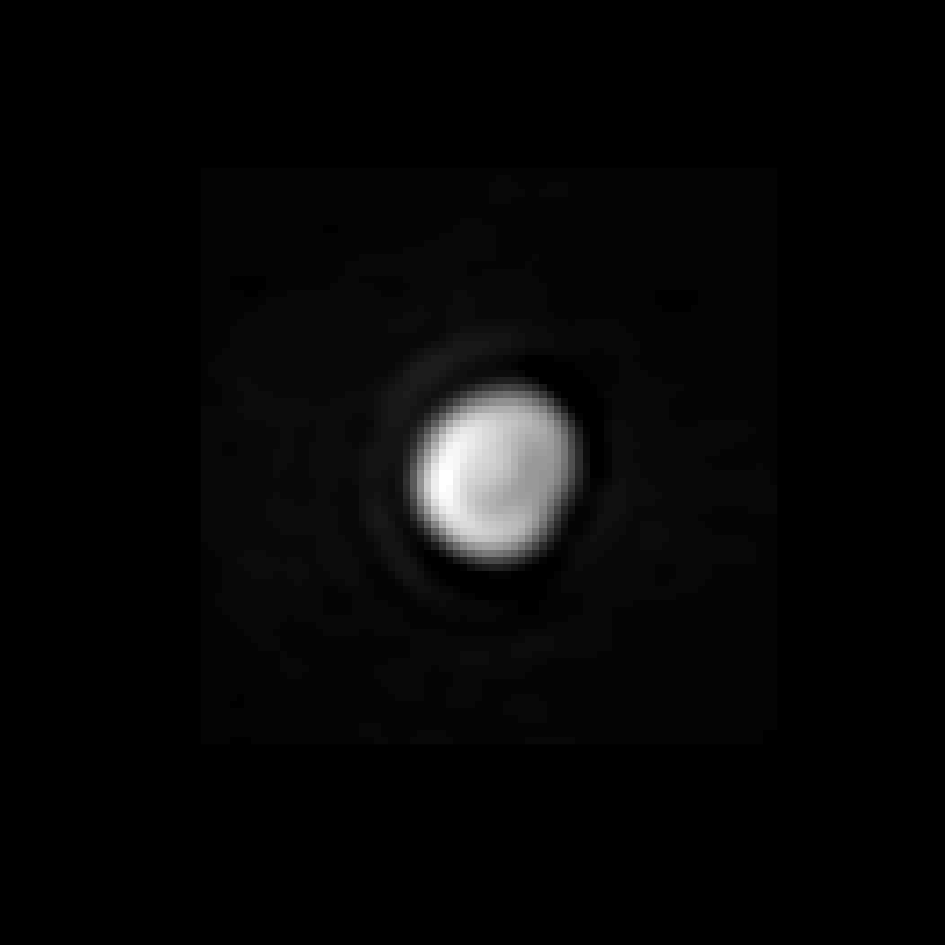} 
    \end{subfigure}%
     \begin{subfigure}[b]{0.16\linewidth}
     \includegraphics[clip=true,trim=65 65 65 65,scale=0.39]{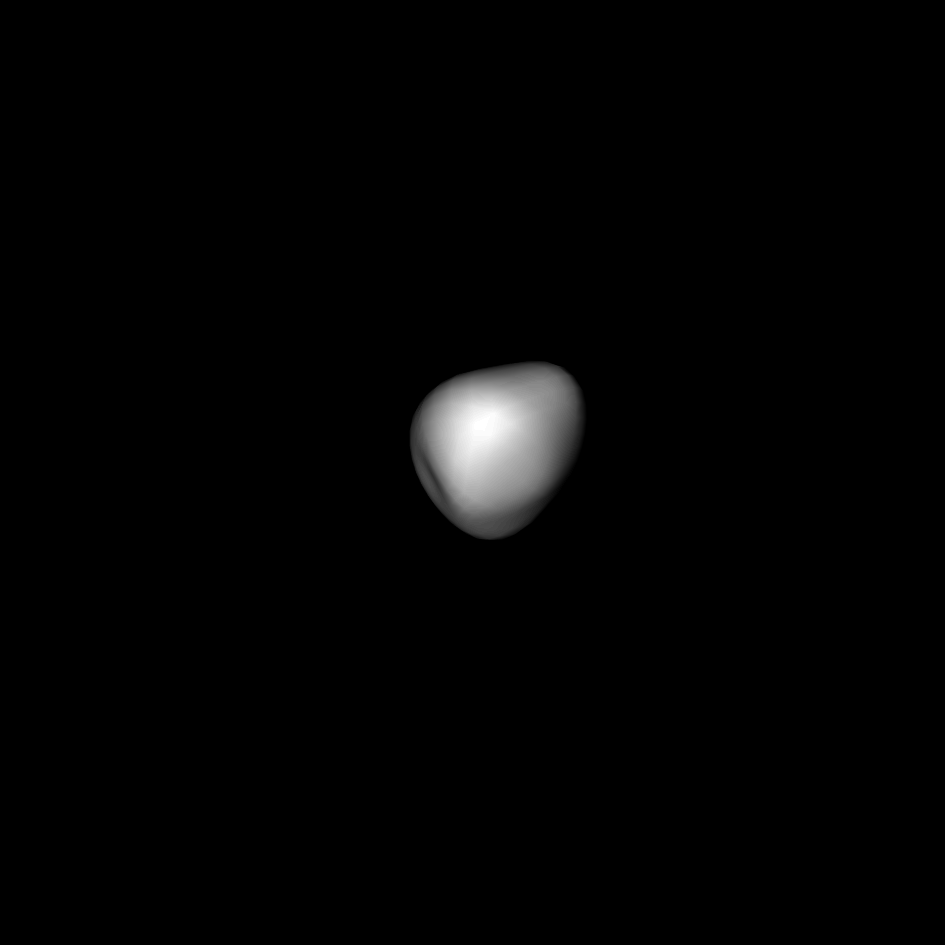}
     \end{subfigure}%
     
     \begin{subfigure}[b]{0.16\linewidth}
     \includegraphics[clip=true,trim=65 65 65 65,scale=0.39]{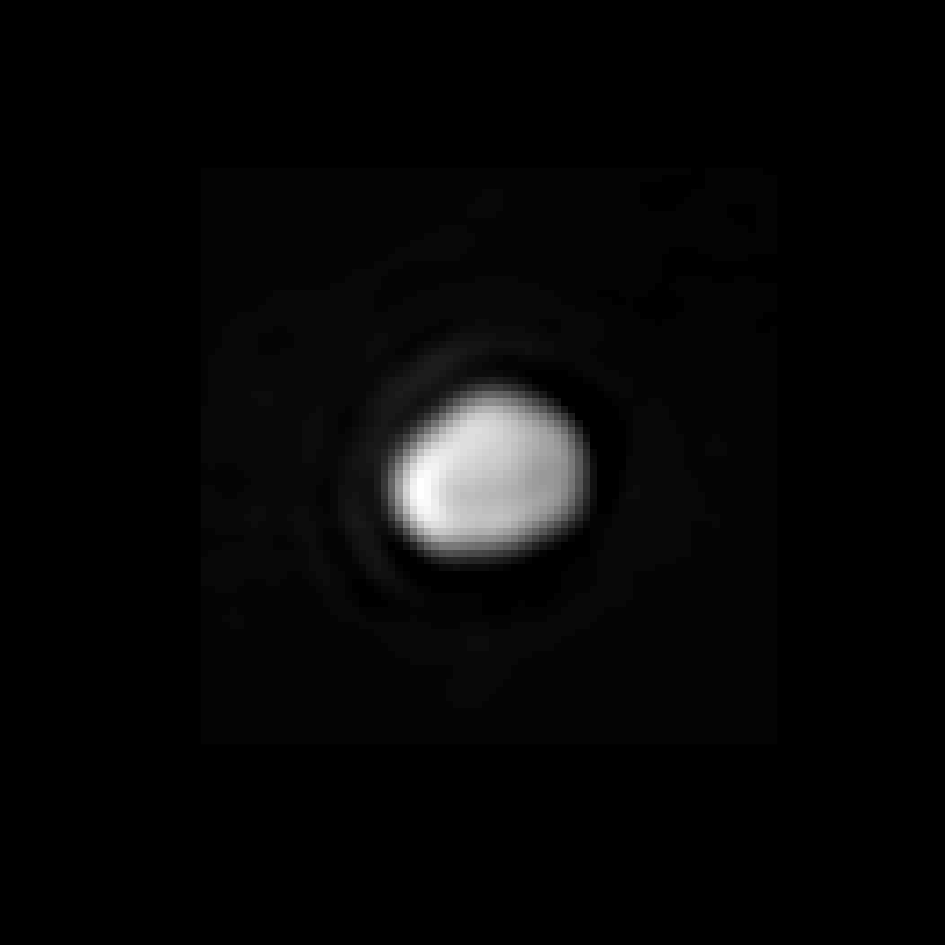}
     \end{subfigure}%
     \begin{subfigure}[b]{0.16\linewidth}
     \includegraphics[clip=true,trim=65 65 65 65,scale=0.39]{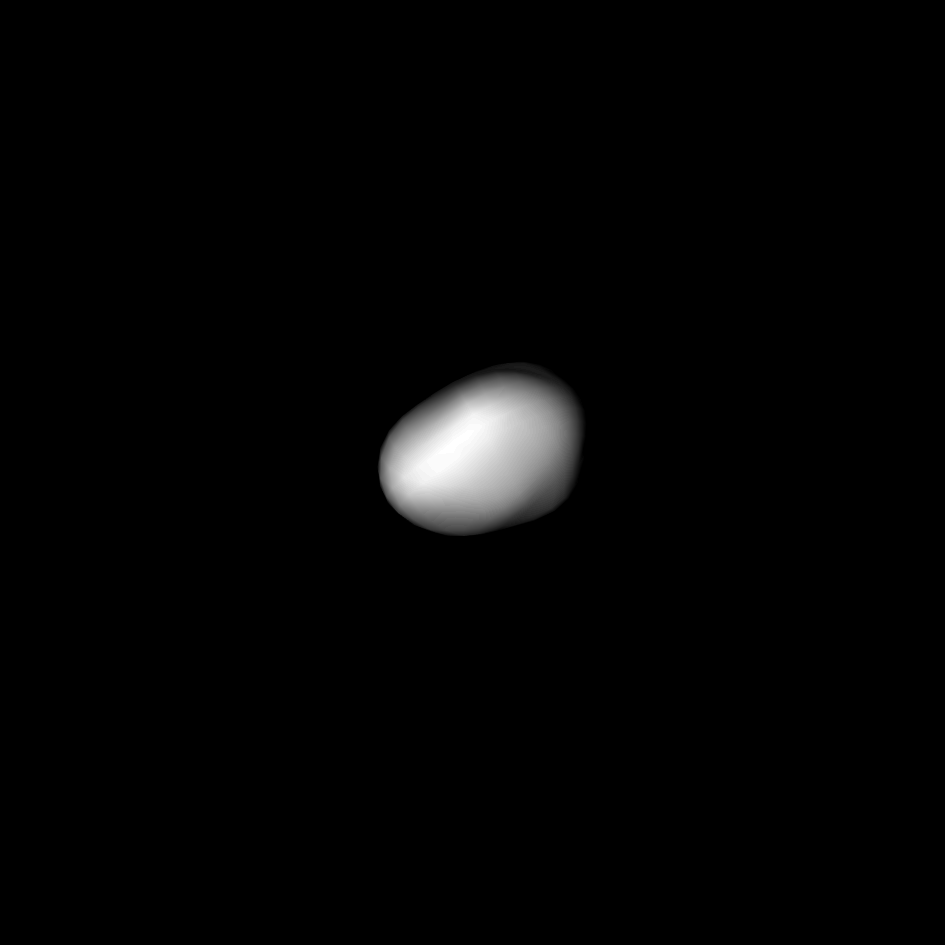}
     \end{subfigure}%
     \begin{subfigure}[b]{0.16\linewidth}
     \includegraphics[clip=true,trim=65 65 65 65,scale=0.39]{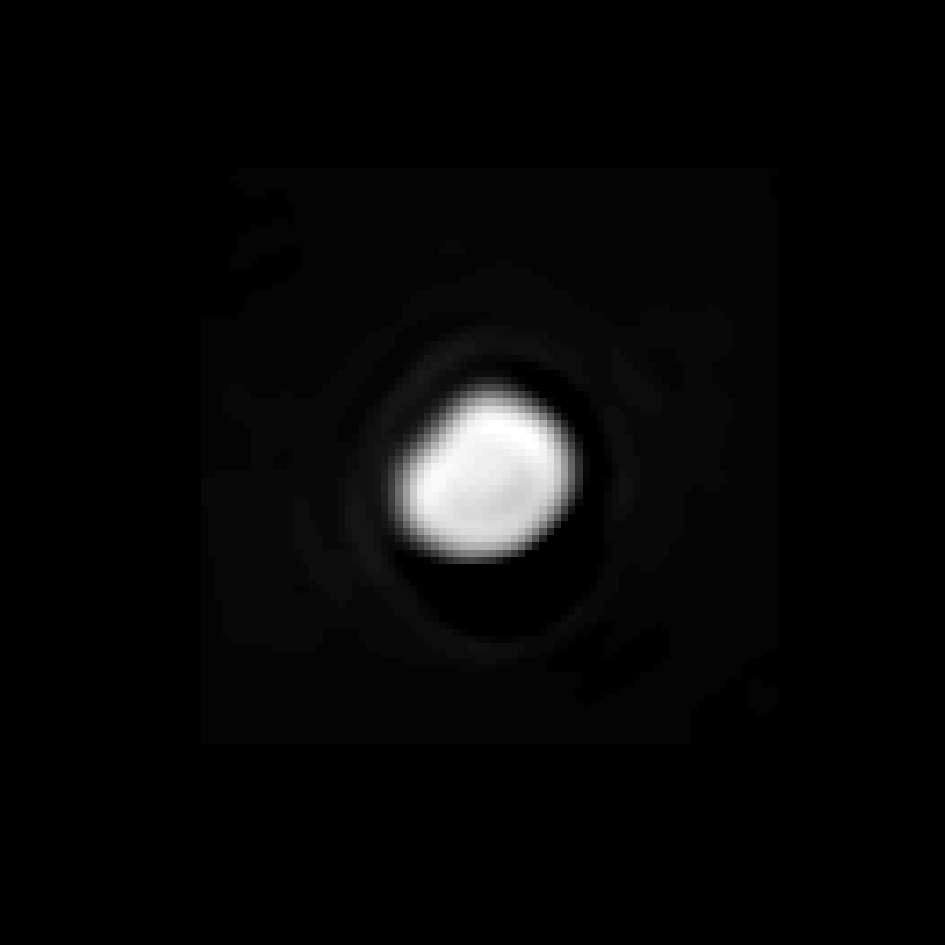}
     \end{subfigure}%
     \begin{subfigure}[b]{0.16\linewidth}
     \includegraphics[clip=true,trim=65 65 65 65,scale=0.39]{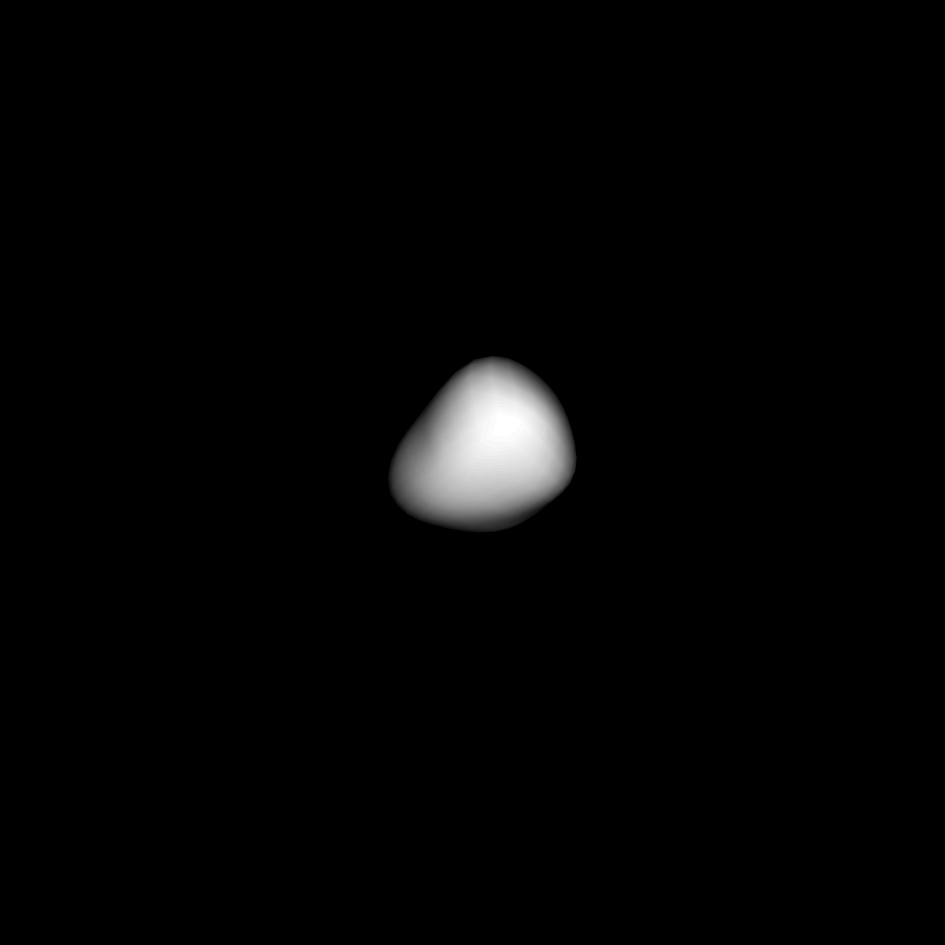}
     \end{subfigure}%
     \caption{\label{fig15}15 Eunomia}
    \end{figure}

\begin{figure}[t]
     \begin{subfigure}[b]{0.16\linewidth}
      \includegraphics[clip=true,trim=85 80 85 90,scale=0.66]{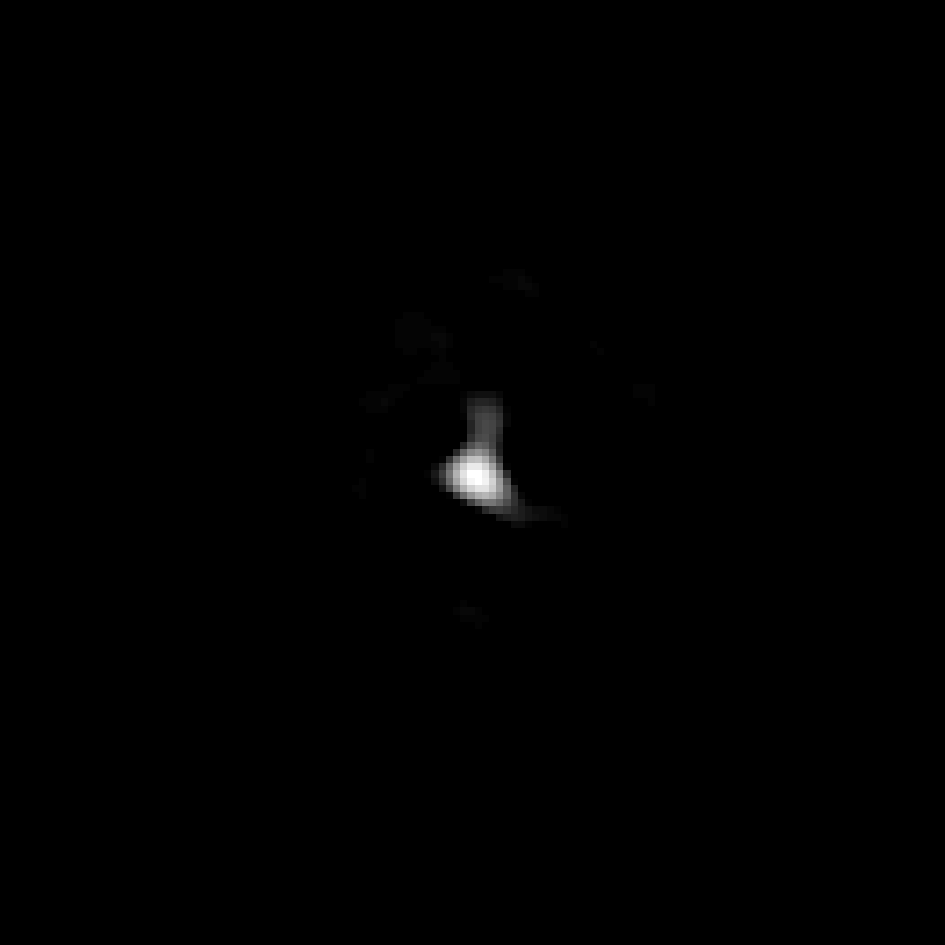} 
    \end{subfigure}%
     \begin{subfigure}[b]{0.16\linewidth}
     \includegraphics[clip=true,trim=90 90 80 80,scale=0.66]{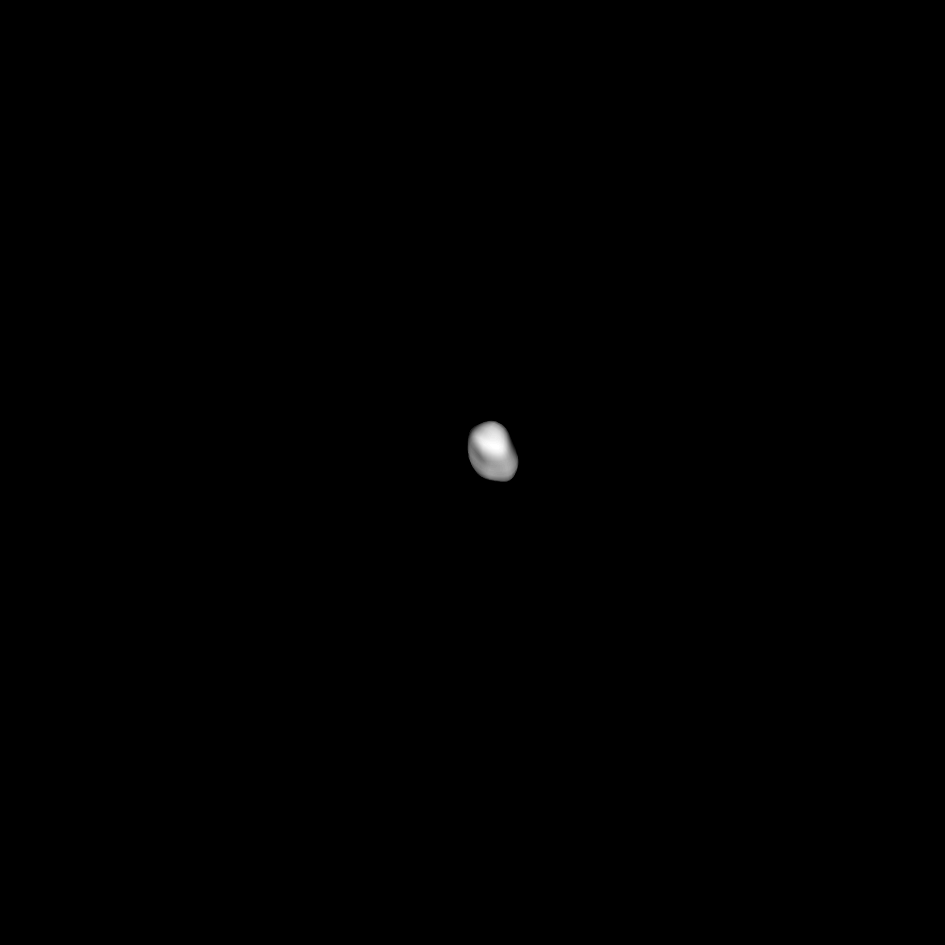}
     \end{subfigure}%
     \begin{subfigure}[b]{0.16\linewidth}
      \includegraphics[clip=true,trim=90 90 80 80,scale=0.66]{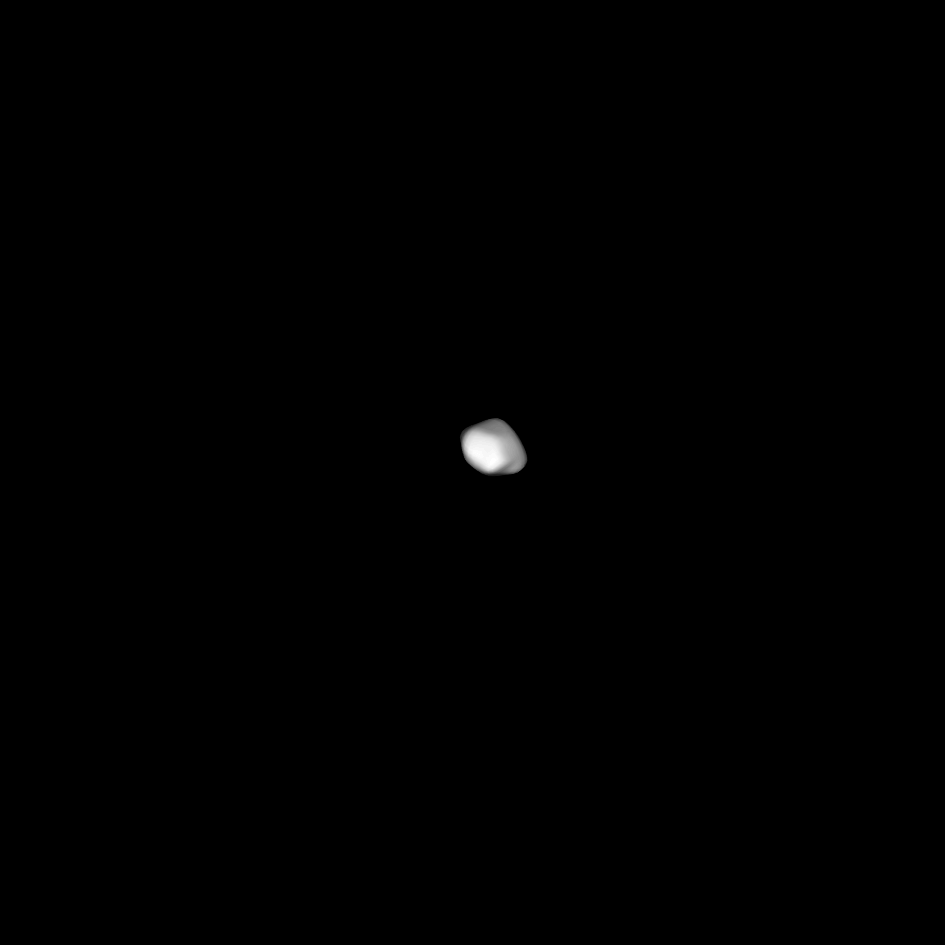}
    \end{subfigure}%
      \begin{subfigure}[b]{0.16\linewidth}
      \includegraphics[clip=true,trim=85 80 85 90,scale=0.66]{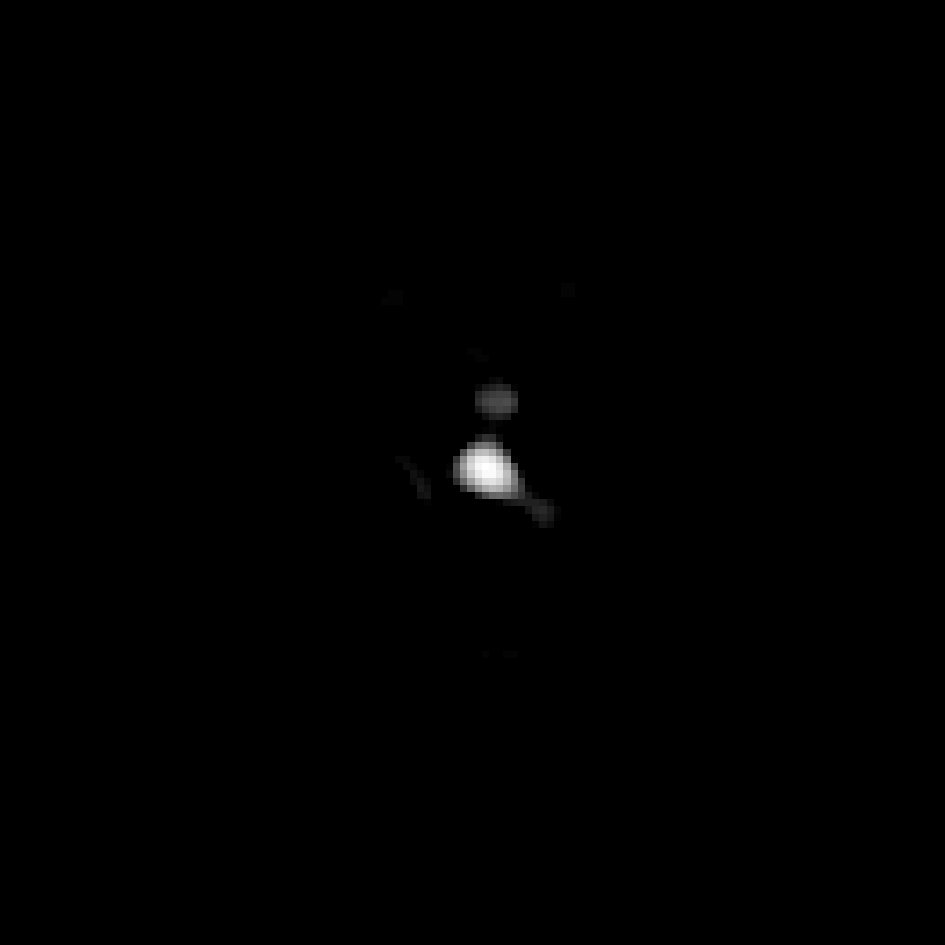}
    \end{subfigure}%
     \begin{subfigure}[b]{0.16\linewidth}
      \includegraphics[clip=true,trim=90 90 80 80,scale=0.66]{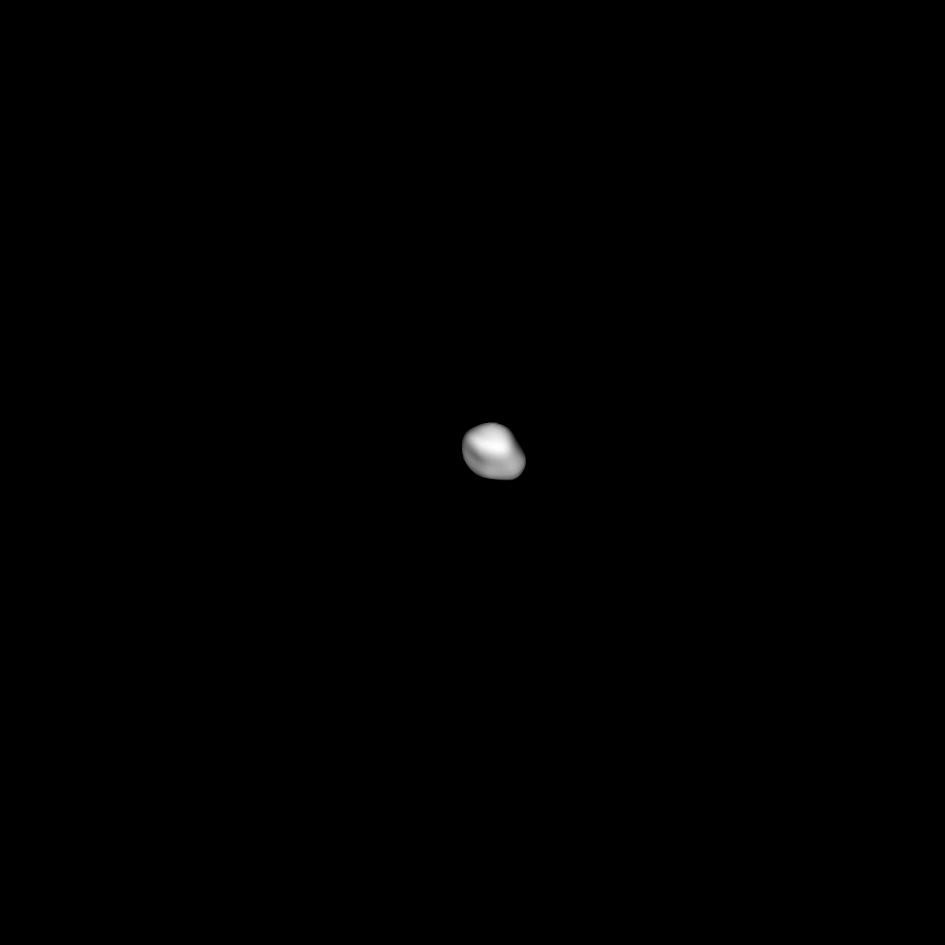}
    \end{subfigure}%
    \begin{subfigure}[b]{0.16\linewidth}
      \includegraphics[clip=true,trim=90 90 80 80,scale=0.66]{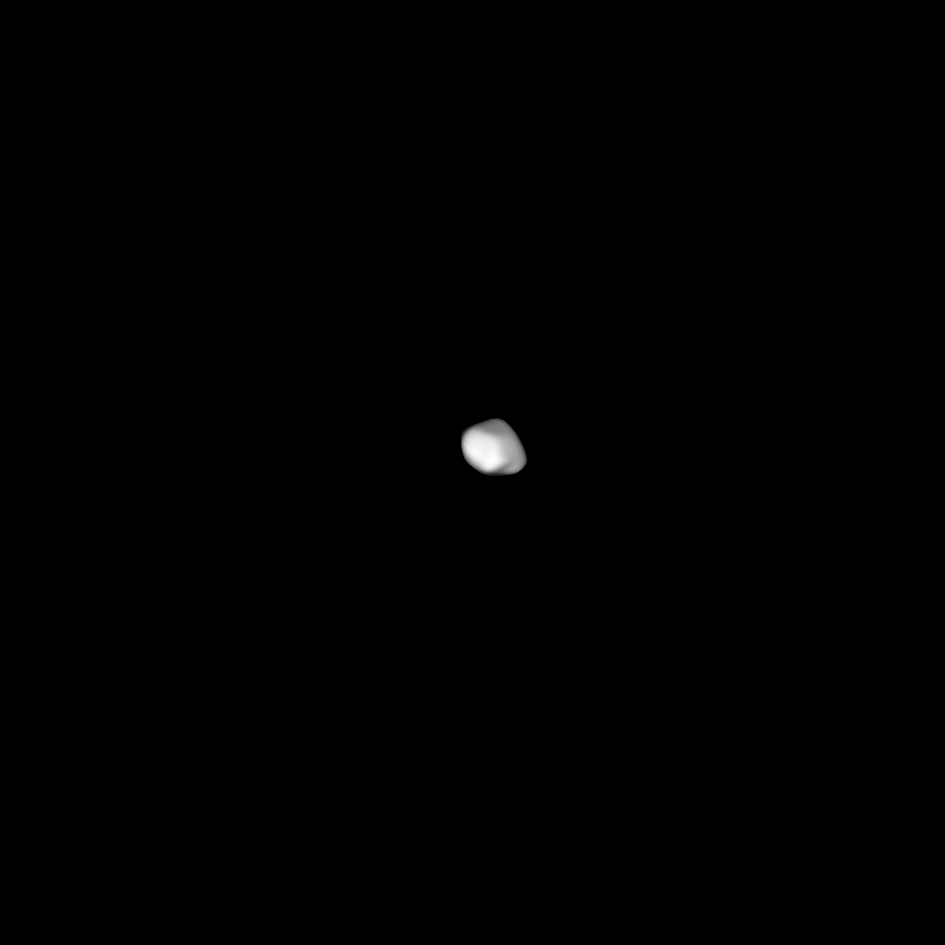}
    \end{subfigure}%
     \caption{\label{fig23}23 Thalia, poles $1$ and $2$.}
\end{figure}

\begin{figure}[t]
     \begin{subfigure}[b]{0.16\linewidth}
      \includegraphics[clip=true,trim=85 85 85 85,scale=0.66]{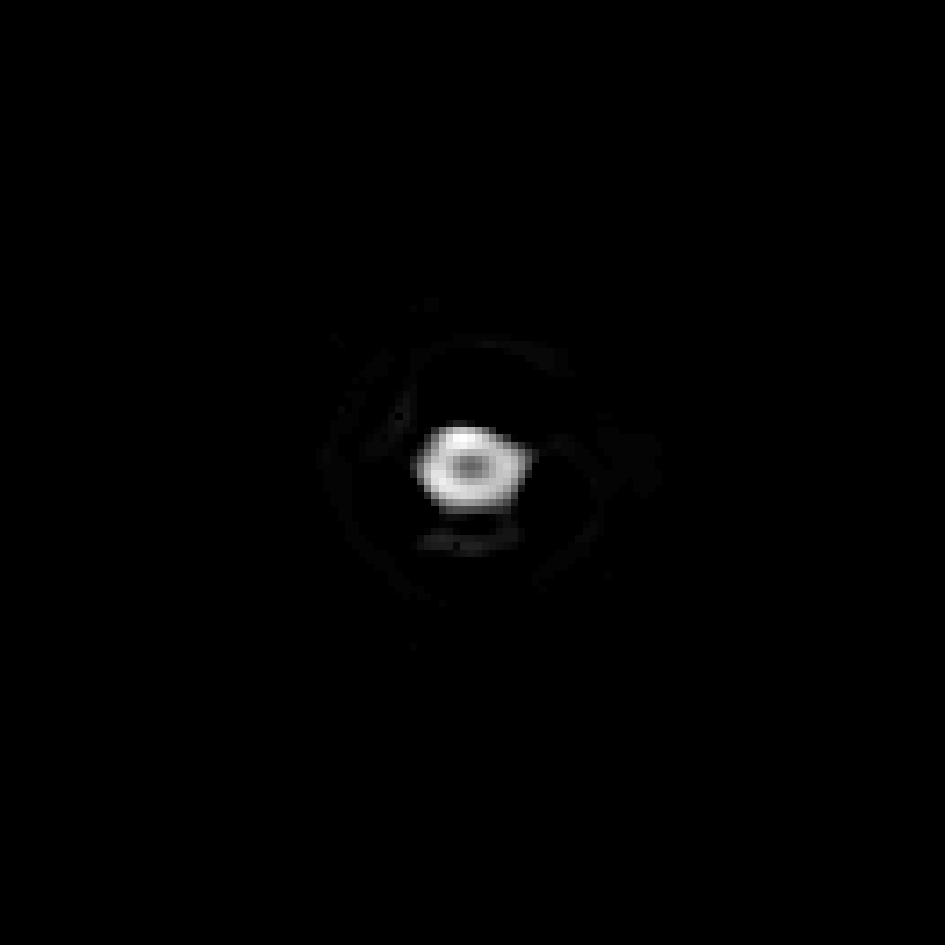} 
    \end{subfigure}%
     \begin{subfigure}[b]{0.16\linewidth}
     \includegraphics[clip=true,trim=90 90 80 80,scale=0.66]{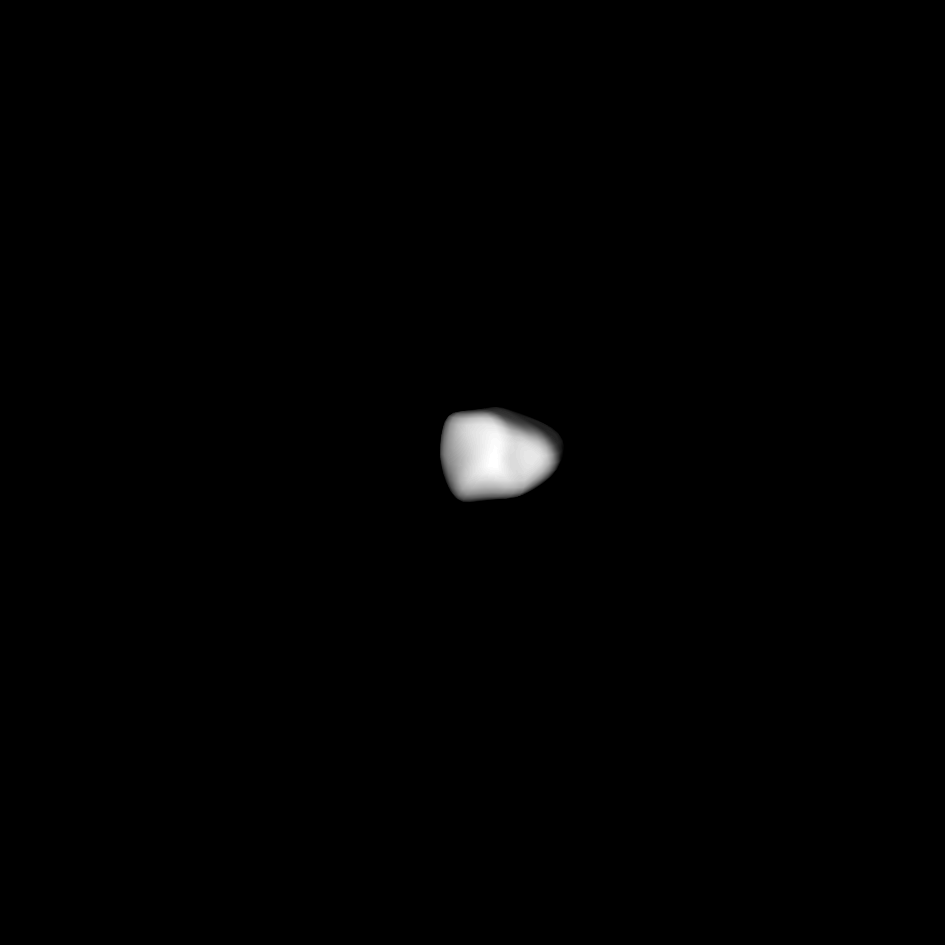}
     \end{subfigure}%
      \begin{subfigure}[b]{0.16\linewidth}
      \includegraphics[clip=true,trim=85 80 85 90,scale=0.66]{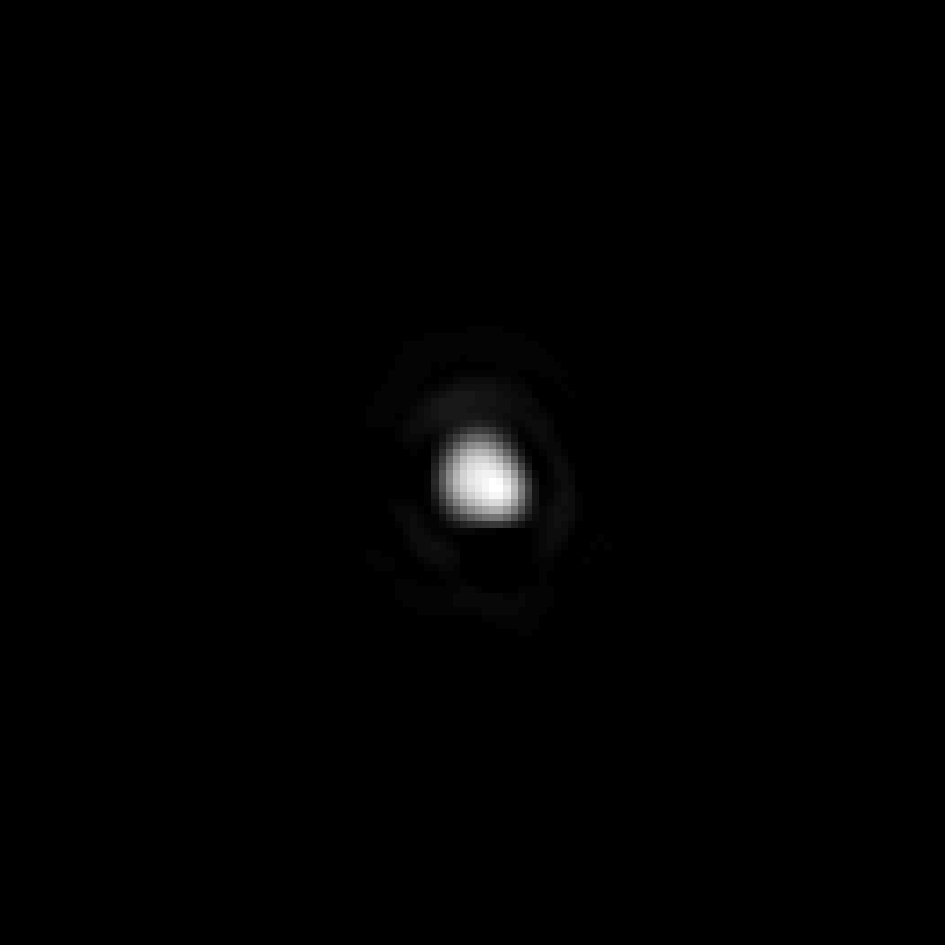}
    \end{subfigure}%
     \begin{subfigure}[b]{0.16\linewidth}
      \includegraphics[clip=true,trim=90 90 80 80,scale=0.66]{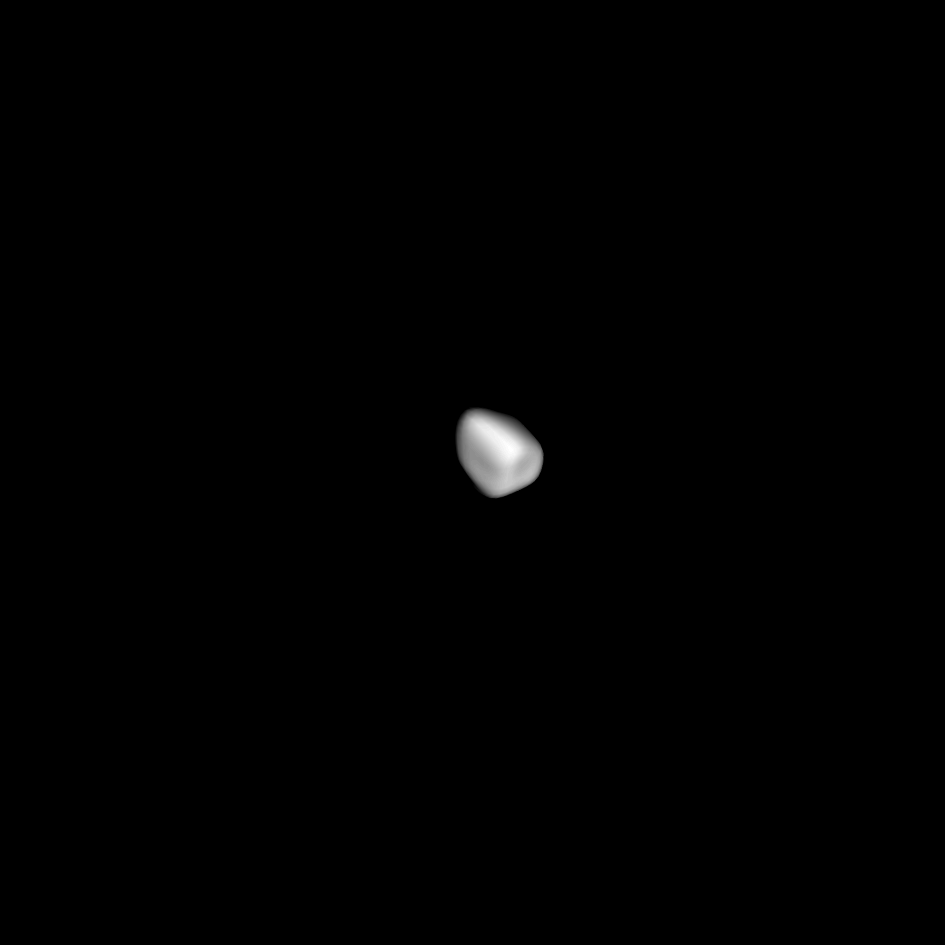}
    \end{subfigure}%
    \begin{subfigure}[b]{0.16\linewidth}
      \includegraphics[clip=true,trim=85 85 85 85,scale=0.66]{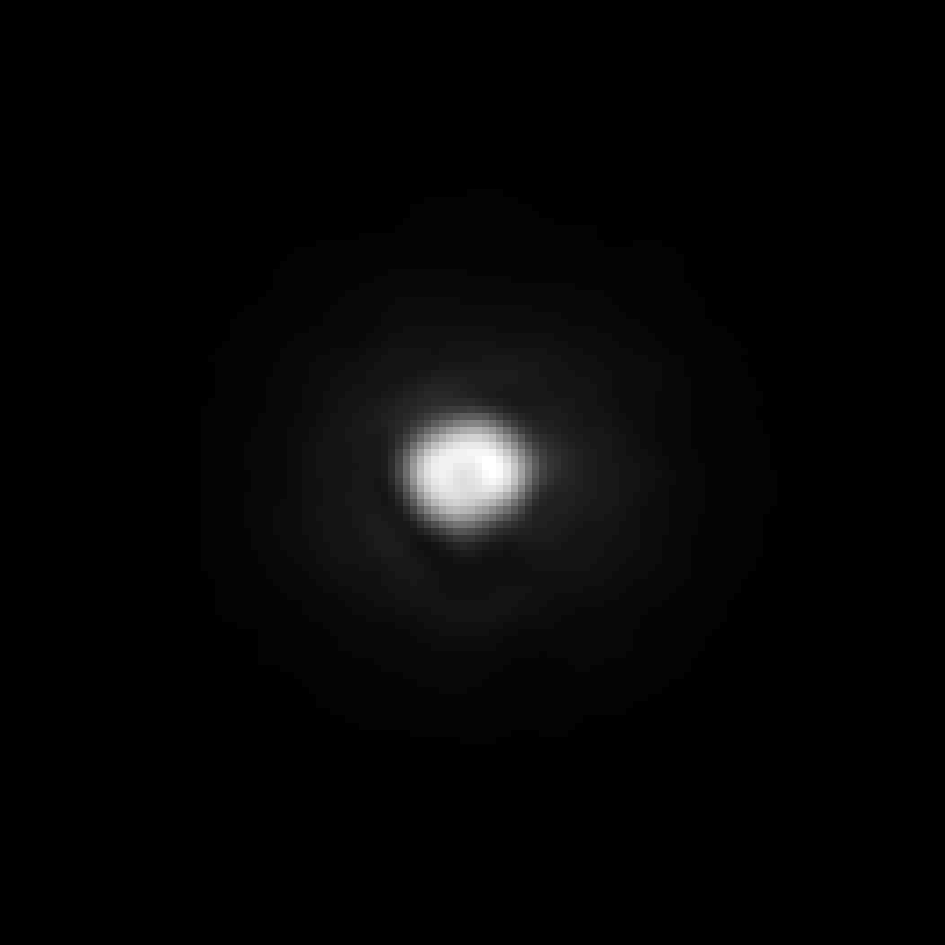} 
    \end{subfigure}%
     \begin{subfigure}[b]{0.16\linewidth}
     \includegraphics[clip=true,trim=90 90 80 80,scale=0.66]{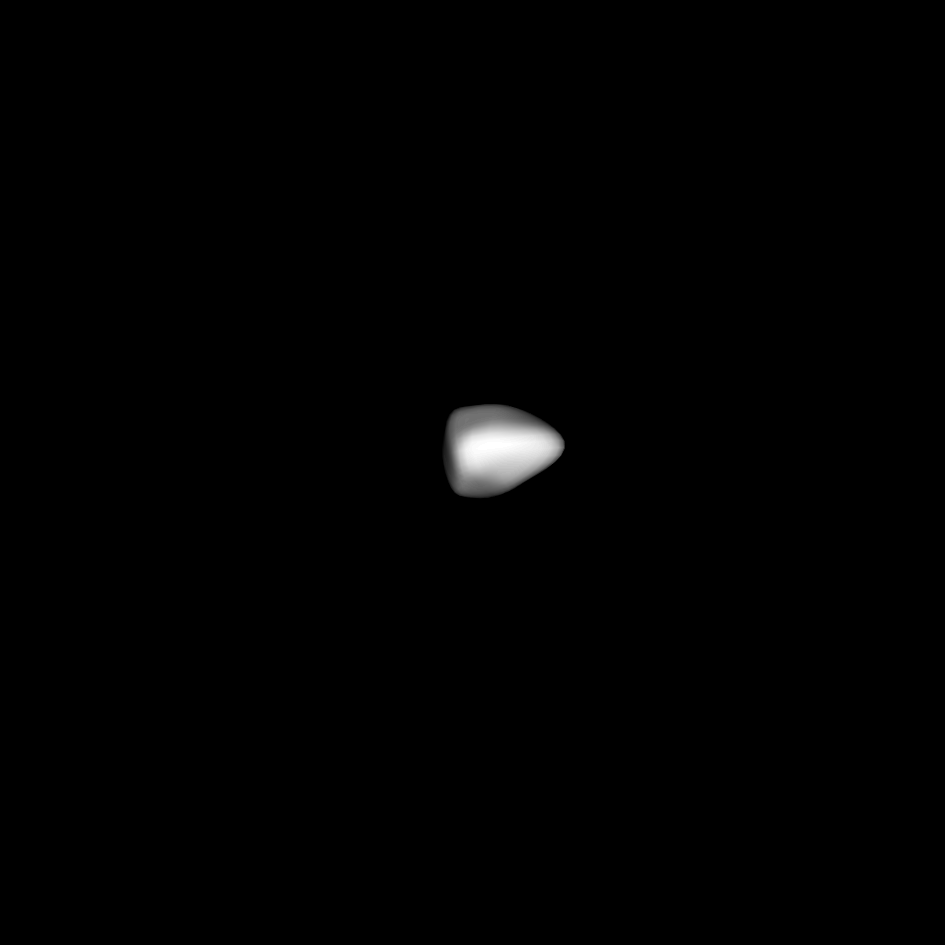}
     \end{subfigure}%
     
     \begin{subfigure}[b]{0.16\linewidth}
      \includegraphics[clip=true,trim=85 85 85 85,scale=0.66]{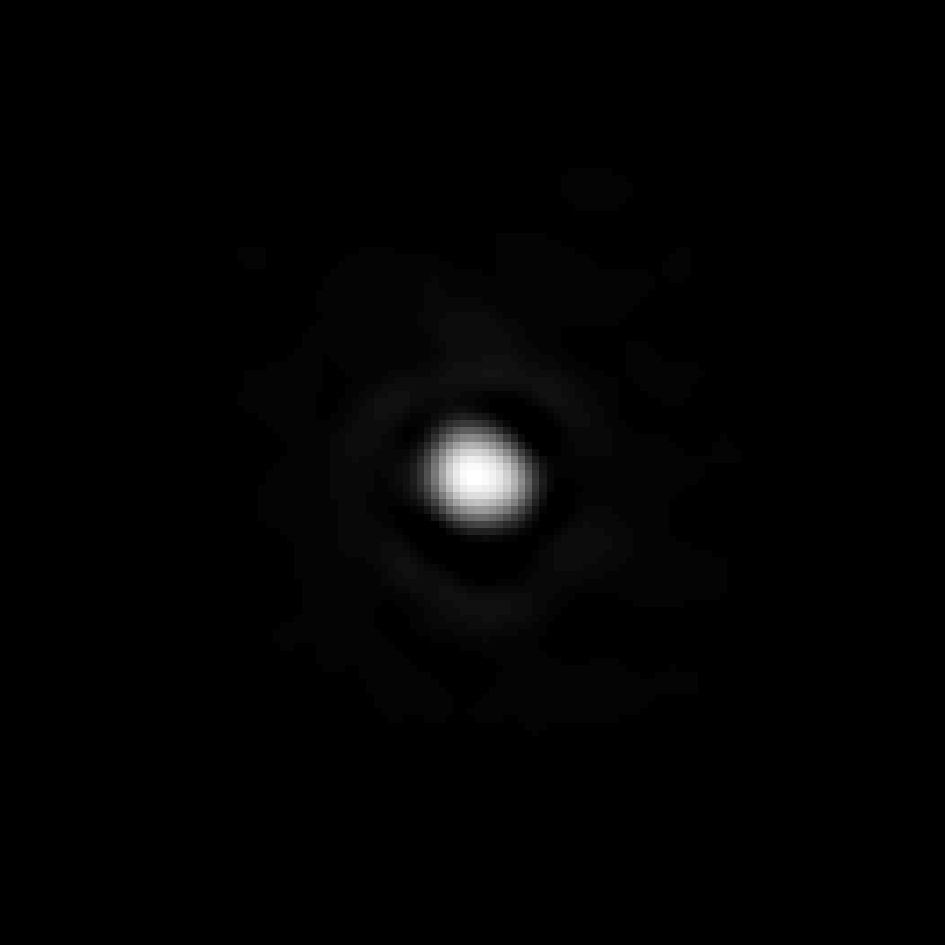} 
    \end{subfigure}%
     \begin{subfigure}[b]{0.16\linewidth}
     \includegraphics[clip=true,trim=90 90 80 80,scale=0.66]{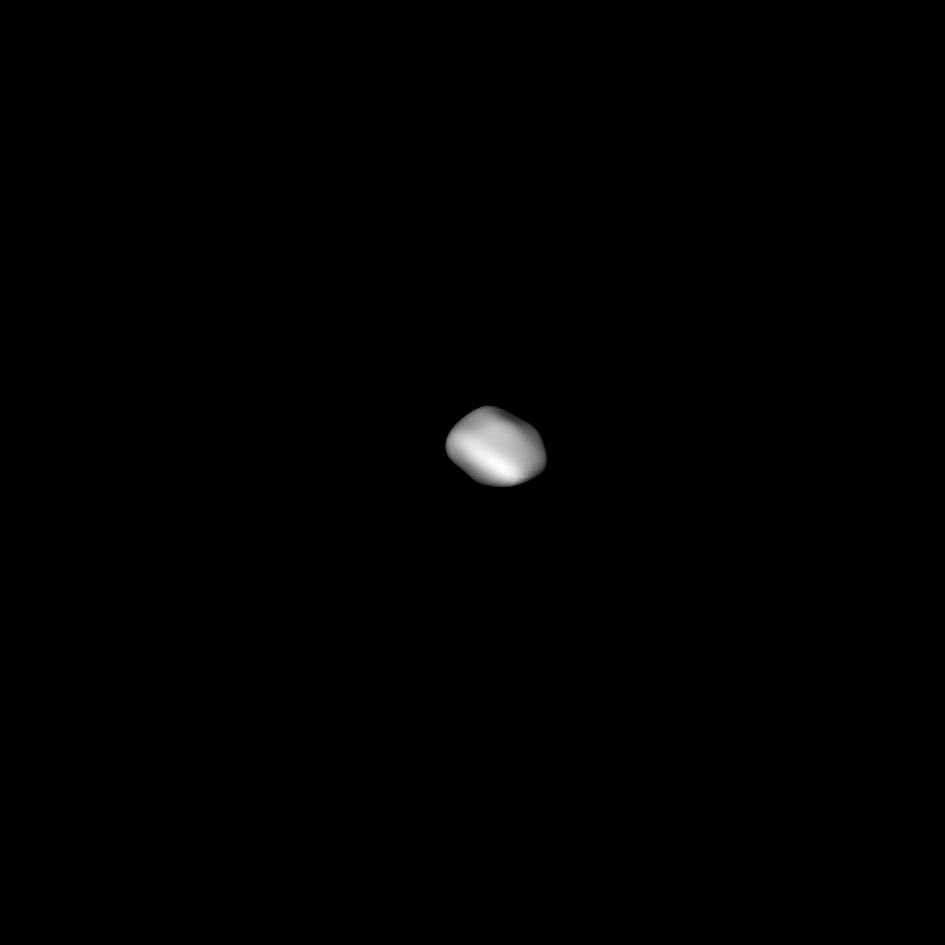}
     \end{subfigure}%
     \caption{\label{fig24}24 Themis}
\end{figure}
    
    \begin{figure}[t]
     \begin{subfigure}[b]{0.16\linewidth}
      \includegraphics[clip=true,trim=85 85 85 85,scale=0.66]{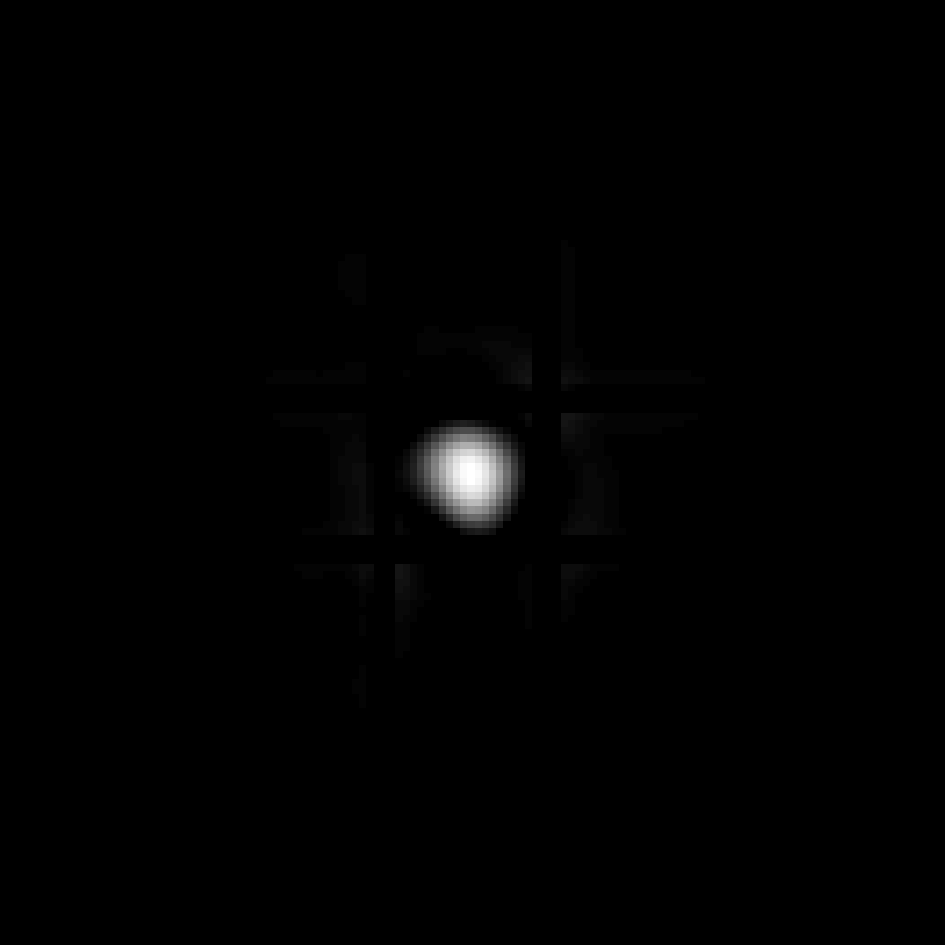} 
    \end{subfigure}%
     \begin{subfigure}[b]{0.16\linewidth}
     \includegraphics[clip=true,trim=90 90 80 80,scale=0.66]{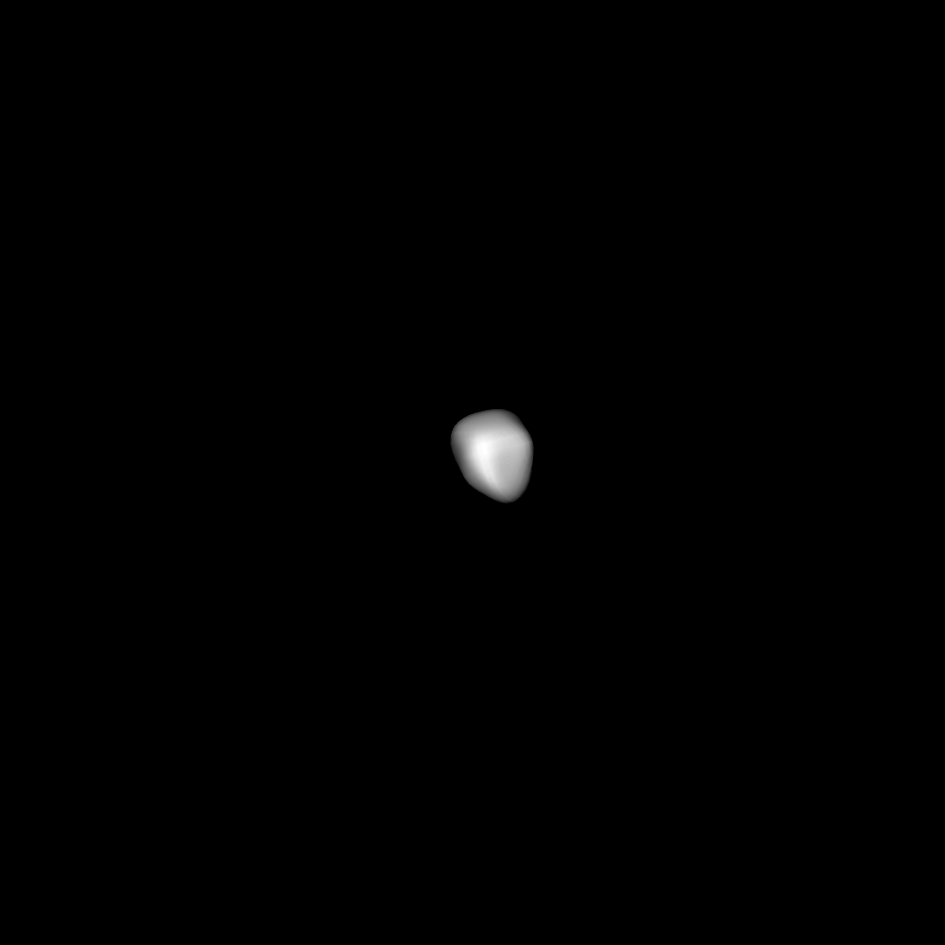}
     \end{subfigure}%
     \caption{\label{fig28}28 Bellona}
     \end{figure}
     
     \begin{figure}[t]
     \begin{subfigure}[b]{0.16\linewidth}
      \includegraphics[clip=true,trim=85 95 85 75,scale=0.66]{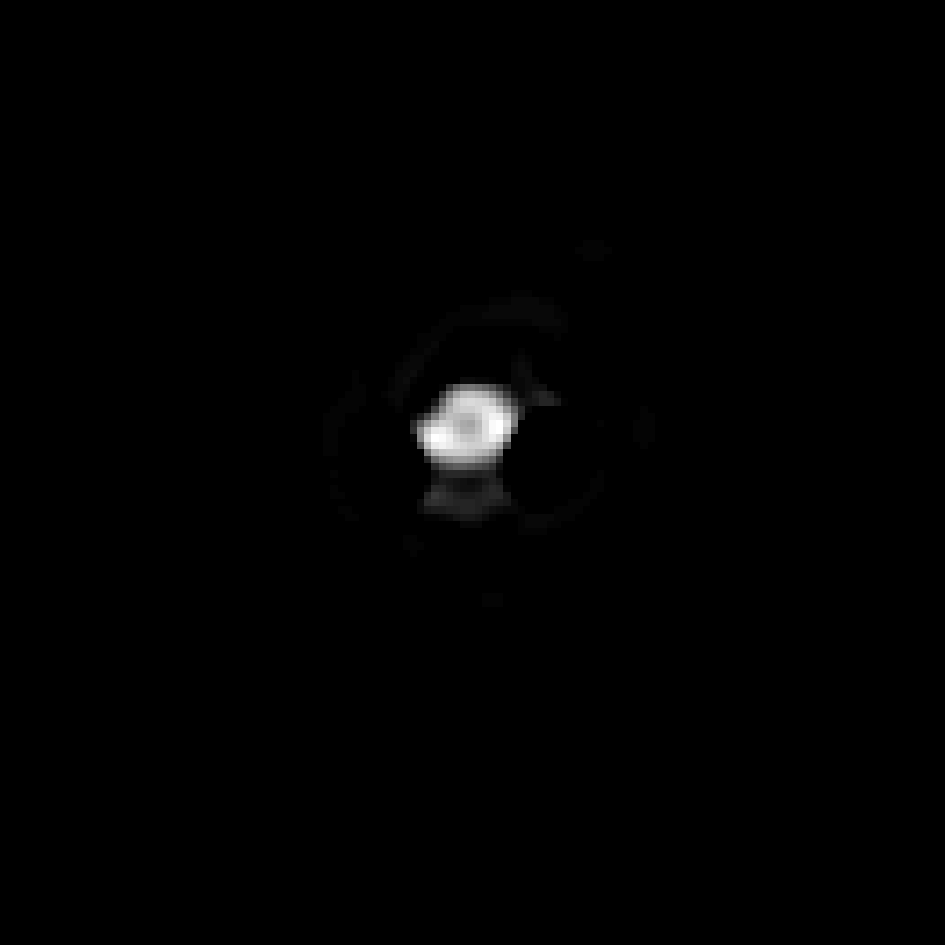} 
    \end{subfigure}%
     \begin{subfigure}[b]{0.16\linewidth}
     \includegraphics[clip=true,trim=90 90 80 80,scale=0.66]{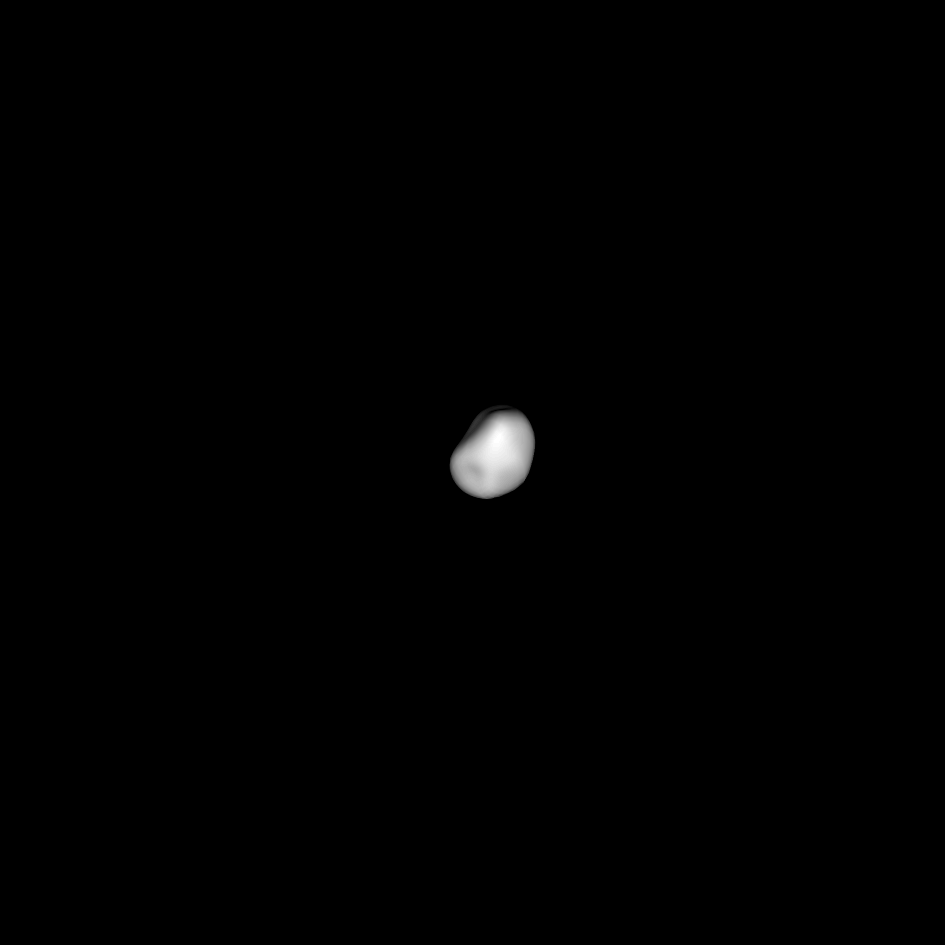}
     \end{subfigure}%
      \begin{subfigure}[b]{0.16\linewidth}
      \includegraphics[clip=true,trim=85 90 85 80,scale=0.66]{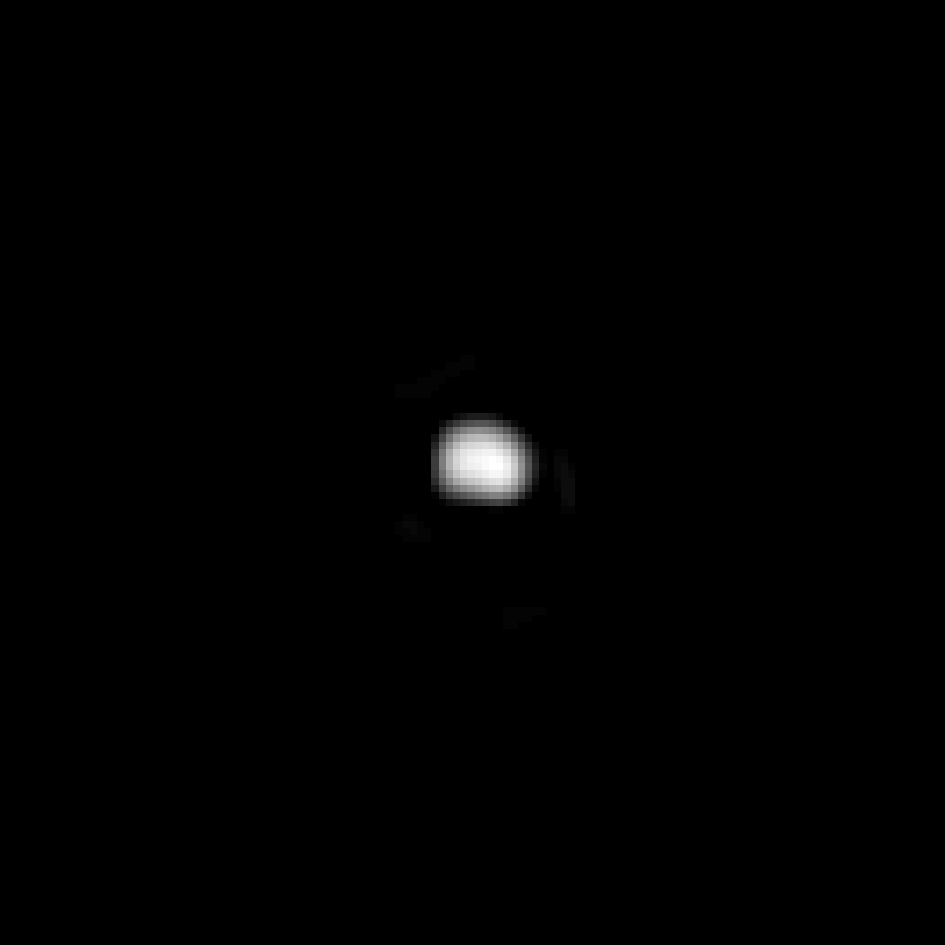}
    \end{subfigure}%
     \begin{subfigure}[b]{0.16\linewidth}
      \includegraphics[clip=true,trim=90 90 80 80,scale=0.66]{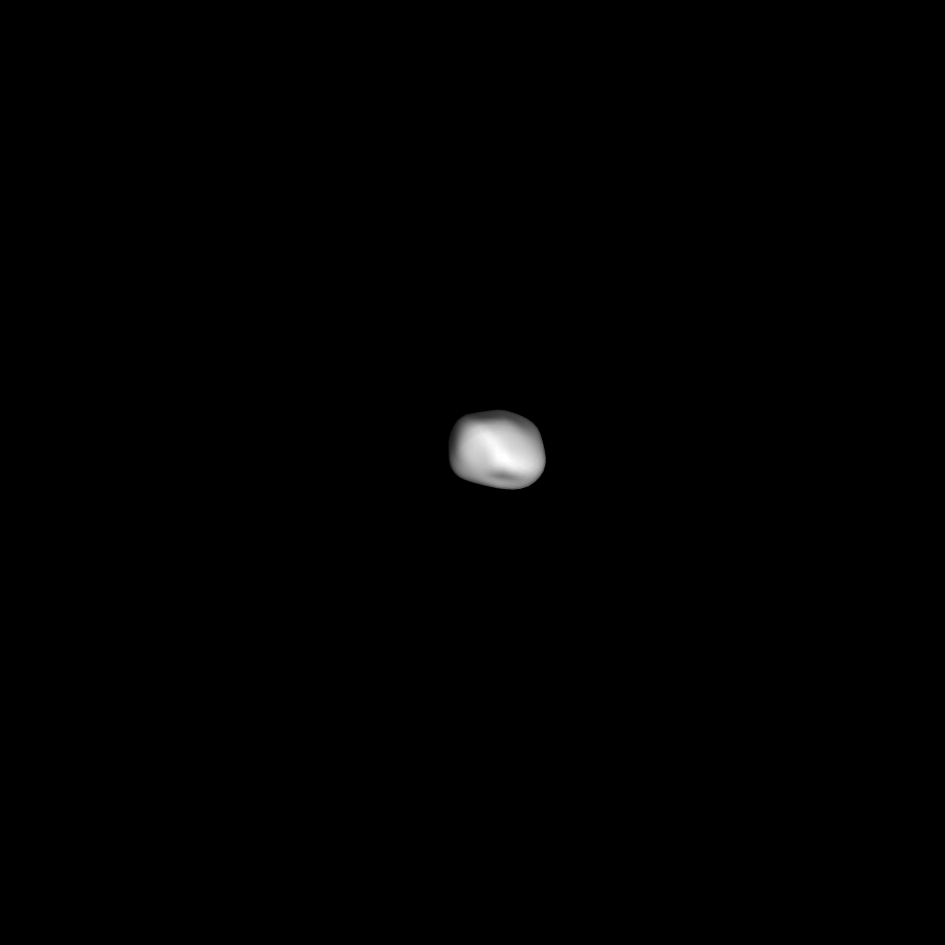}
    \end{subfigure}%
    \begin{subfigure}[b]{0.16\linewidth}
      \includegraphics[clip=true,trim=85 80 85 90,scale=0.66]{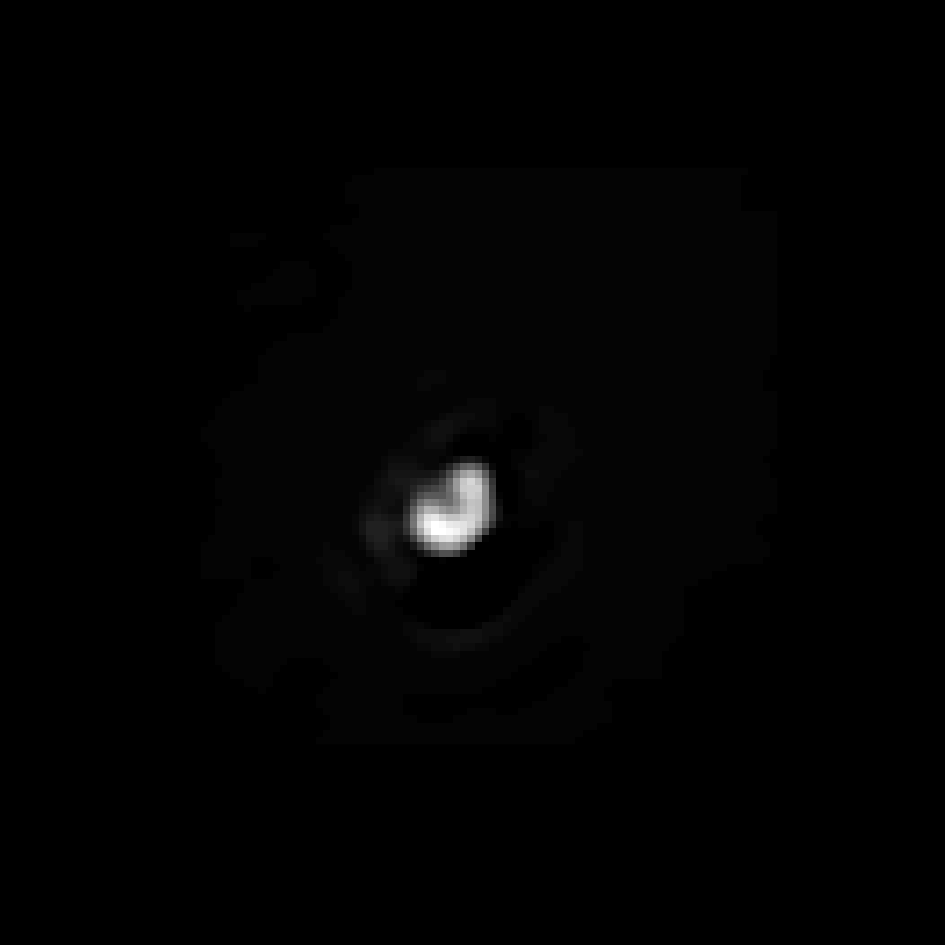} 
    \end{subfigure}%
     \begin{subfigure}[b]{0.16\linewidth}
     \includegraphics[clip=true,trim=90 90 80 80,scale=0.66]{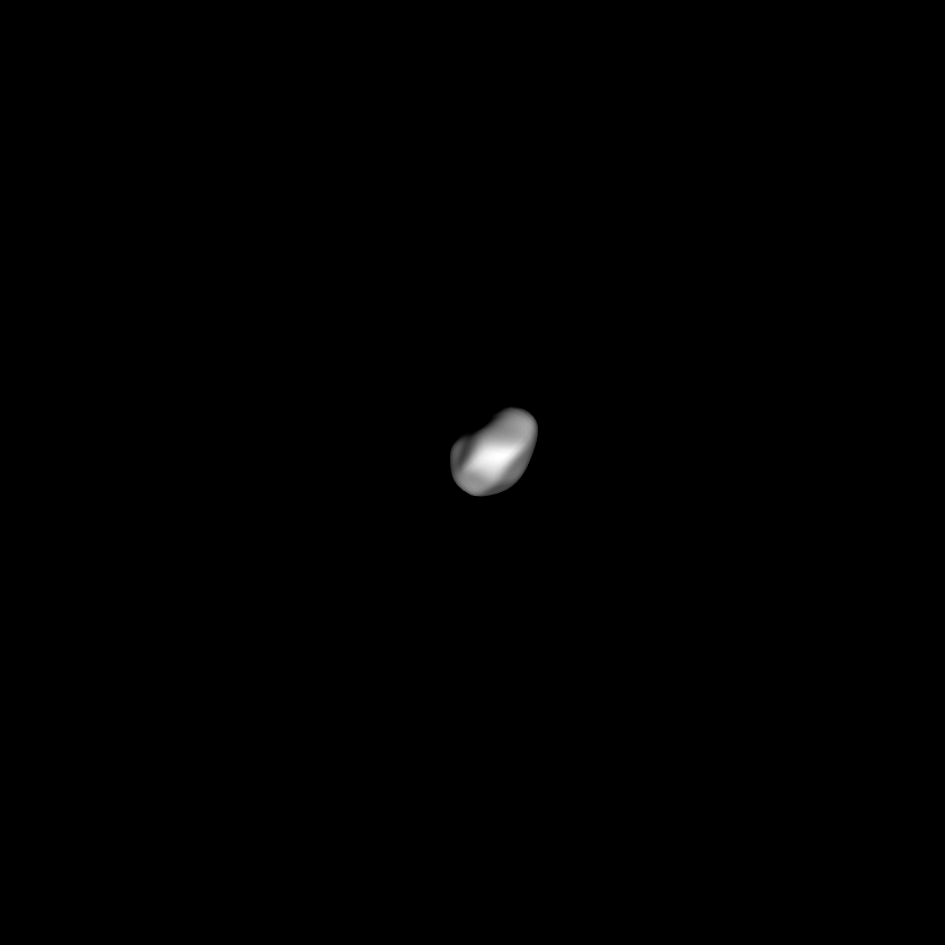}
     \end{subfigure}%
     \caption{\label{fig40}40 Harmonia}
     \end{figure}
     
     \begin{figure}[t]
     \begin{subfigure}[b]{0.16\linewidth}
      \includegraphics[clip=true,trim=75 75 96 95,scale=0.66]{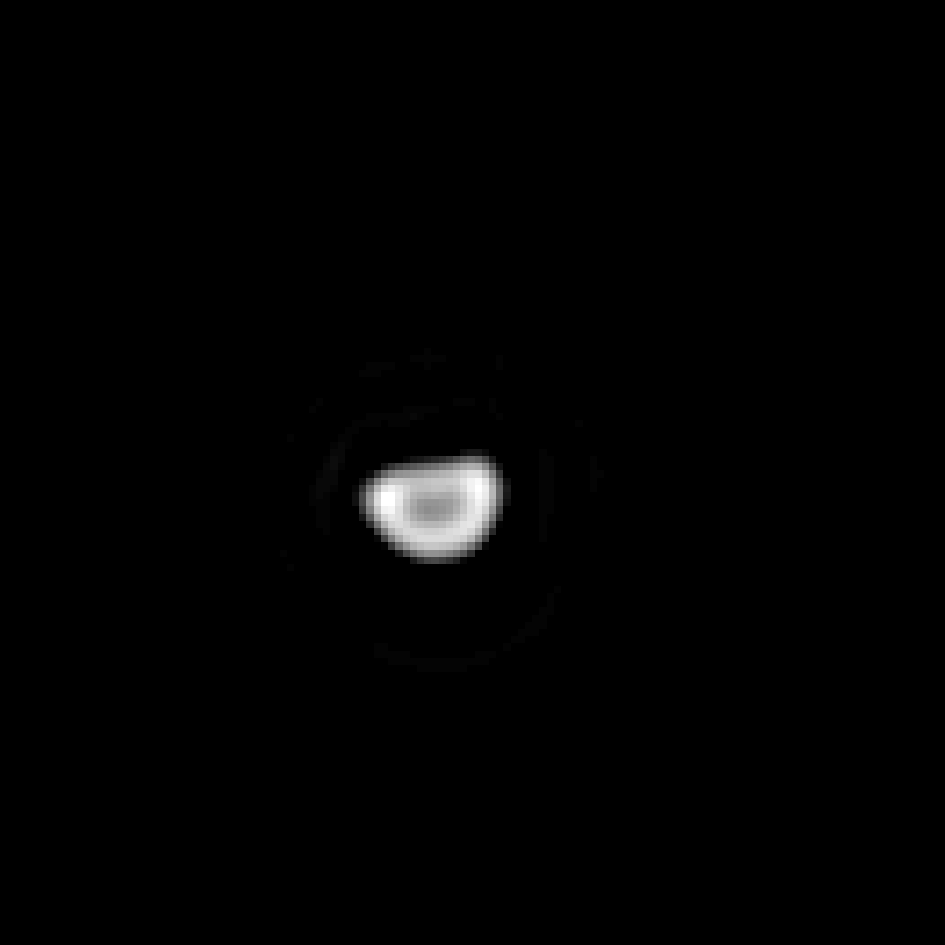} 
    \end{subfigure}%
     \begin{subfigure}[b]{0.16\linewidth}
     \includegraphics[clip=true,trim=90 90 80 80,scale=0.66]{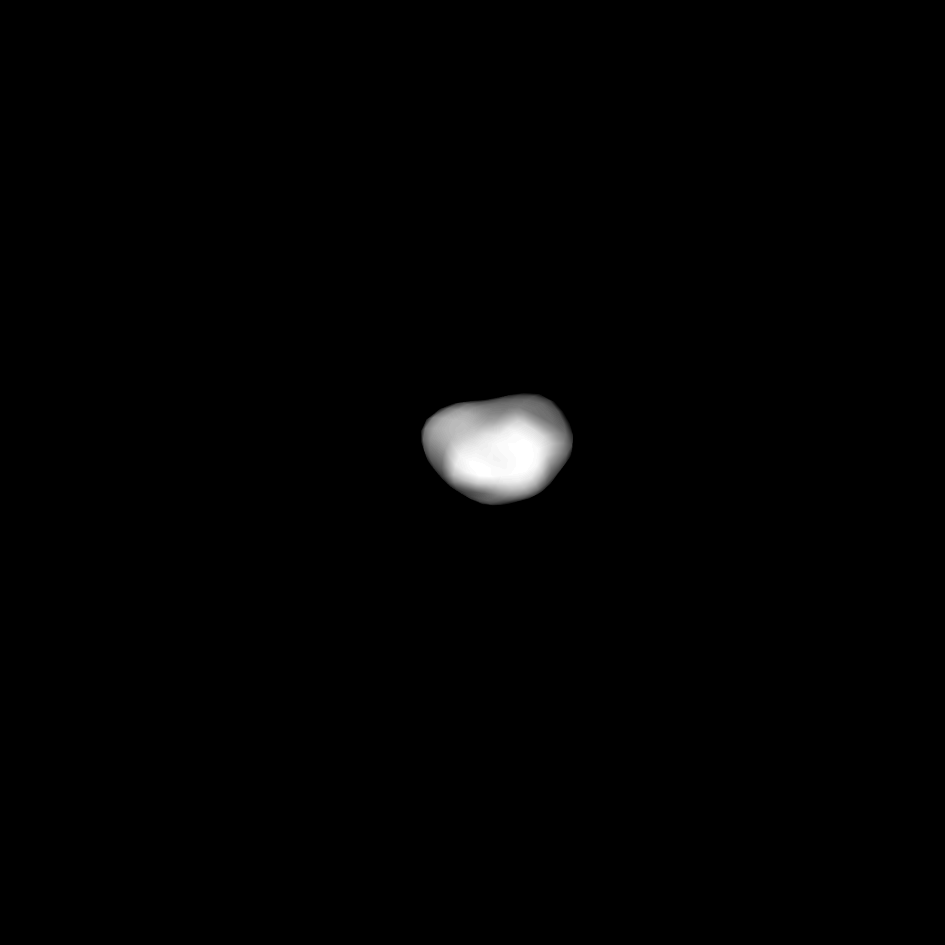}
     \end{subfigure}%
     \caption{\label{fig42}42 Isis}
     \end{figure}
     
\begin{figure}[t]
     \begin{subfigure}[b]{0.16\linewidth}
      \includegraphics[clip=true,trim=85 80 85 90,scale=0.66]{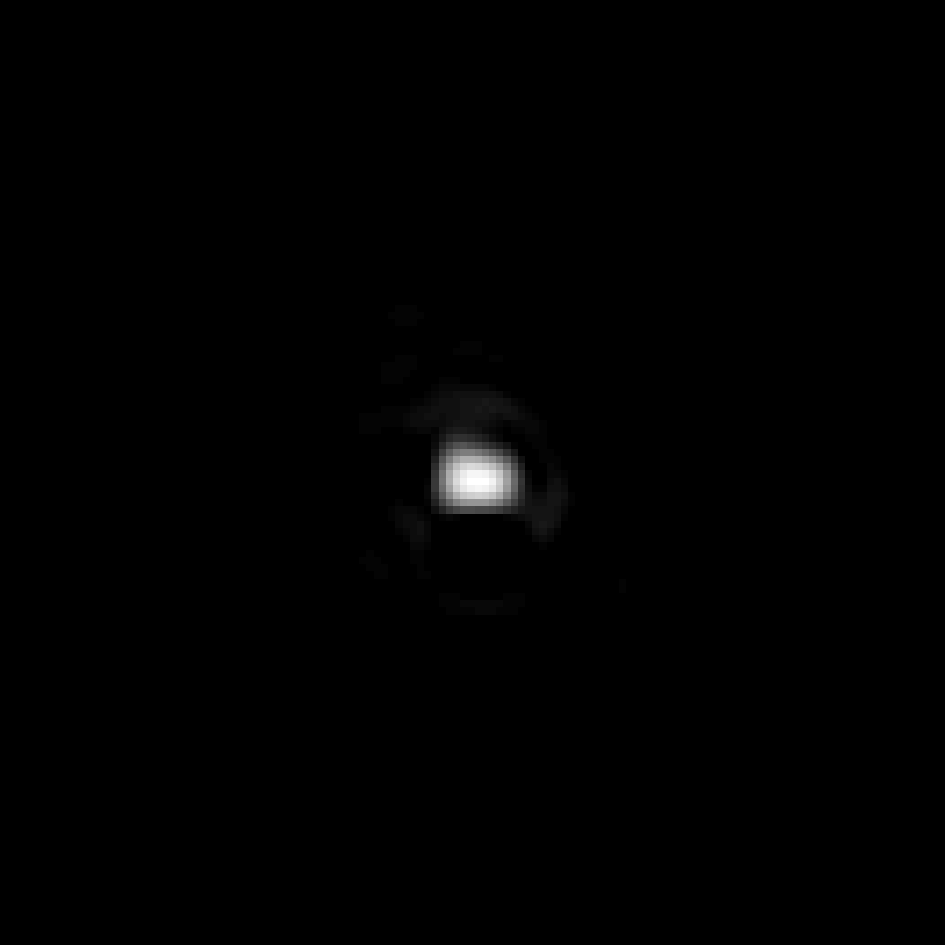} 
    \end{subfigure}%
     \begin{subfigure}[b]{0.16\linewidth}
     \includegraphics[clip=true,trim=90 90 80 80,scale=0.66]{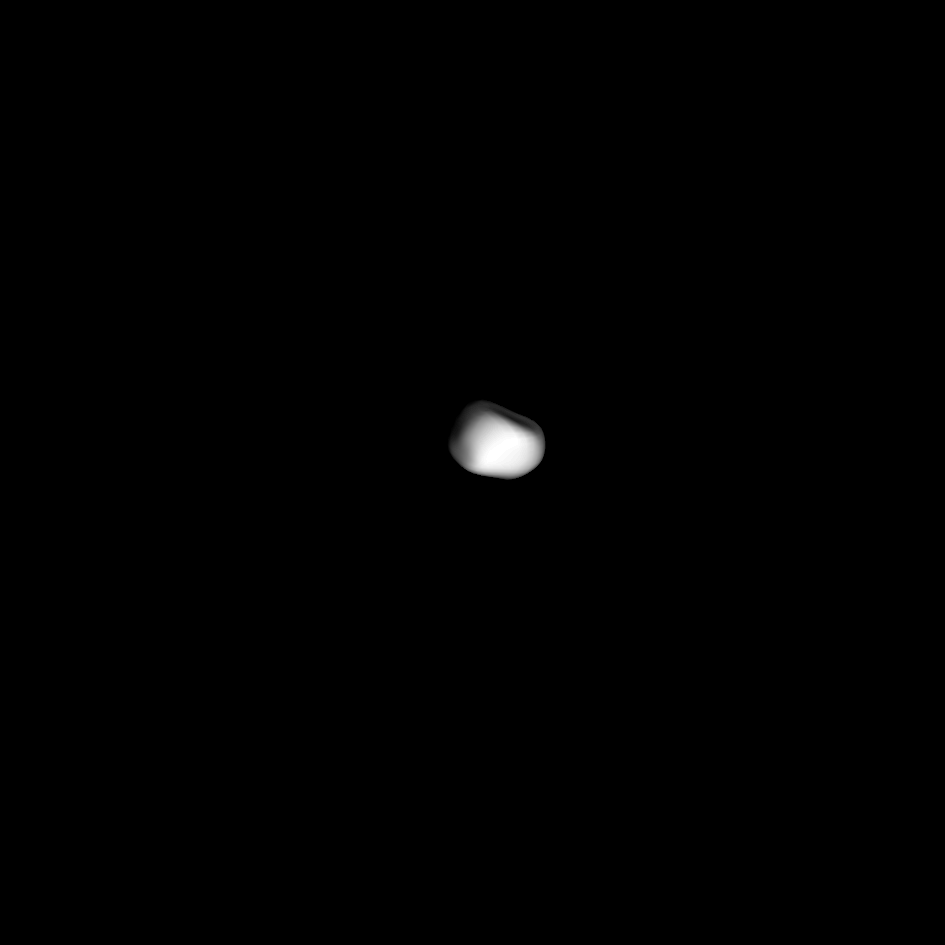}
     \end{subfigure}%
      \begin{subfigure}[b]{0.16\linewidth}
      \includegraphics[clip=true,trim=85 80 85 90,scale=0.66]{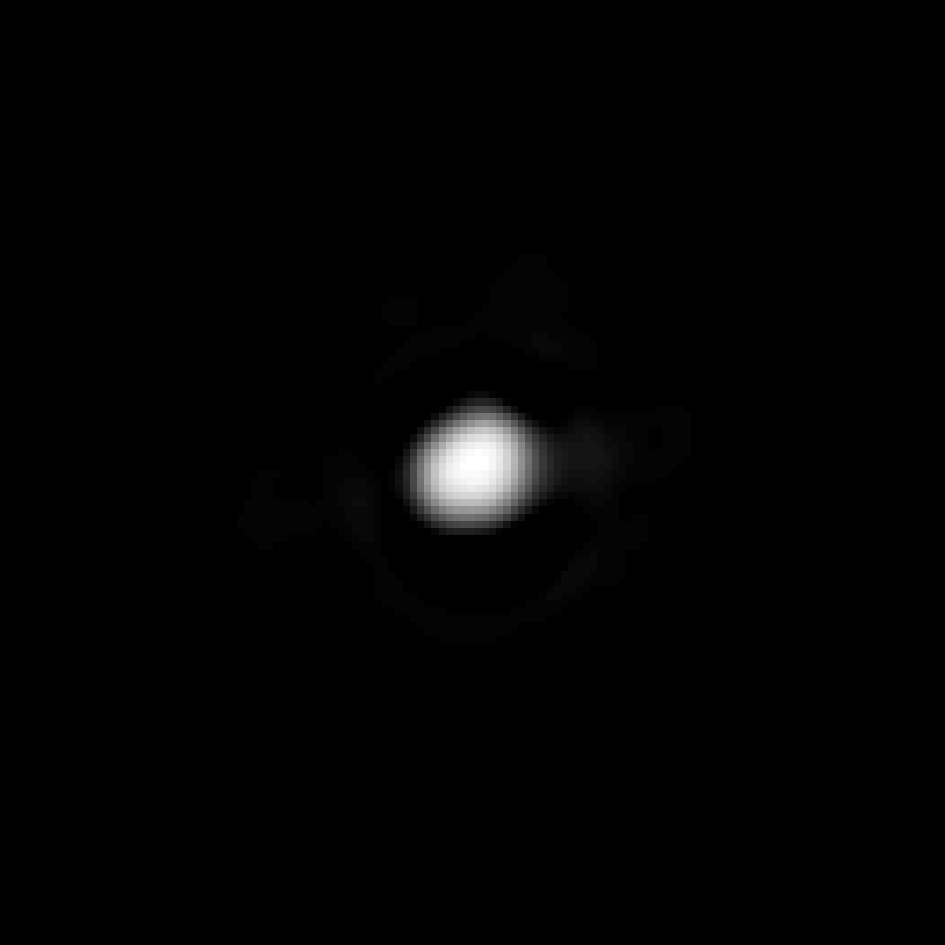}
    \end{subfigure}%
     \begin{subfigure}[b]{0.16\linewidth}
      \includegraphics[clip=true,trim=90 90 80 80,scale=0.66]{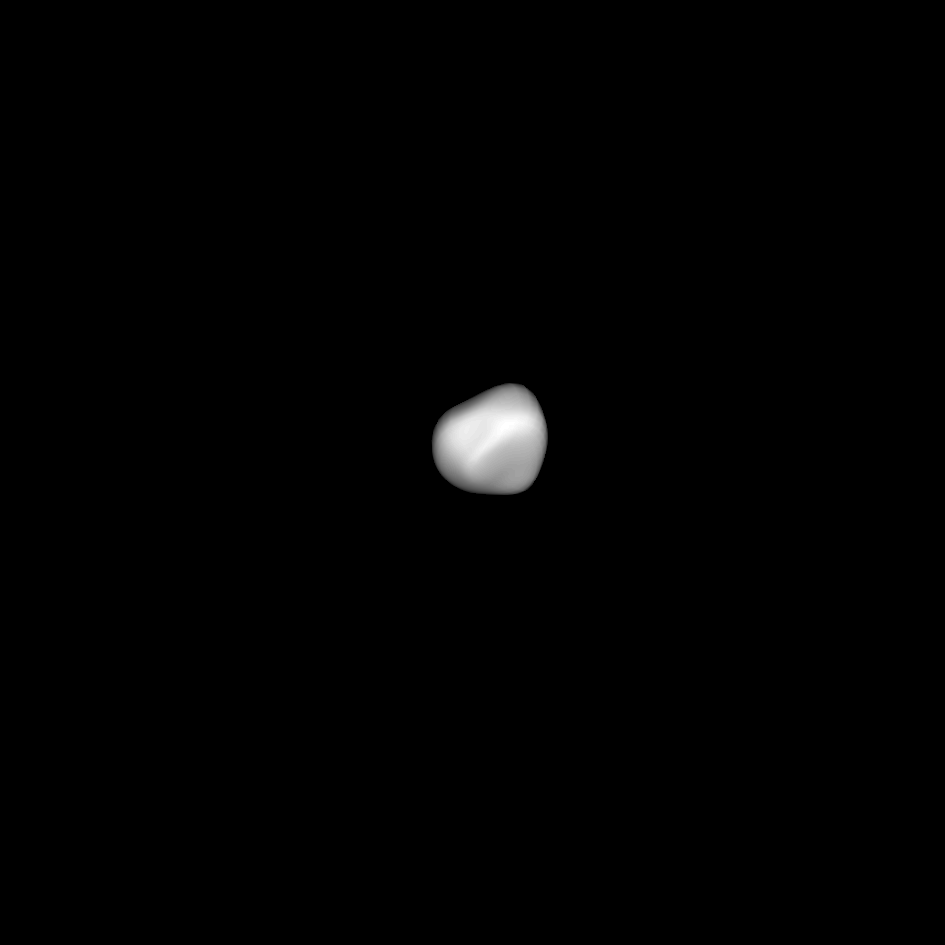}
    \end{subfigure}%
     \caption{\label{fig48}48 Doris}
\end{figure}

\begin{figure}[t]
     \begin{subfigure}[b]{0.16\linewidth}
      \includegraphics[clip=true,trim=85 85 85 85,scale=0.66]{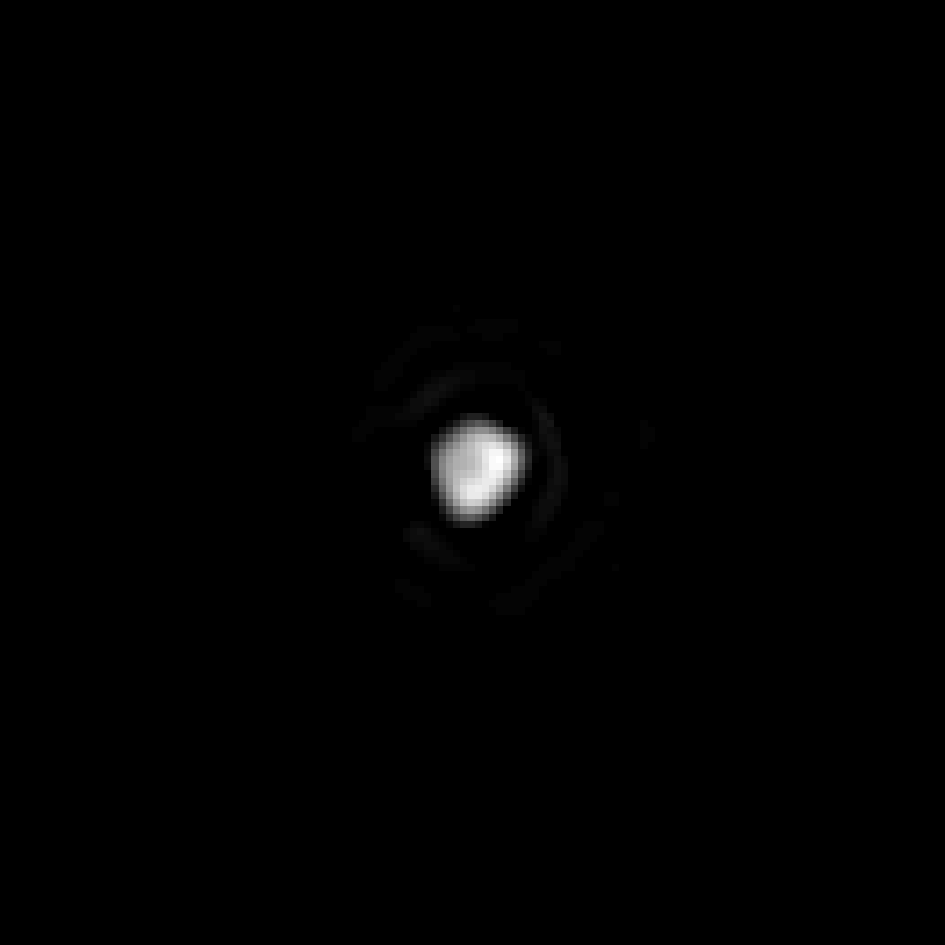} 
    \end{subfigure}%
     \begin{subfigure}[b]{0.16\linewidth}
     \includegraphics[clip=true,trim=90 90 80 80,scale=0.66]{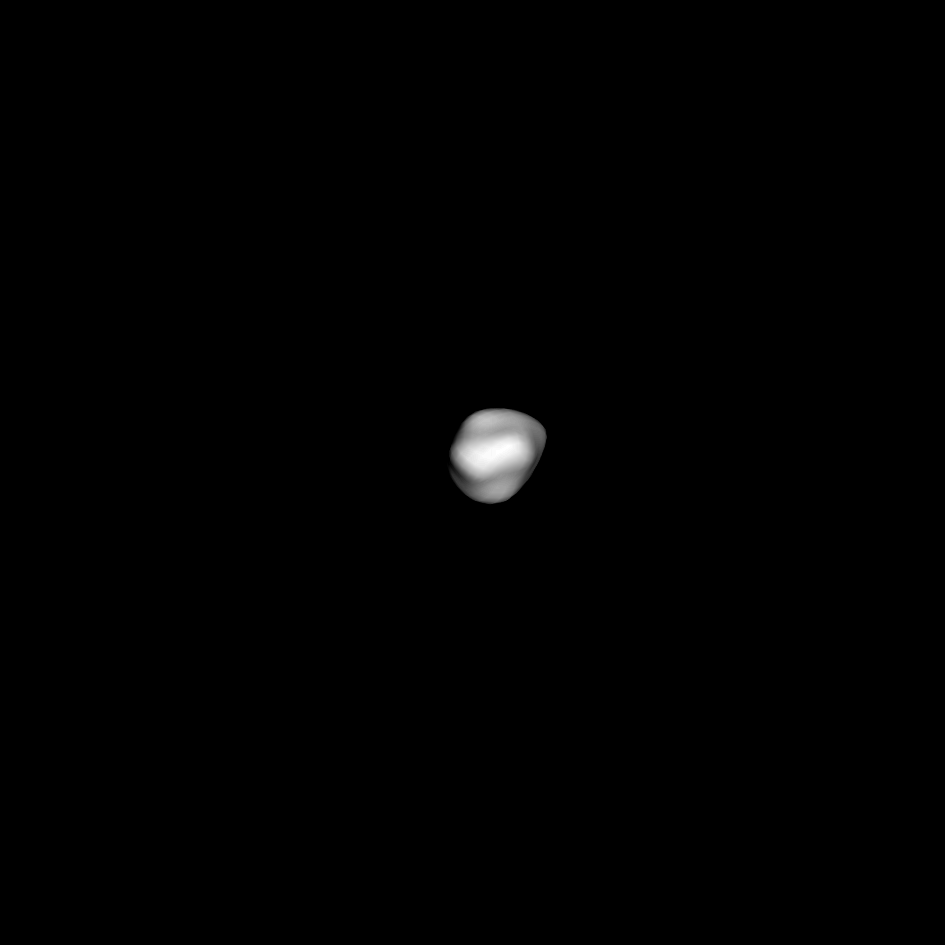}
     \end{subfigure}%
     \begin{subfigure}[b]{0.16\linewidth}
      \includegraphics[clip=true,trim=90 90 80 80,scale=0.66]{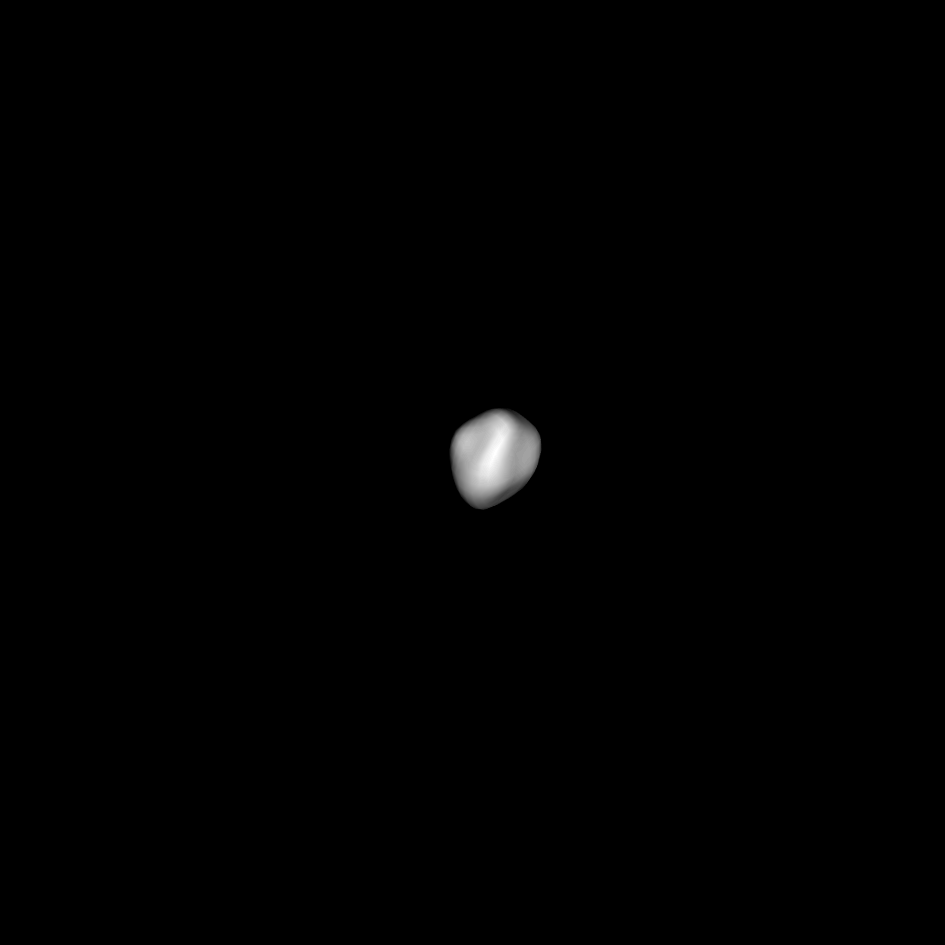}
    \end{subfigure}%
      \begin{subfigure}[b]{0.16\linewidth}
      \includegraphics[clip=true,trim=85 85 85 85,scale=0.66]{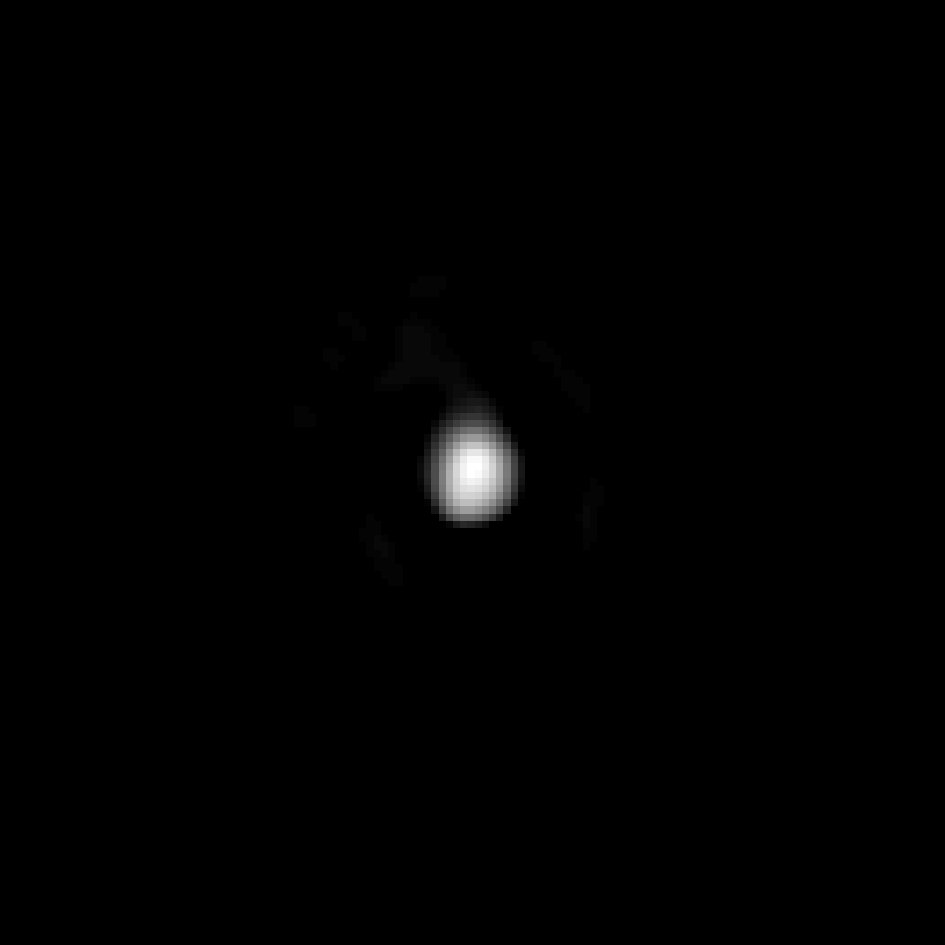}
    \end{subfigure}%
     \begin{subfigure}[b]{0.16\linewidth}
      \includegraphics[clip=true,trim=90 90 80 80,scale=0.66]{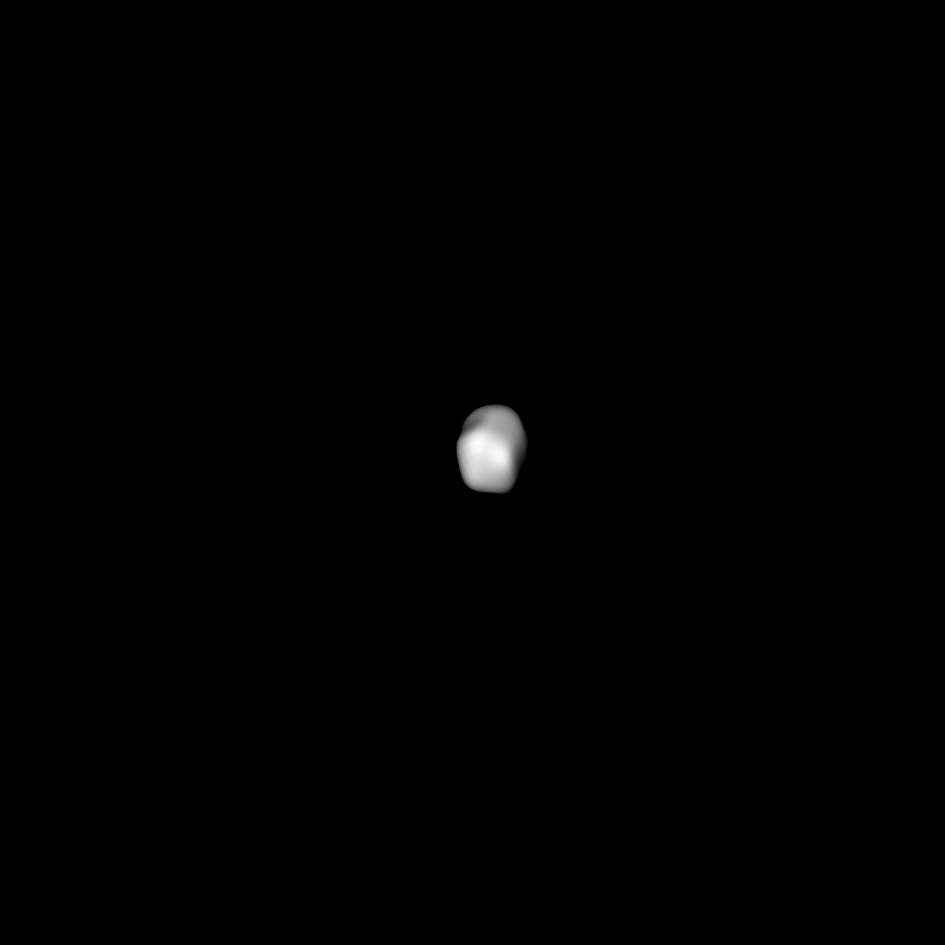}
    \end{subfigure}%
    \begin{subfigure}[b]{0.16\linewidth}
      \includegraphics[clip=true,trim=90 90 80 80,scale=0.66]{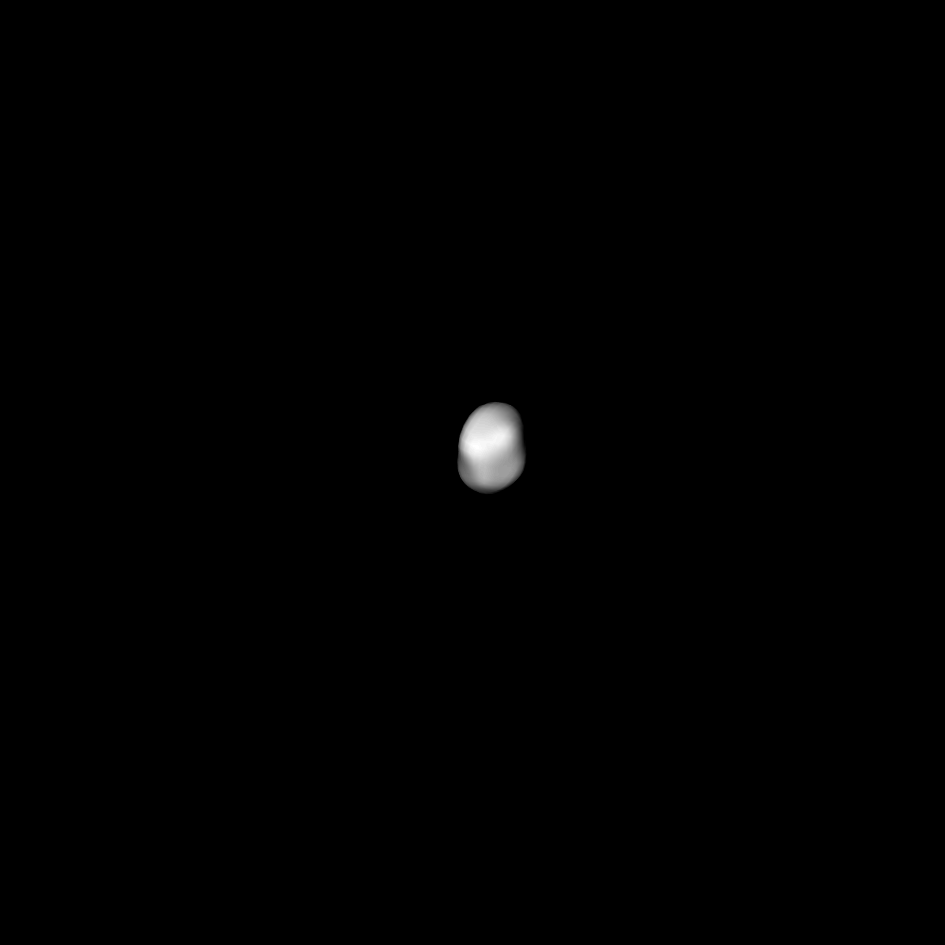}
    \end{subfigure}%
     \caption{\label{fig56}56 Melete, poles $1$ and $2$.}
\end{figure}     
\begin{figure}[t]
     \begin{subfigure}[b]{0.16\linewidth}
      \includegraphics[clip=true,trim=85 85 85 85,scale=0.66]{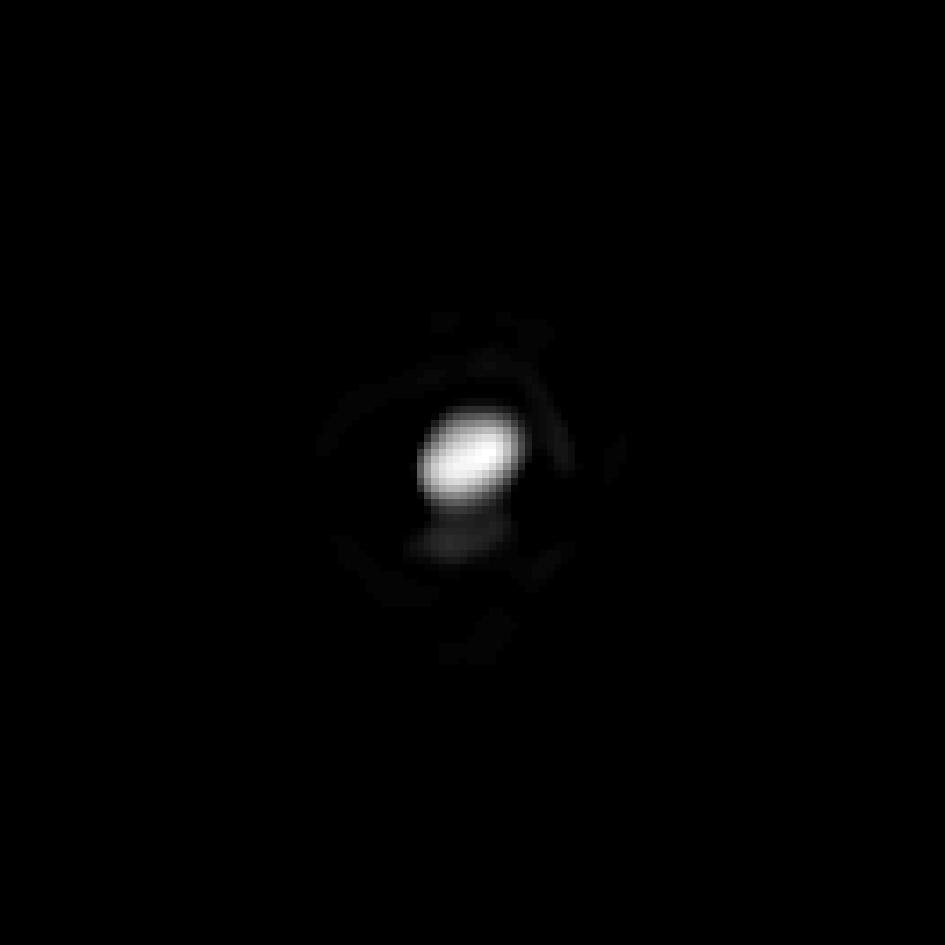} 
    \end{subfigure}%
     \begin{subfigure}[b]{0.16\linewidth}
     \includegraphics[clip=true,trim=90 90 80 80,scale=0.66]{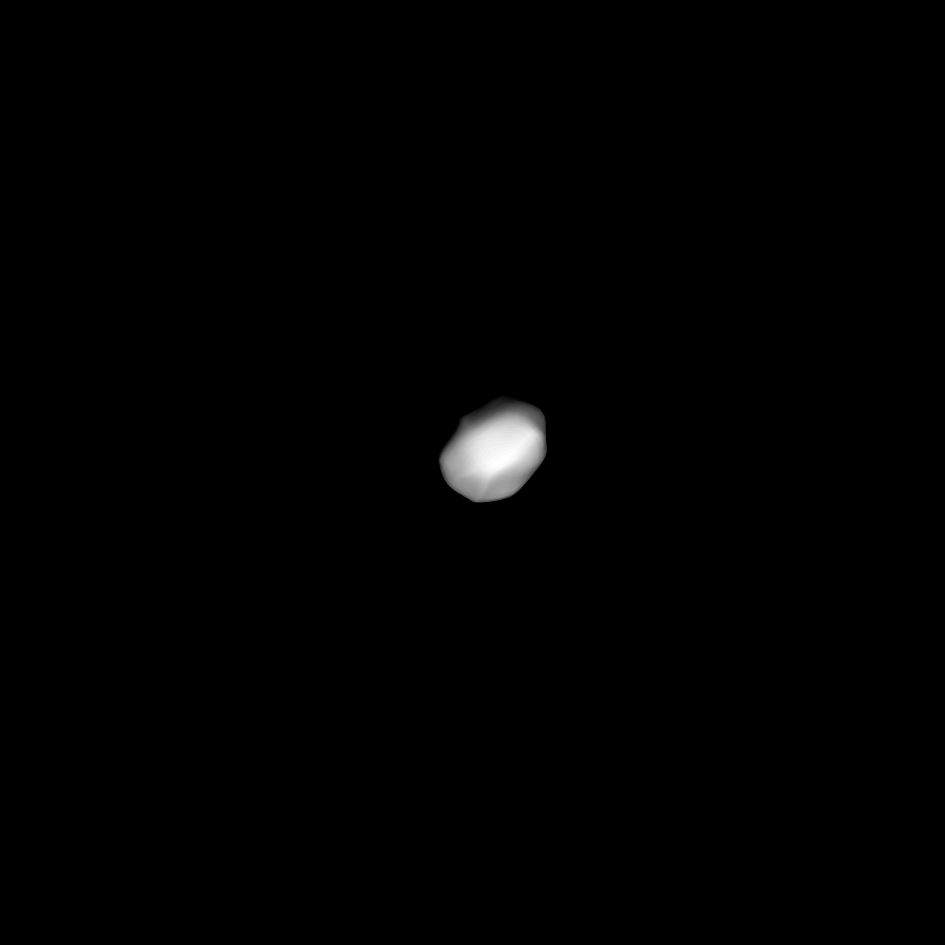}
     \end{subfigure}%
      \begin{subfigure}[b]{0.16\linewidth}
      \includegraphics[clip=true,trim=85 85 85 85,scale=0.66]{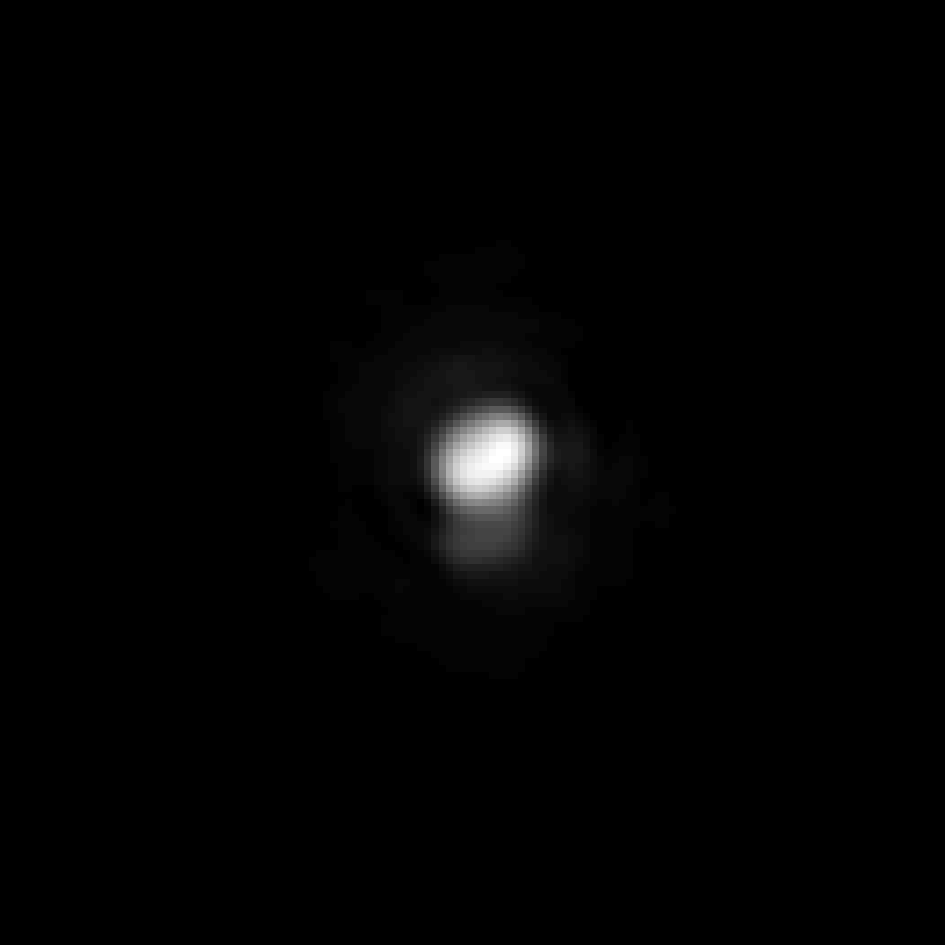}
    \end{subfigure}%
     \begin{subfigure}[b]{0.16\linewidth}
      \includegraphics[clip=true,trim=90 90 80 80,scale=0.66]{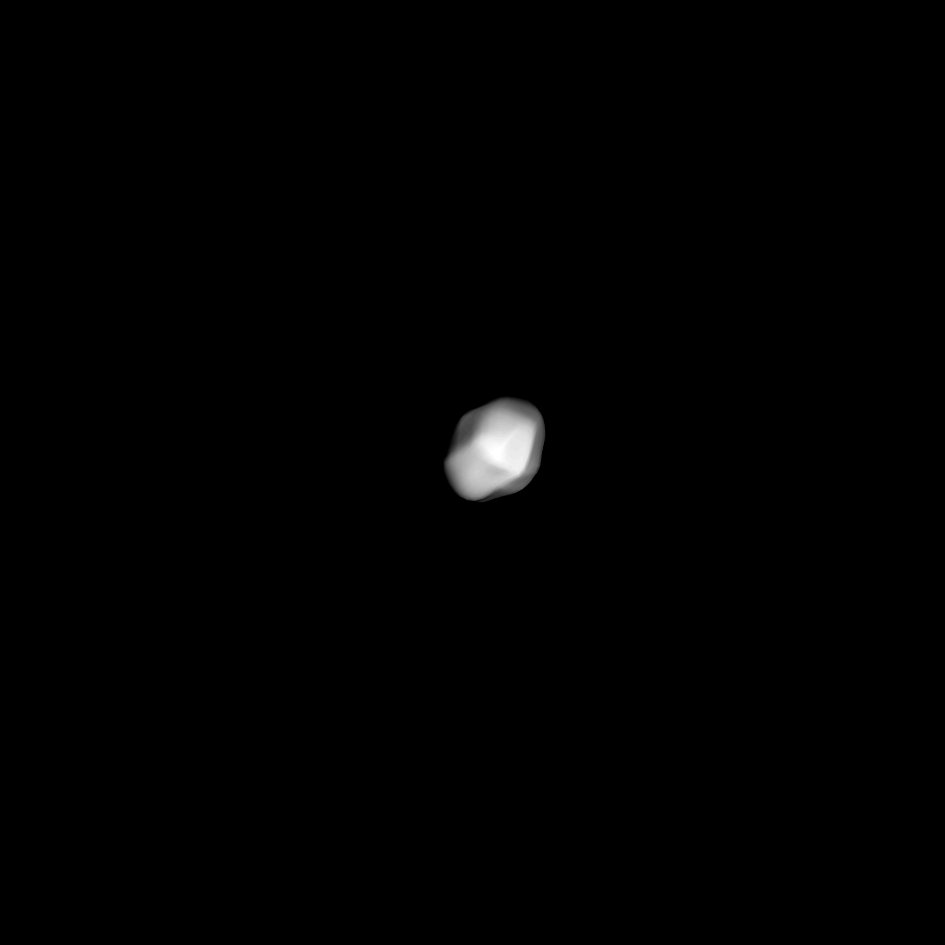}
    \end{subfigure}%
    \begin{subfigure}[b]{0.16\linewidth}
      \includegraphics[clip=true,trim=85 85 85 85,scale=0.66]{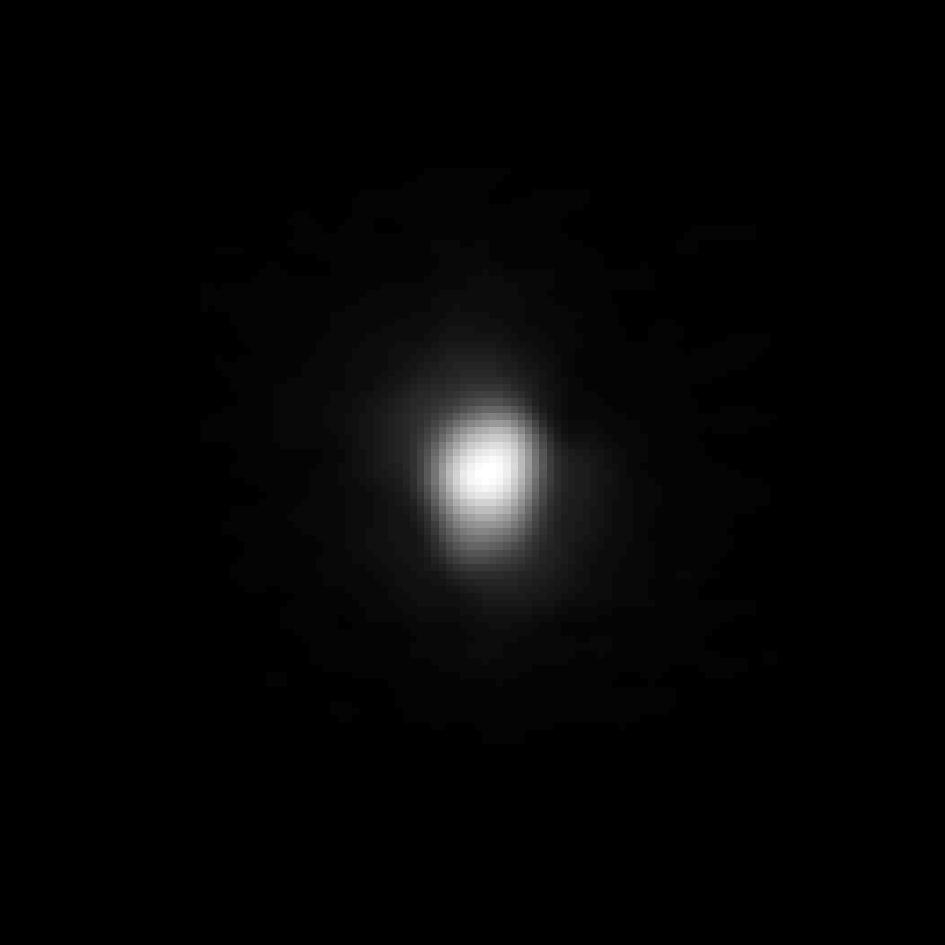} 
    \end{subfigure}%
     \begin{subfigure}[b]{0.16\linewidth}
     \includegraphics[clip=true,trim=90 90 80 80,scale=0.66]{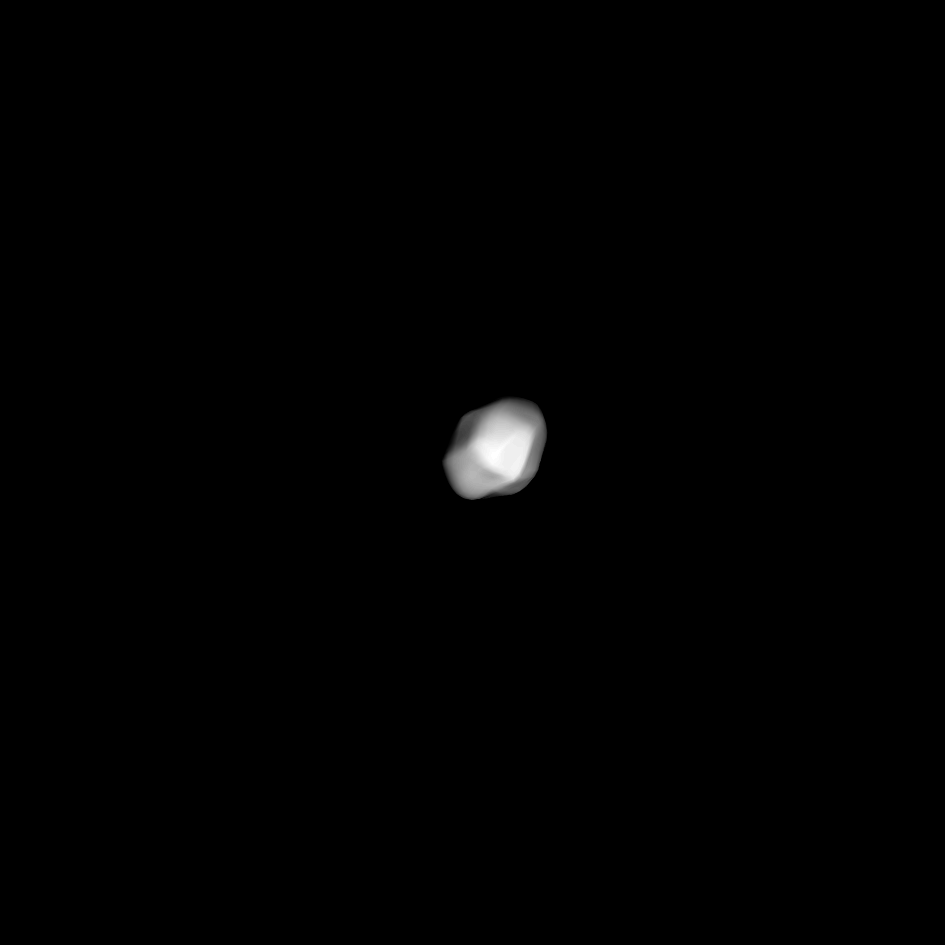}
     \end{subfigure}%
     
      \begin{subfigure}[b]{0.16\linewidth}
      \includegraphics[clip=true,trim=85 85 85 85,scale=0.66]{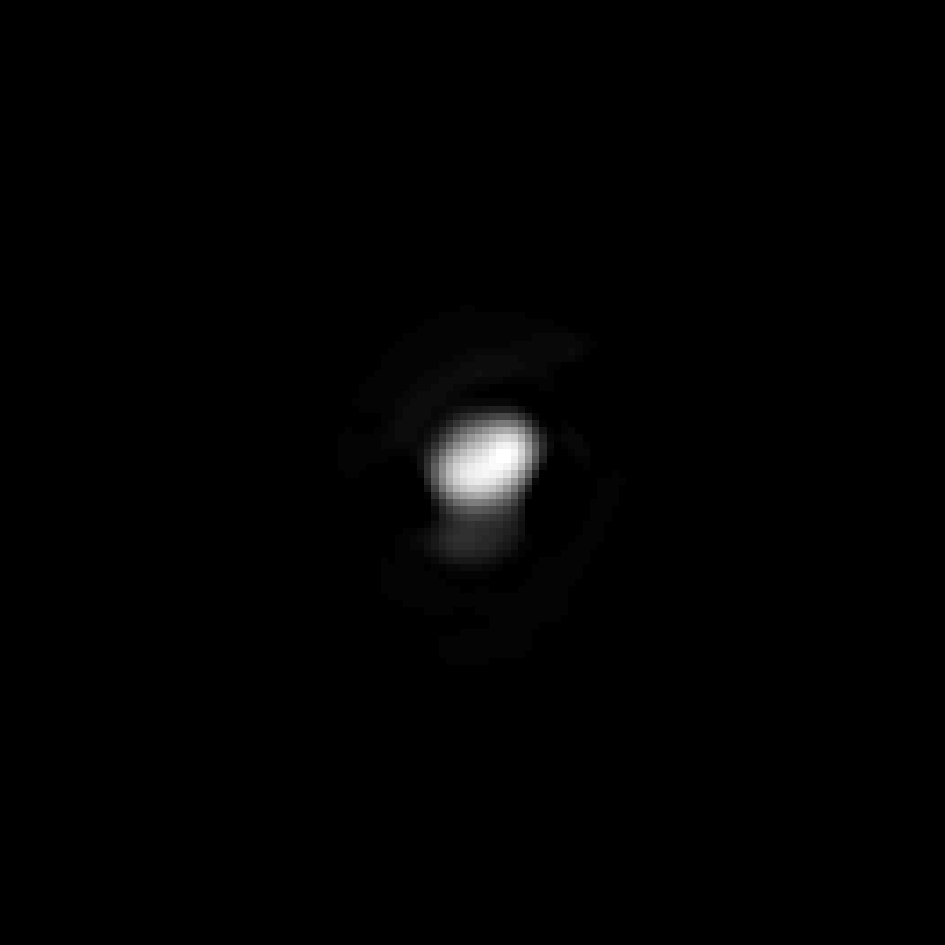} 
    \end{subfigure}%
     \begin{subfigure}[b]{0.16\linewidth}
     \includegraphics[clip=true,trim=90 90 80 80,scale=0.66]{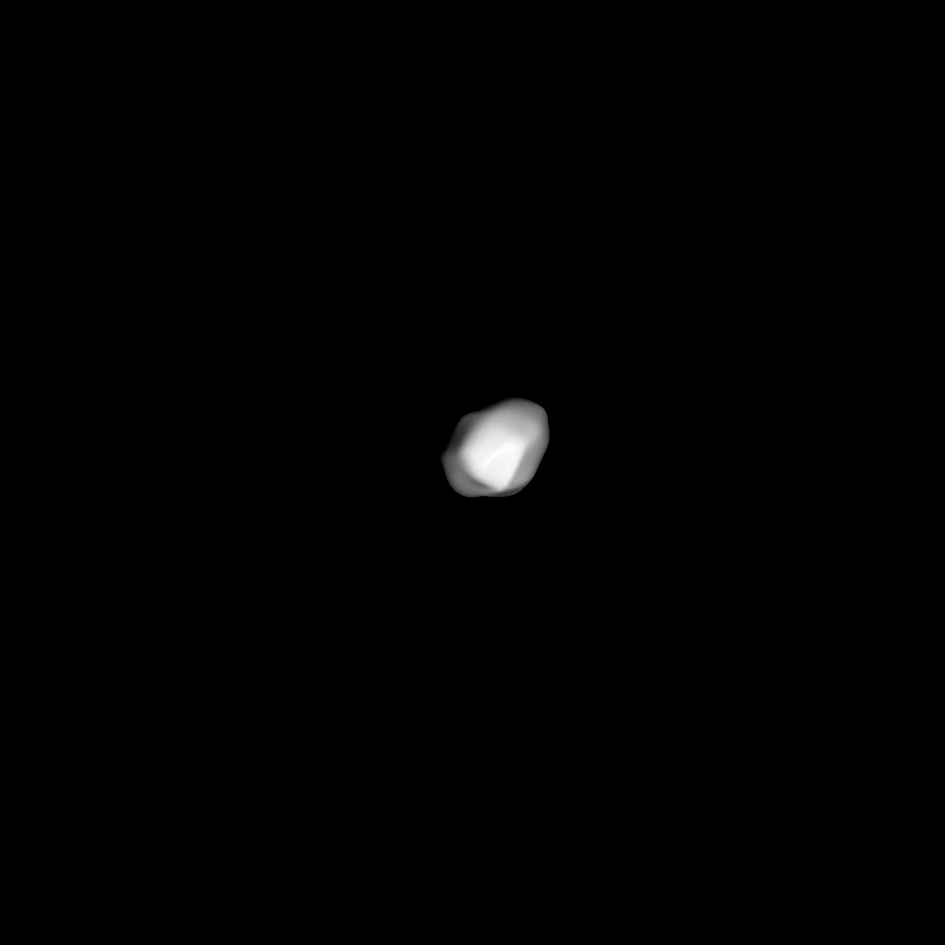}
     \end{subfigure}%
     \begin{subfigure}[b]{0.16\linewidth}
      \includegraphics[clip=true,trim=85 85 85 85,scale=0.66]{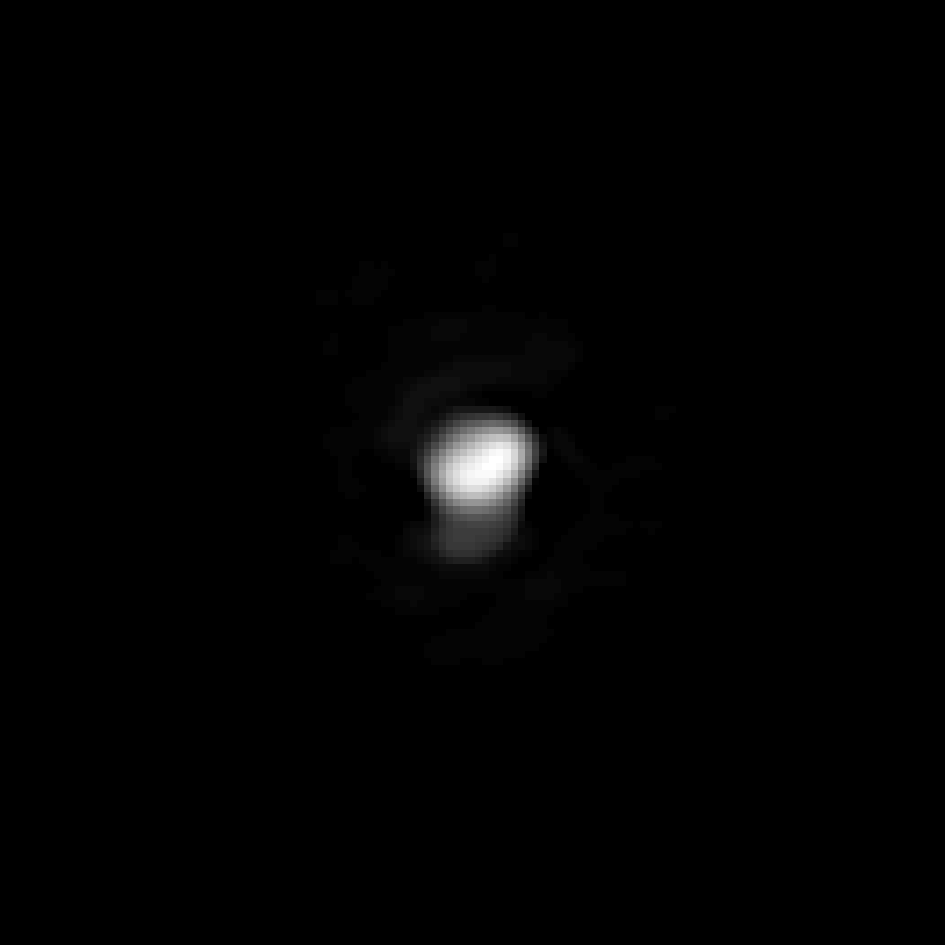} 
    \end{subfigure}%
     \begin{subfigure}[b]{0.16\linewidth}
     \includegraphics[clip=true,trim=90 90 80 80,scale=0.66]{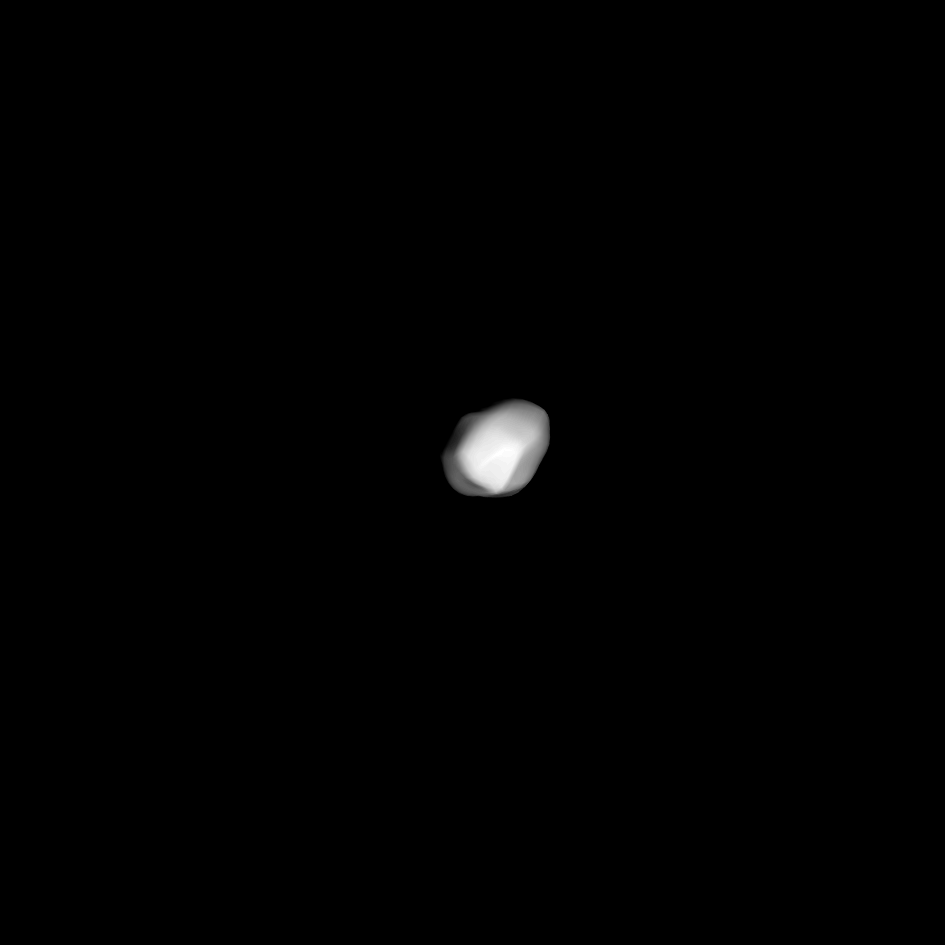}
     \end{subfigure}%
     \begin{subfigure}[b]{0.16\linewidth}
      \includegraphics[clip=true,trim=85 85 85 85,scale=0.66]{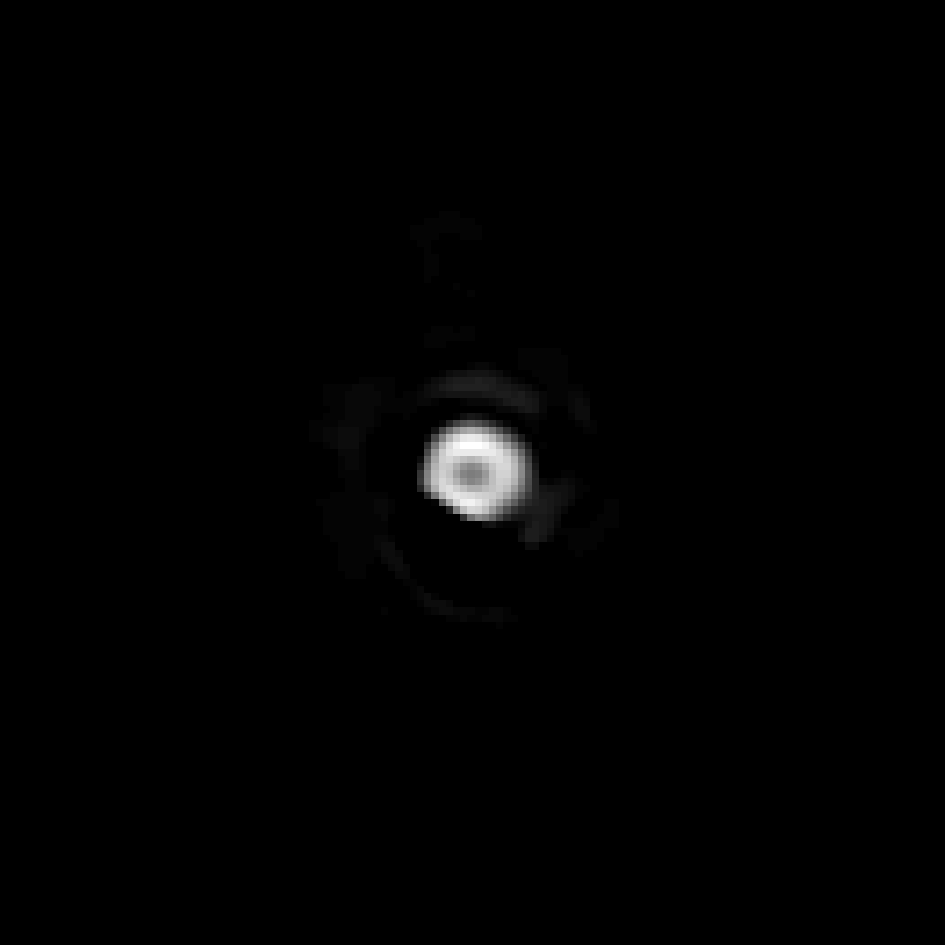} 
    \end{subfigure}%
     \begin{subfigure}[b]{0.16\linewidth}
     \includegraphics[clip=true,trim=90 90 80 80,scale=0.66]{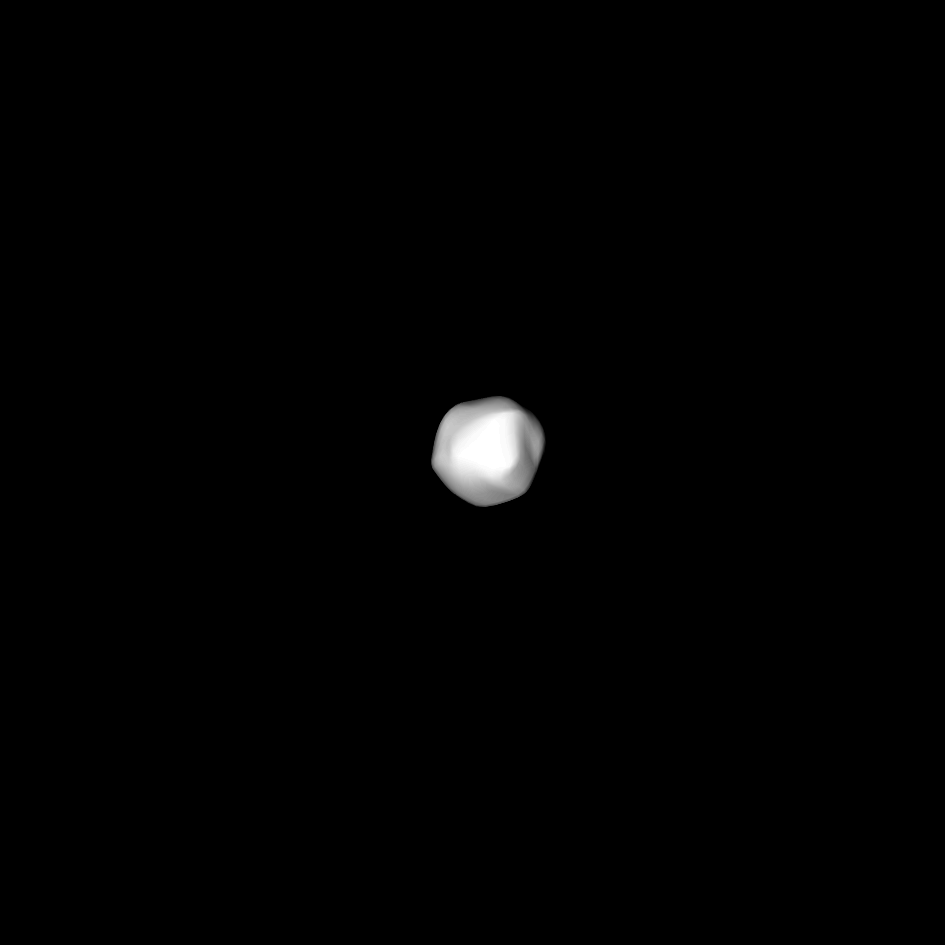}
     \end{subfigure}%
     
     \begin{subfigure}[b]{0.16\linewidth}
     \includegraphics[clip=true,trim=85 85 85 85,scale=0.66]{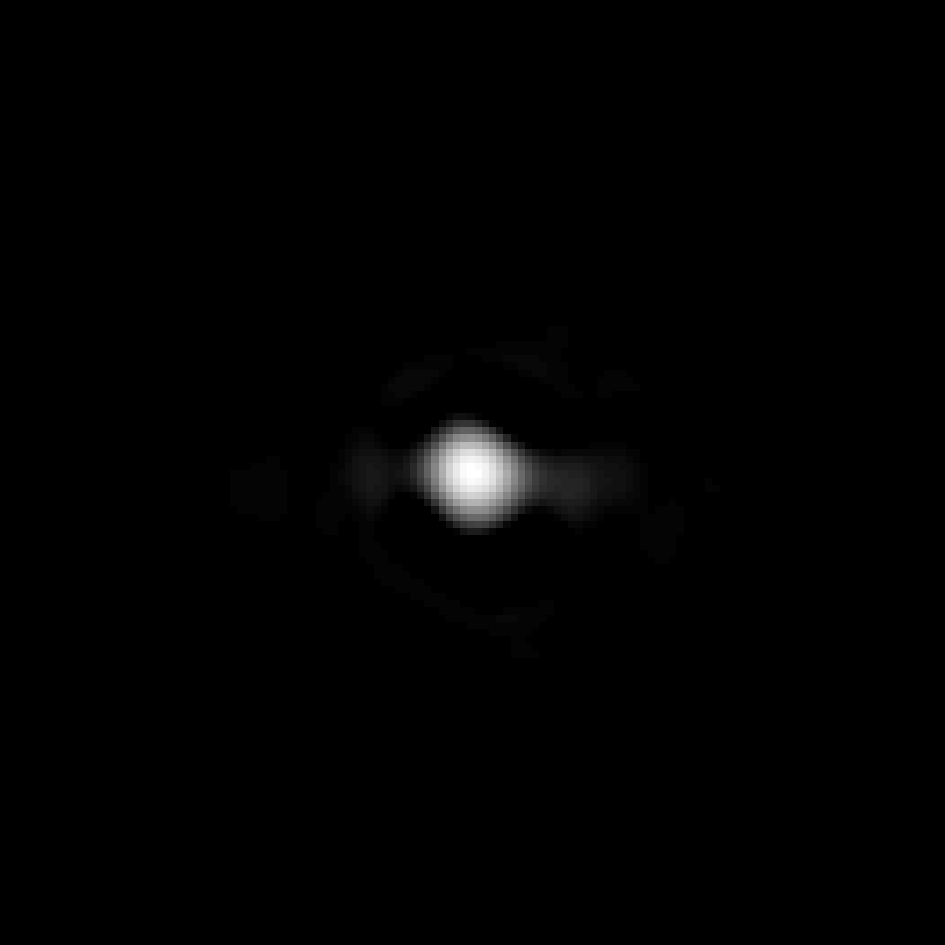}
     \end{subfigure}%
     \begin{subfigure}[b]{0.16\linewidth}
     \includegraphics[clip=true,trim=90 90 80 80,scale=0.66]{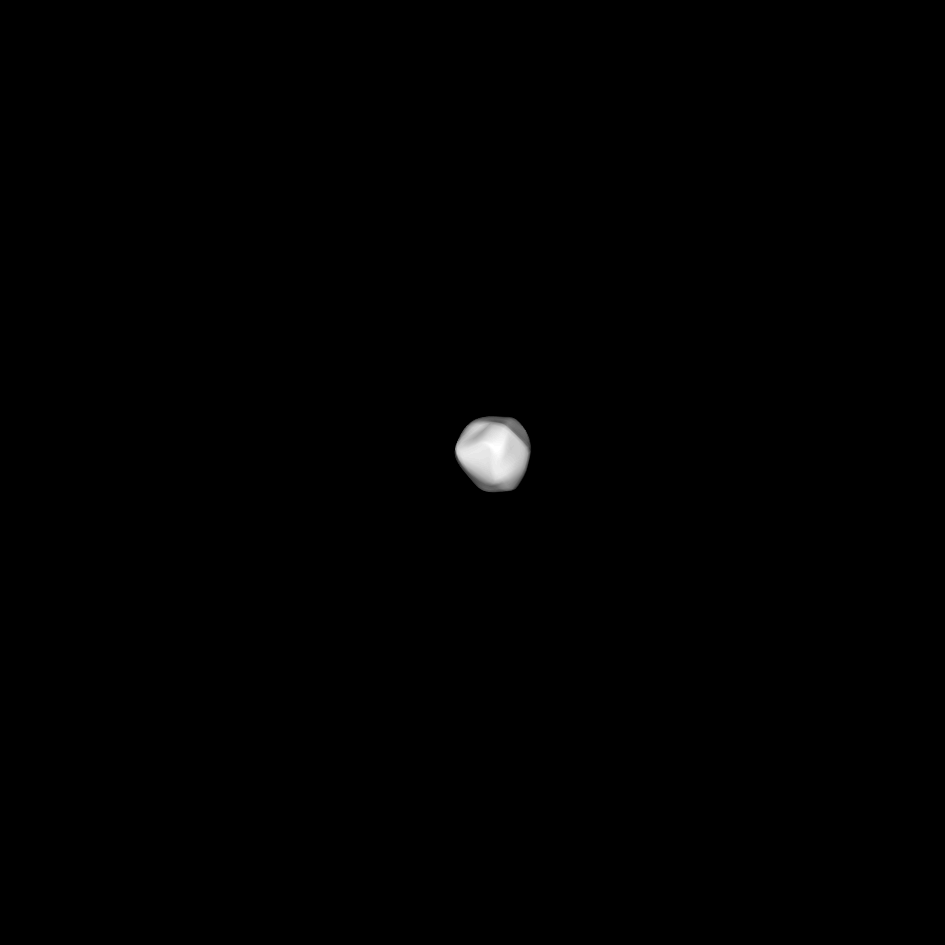}
     \end{subfigure}%
     \caption{\label{fig65}65 Cybele}
    \end{figure}
    \begin{figure}[t]
     \begin{subfigure}[b]{0.16\linewidth}
      \includegraphics[clip=true,trim=75 75 95 95,scale=0.66]{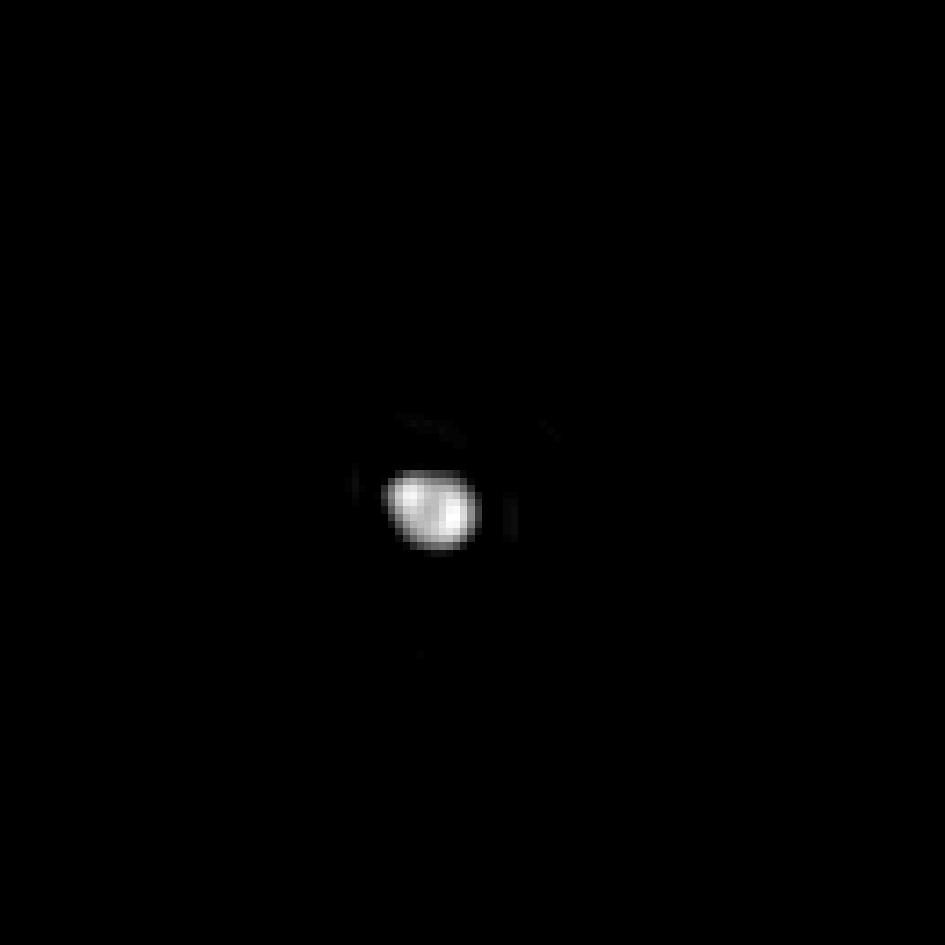} 
    \end{subfigure}%
     \begin{subfigure}[b]{0.16\linewidth}
     \includegraphics[clip=true,trim=90 90 80 80,scale=0.66]{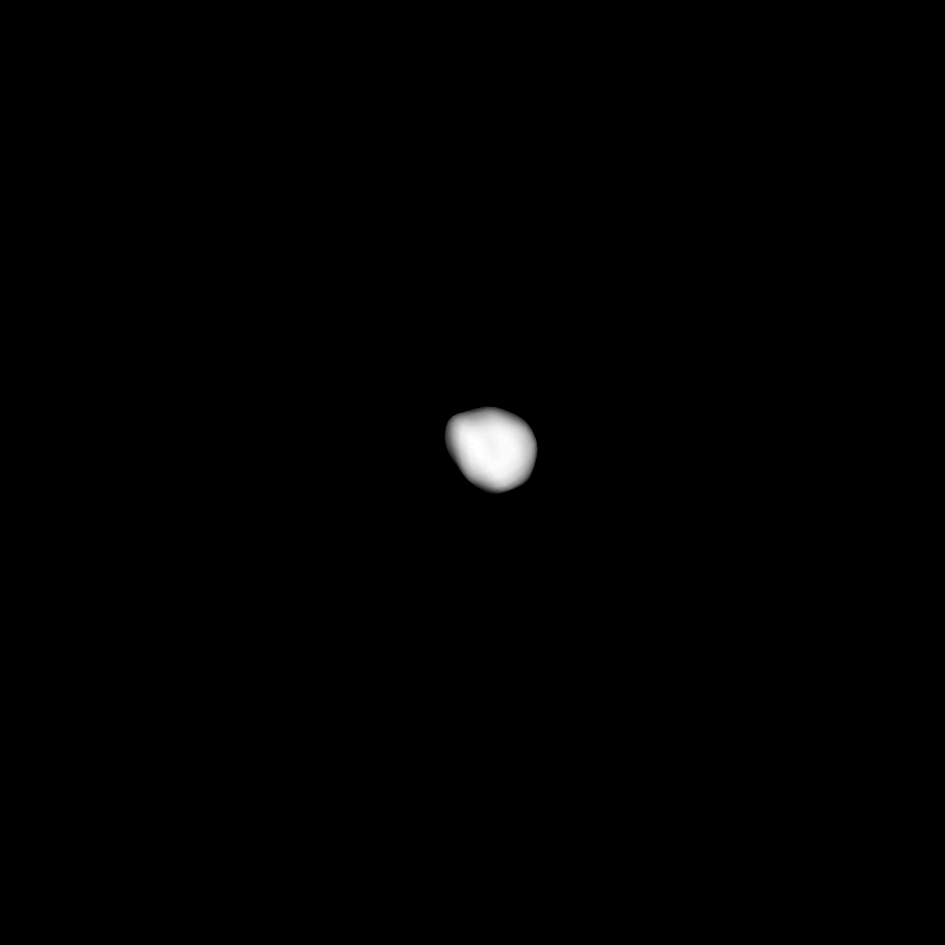}
     \end{subfigure}%
      \begin{subfigure}[b]{0.16\linewidth}
      \includegraphics[clip=true,trim=85 80 85 90,scale=0.66]{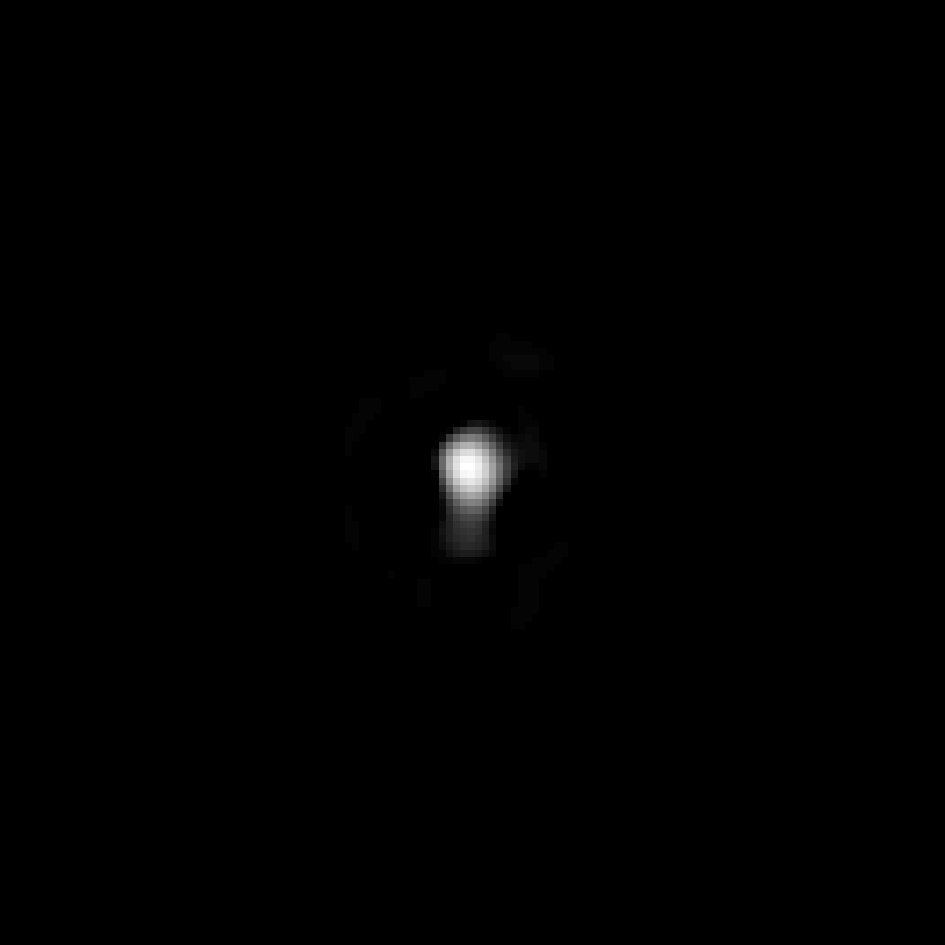}
    \end{subfigure}%
     \begin{subfigure}[b]{0.16\linewidth}
      \includegraphics[clip=true,trim=90 90 80 80,scale=0.66]{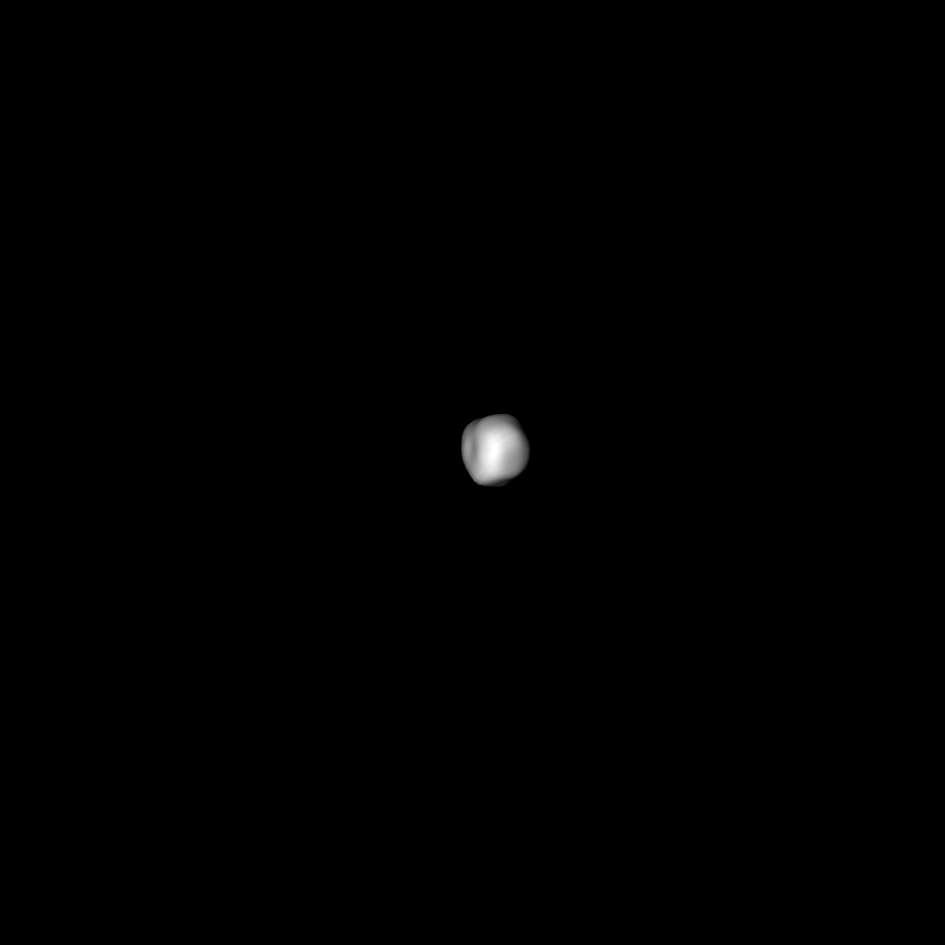}
    \end{subfigure}%
     \caption{\label{fig72}72 Feronia}
\end{figure}

    \begin{figure}[t]
     \begin{subfigure}[b]{0.16\linewidth}
      \includegraphics[clip=true,trim=85 85 85 85,scale=0.66]{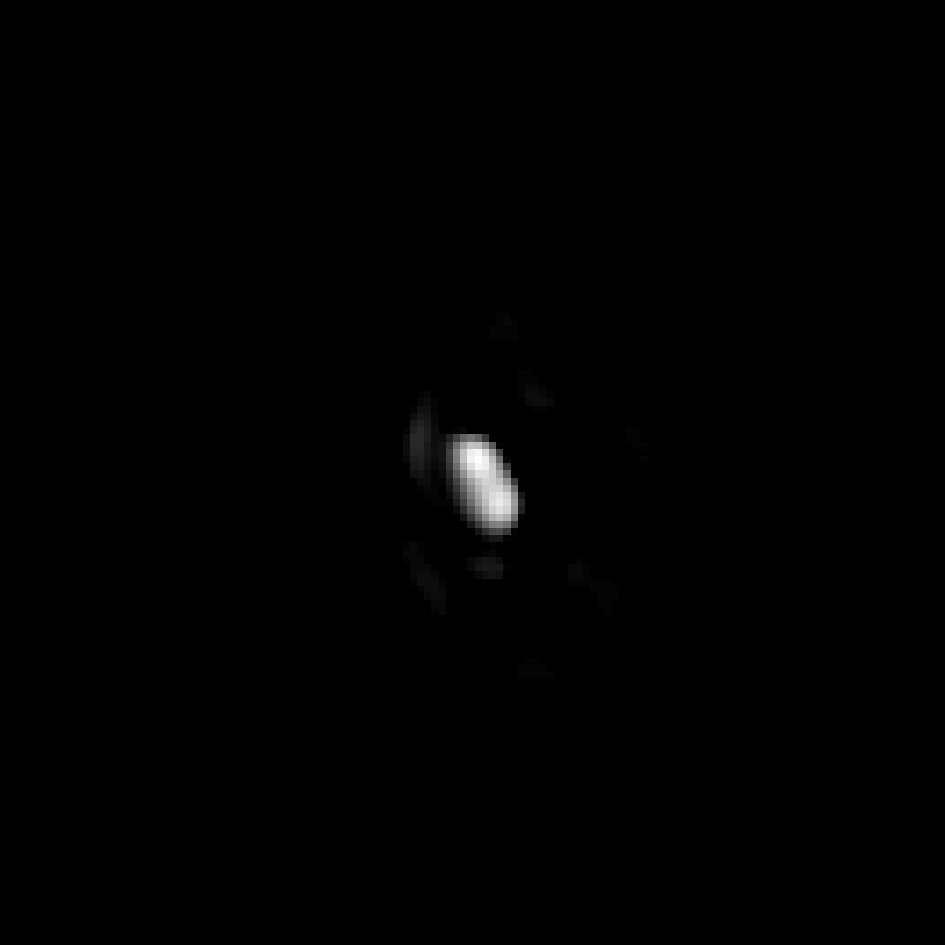} 
    \end{subfigure}%
     \begin{subfigure}[b]{0.16\linewidth}
     \includegraphics[clip=true,trim=90 90 80 80,scale=0.66]{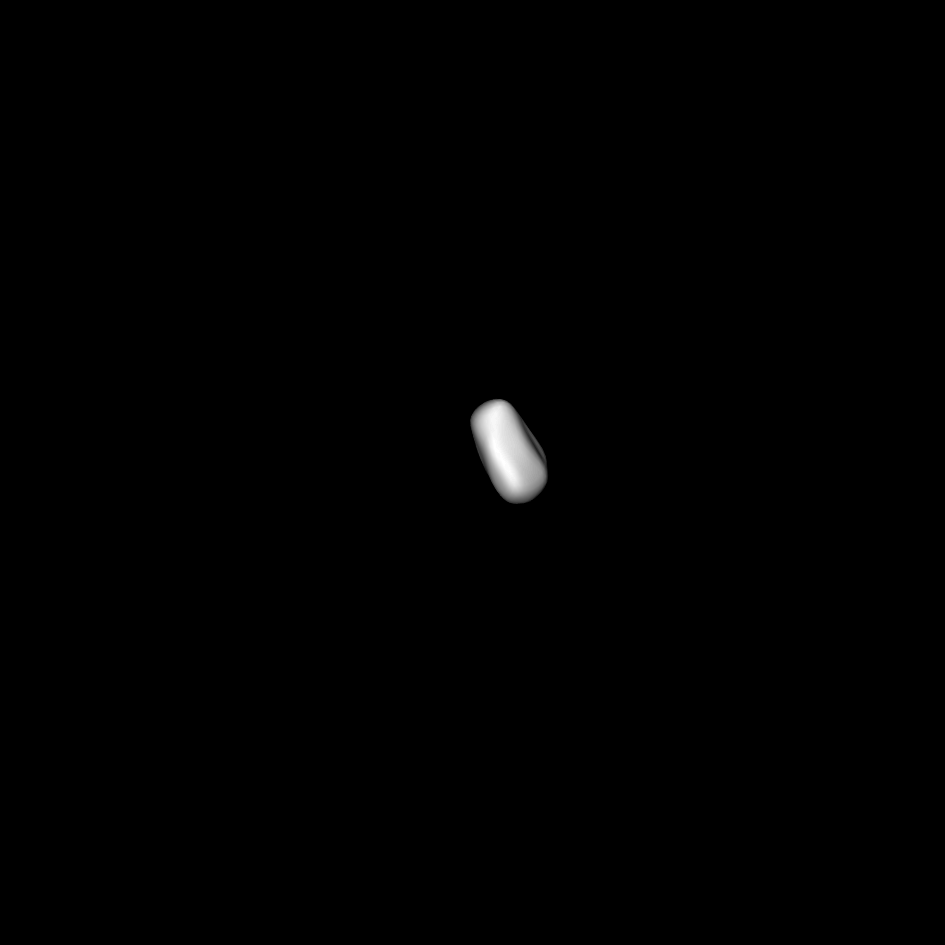}
     \end{subfigure}%
      \begin{subfigure}[b]{0.16\linewidth}
      \includegraphics[clip=true,trim=85 85 85 85,scale=0.66]{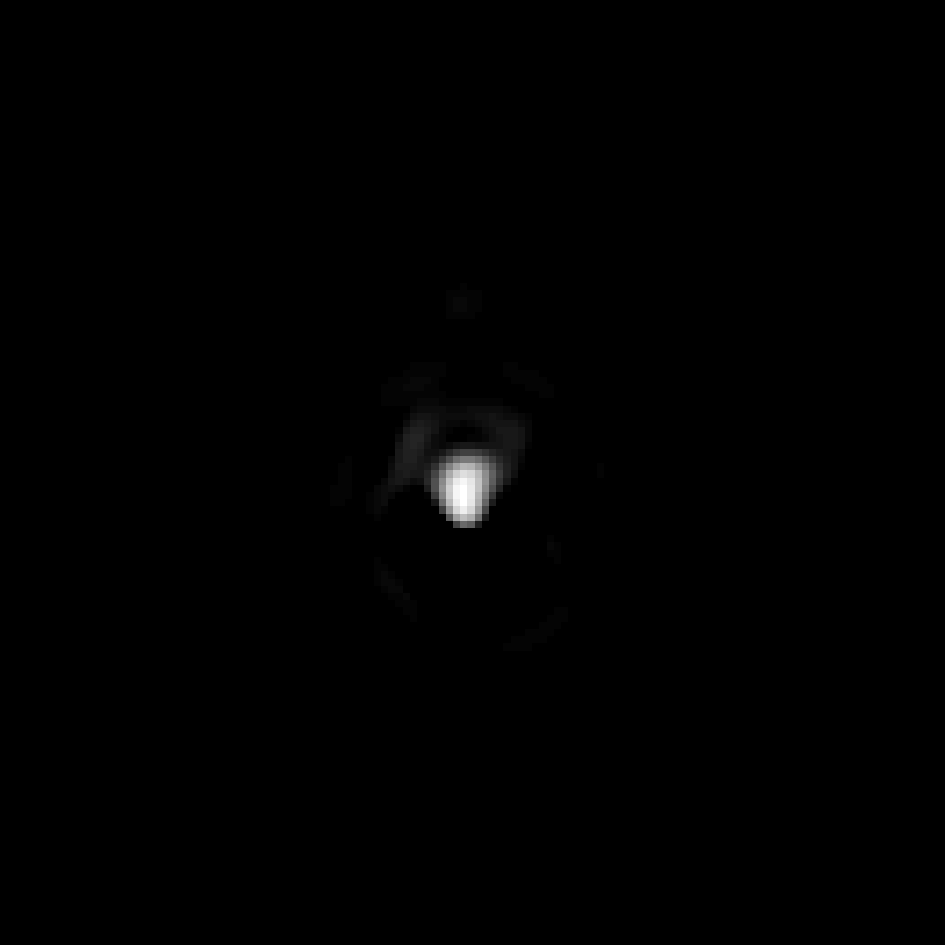}
    \end{subfigure}%
     \begin{subfigure}[b]{0.16\linewidth}
      \includegraphics[clip=true,trim=90 90 80 80,scale=0.66]{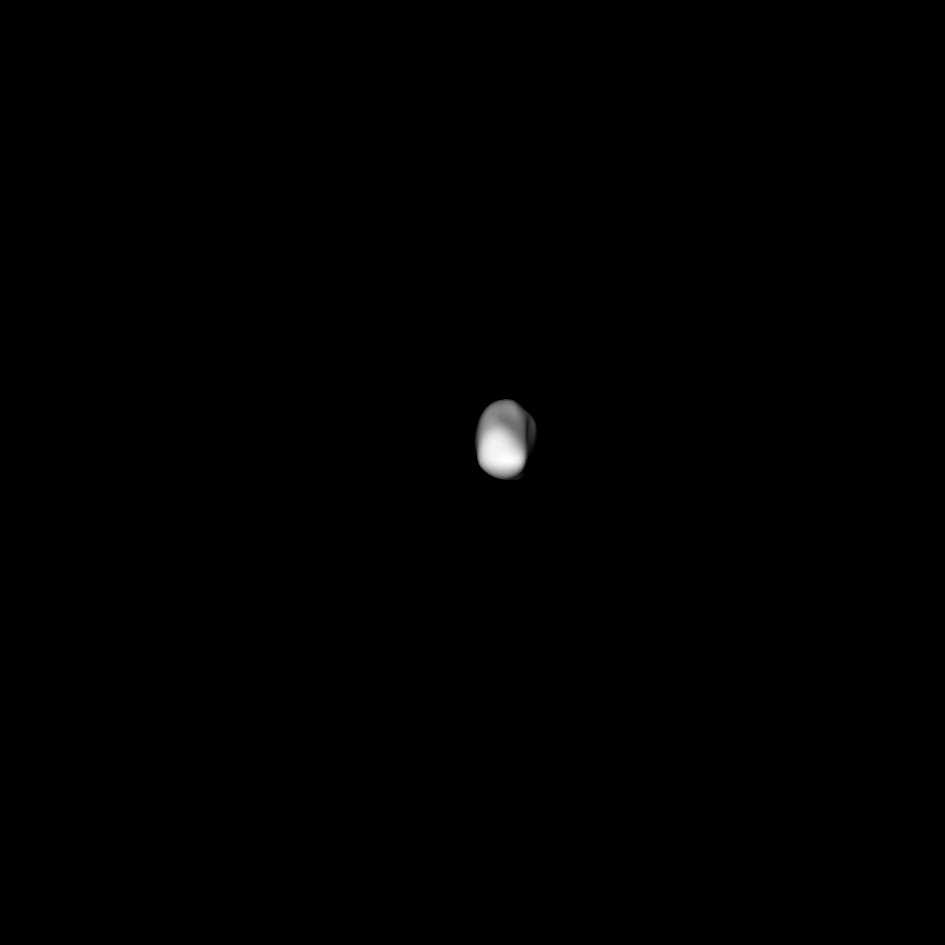}
    \end{subfigure}%
    \begin{subfigure}[b]{0.16\linewidth}
      \includegraphics[clip=true,trim=85 85 85 85,scale=0.66]{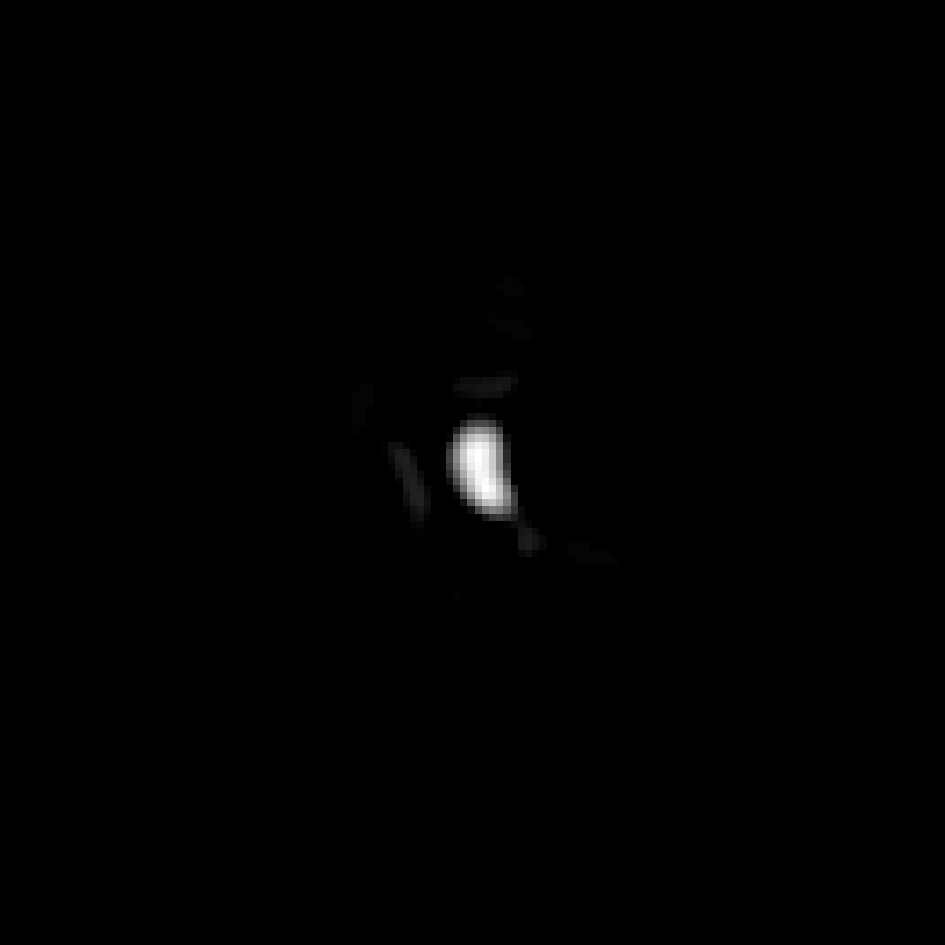} 
    \end{subfigure}%
     \begin{subfigure}[b]{0.16\linewidth}
     \includegraphics[clip=true,trim=90 90 80 80,scale=0.66]{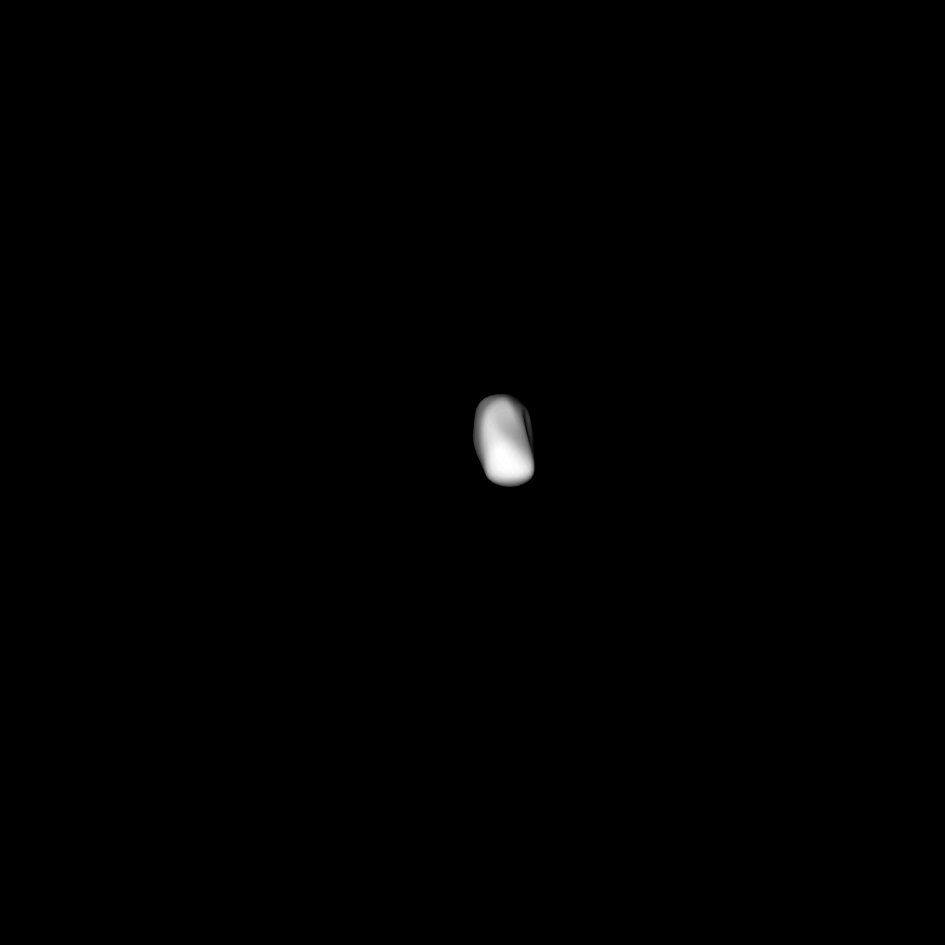}
     \end{subfigure}%
     
      \begin{subfigure}[b]{0.16\linewidth}
      \includegraphics[clip=true,trim=85 85 85 85,scale=0.66]{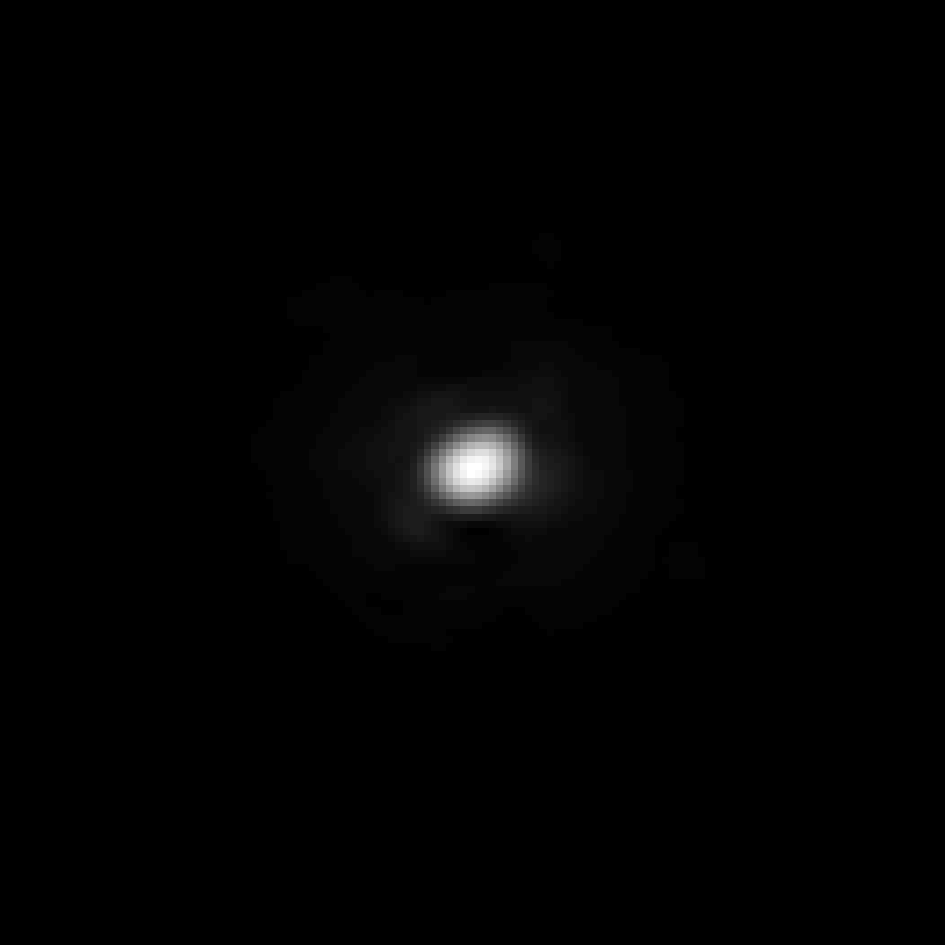} 
    \end{subfigure}%
     \begin{subfigure}[b]{0.16\linewidth}
     \includegraphics[clip=true,trim=90 90 80 80,scale=0.66]{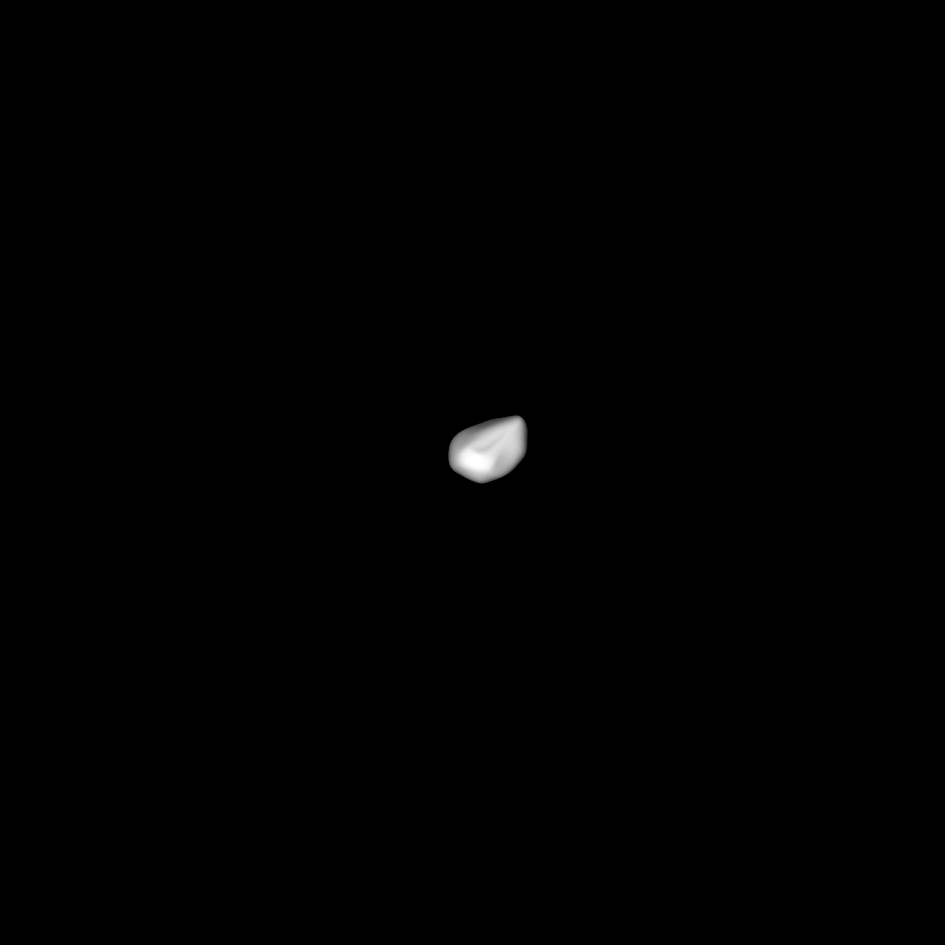}
     \end{subfigure}%
     \begin{subfigure}[b]{0.16\linewidth}
      \includegraphics[clip=true,trim=85 85 85 85,scale=0.66]{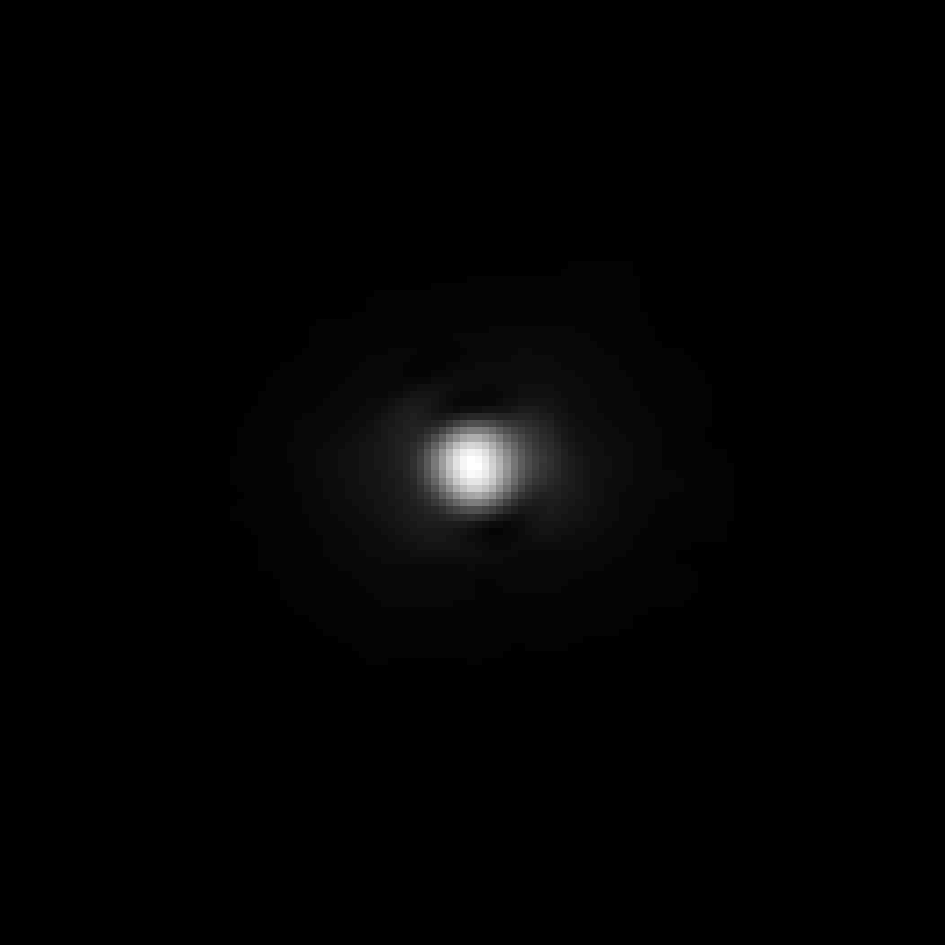} 
    \end{subfigure}%
     \begin{subfigure}[b]{0.16\linewidth}
     \includegraphics[clip=true,trim=90 90 80 80,scale=0.66]{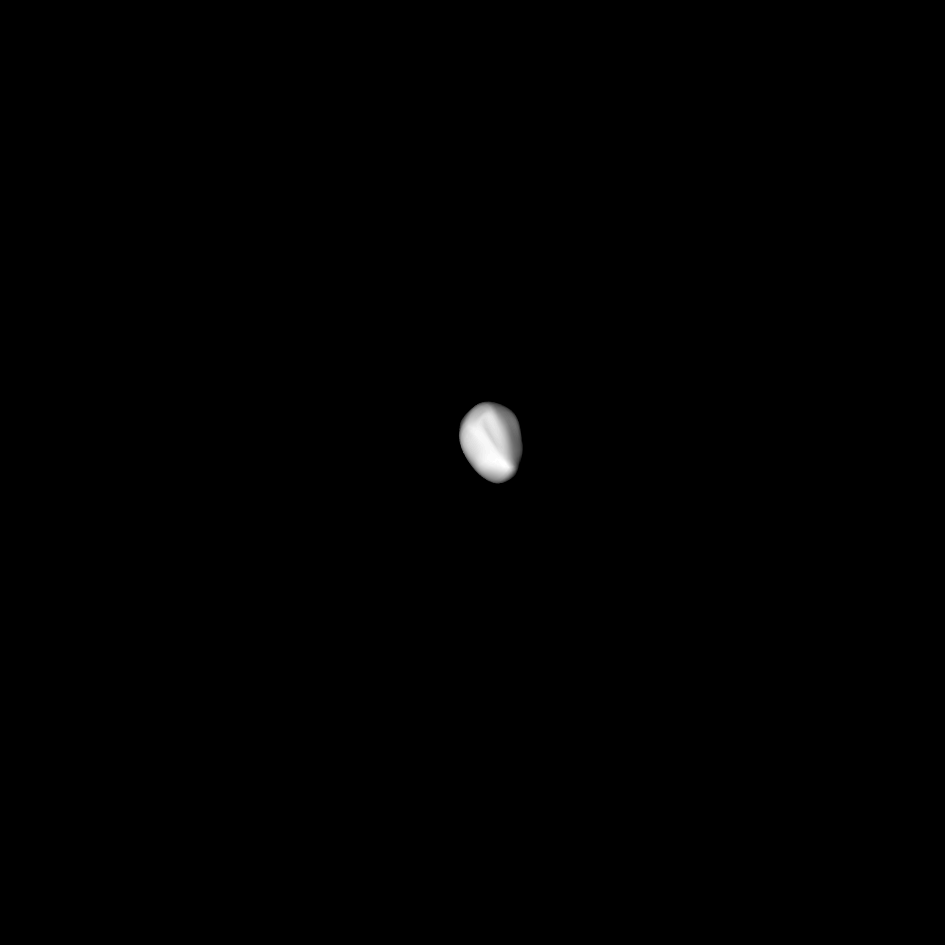}
     \end{subfigure}%
     \begin{subfigure}[b]{0.16\linewidth}
      \includegraphics[clip=true,trim=85 85 85 85,scale=0.66]{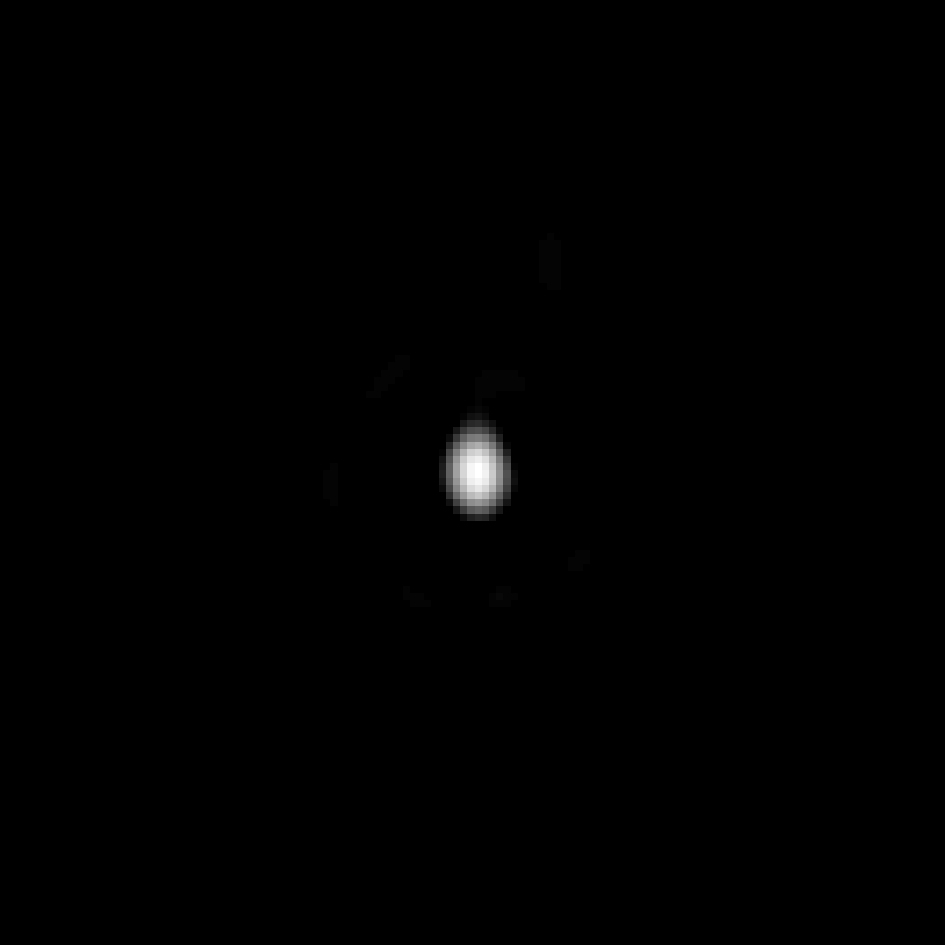} 
    \end{subfigure}%
     \begin{subfigure}[b]{0.16\linewidth}
     \includegraphics[clip=true,trim=90 90 80 80,scale=0.66]{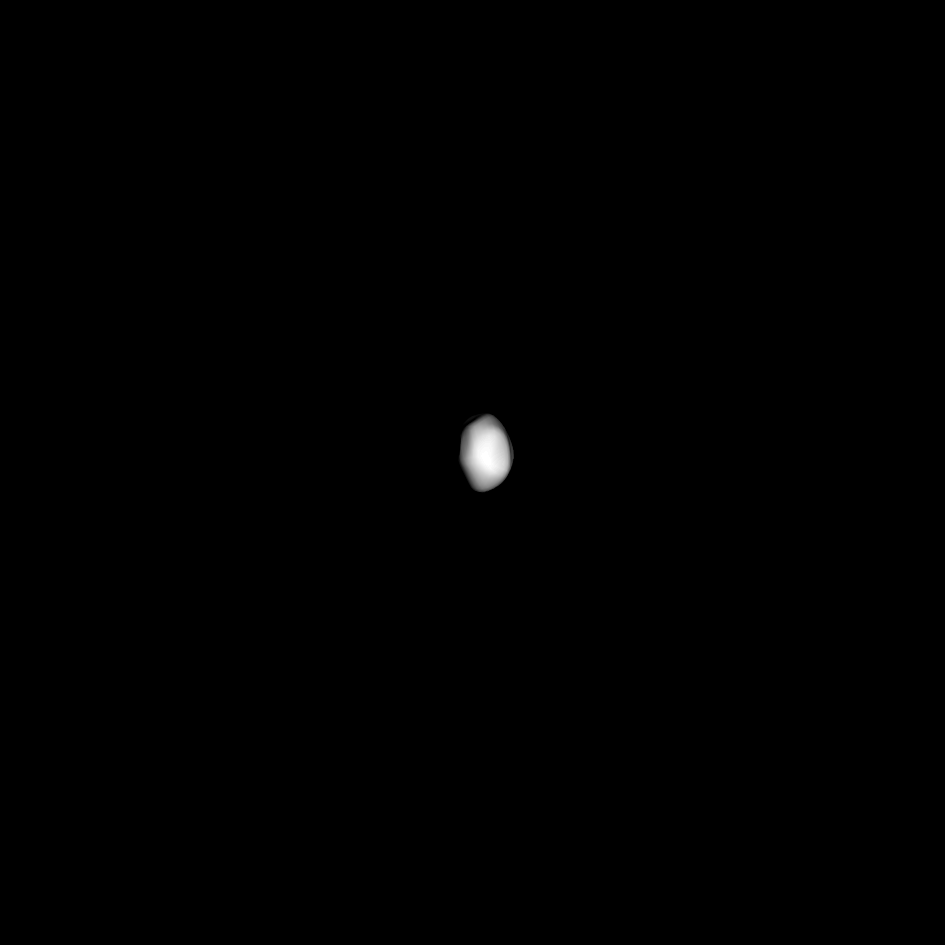}
     \end{subfigure}%
     
     \begin{subfigure}[b]{0.16\linewidth}
     \includegraphics[clip=true,trim=85 85 85 85,scale=0.66]{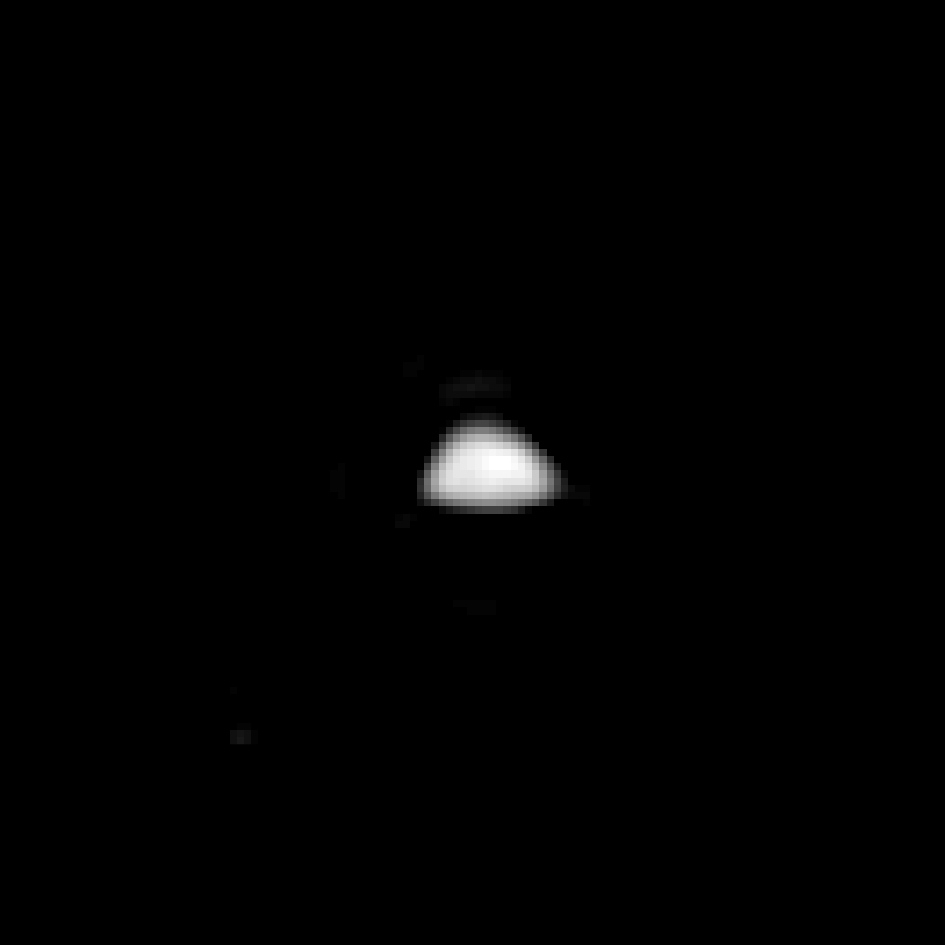}
     \end{subfigure}%
     \begin{subfigure}[b]{0.16\linewidth}
     \includegraphics[clip=true,trim=90 90 80 80,scale=0.66]{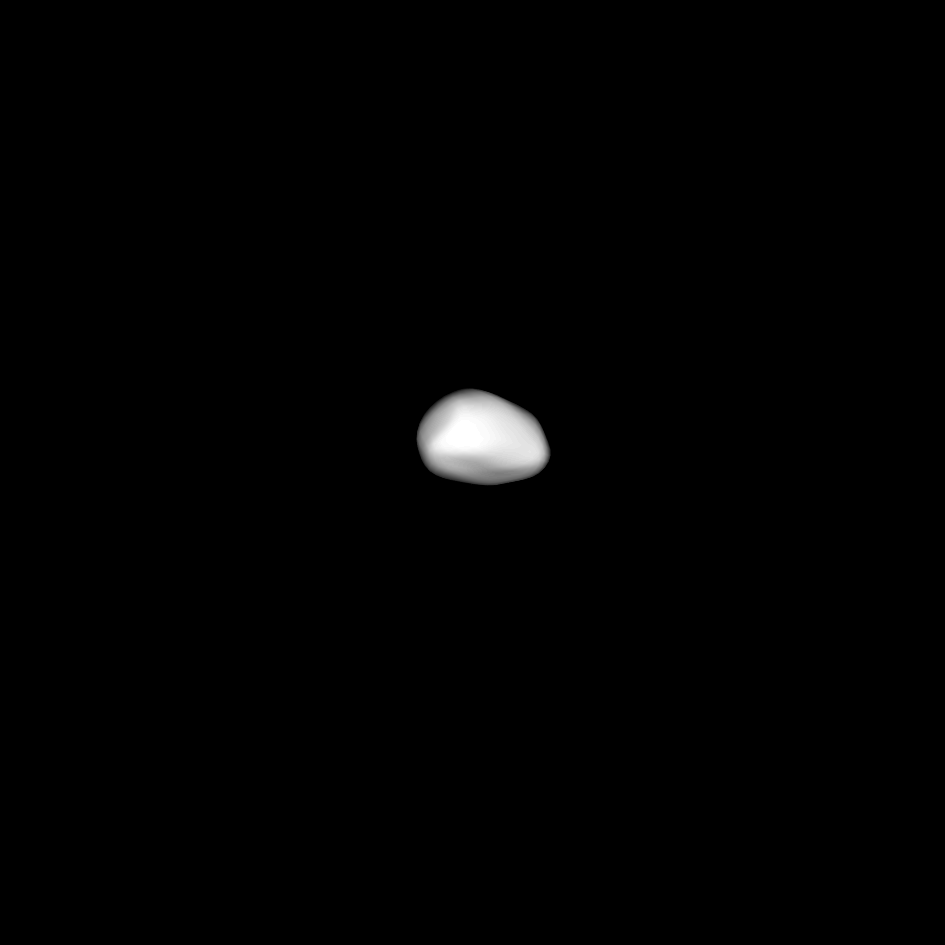}
     \end{subfigure}%
     \begin{subfigure}[b]{0.16\linewidth}
     \includegraphics[clip=true,trim=85 85 85 85,scale=0.66]{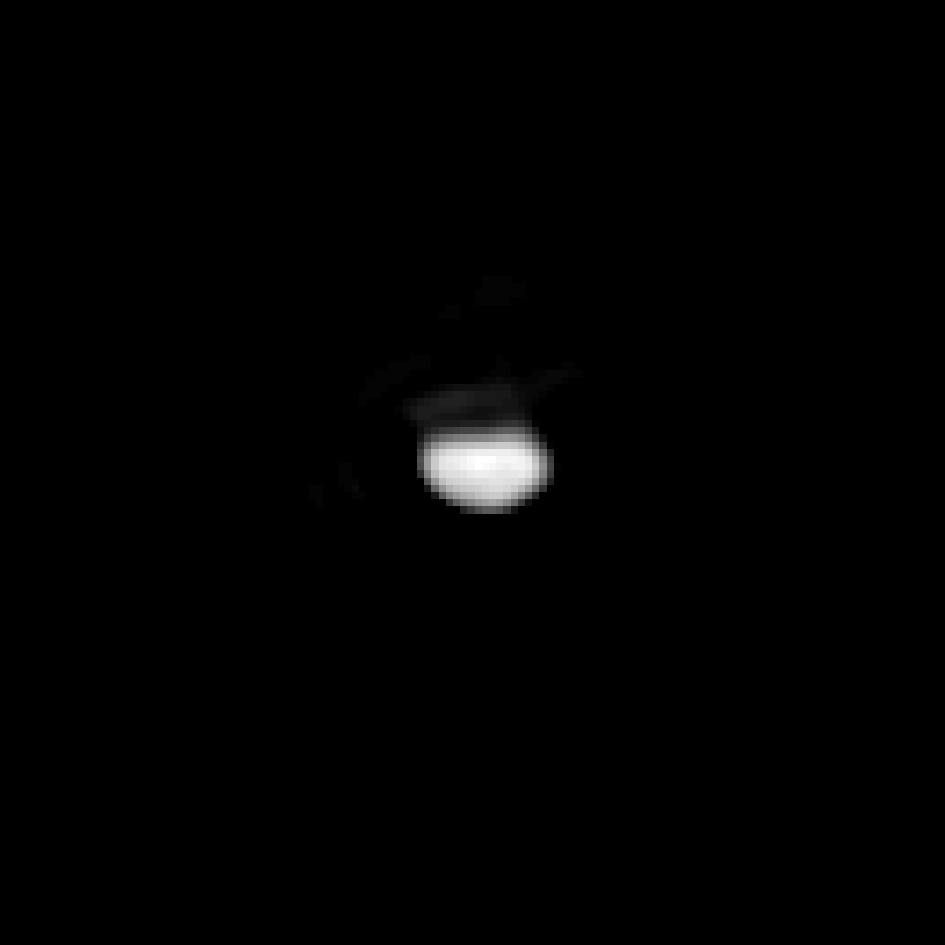}
     \end{subfigure}%
     \begin{subfigure}[b]{0.16\linewidth}
     \includegraphics[clip=true,trim=90 90 80 80,scale=0.66]{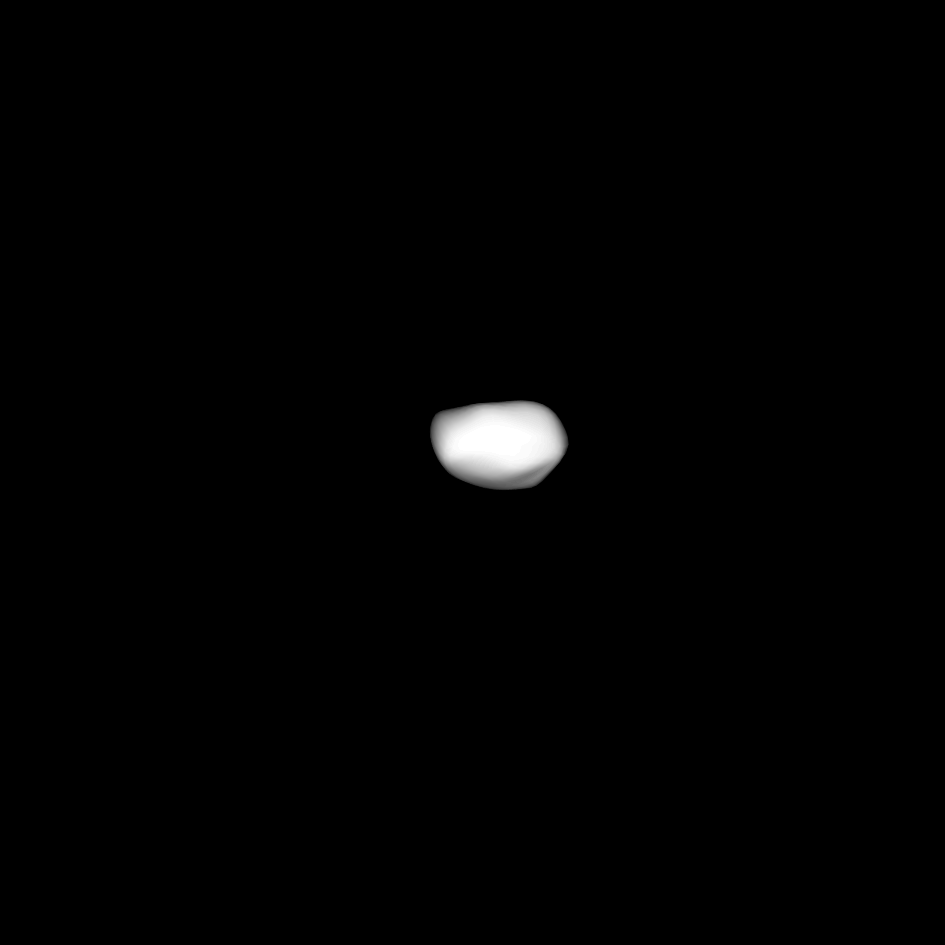}
     \end{subfigure}%
     \begin{subfigure}[b]{0.16\linewidth}
     \includegraphics[clip=true,trim=85 85 85 85,scale=0.66]{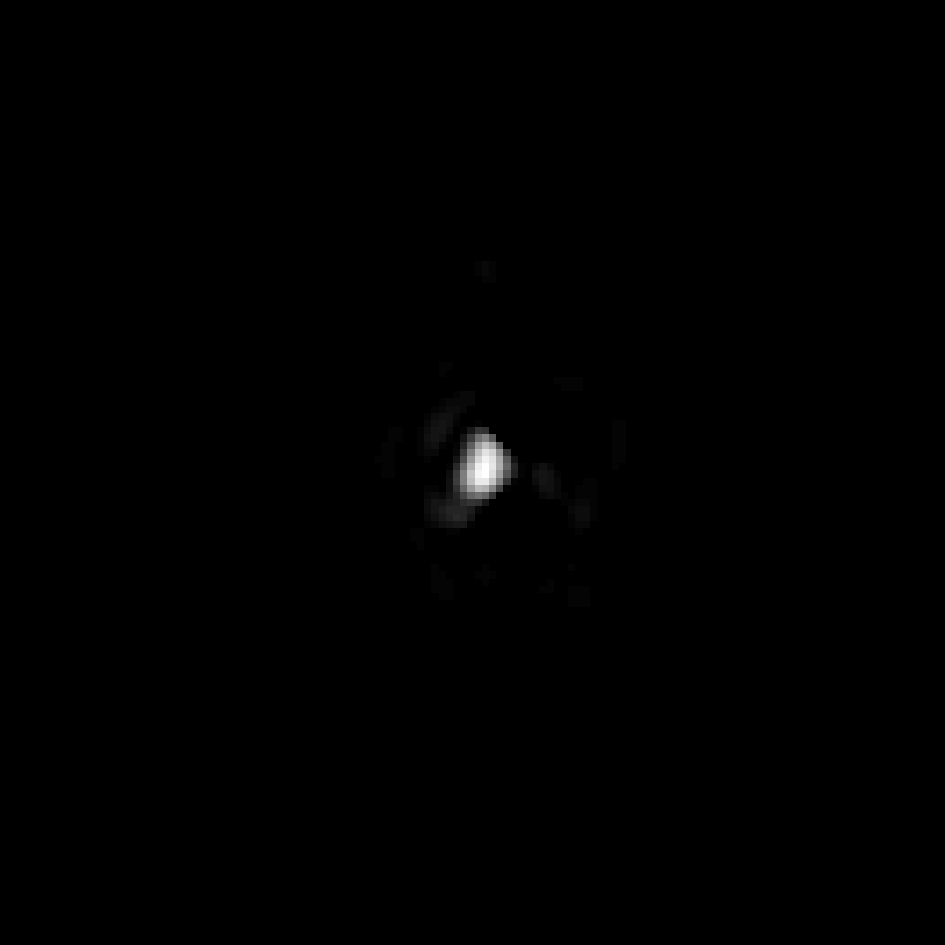}
     \end{subfigure}%
     \begin{subfigure}[b]{0.16\linewidth}
     \includegraphics[clip=true,trim=90 90 80 80,scale=0.66]{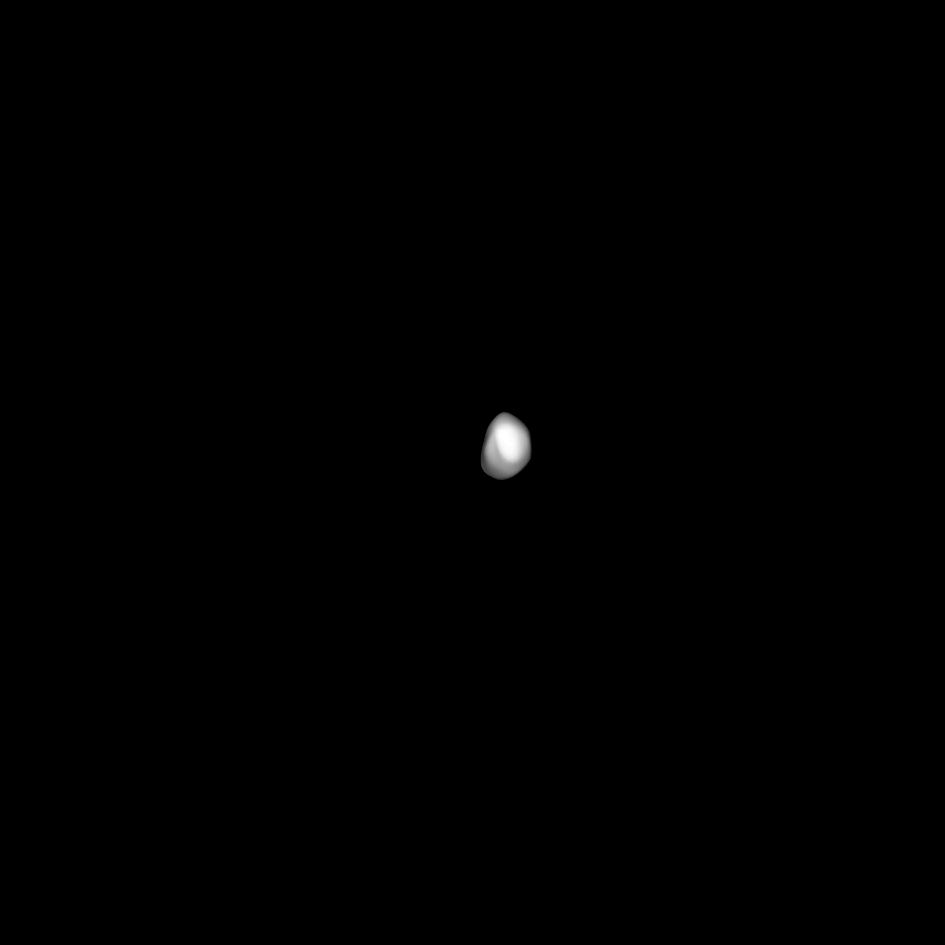}
     \end{subfigure}%
     
     \begin{subfigure}[b]{0.16\linewidth}
     \includegraphics[clip=true,trim=85 85 85 85,scale=0.66]{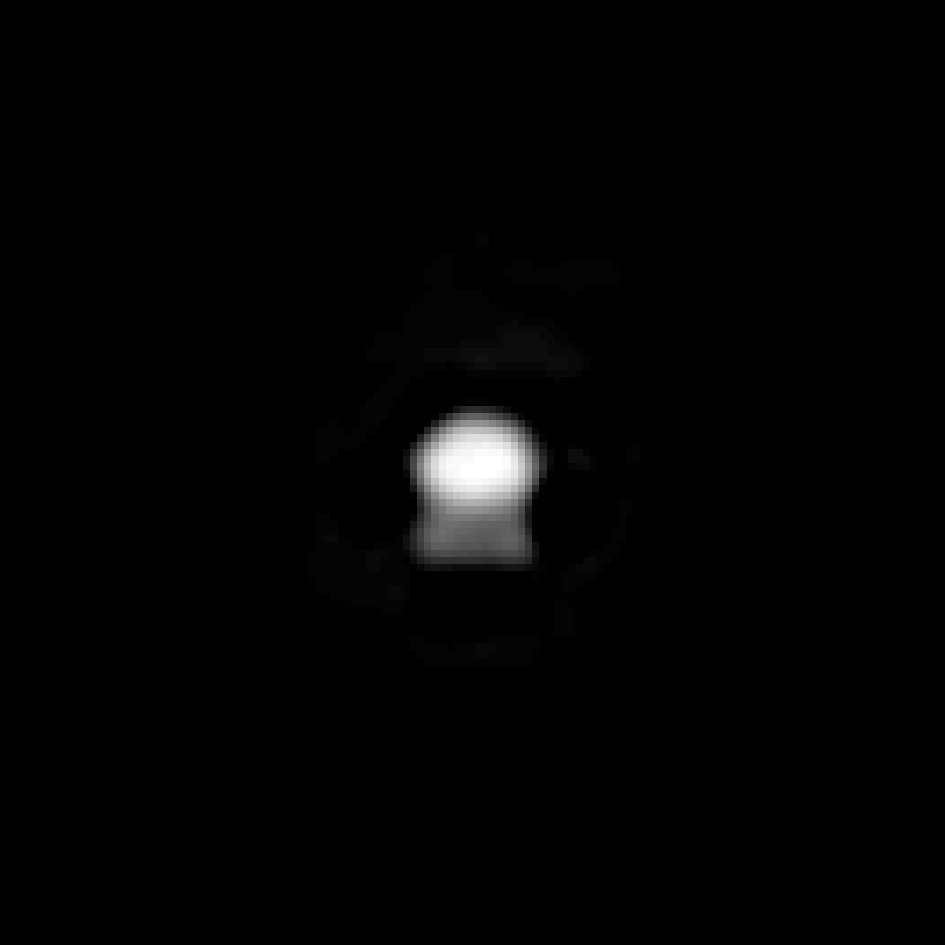}
     \end{subfigure}%
     \begin{subfigure}[b]{0.16\linewidth}
     \includegraphics[clip=true,trim=90 90 80 80,scale=0.66]{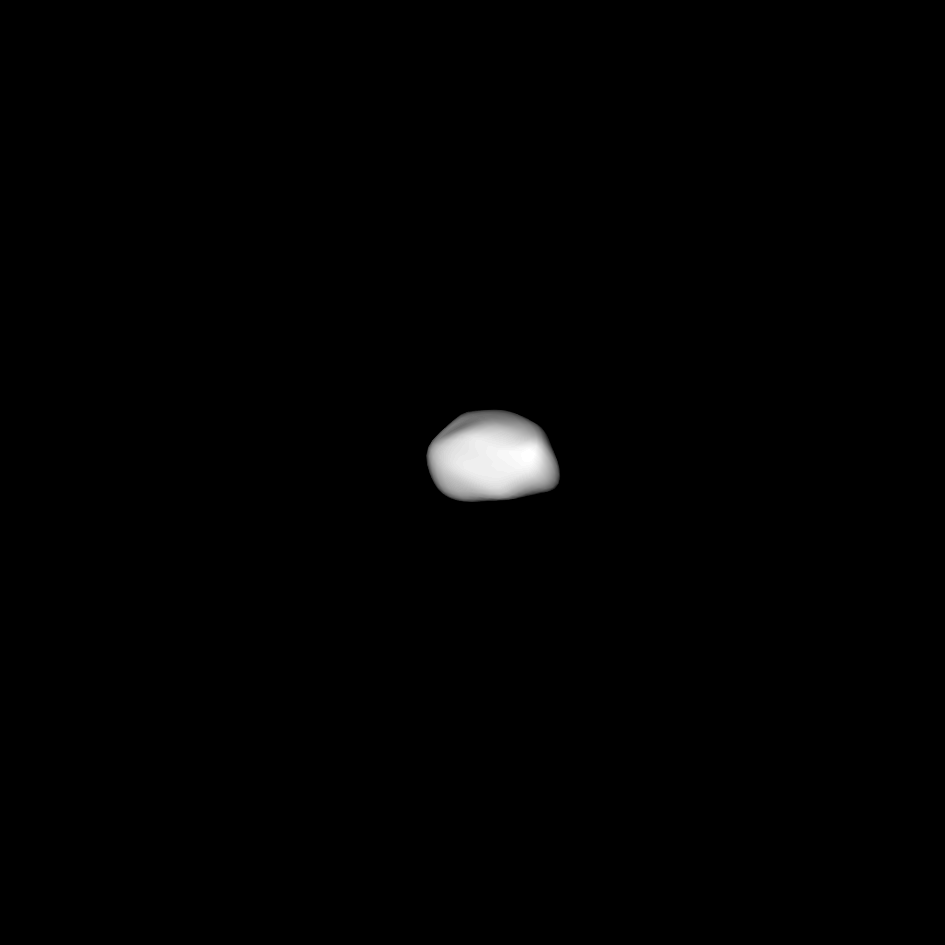}
     \end{subfigure}%
     \begin{subfigure}[b]{0.16\linewidth}
     \includegraphics[clip=true,trim=85 85 85 85,scale=0.66]{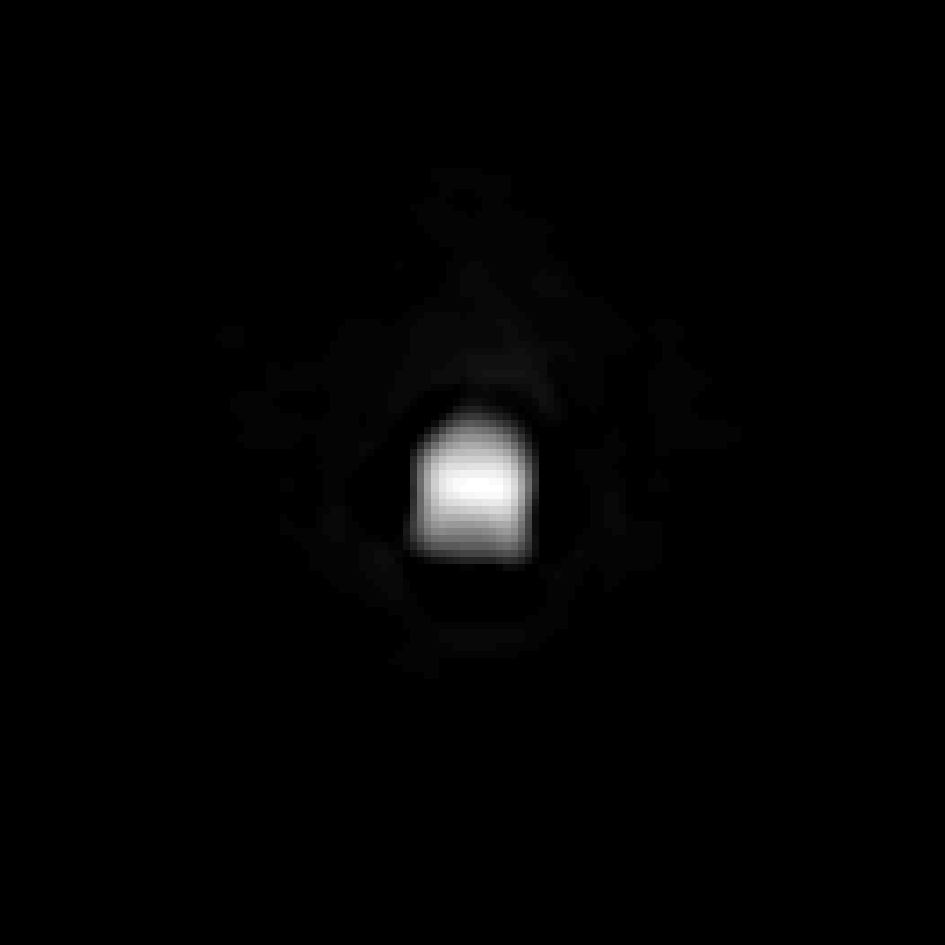}
     \end{subfigure}%
     \begin{subfigure}[b]{0.16\linewidth}
     \includegraphics[clip=true,trim=90 90 80 80,scale=0.66]{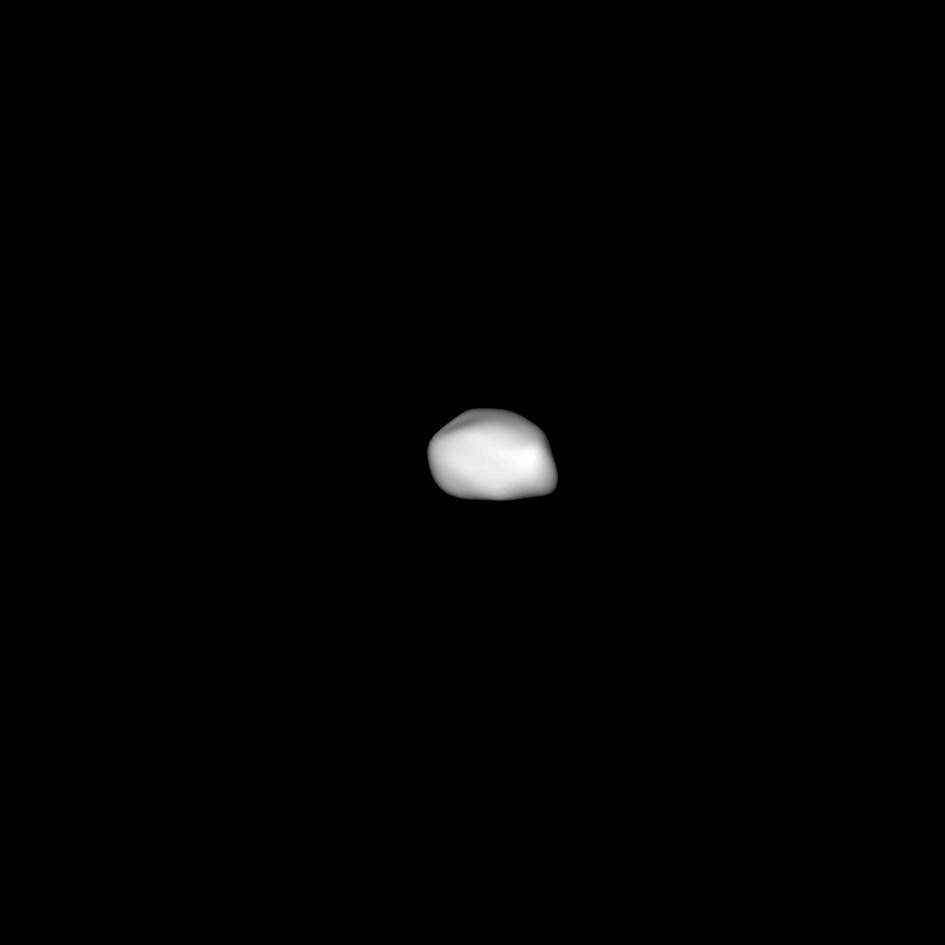}
     \end{subfigure}%
     \begin{subfigure}[b]{0.16\linewidth}
     \includegraphics[clip=true,trim=85 85 85 85,scale=0.66]{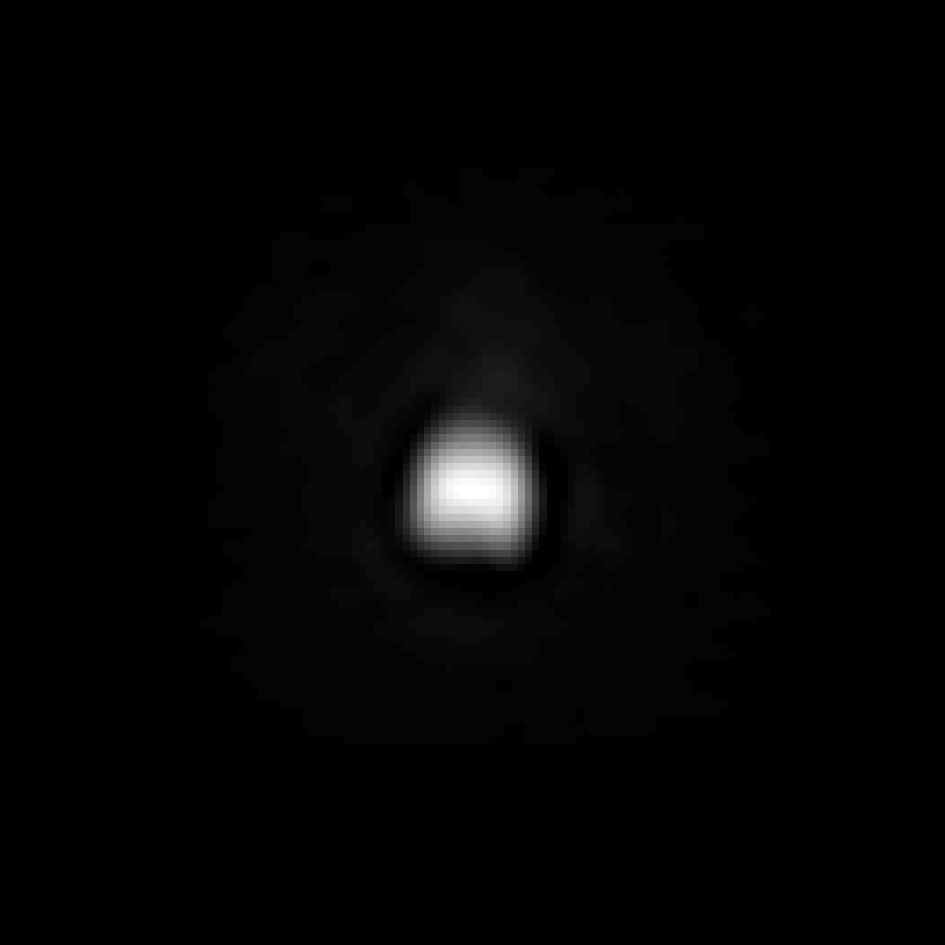}
     \end{subfigure}%
     \begin{subfigure}[b]{0.16\linewidth}
     \includegraphics[clip=true,trim=90 90 80 80,scale=0.66]{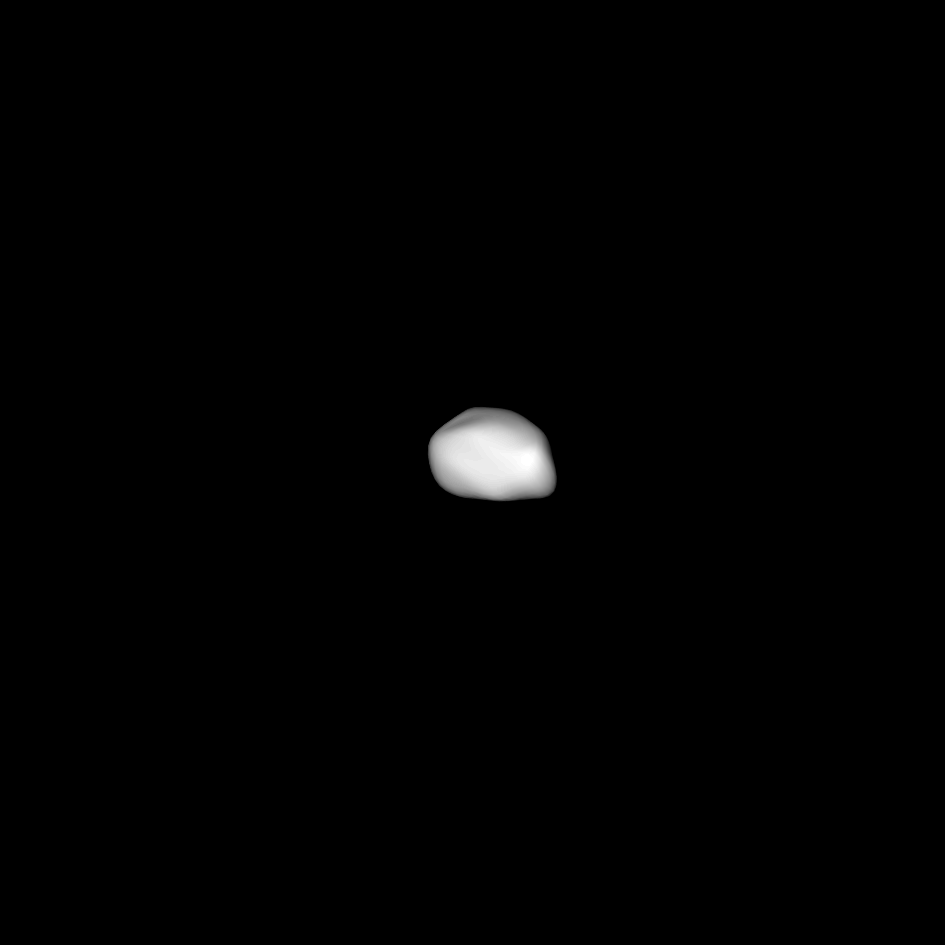}
     \end{subfigure}%
     
     \begin{subfigure}[b]{0.16\linewidth}
     \includegraphics[clip=true,trim=85 85 85 85,scale=0.66]{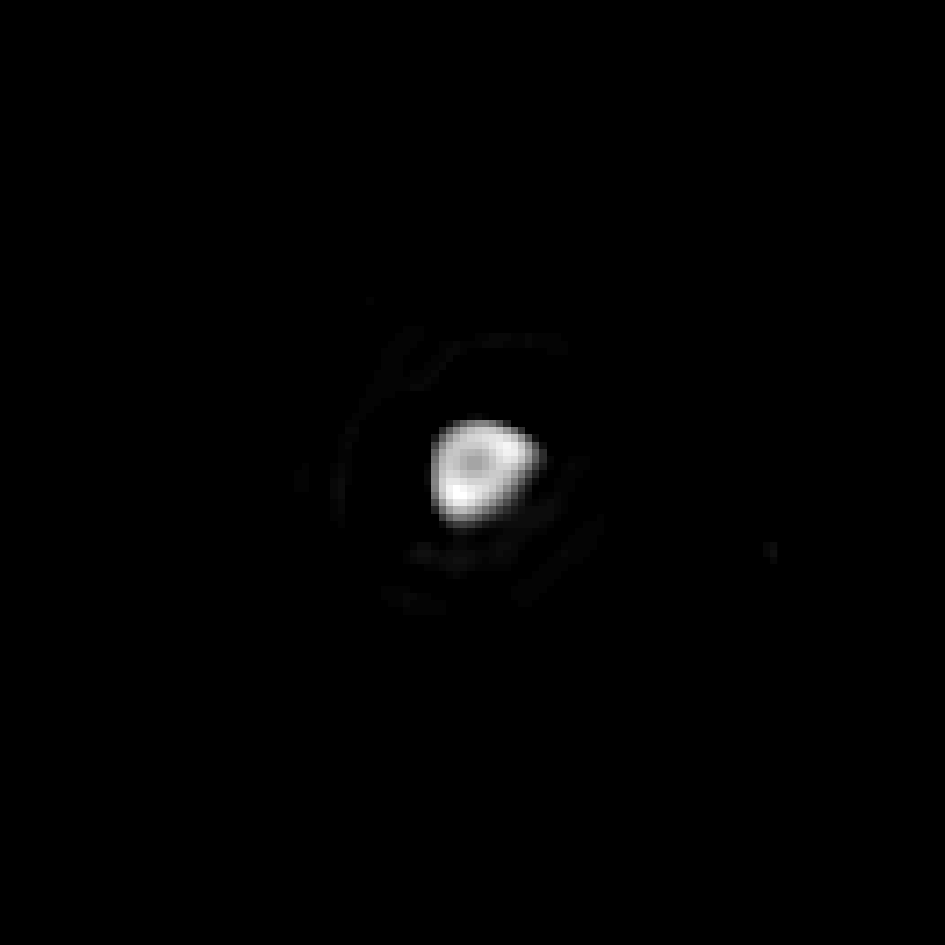}
     \end{subfigure}%
     \begin{subfigure}[b]{0.16\linewidth}
     \includegraphics[clip=true,trim=90 90 80 80,scale=0.66]{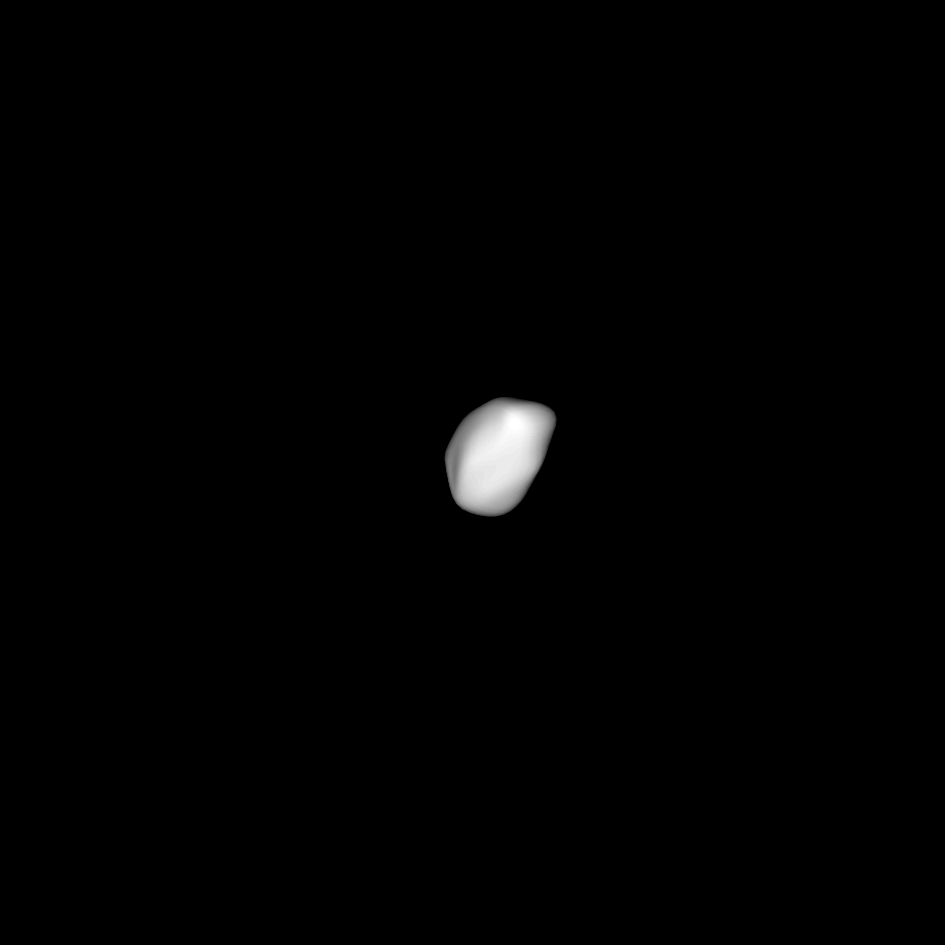}
     \end{subfigure}%
     \caption{\label{fig121}121 Hermione}
    \end{figure}
    
    \begin{figure}[t]
     \begin{subfigure}[b]{0.16\linewidth}
      \includegraphics[clip=true,trim=75 75 95 95,scale=0.66]{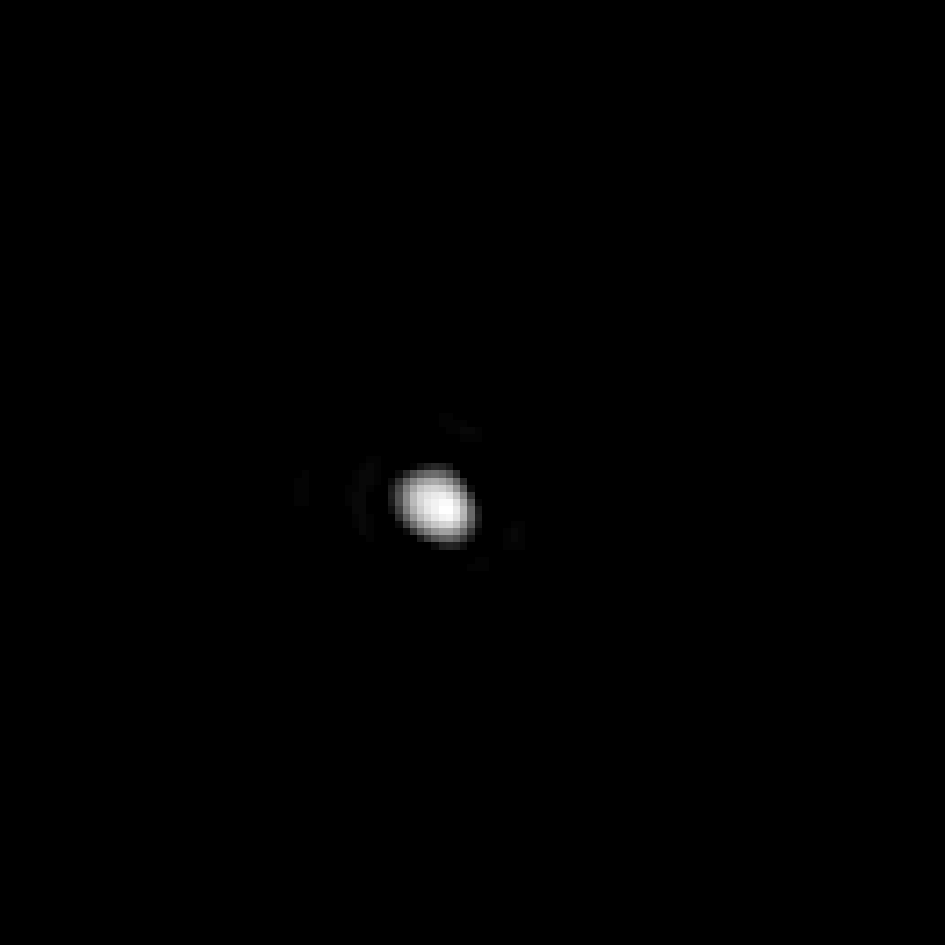} 
    \end{subfigure}%
     \begin{subfigure}[b]{0.16\linewidth}
     \includegraphics[clip=true,trim=90 90 80 80,scale=0.66]{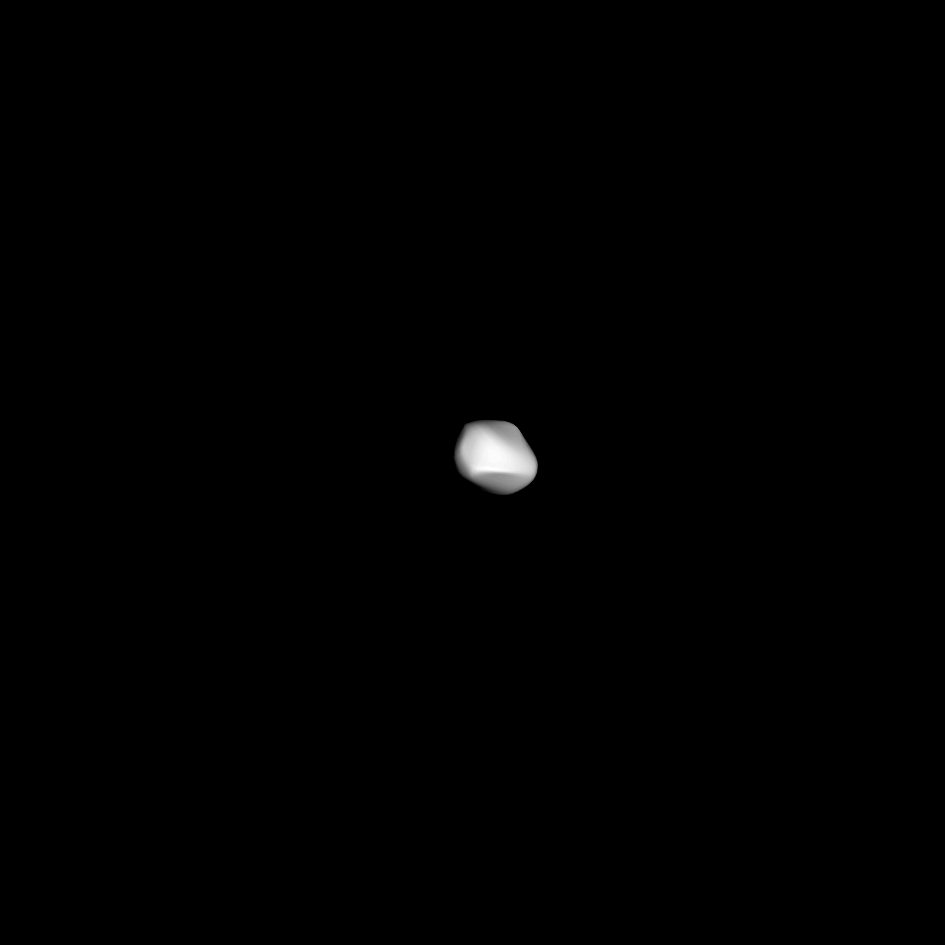}
     \end{subfigure}%
      \begin{subfigure}[b]{0.16\linewidth}
      \includegraphics[clip=true,trim=85 80 85 90,scale=0.66]{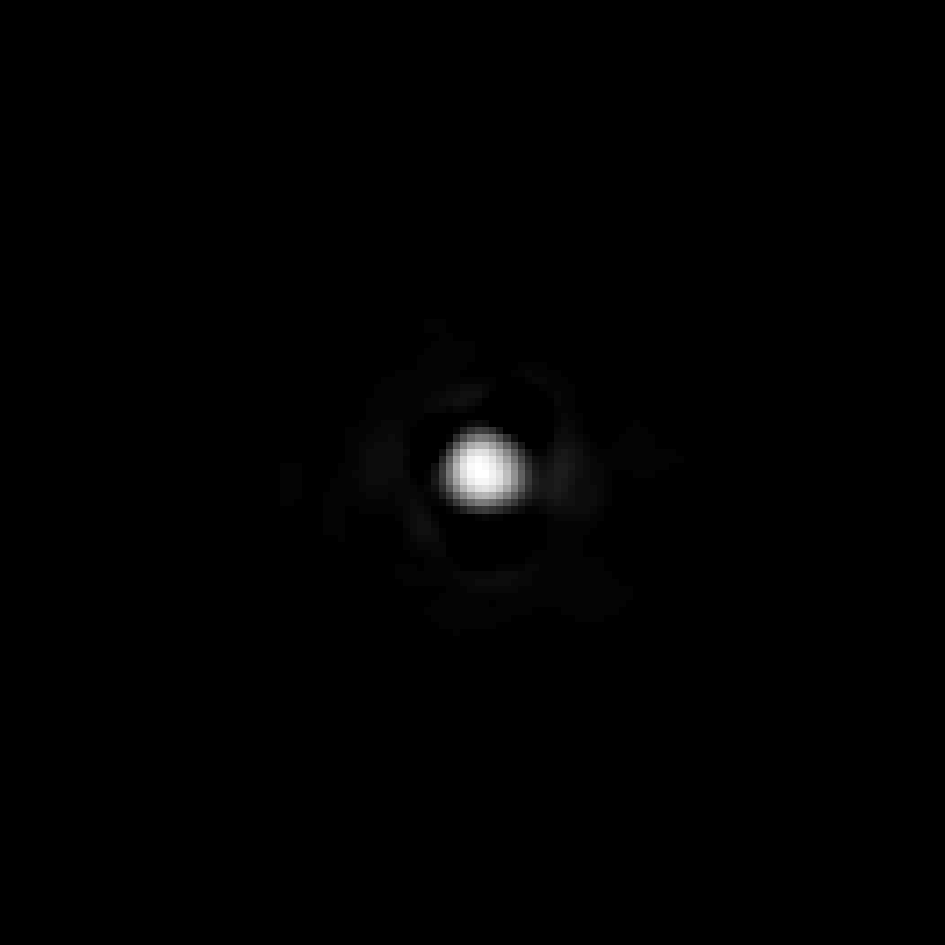}
    \end{subfigure}%
     \begin{subfigure}[b]{0.16\linewidth}
      \includegraphics[clip=true,trim=90 90 80 80,scale=0.66]{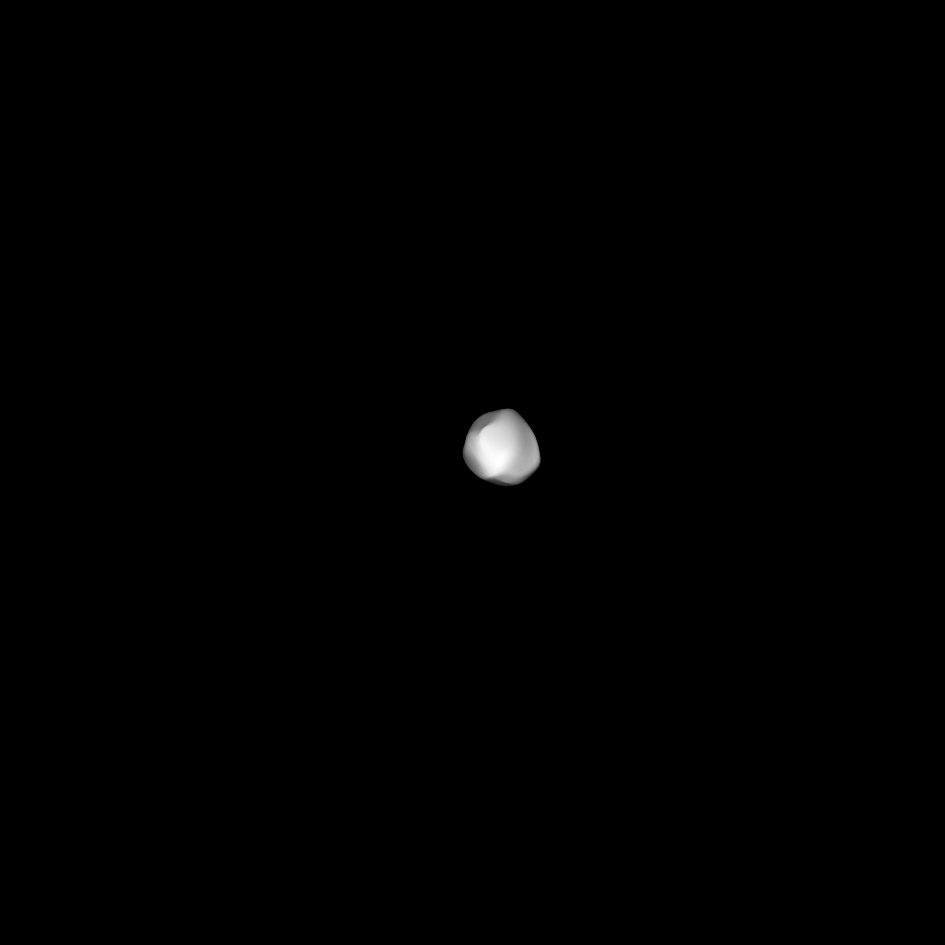}
    \end{subfigure}%
     \caption{\label{fig146}146 Lucina}
\end{figure}
\begin{figure}[t]
     \begin{subfigure}[b]{0.16\linewidth}
      \includegraphics[clip=true,trim=85 85 85 85,scale=0.66]{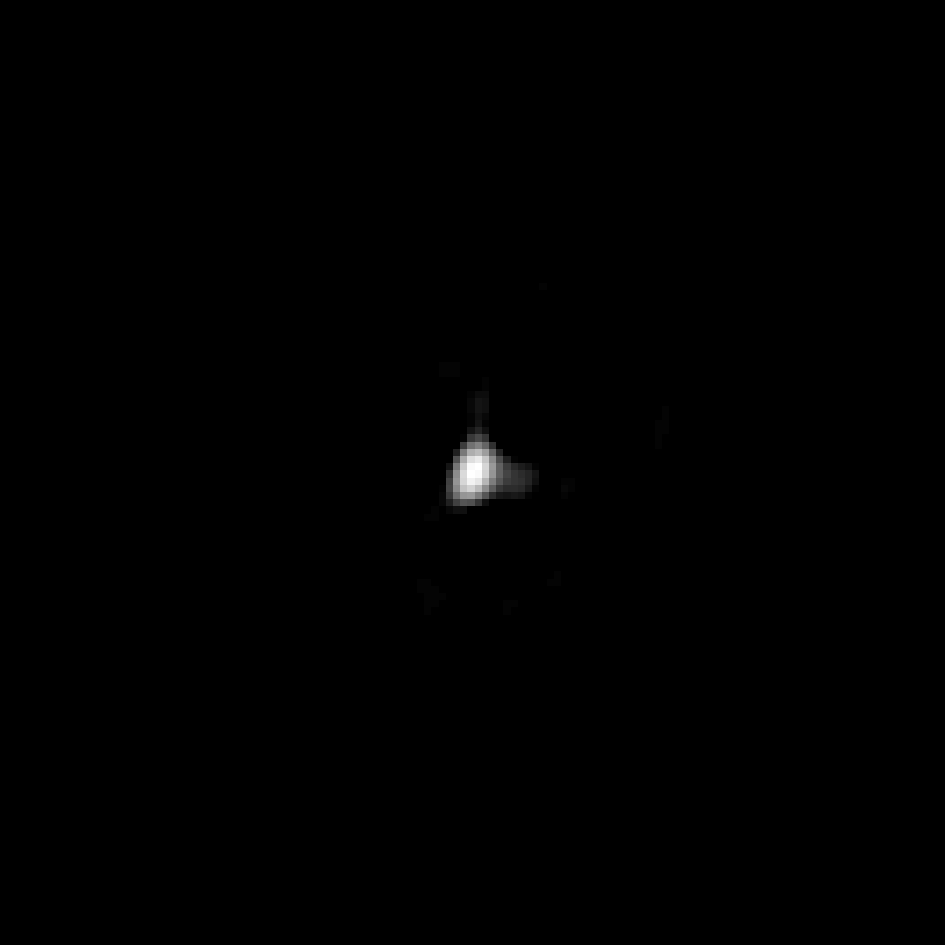} 
    \end{subfigure}%
     \begin{subfigure}[b]{0.16\linewidth}
     \includegraphics[clip=true,trim=90 90 80 80,scale=0.66]{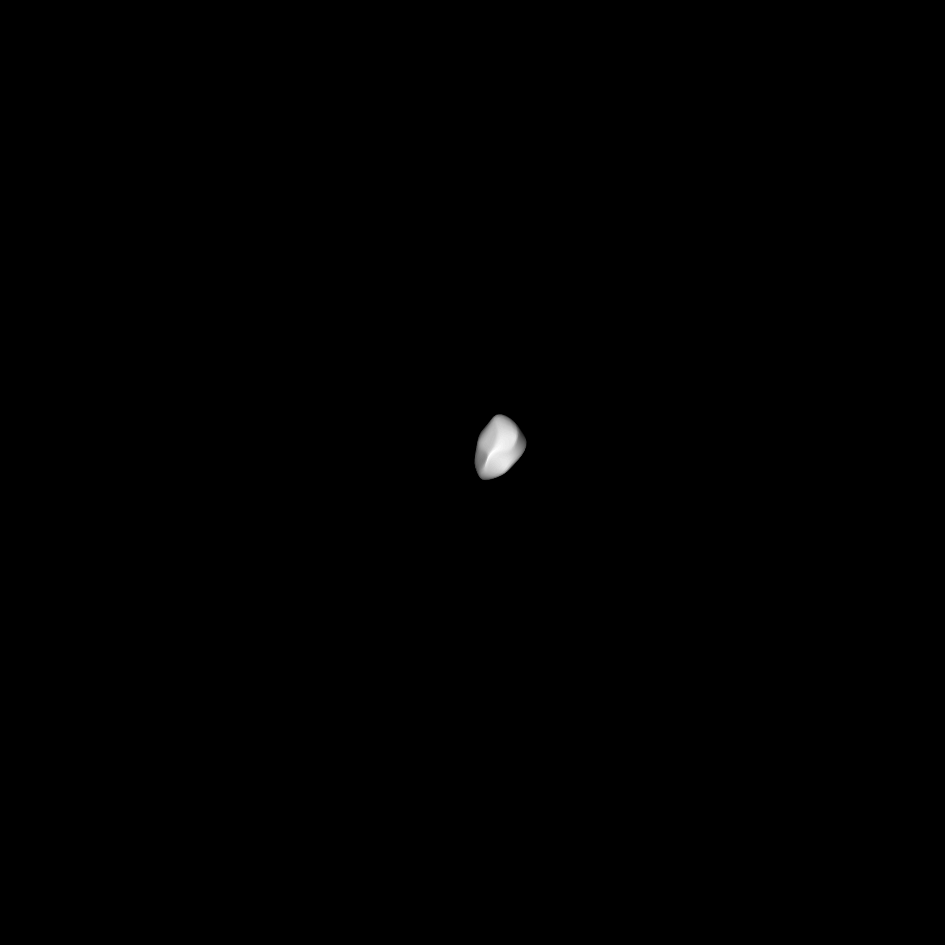}
     \end{subfigure}%
     \caption{\label{fig250}250 Bettina}
\end{figure}

\begin{figure}[t]
     \begin{subfigure}[b]{0.16\linewidth}
      \includegraphics[clip=true,trim=85 85 85 85,scale=0.66]{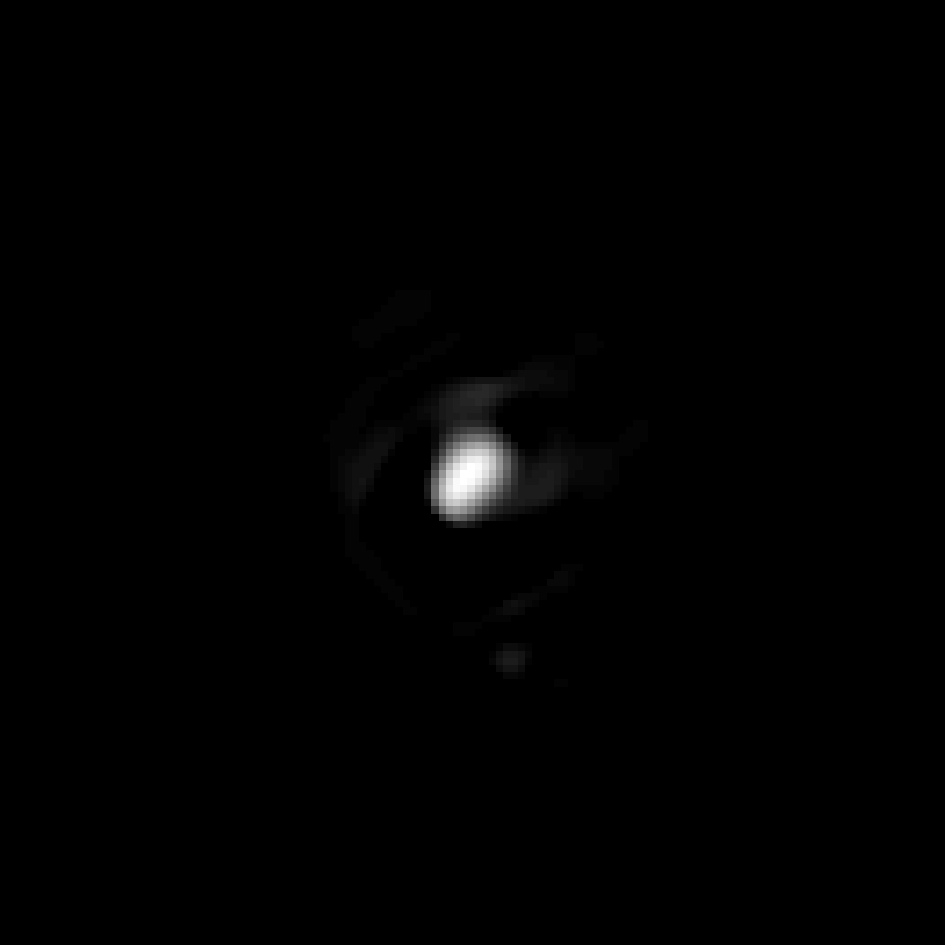} 
    \end{subfigure}%
     \begin{subfigure}[b]{0.16\linewidth}
     \includegraphics[clip=true,trim=90 90 80 80,scale=0.66]{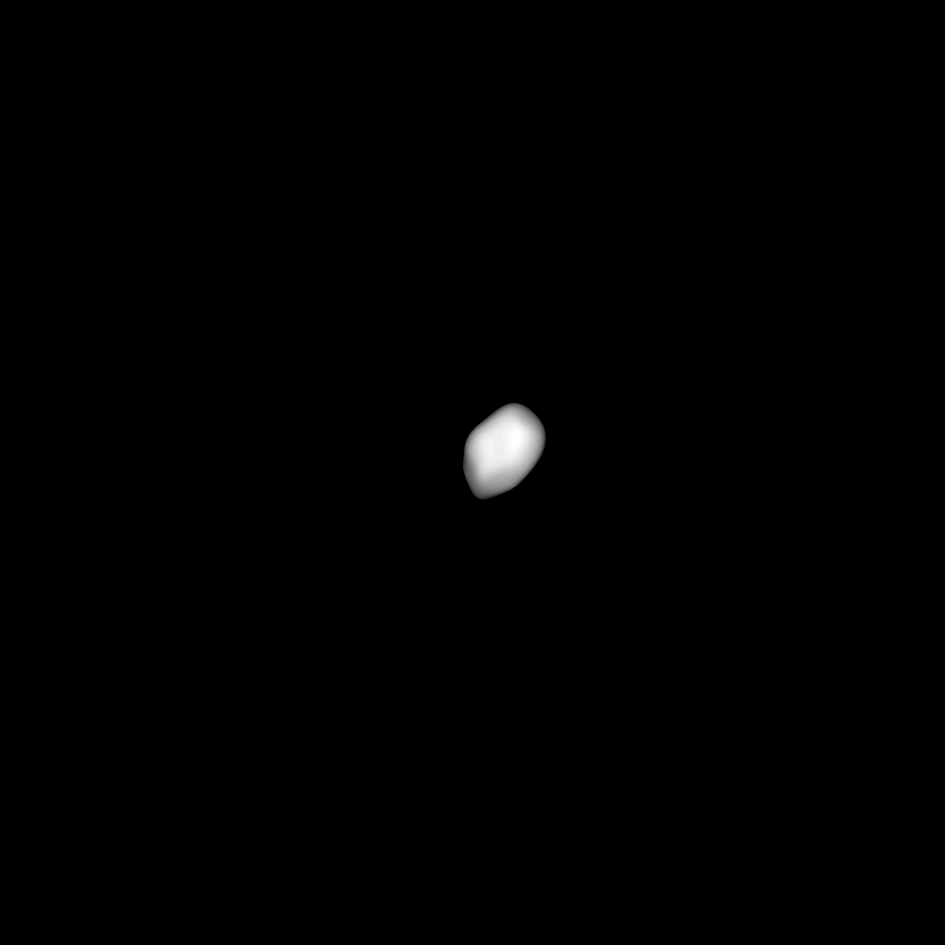}
     \end{subfigure}%
      \begin{subfigure}[b]{0.16\linewidth}
      \includegraphics[clip=true,trim=85 85 85 85,scale=0.66]{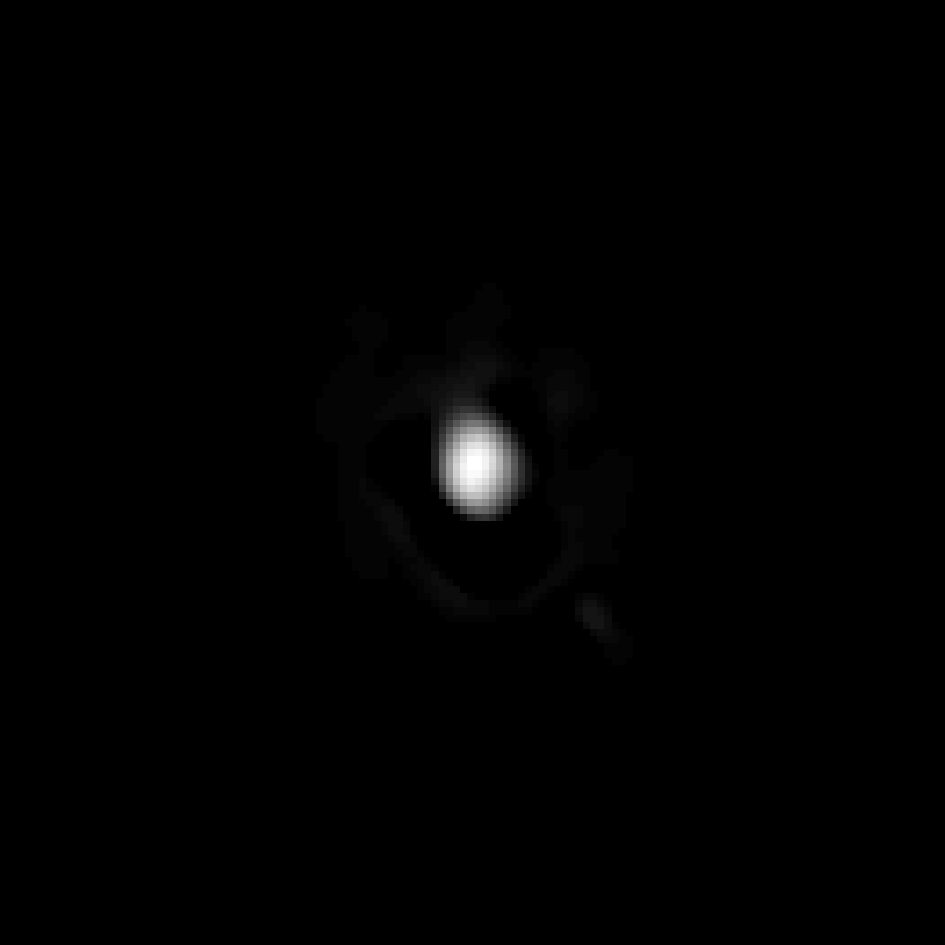}
    \end{subfigure}%
     \begin{subfigure}[b]{0.16\linewidth}
      \includegraphics[clip=true,trim=90 90 80 80,scale=0.66]{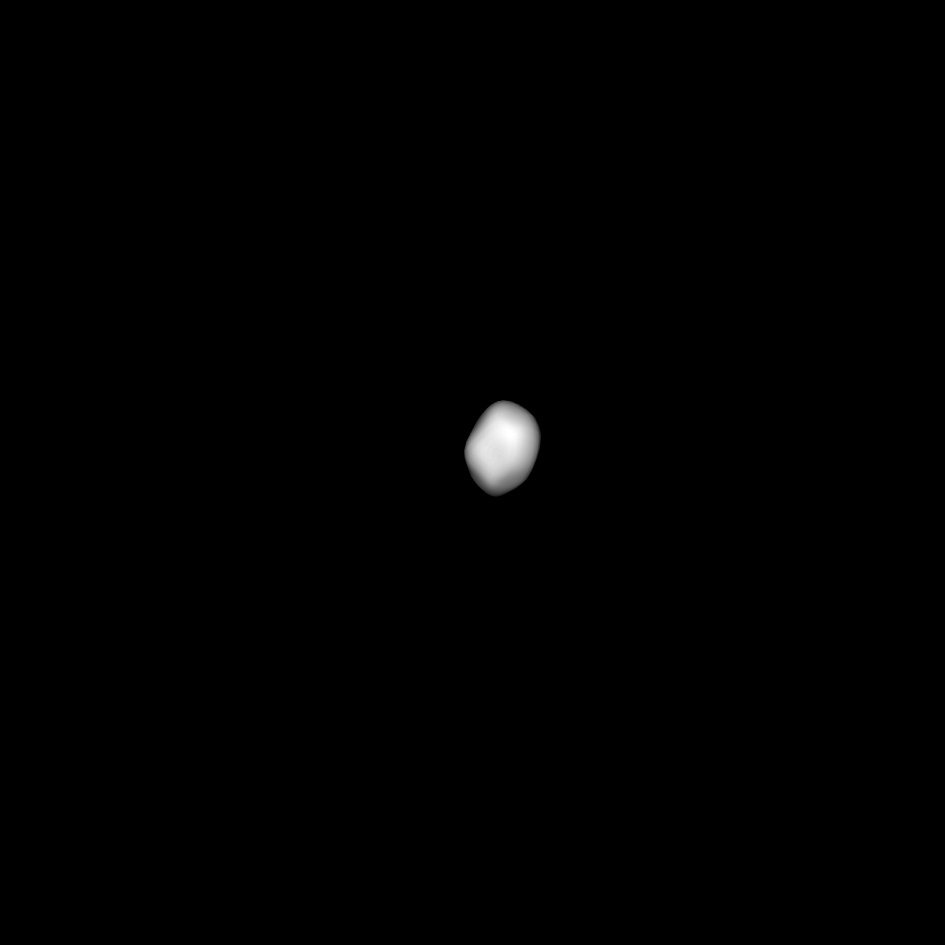}
    \end{subfigure}%
    \begin{subfigure}[b]{0.16\linewidth}
      \includegraphics[clip=true,trim=85 85 85 85,scale=0.66]{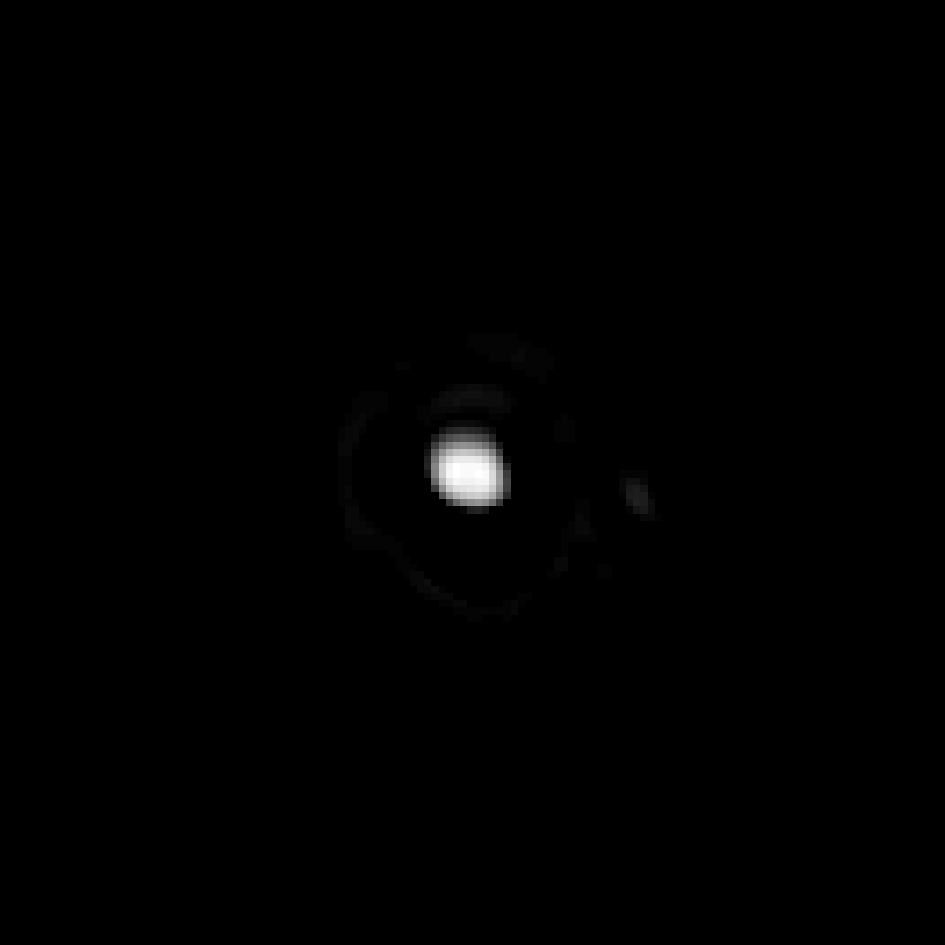} 
    \end{subfigure}%
     \begin{subfigure}[b]{0.16\linewidth}
     \includegraphics[clip=true,trim=90 90 80 80,scale=0.66]{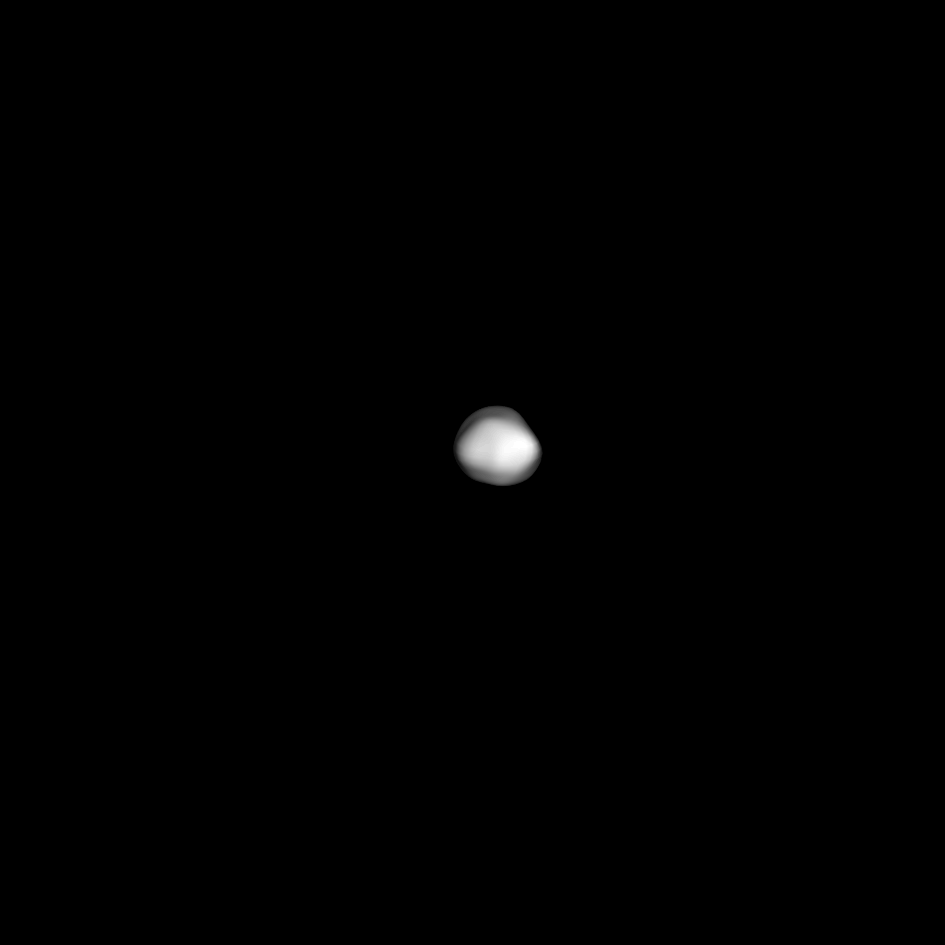}
     \end{subfigure}%
     
     \begin{subfigure}[b]{0.16\linewidth}
      \includegraphics[clip=true,trim=85 85 85 85,scale=0.66]{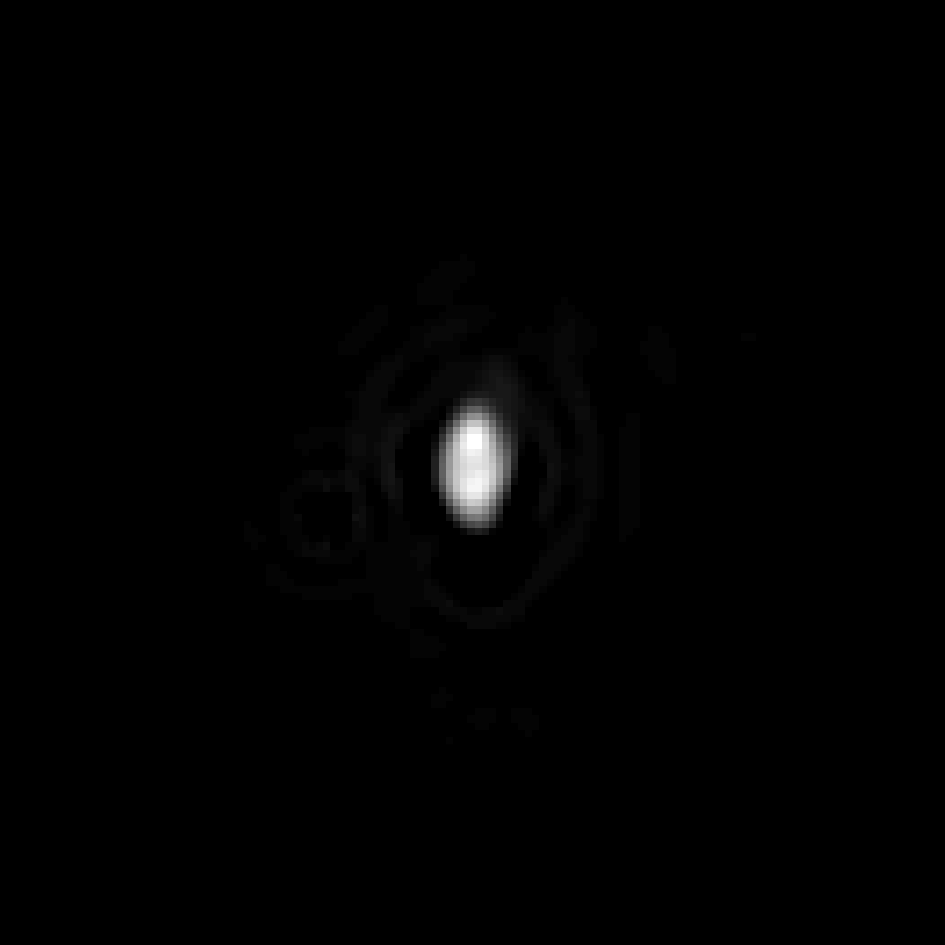} 
    \end{subfigure}%
     \begin{subfigure}[b]{0.16\linewidth}
     \includegraphics[clip=true,trim=90 90 80 80,scale=0.66]{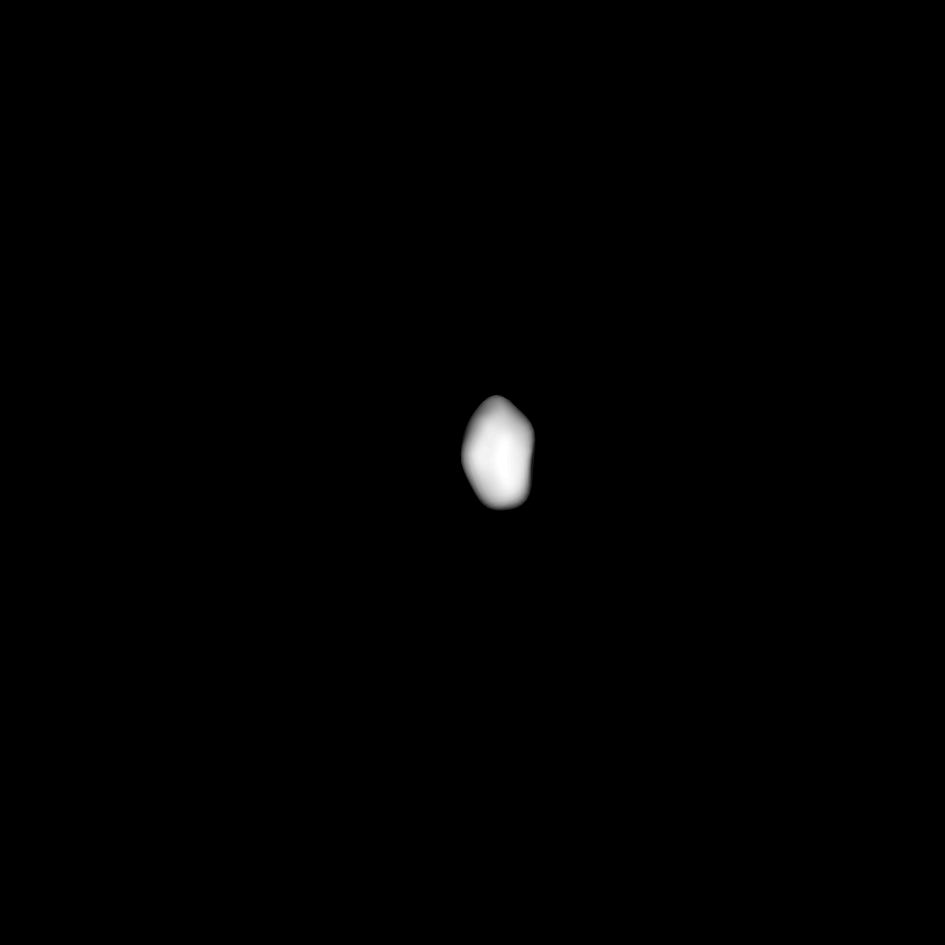}
     \end{subfigure}%
     \begin{subfigure}[b]{0.16\linewidth}
      \includegraphics[clip=true,trim=85 85 85 85,scale=0.66]{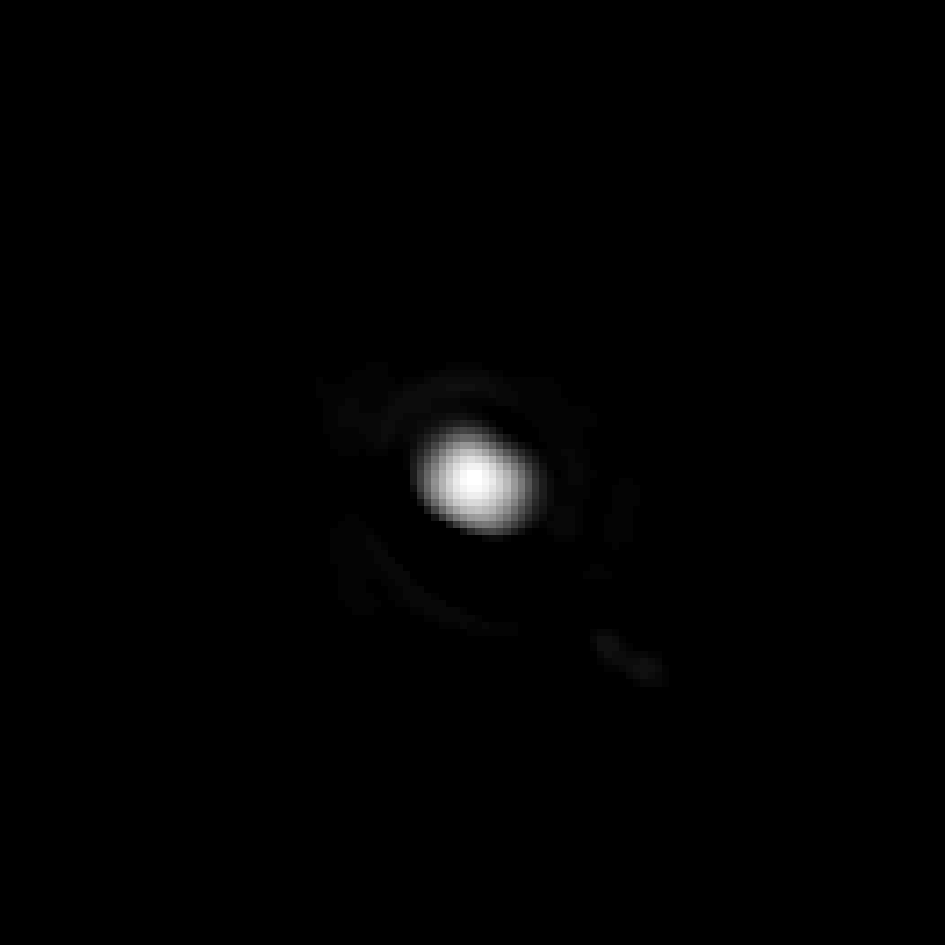} 
    \end{subfigure}%
     \begin{subfigure}[b]{0.16\linewidth}
     \includegraphics[clip=true,trim=90 90 80 80,scale=0.66]{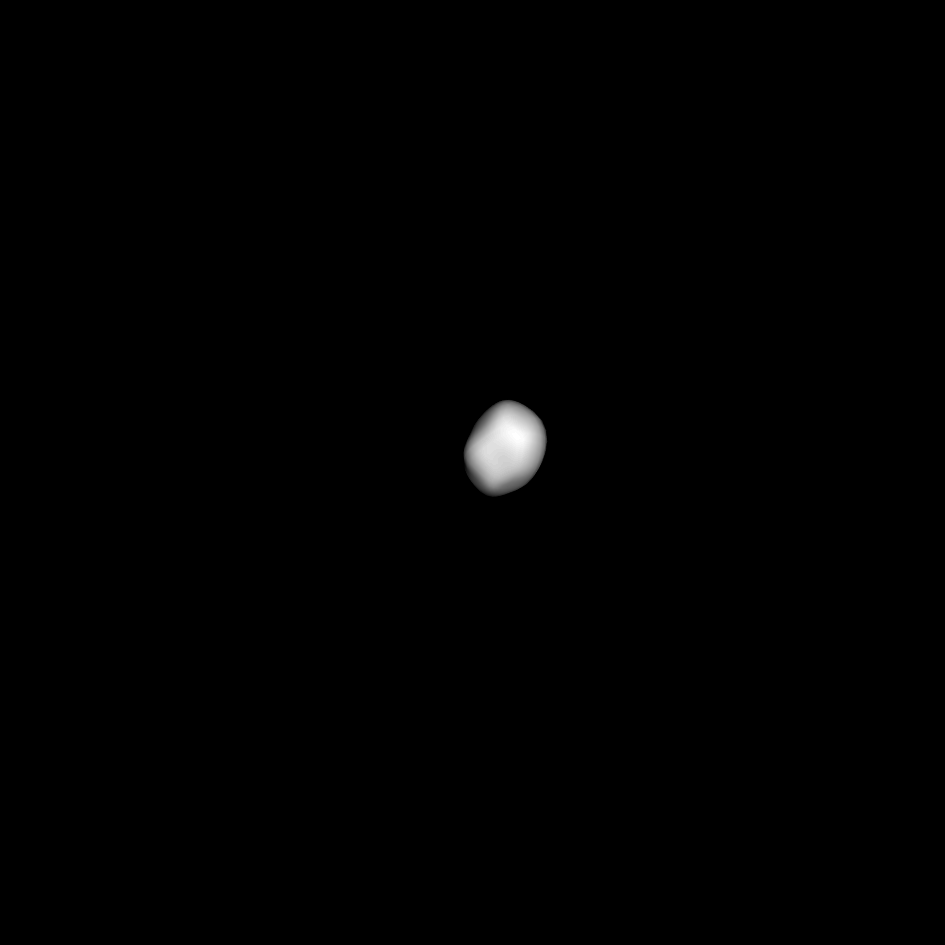}
     \end{subfigure}%
     \caption{\label{fig283}283 Emma}
\end{figure}

\begin{figure}[t]
     \begin{subfigure}[b]{0.16\linewidth}
      \includegraphics[clip=true,trim=85 85 85 85,scale=0.66]{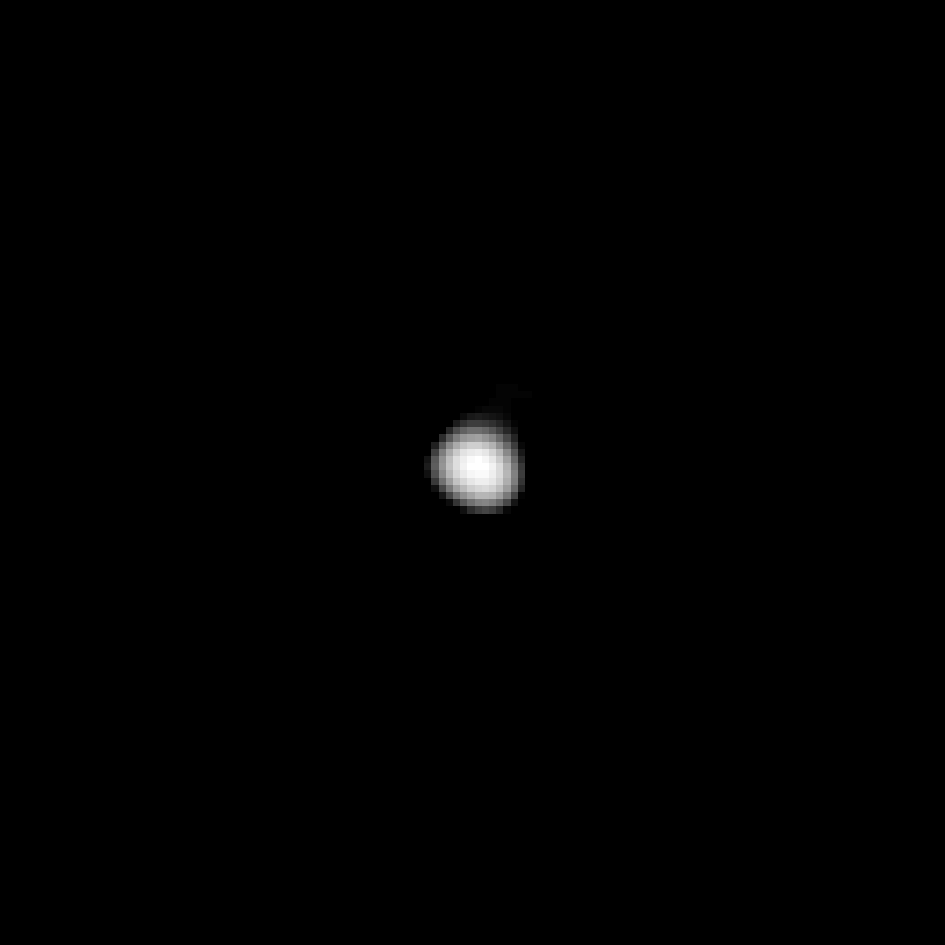} 
    \end{subfigure}%
     \begin{subfigure}[b]{0.16\linewidth}
     \includegraphics[clip=true,trim=90 90 80 80,scale=0.66]{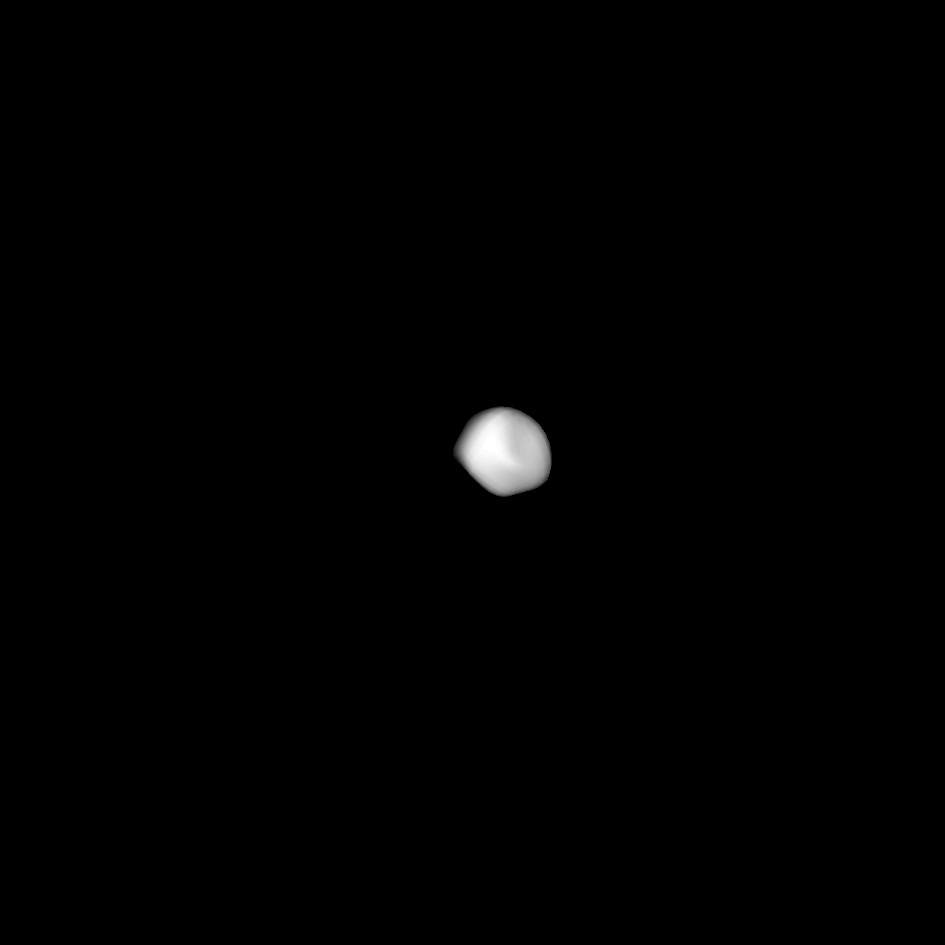}
     \end{subfigure}%
      \begin{subfigure}[b]{0.16\linewidth}
      \includegraphics[clip=true,trim=85 85 85 85,scale=0.66]{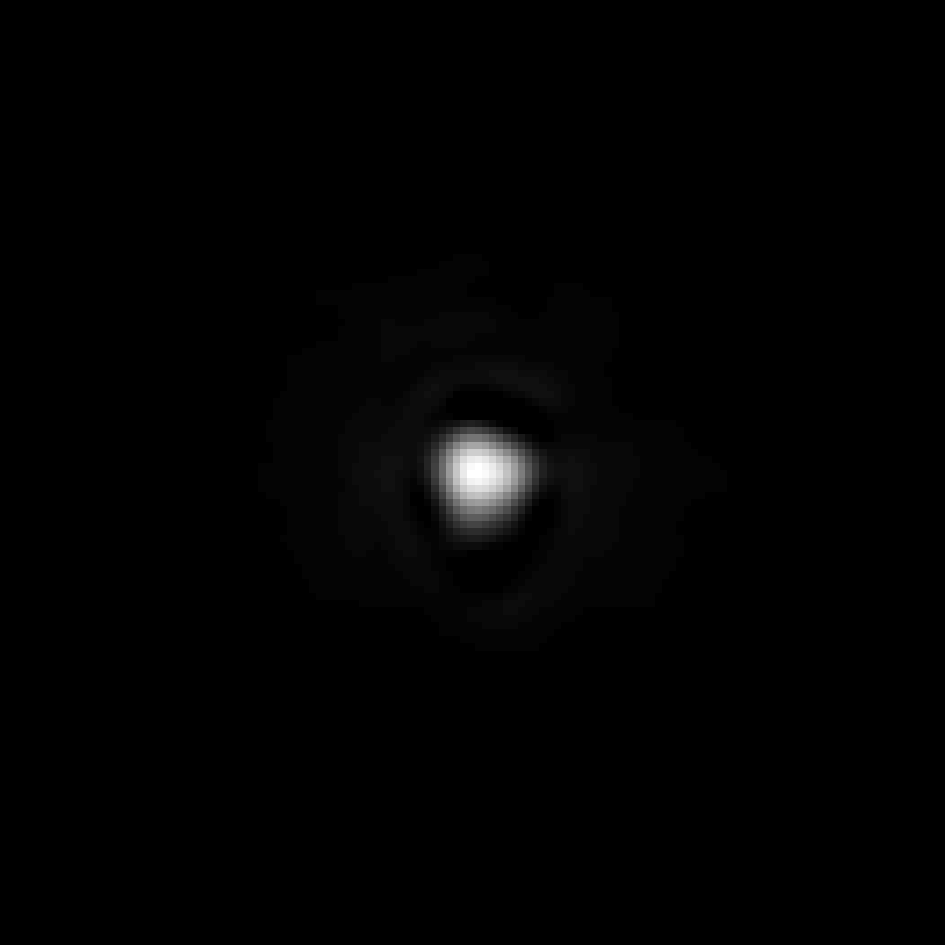}
    \end{subfigure}%
     \begin{subfigure}[b]{0.16\linewidth}
      \includegraphics[clip=true,trim=90 90 80 80,scale=0.66]{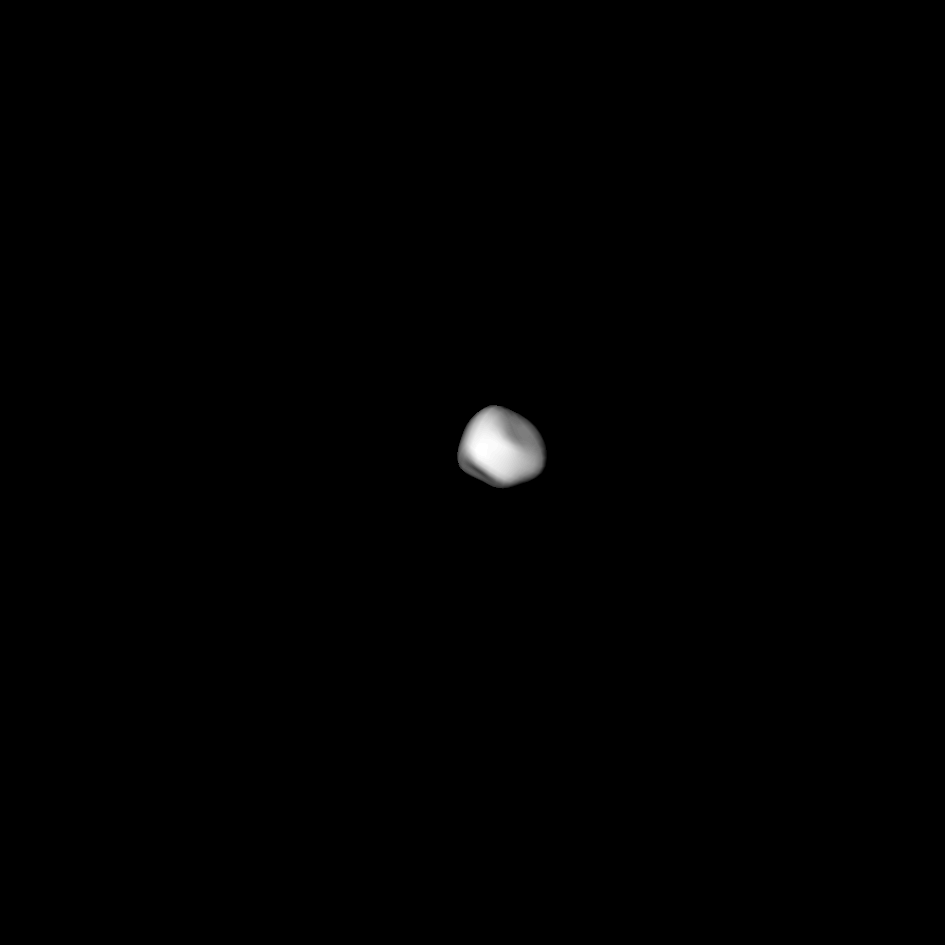}
    \end{subfigure}%
    \begin{subfigure}[b]{0.16\linewidth}
      \includegraphics[clip=true,trim=85 85 85 85,scale=0.66]{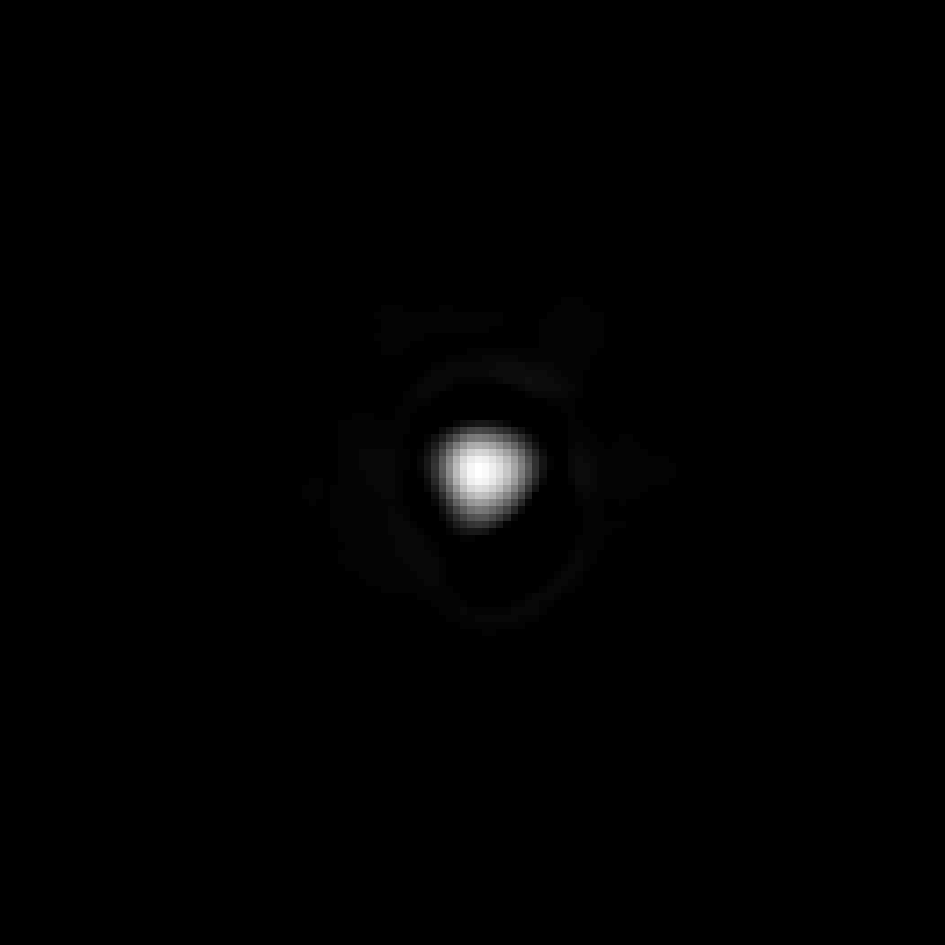} 
    \end{subfigure}%
     \begin{subfigure}[b]{0.16\linewidth}
     \includegraphics[clip=true,trim=90 90 80 80,scale=0.66]{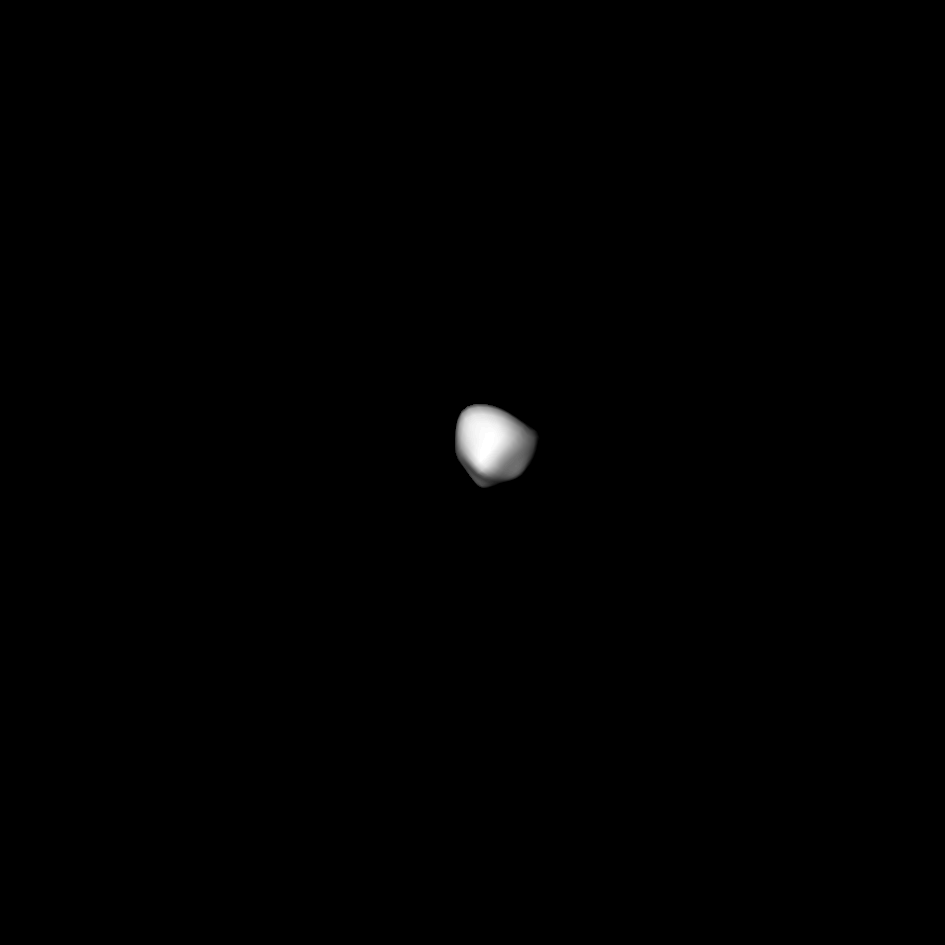}
     \end{subfigure}%
     \caption{\label{fig354}354 Eleonora}
\end{figure}
\begin{figure}[t]
     \begin{subfigure}[b]{0.16\linewidth}
      \includegraphics[clip=true,trim=65 65 65 65,scale=0.39]{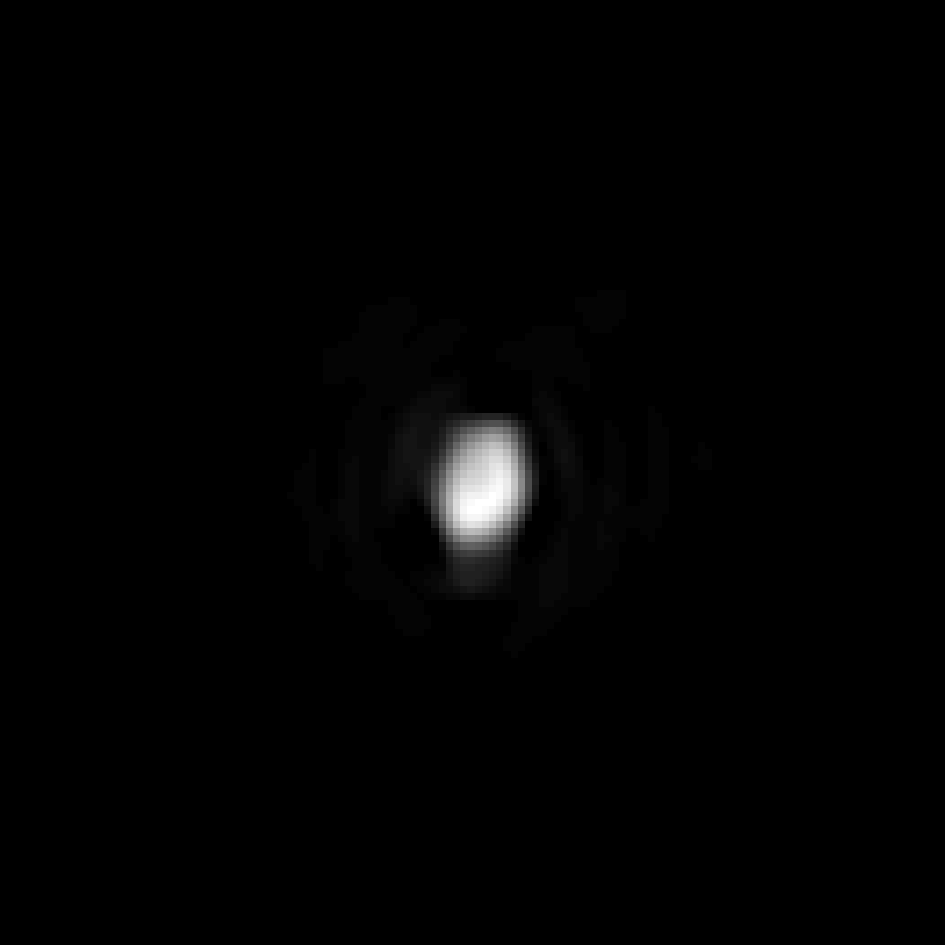} 
    \end{subfigure}%
     \begin{subfigure}[b]{0.16\linewidth}
     \includegraphics[clip=true,trim=70 75 60 55,scale=0.39]{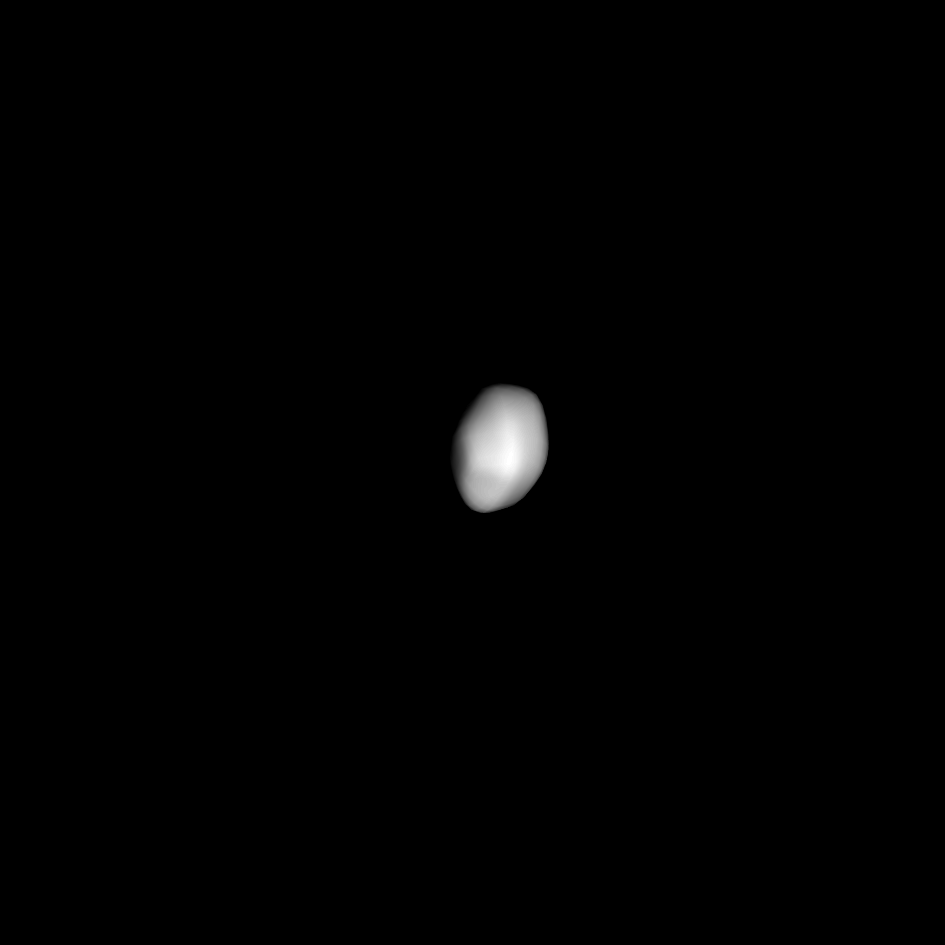}
     \end{subfigure}%
      \begin{subfigure}[b]{0.16\linewidth}
      \includegraphics[clip=true,trim=65 65 65 65,scale=0.39]{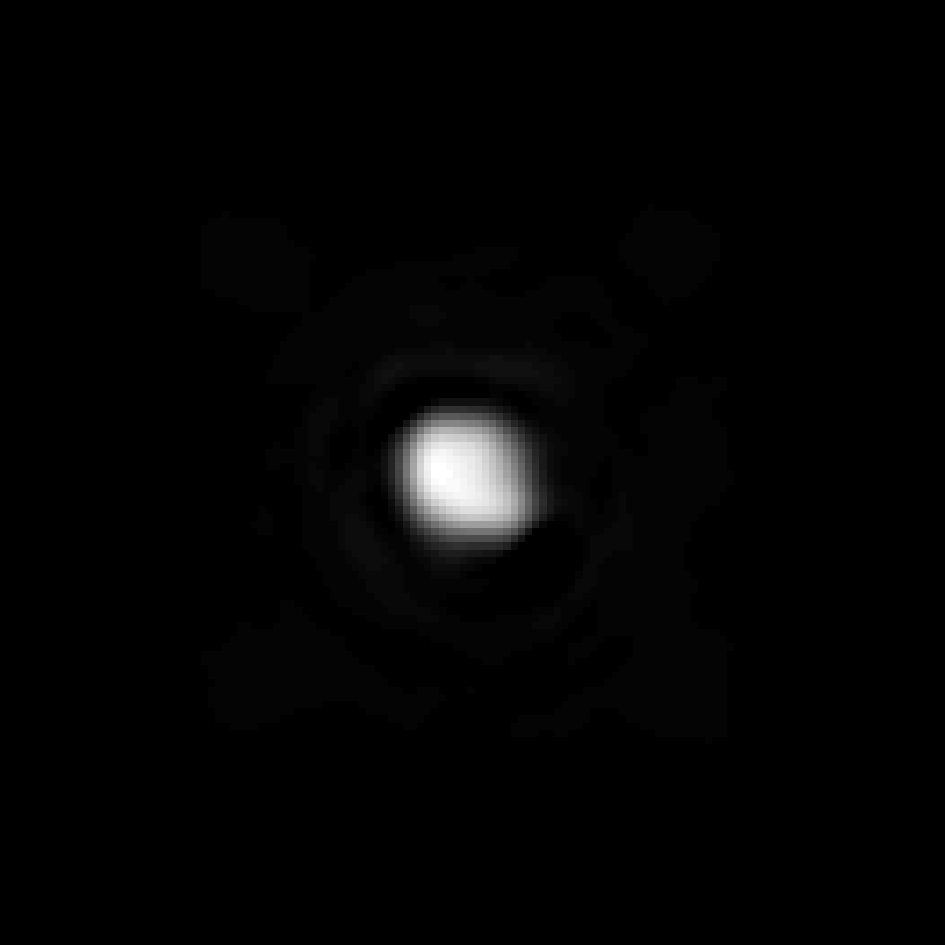}
    \end{subfigure}%
     \begin{subfigure}[b]{0.16\linewidth}
      \includegraphics[clip=true,trim=70 75 60 55,scale=0.39]{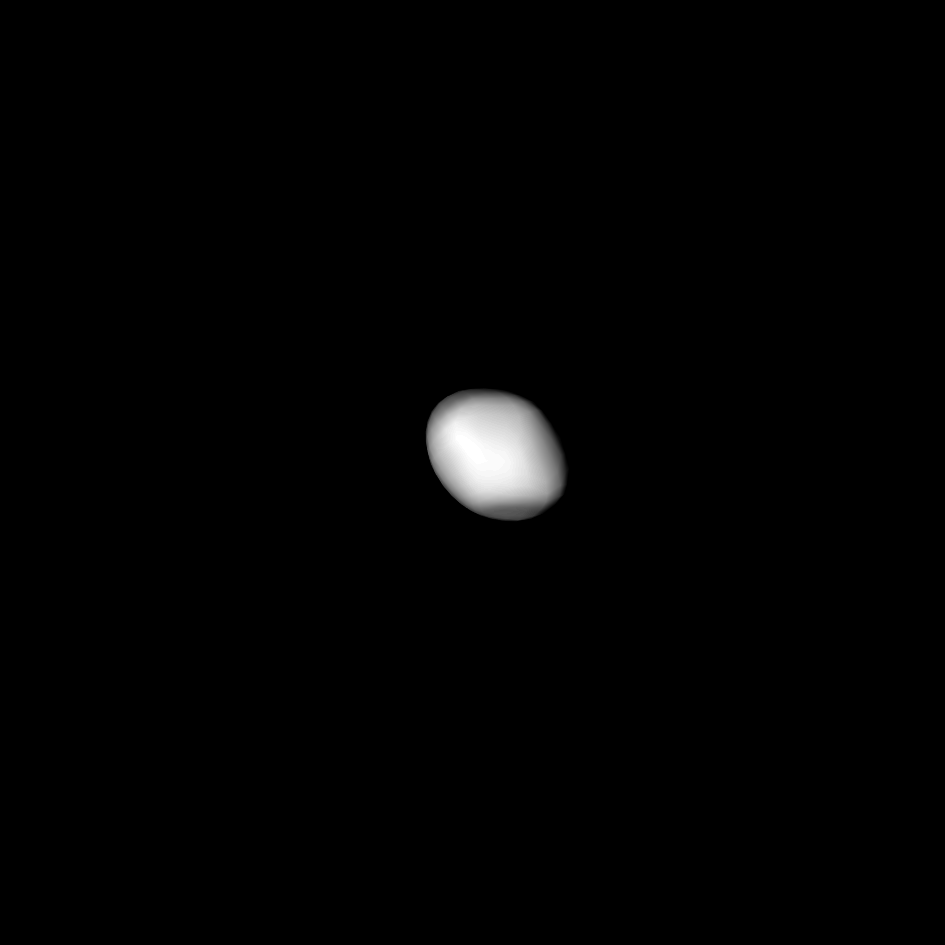}
    \end{subfigure}%
    \begin{subfigure}[b]{0.16\linewidth}
      \includegraphics[clip=true,trim=65 65 65 65,scale=0.39]{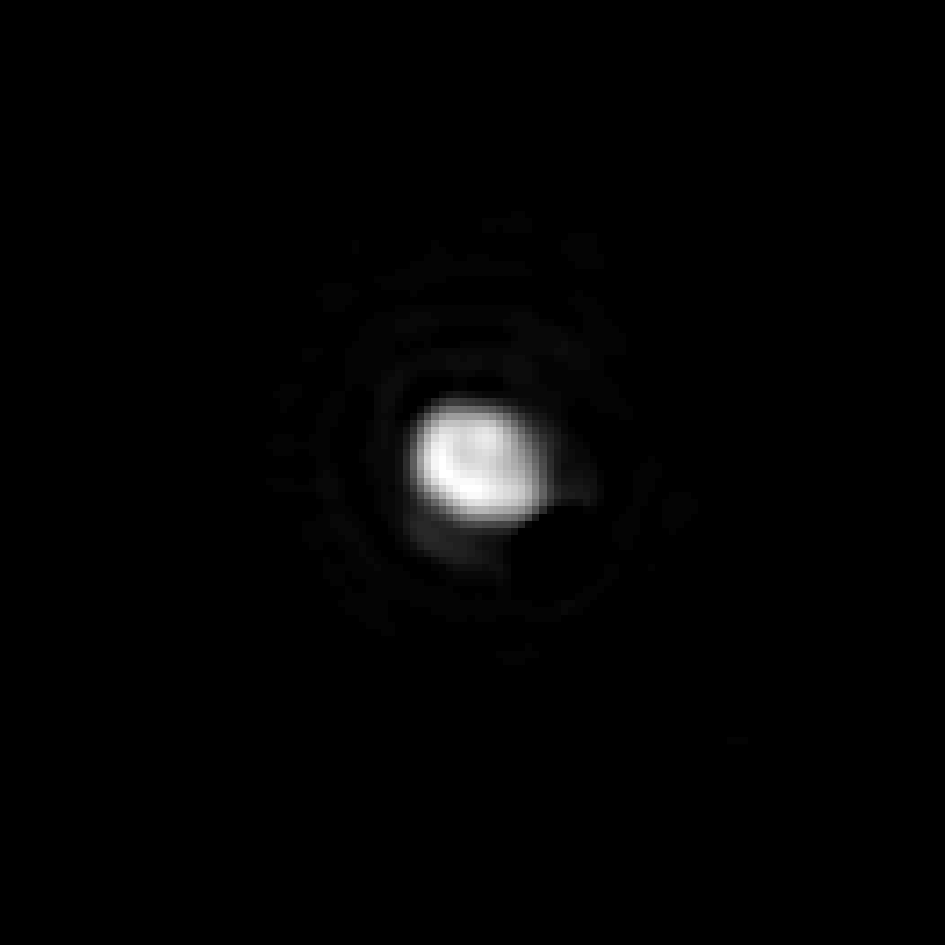} 
    \end{subfigure}%
     \begin{subfigure}[b]{0.16\linewidth}
     \includegraphics[clip=true,trim=70 75 60 55,scale=0.39]{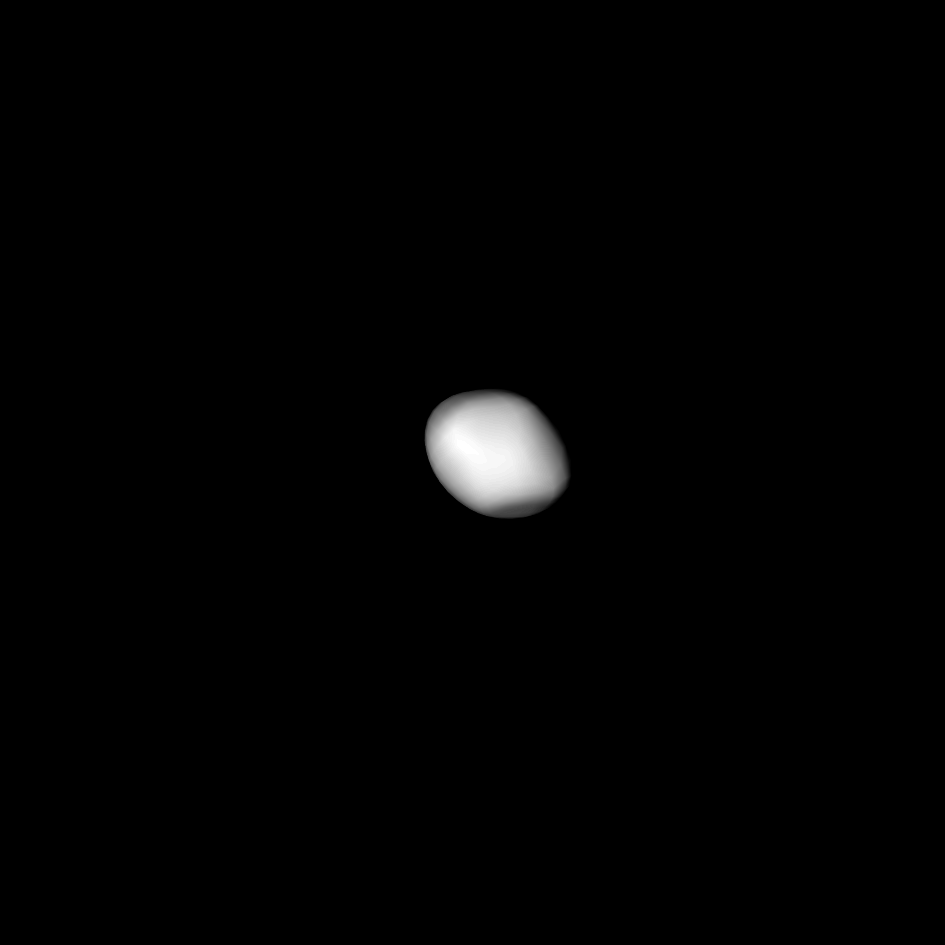}
     \end{subfigure}%
     
      \begin{subfigure}[b]{0.16\linewidth}
      \includegraphics[clip=true,trim=65 65 65 65,scale=0.39]{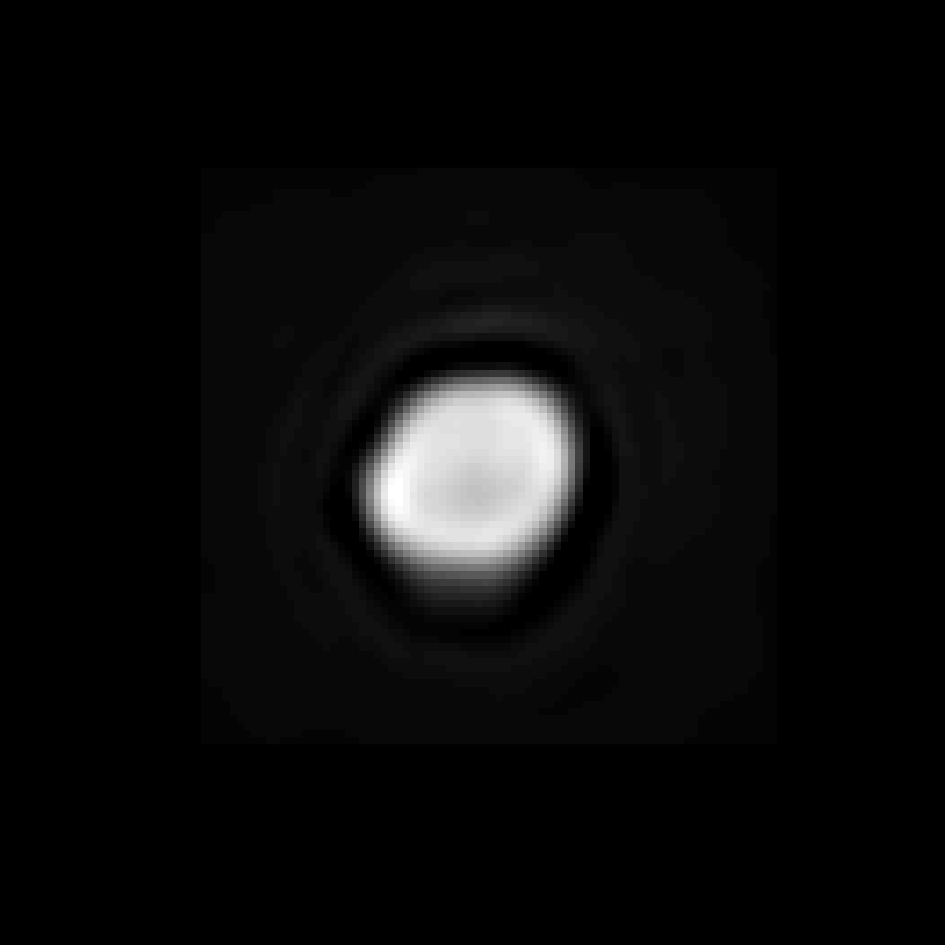} 
    \end{subfigure}%
     \begin{subfigure}[b]{0.16\linewidth}
     \includegraphics[clip=true,trim=70 75 60 55,scale=0.39]{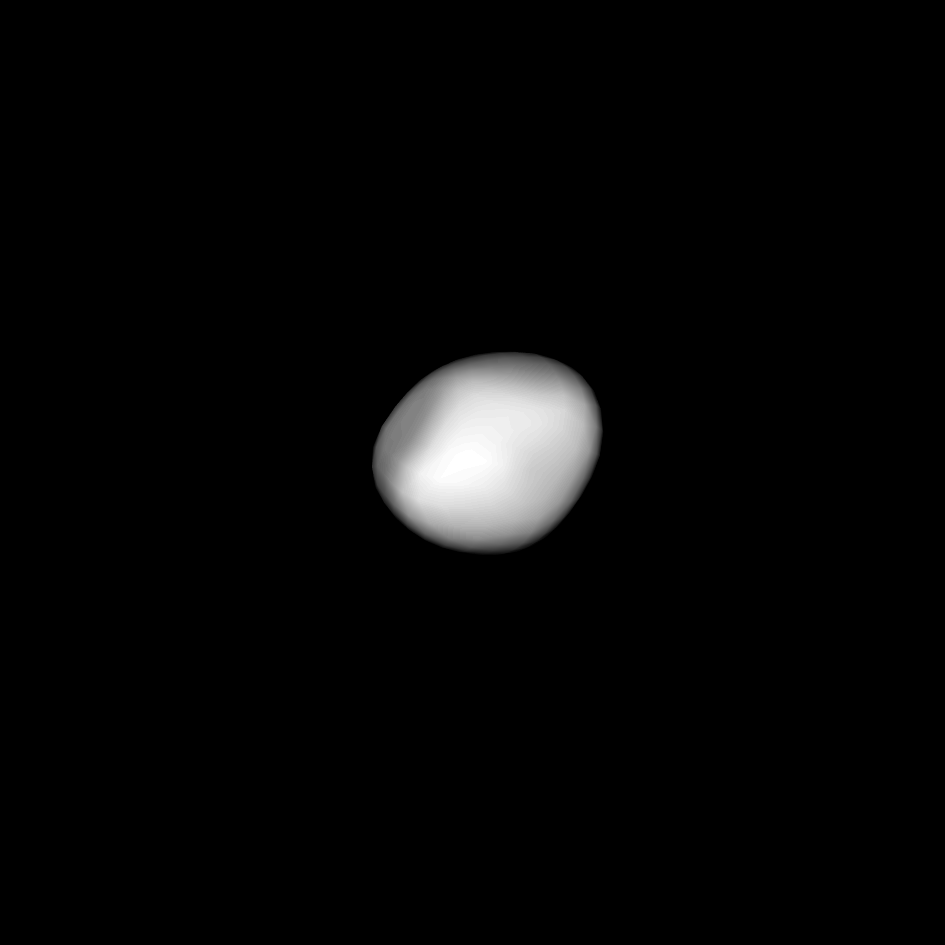}
     \end{subfigure}%
     \begin{subfigure}[b]{0.16\linewidth}
      \includegraphics[clip=true,trim=65 65 65 65,scale=0.39]{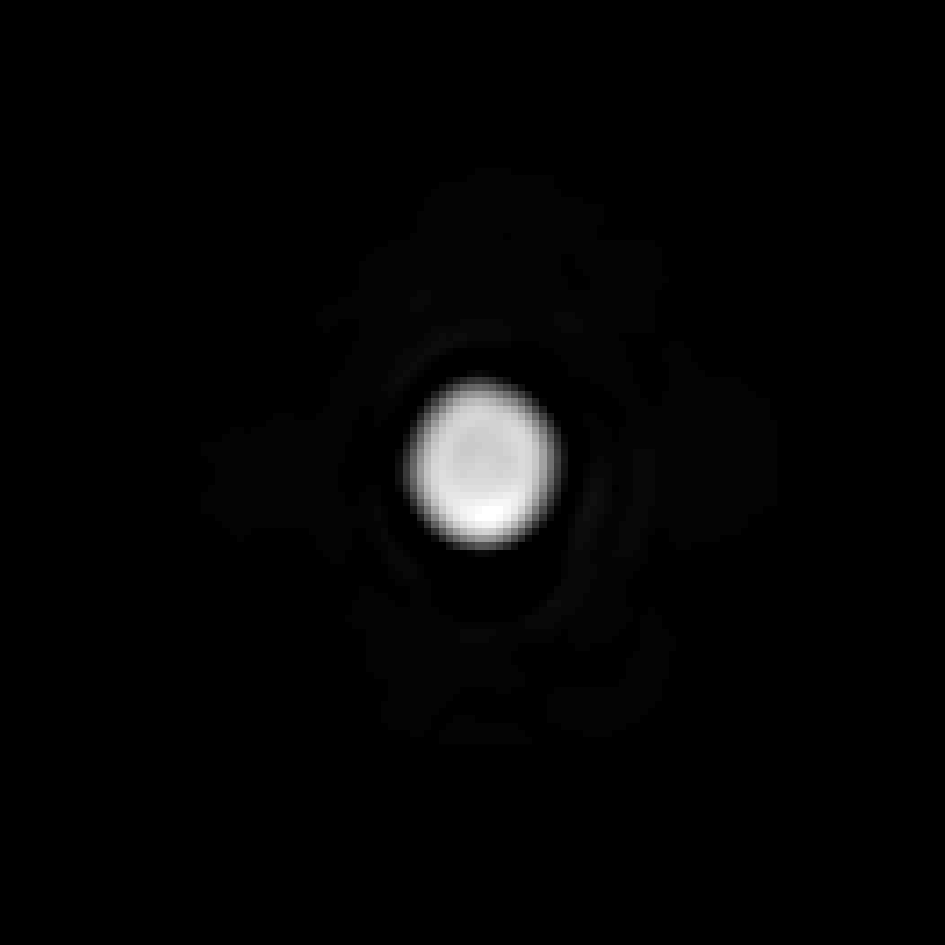} 
    \end{subfigure}%
     \begin{subfigure}[b]{0.16\linewidth}
     \includegraphics[clip=true,trim=70 75 60 55,scale=0.39]{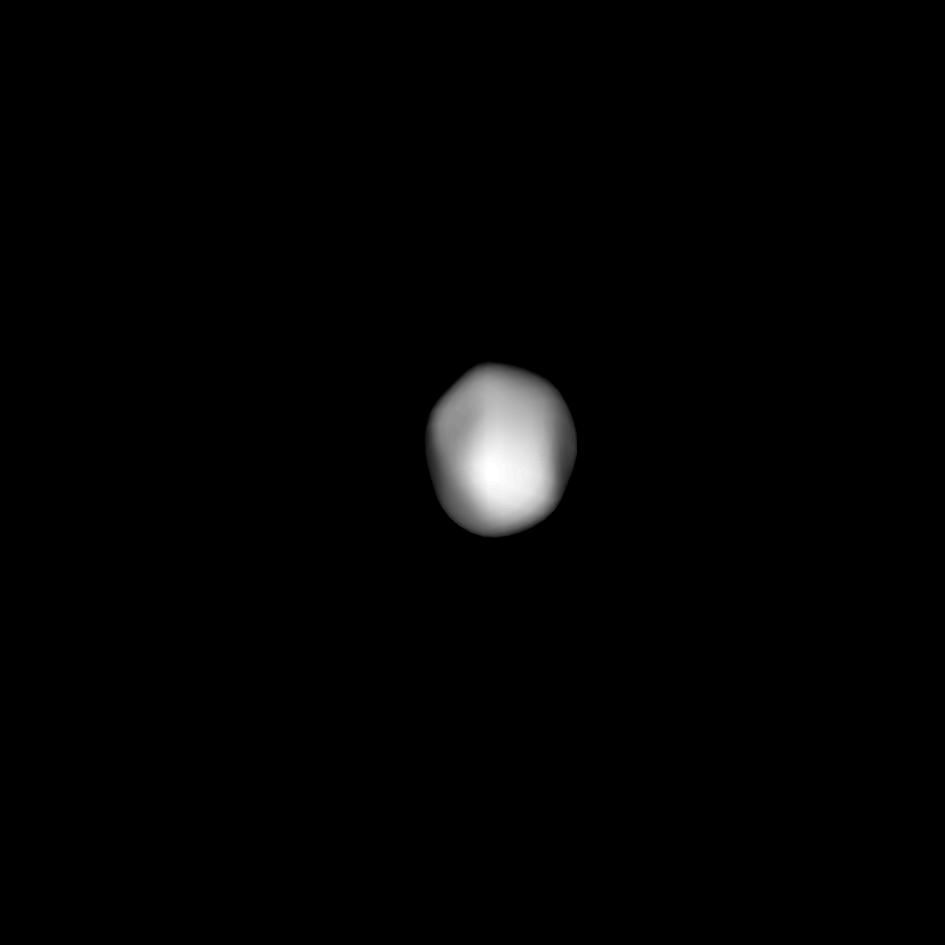}
     \end{subfigure}%
     \begin{subfigure}[b]{0.16\linewidth}
      \includegraphics[clip=true,trim=65 65 65 65,scale=0.39]{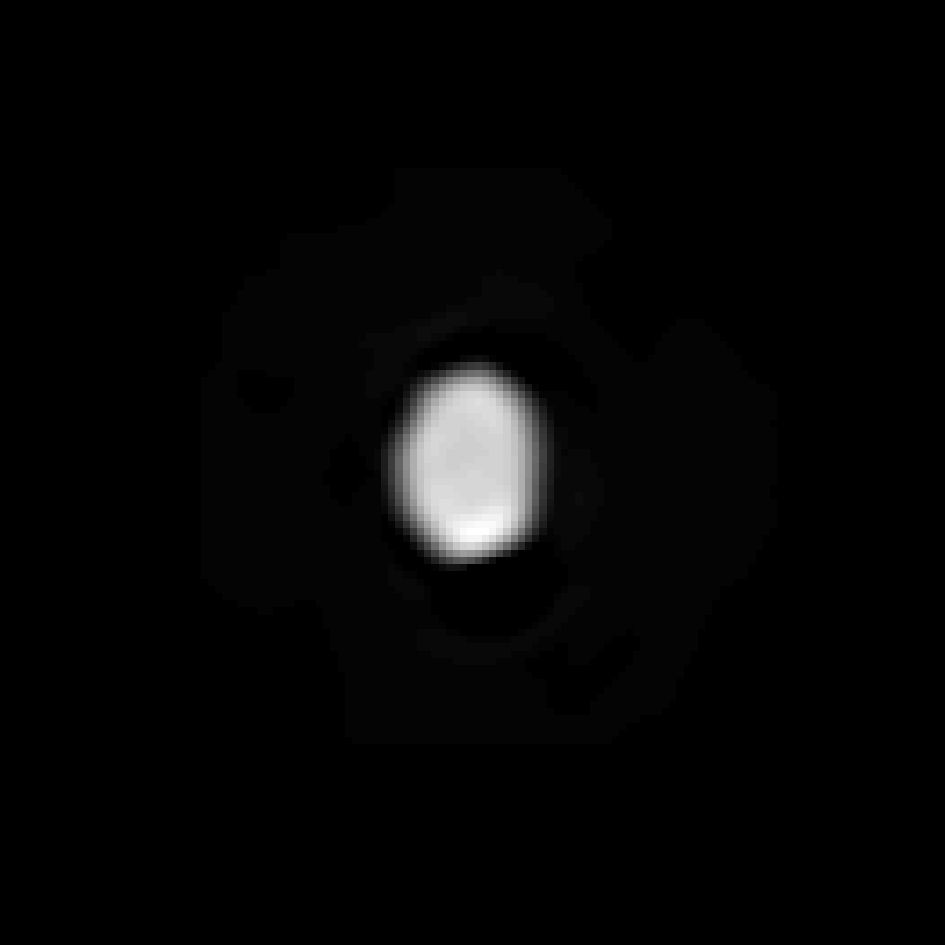} 
    \end{subfigure}%
     \begin{subfigure}[b]{0.16\linewidth}
     \includegraphics[clip=true,trim=70 75 60 55,scale=0.39]{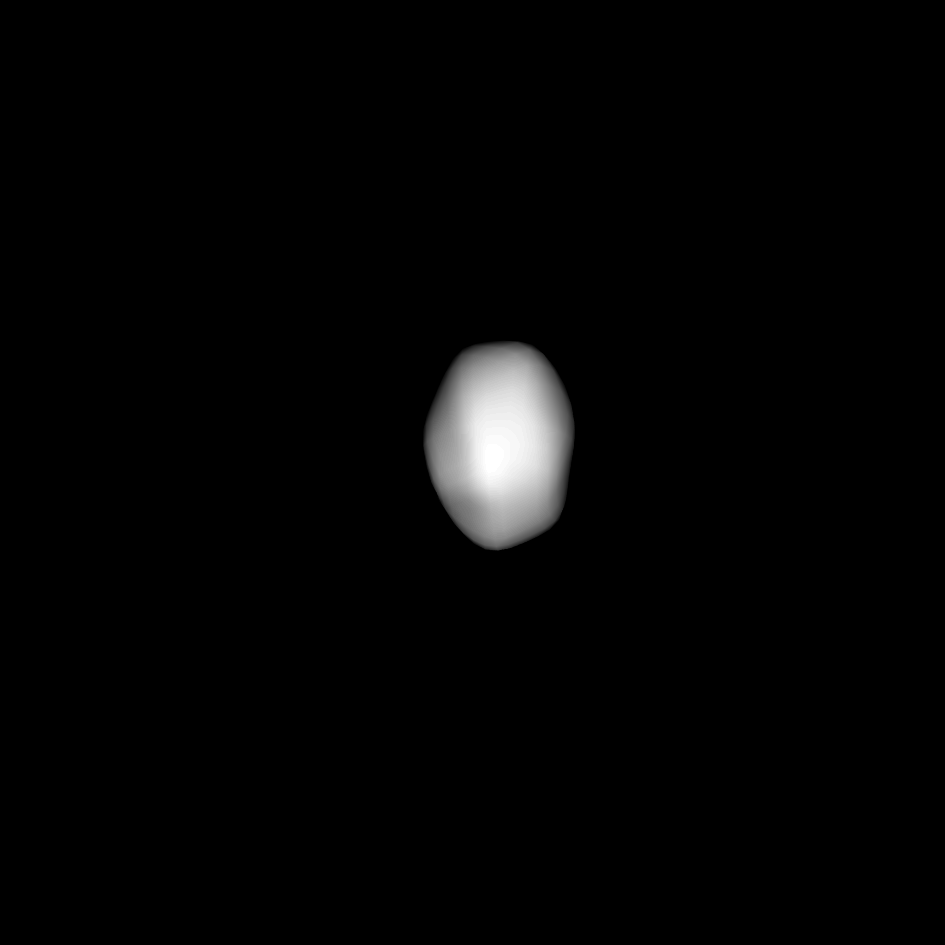}
     \end{subfigure}%
     
     \begin{subfigure}[b]{0.16\linewidth}
     \includegraphics[clip=true,trim=65 65 65 65,scale=0.39]{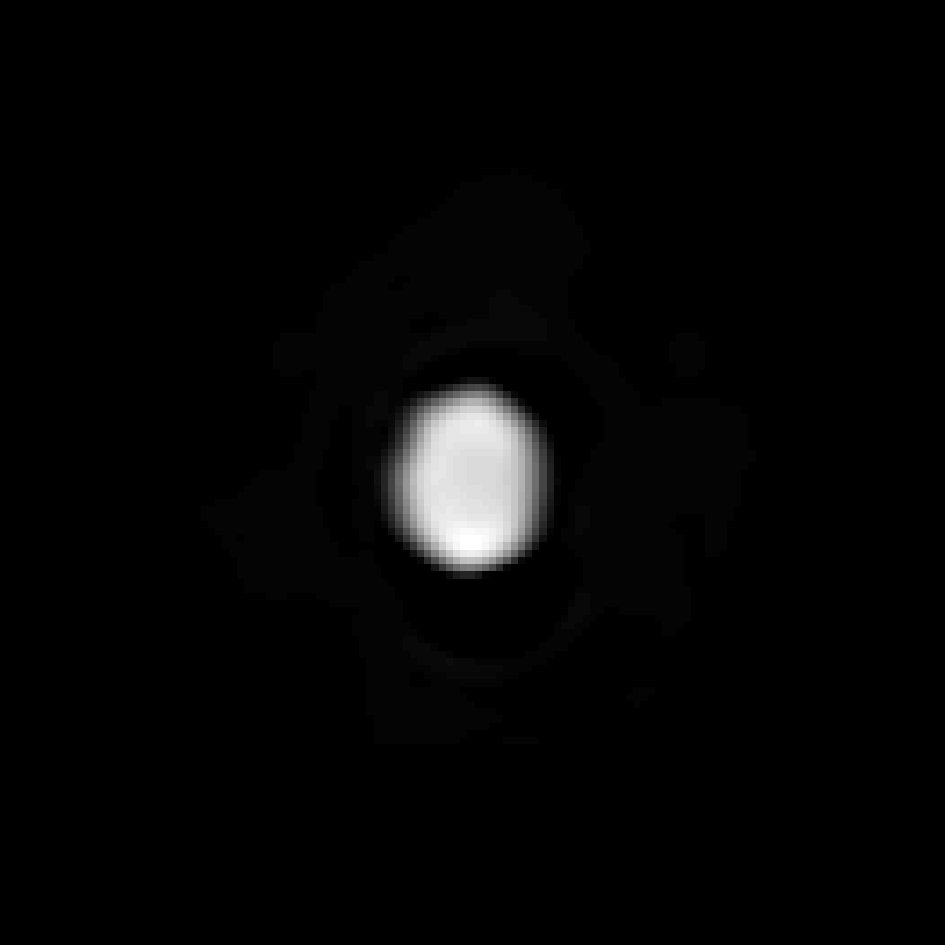}
     \end{subfigure}%
     \begin{subfigure}[b]{0.16\linewidth}
     \includegraphics[clip=true,trim=70 75 60 55,scale=0.39]{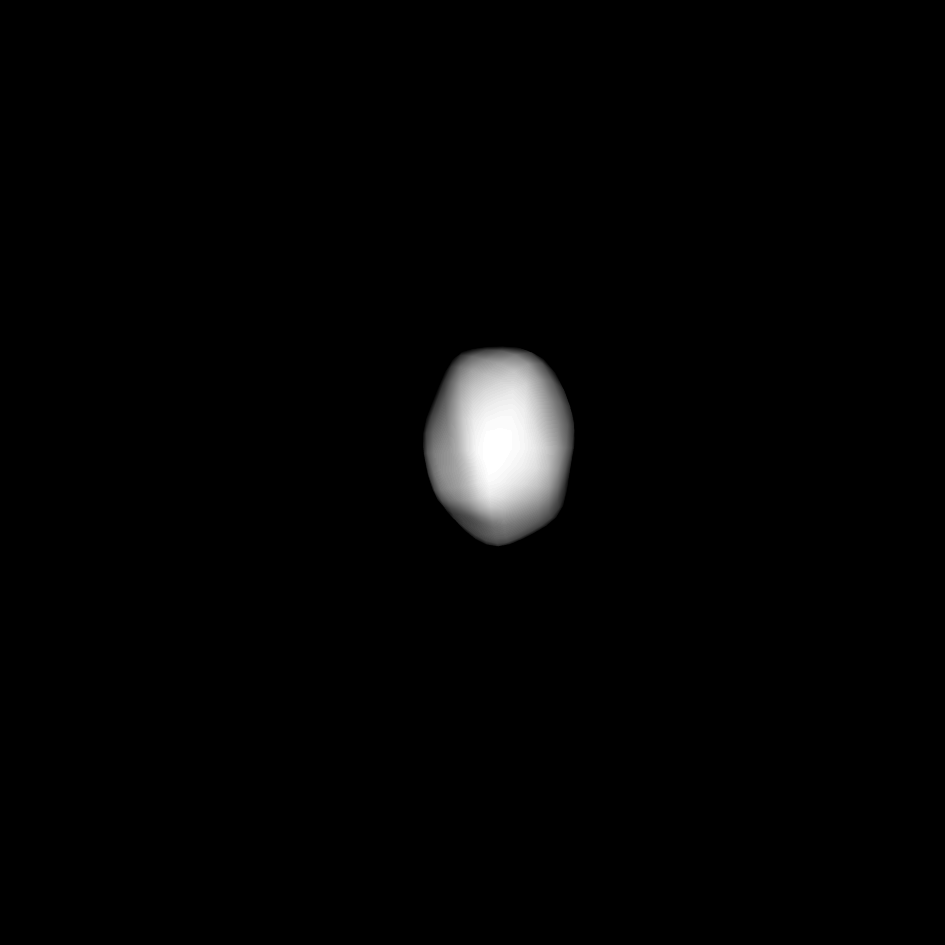}
     \end{subfigure}%
     \begin{subfigure}[b]{0.16\linewidth}
     \includegraphics[clip=true,trim=65 65 65 65,scale=0.39]{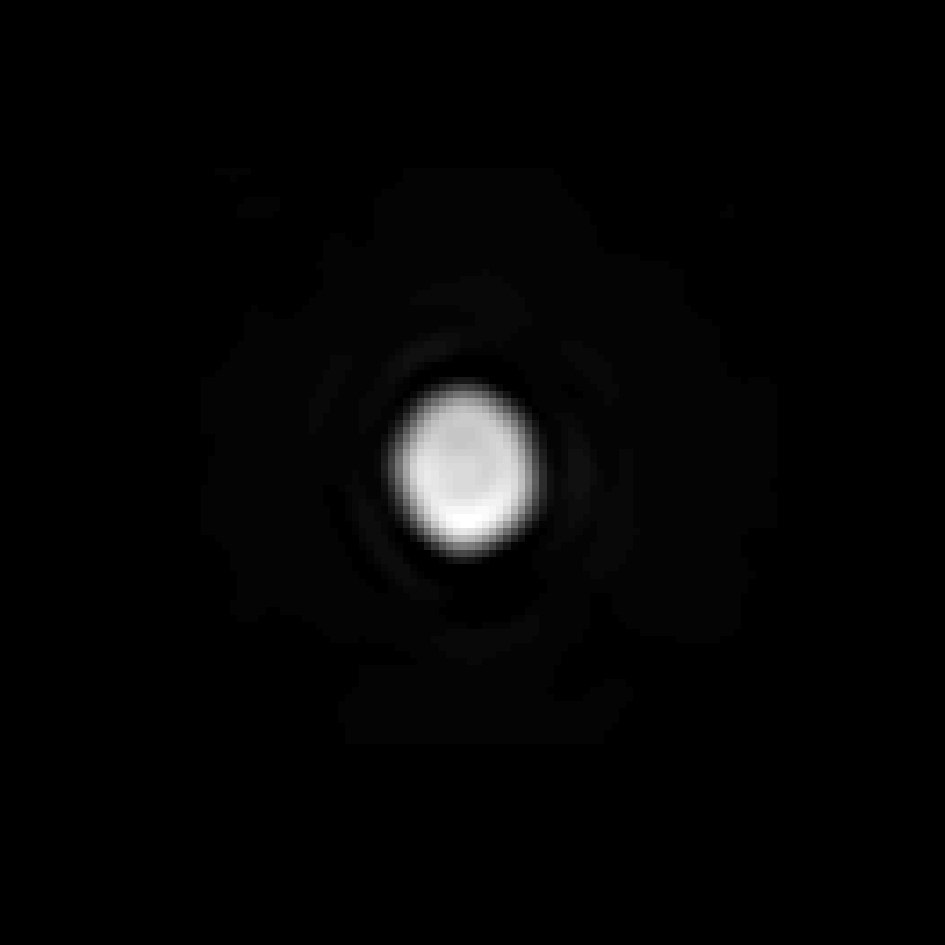}
     \end{subfigure}%
     \begin{subfigure}[b]{0.16\linewidth}
     \includegraphics[clip=true,trim=70 75 60 55,scale=0.39]{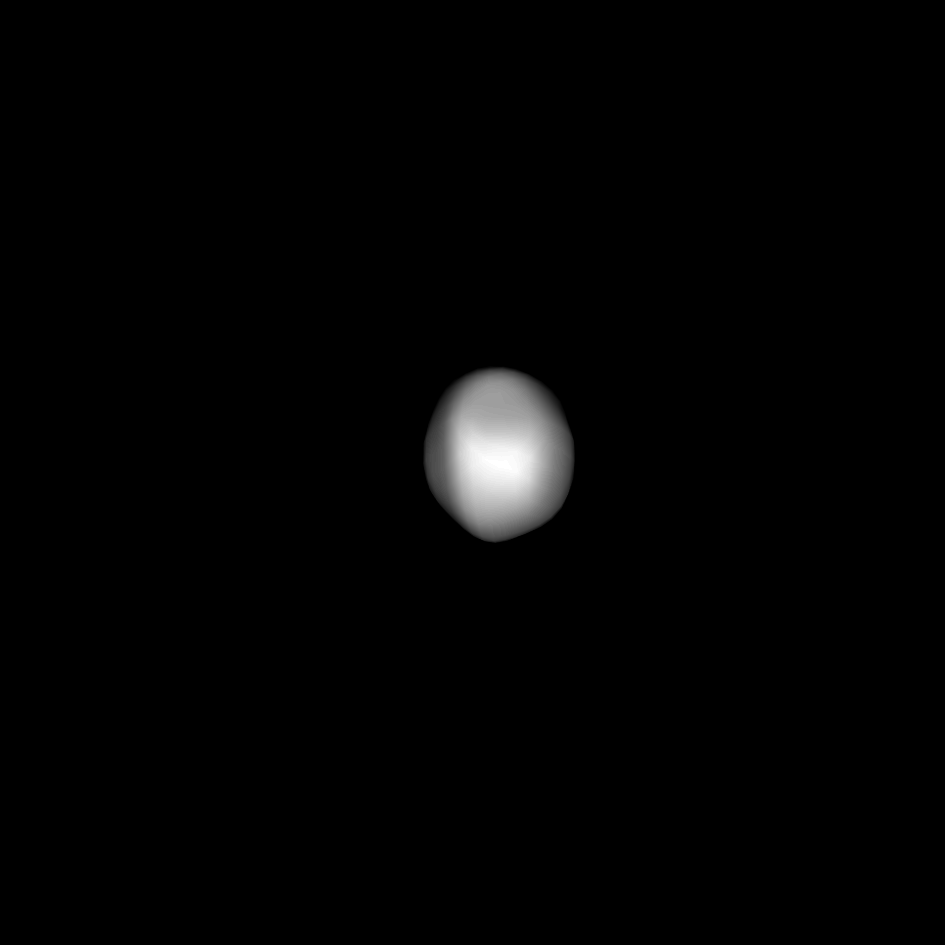}
     \end{subfigure}%
     \begin{subfigure}[b]{0.16\linewidth}
     \includegraphics[clip=true,trim=65 65 65 65,scale=0.39]{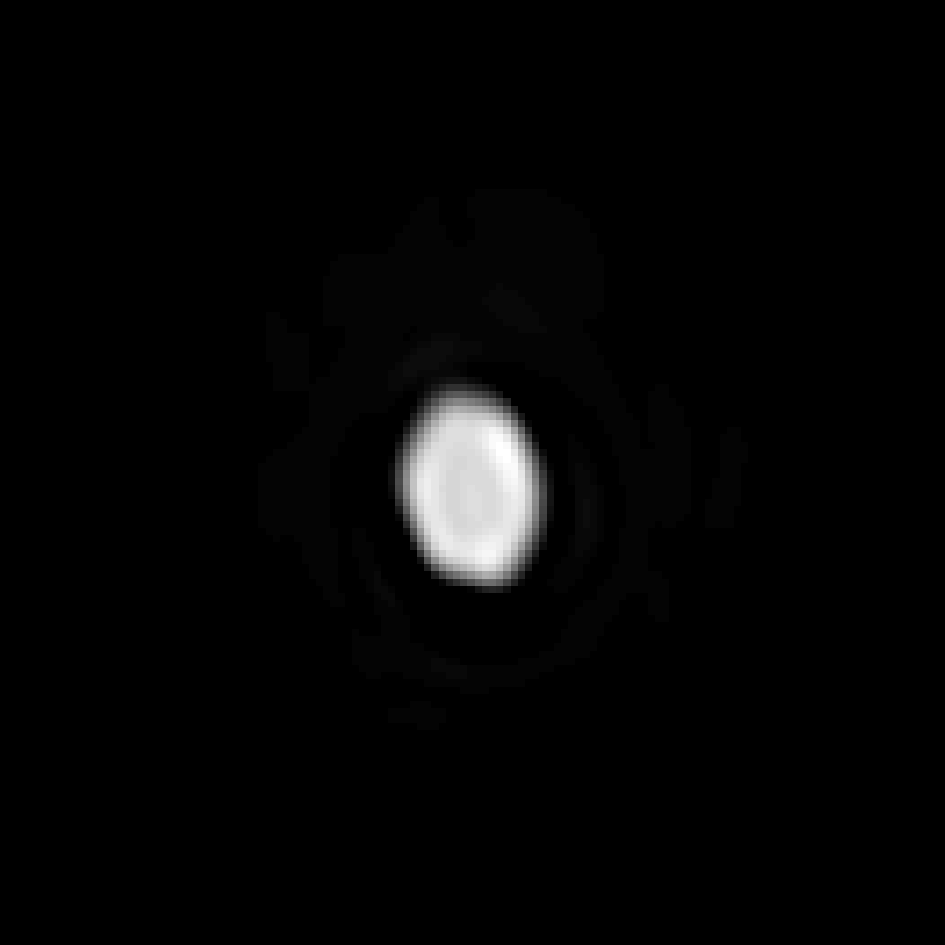}
     \end{subfigure}%
     \begin{subfigure}[b]{0.16\linewidth}
     \includegraphics[clip=true,trim=70 75 60 55,scale=0.39]{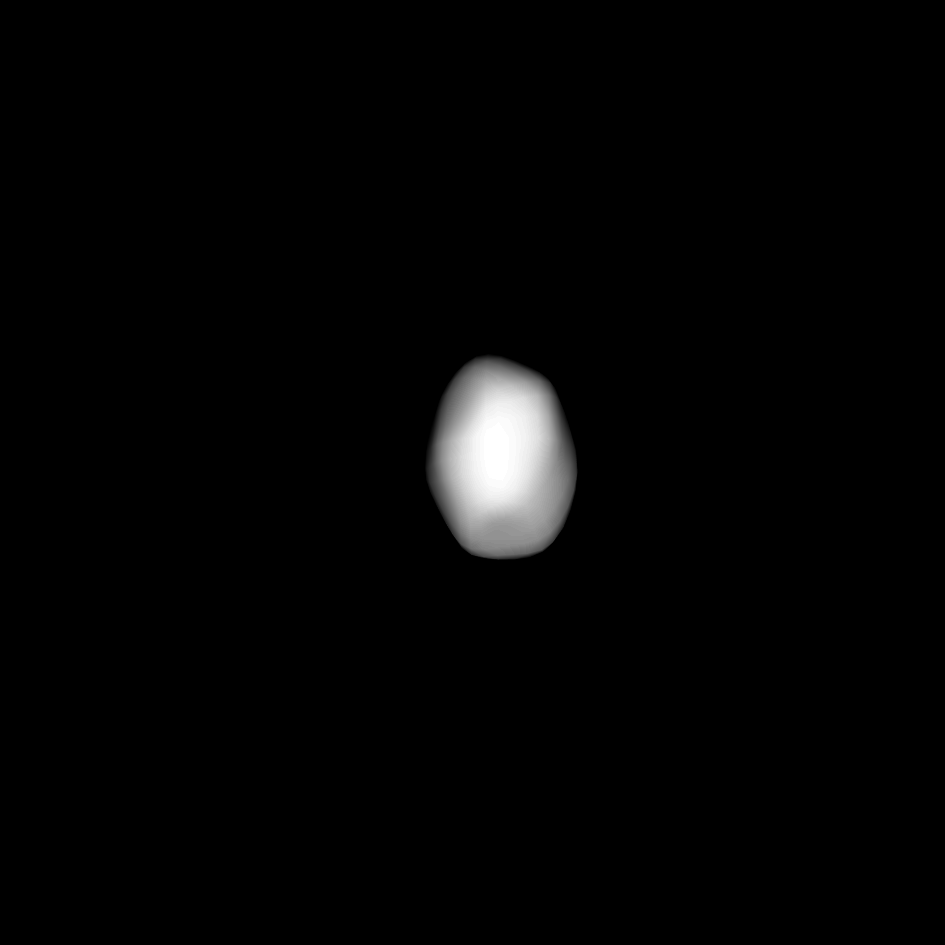}
     \end{subfigure}%
     
     \begin{subfigure}[b]{0.16\linewidth}
     \includegraphics[clip=true,trim=65 65 65 65,scale=0.39]{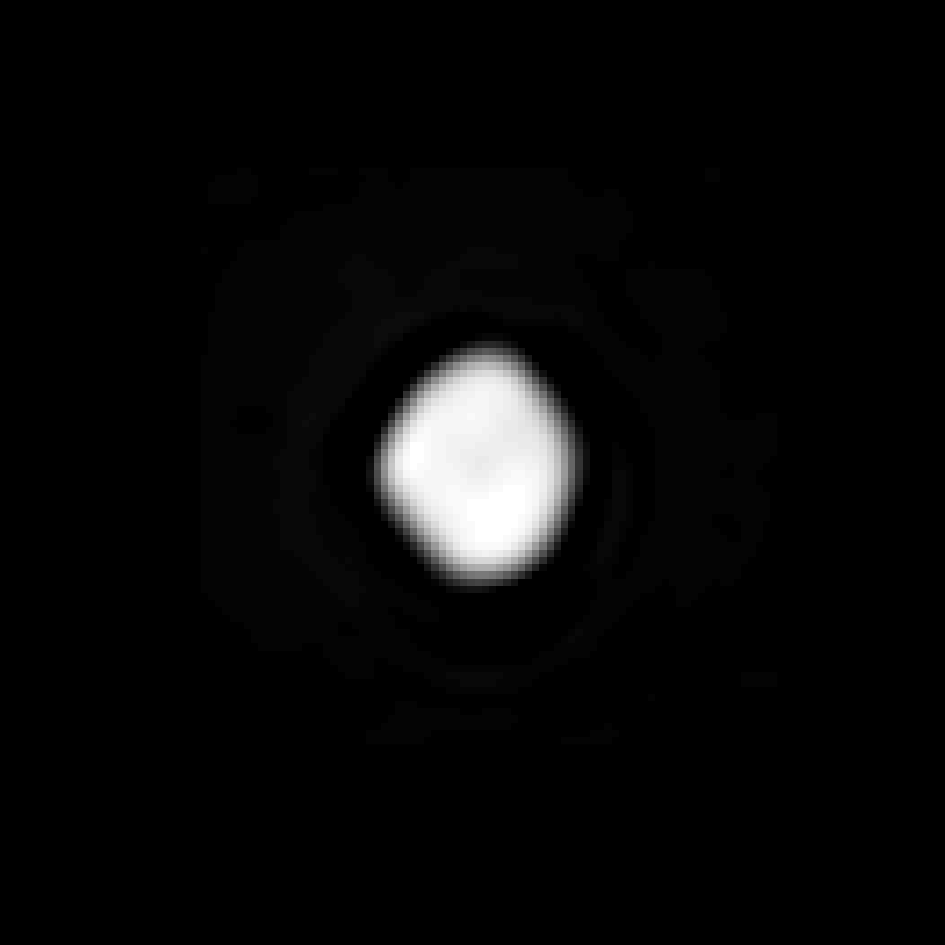}
     \end{subfigure}%
     \begin{subfigure}[b]{0.16\linewidth}
     \includegraphics[clip=true,trim=70 75 60 55,scale=0.39]{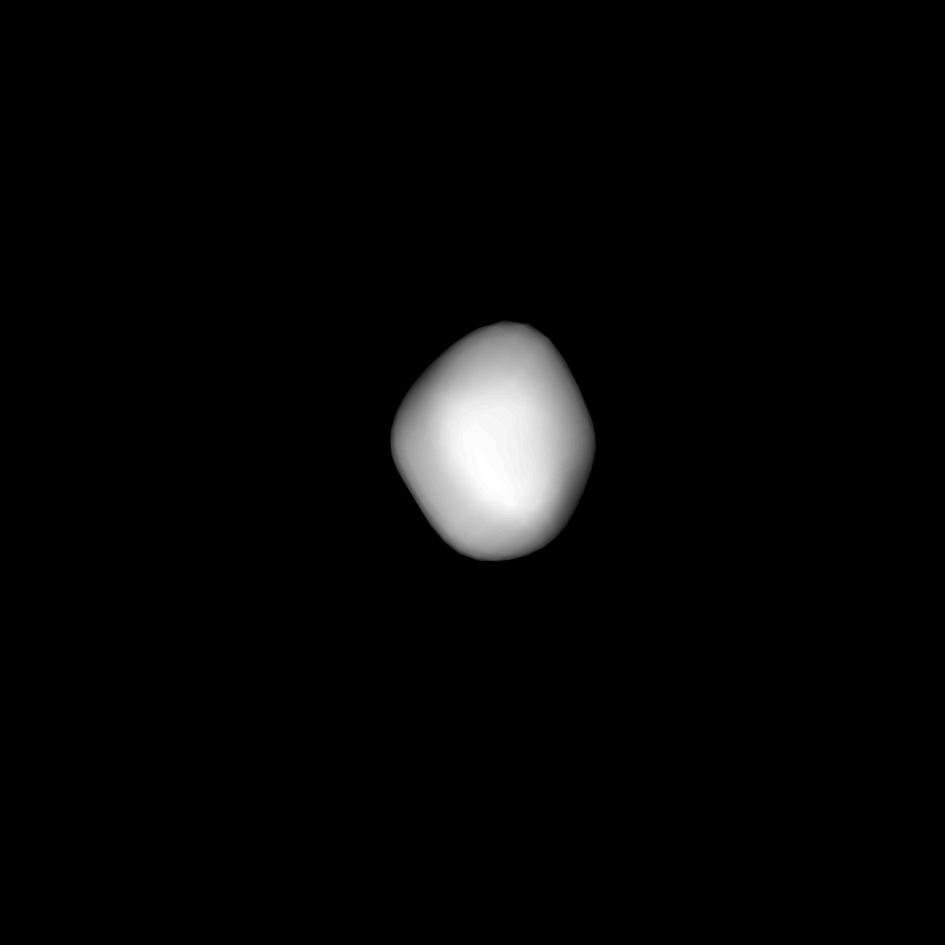}
     \end{subfigure}%
     \caption{\label{fig511}511 Davida}
     \end{figure}
     
     \begin{figure}[t]
     \begin{subfigure}[b]{0.16\linewidth}
      \includegraphics[clip=true,trim=85 85 85 85,scale=0.66]{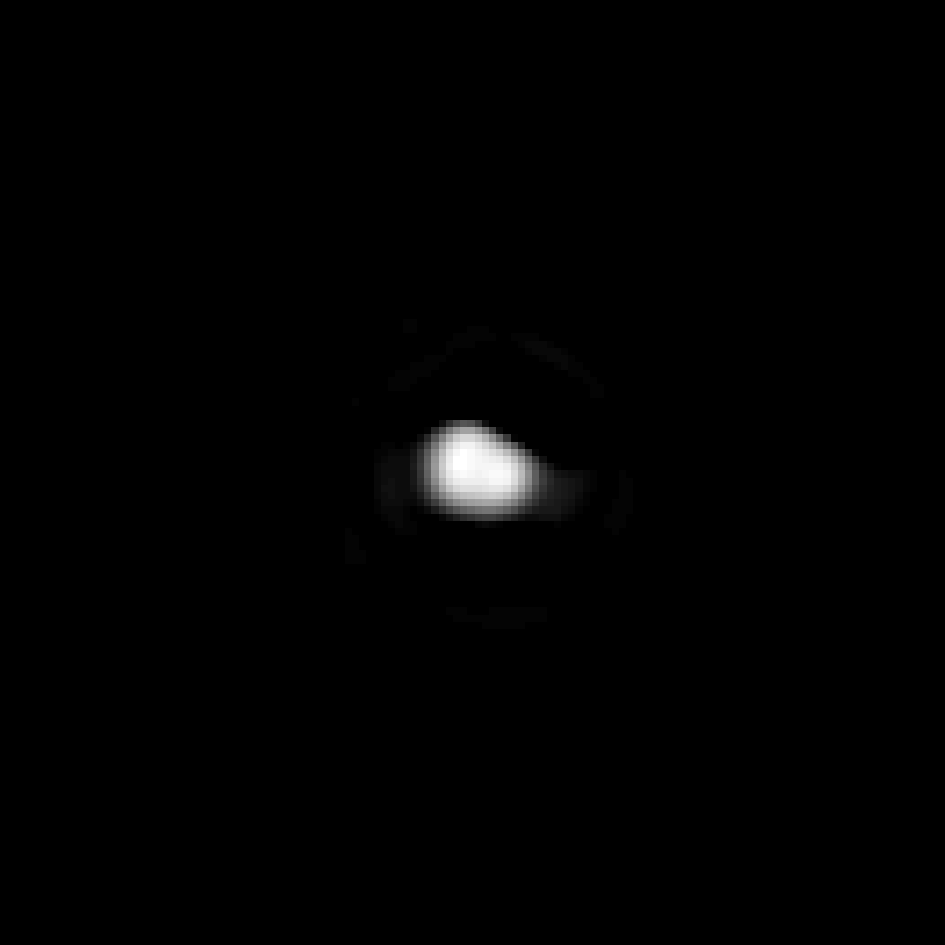} 
    \end{subfigure}%
     \begin{subfigure}[b]{0.16\linewidth}
     \includegraphics[clip=true,trim=90 90 80 80,scale=0.66]{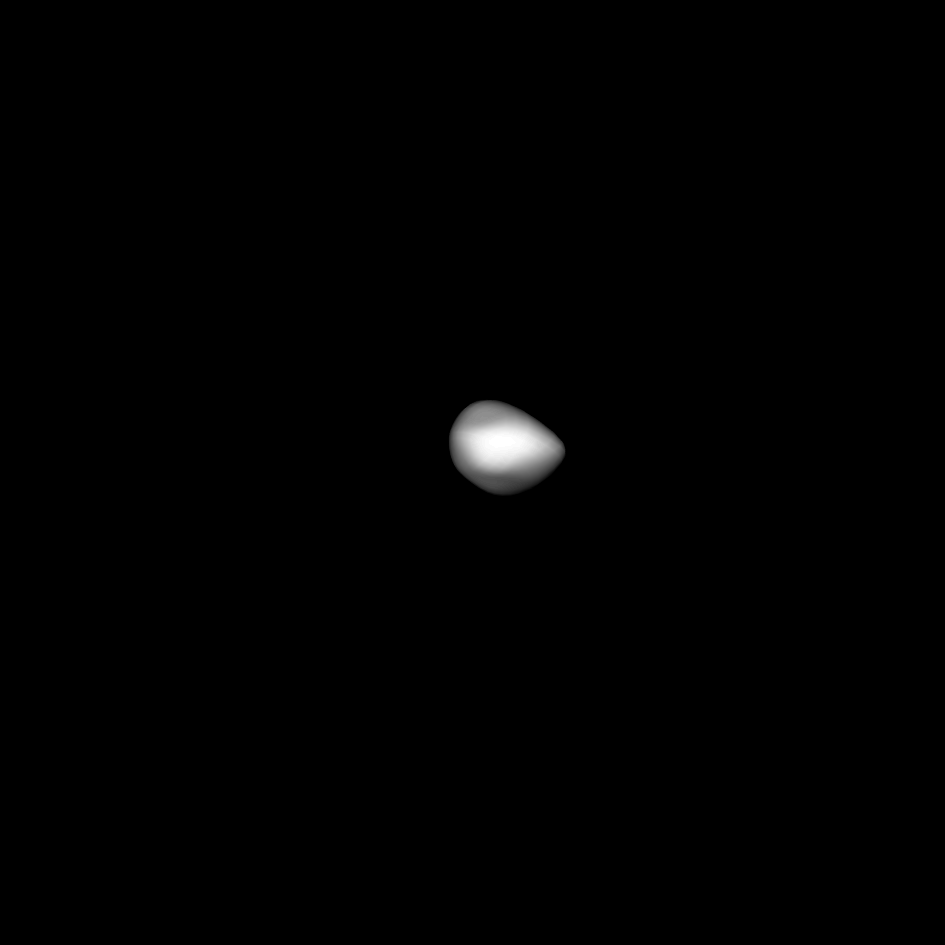}
     \end{subfigure}%
      \begin{subfigure}[b]{0.16\linewidth}
      \includegraphics[clip=true,trim=85 85 85 85,scale=0.66]{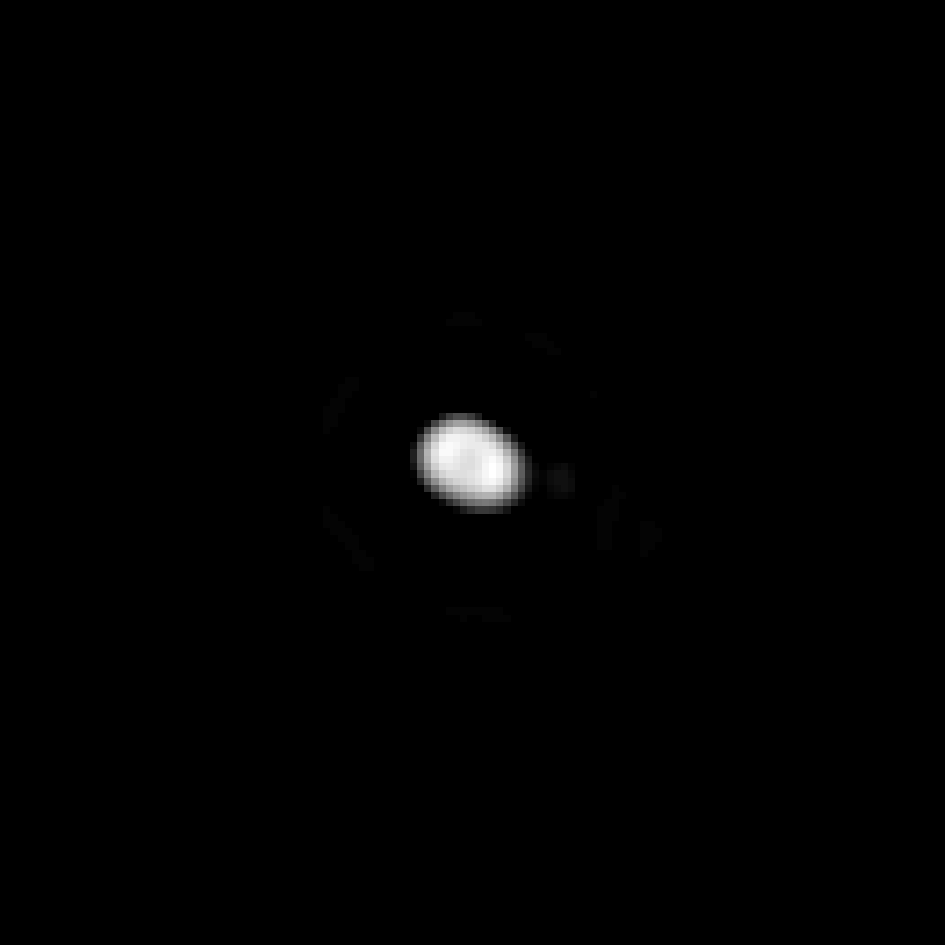}
    \end{subfigure}%
     \begin{subfigure}[b]{0.16\linewidth}
      \includegraphics[clip=true,trim=90 90 80 80,scale=0.66]{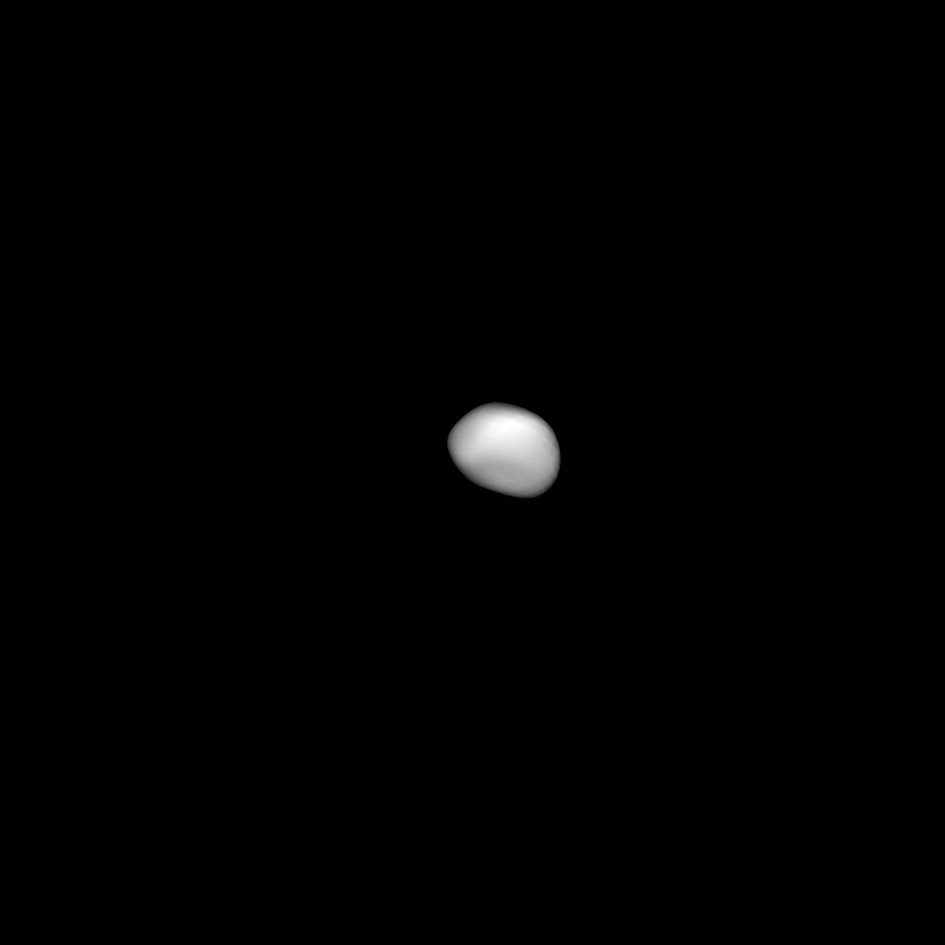}
    \end{subfigure}%
    \begin{subfigure}[b]{0.16\linewidth}
      \includegraphics[clip=true,trim=85 85 85 85,scale=0.66]{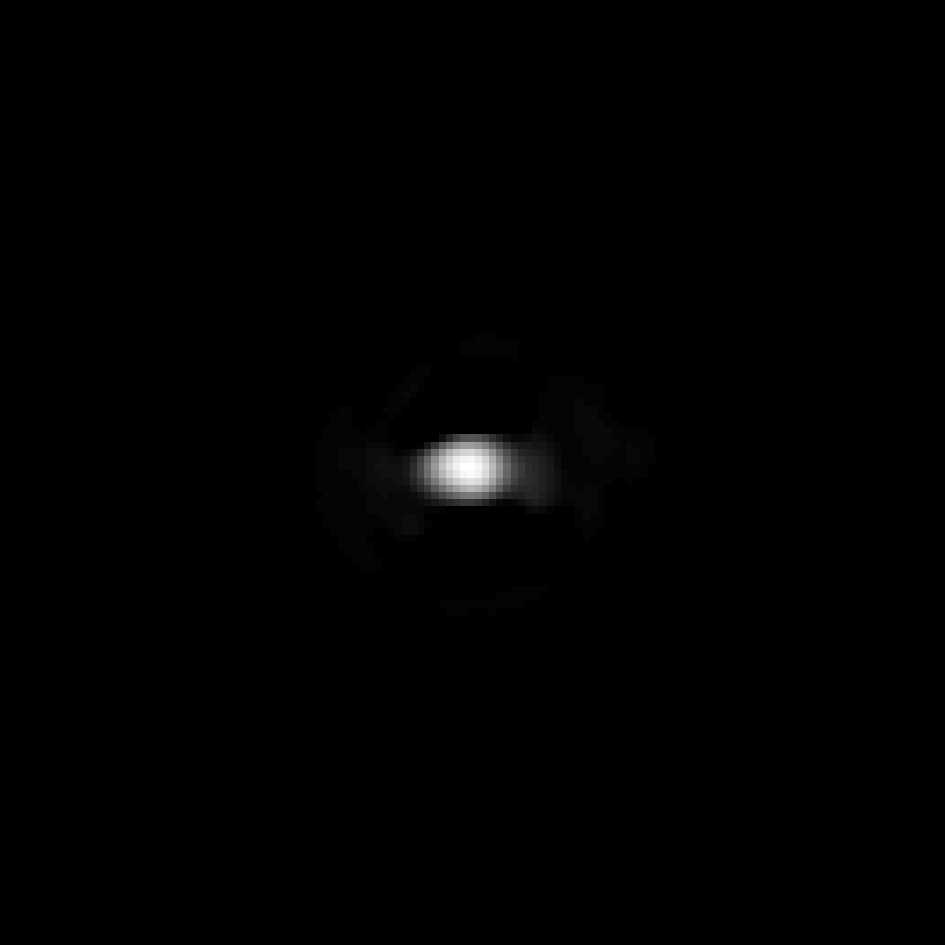} 
    \end{subfigure}%
     \begin{subfigure}[b]{0.16\linewidth}
     \includegraphics[clip=true,trim=90 90 80 80,scale=0.66]{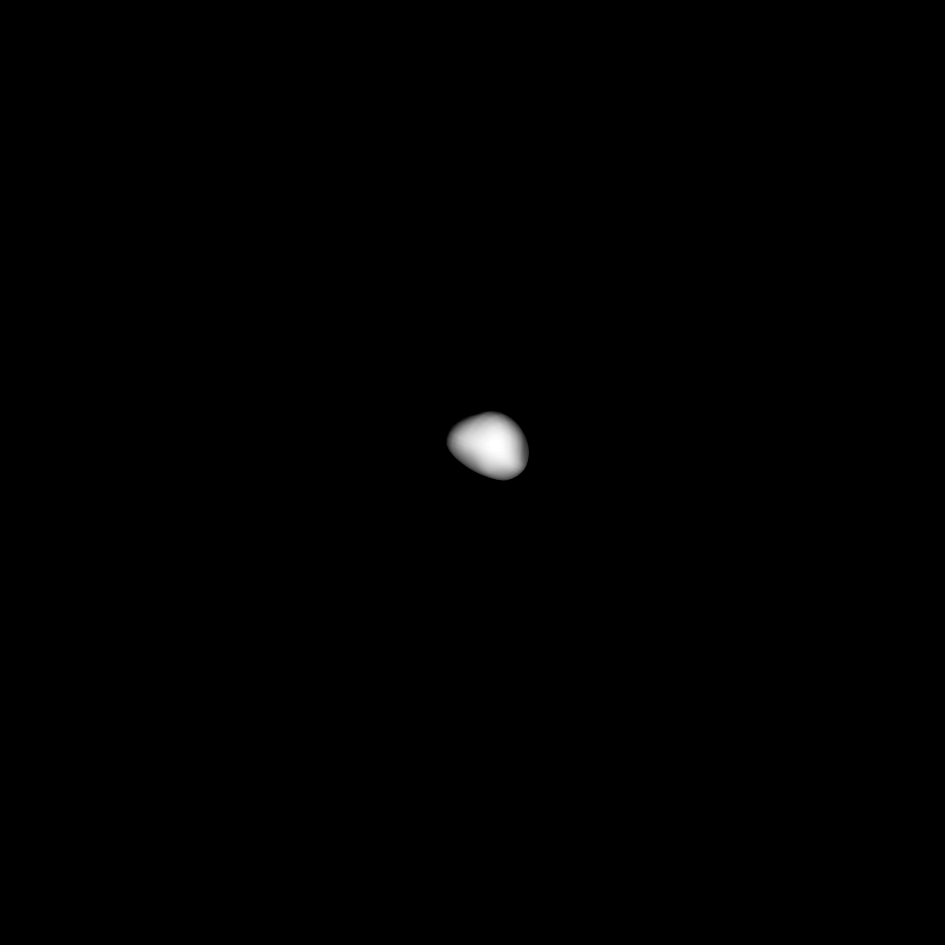}
     \end{subfigure}%
     
     \begin{subfigure}[b]{0.16\linewidth}
      \includegraphics[clip=true,trim=85 85 85 85,scale=0.66]{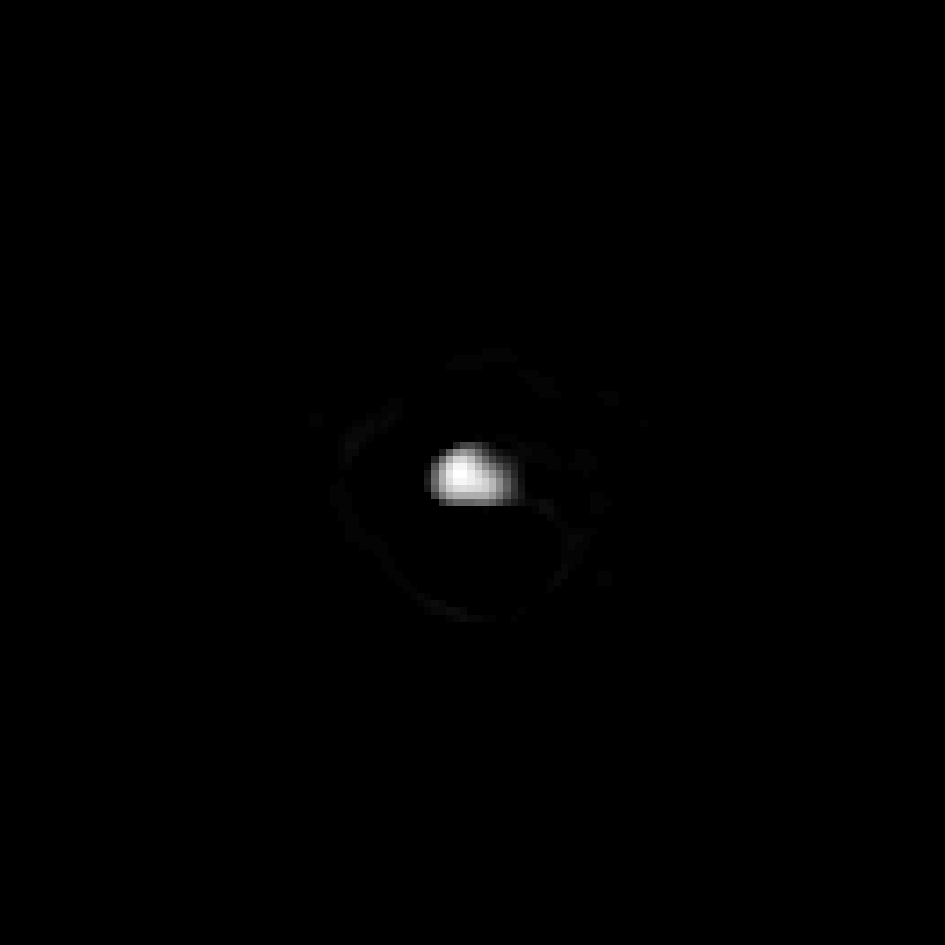} 
    \end{subfigure}%
     \begin{subfigure}[b]{0.16\linewidth}
     \includegraphics[clip=true,trim=90 90 80 80,scale=0.66]{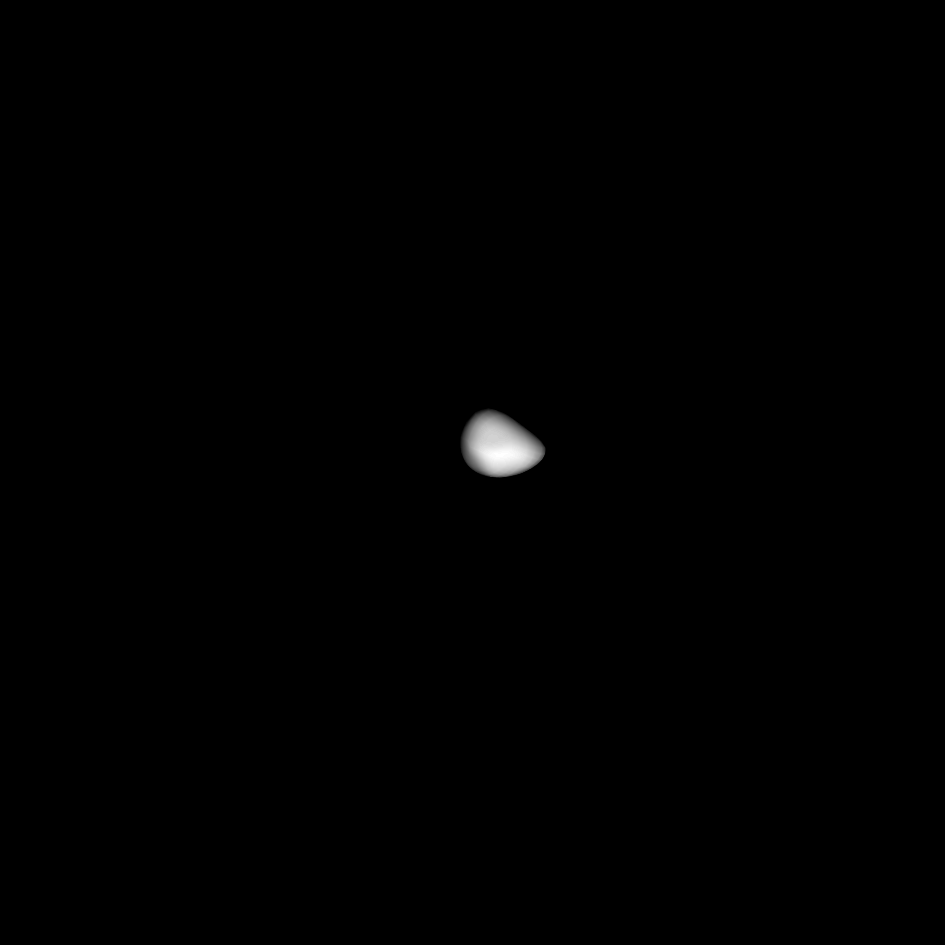}
     \end{subfigure}%
     \begin{subfigure}[b]{0.16\linewidth}
      \includegraphics[clip=true,trim=85 85 85 85,scale=0.66]{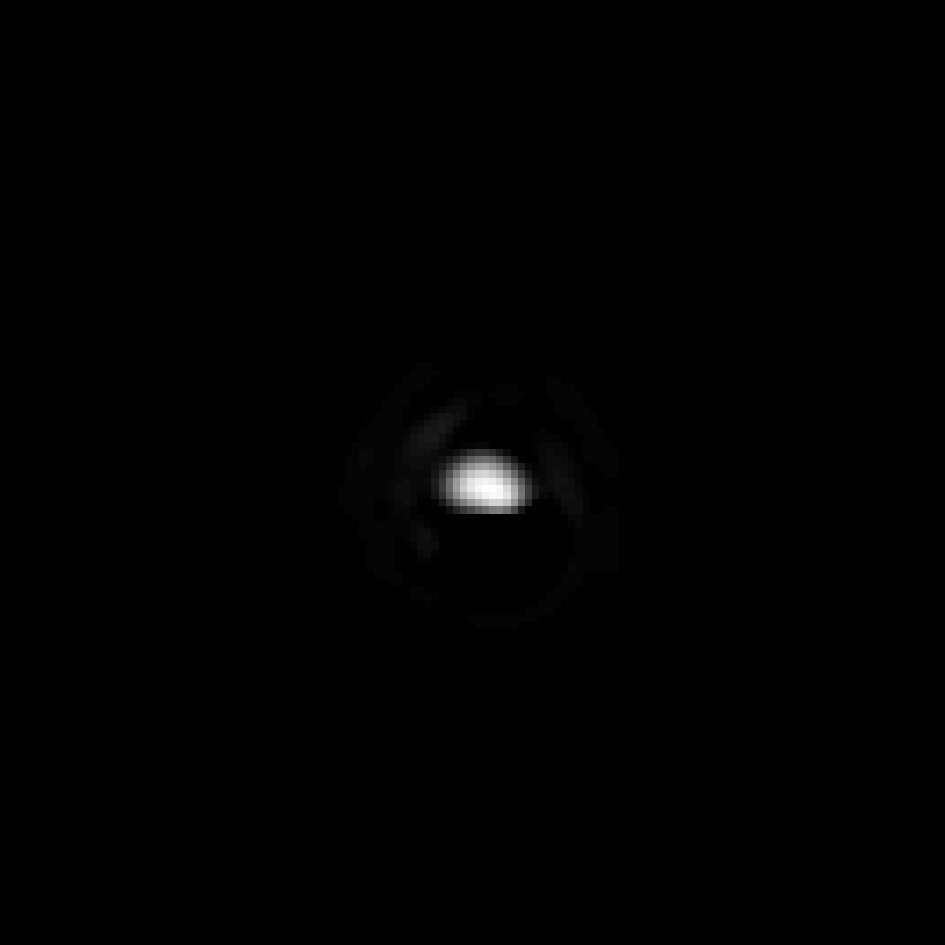} 
    \end{subfigure}%
     \begin{subfigure}[b]{0.16\linewidth}
     \includegraphics[clip=true,trim=90 90 80 80,scale=0.66]{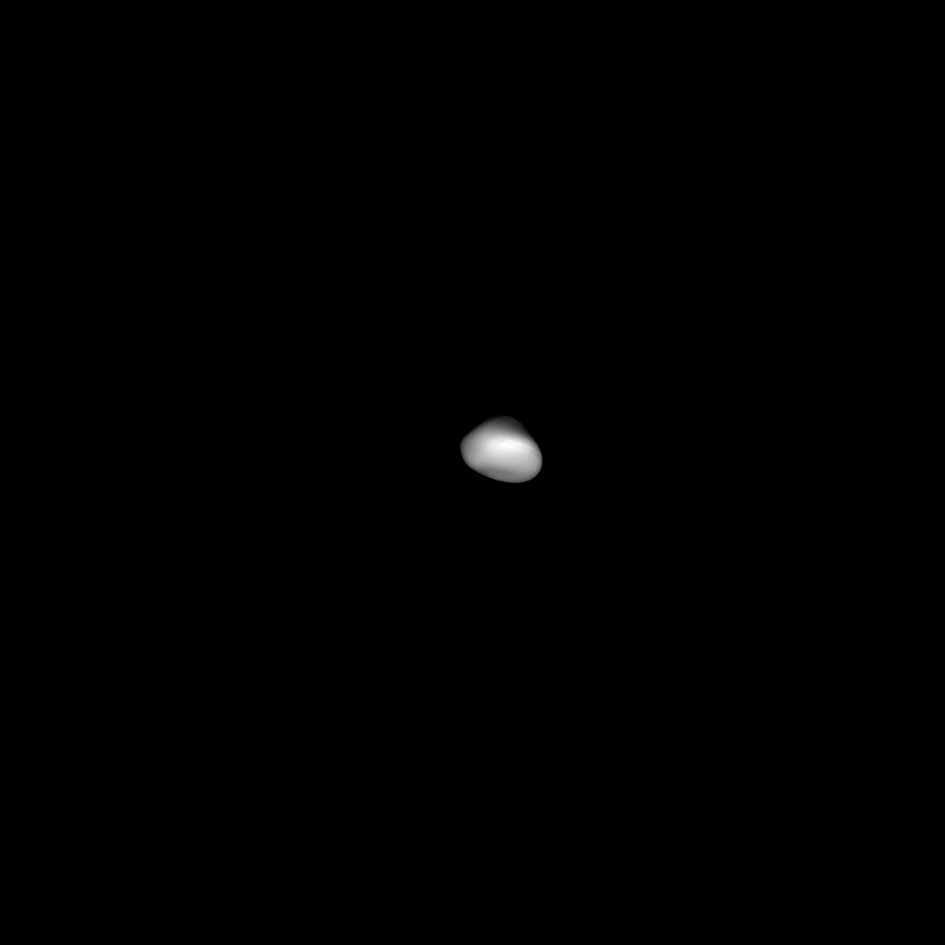}
     \end{subfigure}%
     \caption{\label{fig1036}1036 Ganymed}
\end{figure}

%% file: tabs/tab4.tex
\onecolumn
\scriptsize{
\begin{longtable}{rr rr rr rrr}
\caption{\label{tab:ao}List of disk-resolve images obtained by the NIRC2 at Keck II telescope used for the shape modeling with ADAM. For each observation, the table gives the epoch, filter, exposure time, airmass, R.A. and Dec of the asteroid, distance to the Earth $r$ and the reference or the PI of the project within which were the data obtained.}\\
\hline
\multicolumn{1}{c} {Date} & \multicolumn{1}{c} {UT} & \multicolumn{1}{c} {Filter} & \multicolumn{1}{c} {Exp} & \multicolumn{1}{c} {Airmass} & \multicolumn{1}{c} {R.A.} & \multicolumn{1}{c} {Dec} & \multicolumn{1}{c} {$r$} & Reference or PI \\ \hline\hline
\endfirsthead
\caption{continued.}\\

\hline
\multicolumn{1}{c} {Date} & \multicolumn{1}{c} {UT} & \multicolumn{1}{c} {Filter} & \multicolumn{1}{c} {Exp} & \multicolumn{1}{c} {Airmass} & \multicolumn{1}{c} {R.A.} & \multicolumn{1}{c} {Dec} & \multicolumn{1}{c} {$r$} & Reference or PI \\
\hline\hline
\endhead
\hline
\endfoot
\multicolumn{9}{c}  {      7         Iris}      \\
2005-07-17          &      07:53:59  &          Kp           &  0.6    &  1.37  &  16  13  01  &  $-$20  56  49  &  2.01  &  Marchis      \\
2009-08-16          &      07:50:06  &          PK50$\_$1.5  &  20.0   &  1.32  &  18  17  57  &  $-$19  00  54  &  1.70  &  Marchis      \\
2009-08-16          &      08:15:57  &          PK50$\_$1.5  &  5.0    &  1.36  &  18  17  57  &  $-$19  00  54  &  1.70  &  Marchis      \\
2002-09-27          &      09:54:15  &          Kp           &  0.36   &  1.29  &  21  45  43  &  $-$3   03  48  &  1.14  &  Merline      \\
2002-12-29          &      04:35:18  &          H            &  0.845  &  1.13  &  23  19  49  &  00     44  34  &  1.88  &  Margot       \\
2006-11-17          &      07:06:23  &          K            &  0.181  &  1.26  &  03  10  48  &  23     19  01  &  0.85  &  Engineering  \\
2006-11-17          &      07:13:20  &          H            &  0.1    &  1.24  &  03  10  48  &  23     19  00  &  0.85  &  Engineering  \\
2006-11-17          &      07:18:58  &          J            &  0.2    &  1.22  &  03  10  48  &  23     18  59  &  0.85  &  Engineering  \\
2006-11-17          &      07:53:59  &          Kp           &  0.11   &  1.12  &  03  10  47  &  23     18  45  &  0.85  &  Engineering  \\
2006-11-17          &      07:57:52  &          H            &  0.1    &  1.11  &  03  10  47  &  23     18  46  &  0.85  &  Engineering  \\
2006-11-17          &      08:02:23  &          J            &  0.12   &  1.10  &  03  10  47  &  23     18  45  &  0.85  &  Engineering  \\
2006-11-17          &      08:24:30  &          Kp           &  0.08   &  1.06  &  03  10  45  &  23     18  32  &  0.85  &  Engineering  \\
2006-11-17          &      08:27:22  &          H            &  0.08   &  1.06  &  03  10  45  &  23     18  32  &  0.85  &  Engineering  \\
2006-11-17          &      08:30:57  &          J            &  0.1    &  1.05  &  03  10  45  &  23     18  32  &  0.85  &  Engineering  \\
\multicolumn{9}{c}  {      12        Victoria}  \\
2003-06-05          &      12:41:58  &          Ks           &  0.4    &  1.44  &  17  18  22  &  $-$16  37  28  &  0.93  &  Merline      \\
2003-06-11          &      06:17:31  &          Kp           &  0.2    &  2.78  &  17  13  05  &  $-$15  49  31  &  0.91  &  Engineering  \\
2003-06-11          &      07:31:21  &          Kp           &  0.2    &  1.71  &  17  13  02  &  $-$15  49  02  &  0.91  &  Engineering  \\
2003-06-11          &      09:22:35  &          Kp           &  0.2    &  1.27  &  17  12  57  &  $-$15  48  37  &  0.91  &  Engineering  \\
2003-06-11          &      10:17:28  &          Kp           &  0.2    &  1.23  &  17  12  55  &  $-$15  48  10  &  0.91  &  Engineering  \\
2003-06-11          &      11:49:42  &          Kp           &  0.2    &  1.35  &  17  12  51  &  $-$15  47  39  &  0.91  &  Engineering  \\
2003-06-11          &      12:41:43  &          Kp           &  0.2    &  1.56  &  17  12  48  &  $-$15  47  14  &  0.91  &  Engineering  \\
2010-06-28          &      06:02:11  &          PK50$\_$1.5  &  2.903  &  1.21  &  14  46  41  &  $-$13  05  01  &  1.20  &  Marchis      \\
\multicolumn{9}{c}  {      14        Irene}     \\
2005-07-17          &      07:08:51  &          Kp           &  2.0    &  1.29  &  15  21  45  &  $-$15  45  58  &  1.74  &  Marchis      \\
2005-07-17          &      07:16:40  &          Kp           &  0.544  &  1.30  &  15  21  45  &  $-$15  45  58  &  1.74  &  Marchis      \\
2009-06-07          &      09:33:36  &          Kp           &  0.5    &  1.35  &  13  45  52  &  $-$1   12  34  &  1.47  &  Merline      \\
\multicolumn{9}{c}  {      15        Eunomia}   \\
2002-09-28          &      12:00:05  &          Kp           &  0.4    &  1.50  &  22  46  39  &  12     14  45  &  1.27  &  Merline      \\
2007-12-15          &      13:49:04  &          Kp           &  0.181  &  1.04  &  07  53  06  &  24     21  00  &  1.52  &  Engineering  \\
2007-12-15          &      14:06:07  &          Kp           &  0.181  &  1.07  &  07  53  05  &  24     20  58  &  1.52  &  Engineering  \\
2007-12-15          &      14:21:18  &          Kp           &  0.181  &  1.09  &  07  53  04  &  24     20  54  &  1.52  &  Engineering  \\
2007-12-15          &      14:36:30  &          Kp           &  0.181  &  1.12  &  07  53  04  &  24     20  53  &  1.52  &  Engineering  \\
2007-12-15          &      15:04:43  &          Kp           &  0.181  &  1.20  &  07  53  03  &  24     20  49  &  1.52  &  Engineering  \\
2007-12-15          &      16:05:04  &          Kp           &  0.181  &  1.48  &  07  53  01  &  24     20  43  &  1.52  &  Engineering  \\
2008-01-21          &      12:07:00  &          Kp           &  0.181  &  1.22  &  07  14  12  &  22     37  15  &  1.51  &  Engineering  \\
\multicolumn{9}{c}  {      23        Thalia}    \\
2009-08-16          &      12:08:31  &          PK50$\_$1.5  &  10.0   &  1.24  &  00  02  49  &  $-$15  15  28  &  2.16  &  Marchis      \\
2009-08-16          &      12:15:19  &          PK50$\_$1.5  &  10.0   &  1.23  &  00  02  49  &  $-$15  15  28  &  2.16  &  Marchis      \\
\multicolumn{9}{c}  {      24        Themis}    \\
2001-12-27          &      08:39:34  &          Kp           &  2.0    &  1.00  &  04  43  15  &  23     03  15  &  1.98  &  Merline      \\
2003-06-05          &      07:44:11  &          Ks           &  5.0    &  1.29  &  11  53  34  &  00     59  26  &  2.43  &  Merline      \\
2012-12-24          &      06:21:00  &          H            &  60.0   &  1.14  &  04  14  45  &  21     55  34  &  2.04  &  Margot       \\
2010-06-28          &      11:04:11  &          PK50$\_$1.5  &  20.0   &  1.38  &  20  14  03  &  $-$20  55  24  &  2.52  &  Marchis      \\
\multicolumn{9}{c}  {      28        Bellona}   \\
2007-04-03          &      13:31:47  &          Kp           &  3.0    &  1.25  &  13  40  15  &  02     49  36  &  1.62  &  Marchis      \\
\multicolumn{9}{c}  {      30        Urania}    \\
2010-06-28          &      08:02:14  &          PK50$\_$1.5  &  5.0    &  1.40  &  16  27  30  &  $-$24  15  12  &  1.61  &  Marchis      \\
\multicolumn{9}{c}  {      37        Fides}     \\
2009-08-16          &      07:03:28  &          PK50$\_$1.5  &  20.0   &  1.53  &  19  36  55  &  $-$25  42  35  &  1.96  &  Marchis      \\
\multicolumn{9}{c}  {      40        Harmonia}  \\
2001-12-27          &      11:21:54  &          Kp           &  0.5    &  1.10  &  05  37  33  &  22     43  02  &  1.28  &  Merline      \\
2003-06-05          &      10:14:51  &          Ks           &  1.0    &  1.44  &  14  21  39  &  $-$9   12  28  &  1.45  &  Merline      \\
2009-02-10          &      07:28:45  &          Kp           &  30.0   &  1.02  &  07  09  06  &  25     29  43  &  1.44  &  Engineering  \\
\multicolumn{9}{c}  {      42        Isis}      \\
2005-07-17          &      10:51:46  &          Kp           &  0.8    &  1.54  &  20  07  09  &  $-$29  56  29  &  0.92  &  Marchis      \\
\multicolumn{9}{c}  {      48        Doris}     \\
2003-06-05          &      07:34:20  &          Ks           &  5.0    &  1.31  &  11  34  41  &  03     45  38  &  2.86  &  Merline      \\
2010-06-28          &      08:19:01  &          PK50$\_$1.5  &  8.0    &  1.63  &  19  06  53  &  $-$13  13  48  &  2.32  &  Marchis      \\
\multicolumn{9}{c}  {      53        Kalypso}   \\
2002-03-07          &      13:02:56  &          Kp           &  5.0    &  1.27  &  11  10  11  &  08     58  21  &  1.35  &  Merline      \\
\multicolumn{9}{c}  {      56        Melete}    \\
2003-06-05          &      10:26:26  &          Ks           &  4.0    &  1.35  &  14  42  12  &  $-$6   05  54  &  1.29  &  Merline      \\
2008-09-19          &      13:05:41  &          PK50$\_$1.5  &  20.0   &  1.02  &  02  08  24  &  11     37  15  &  1.49  &  Marchis      \\
\multicolumn{9}{c}  {      65        Cybele}    \\
2002-09-27          &      07:37:48  &          Kp           &  10.0   &  1.58  &  19  30  25  &  $-$19  54  47  &  2.73  &  Merline      \\
2002-09-28          &      06:45:48  &          Kp           &  10.0   &  1.40  &  19  30  52  &  $-$19  54  59  &  2.74  &  Merline      \\
2002-09-28          &      06:55:36  &          H            &  12.0   &  1.42  &  19  30  52  &  $-$19  55  00  &  2.74  &  Merline      \\
2002-09-28          &      07:11:00  &          Kp           &  12.0   &  1.47  &  19  30  52  &  $-$19  55  06  &  2.75  &  Merline      \\
2002-09-28          &      07:18:40  &          Kp           &  8.0    &  1.51  &  19  30  52  &  $-$19  55  06  &  2.75  &  Merline      \\
2003-08-10          &      14:53:57  &          Kp           &  8.0    &  1.06  &  01  01  02  &  04     28  28  &  2.82  &  Merline      \\
2010-06-28          &      14:45:28  &          PK50$\_$1.5  &  40.0   &  1.76  &  02  40  04  &  12     56  53  &  4.13  &  Marchis      \\
\multicolumn{9}{c}  {      72        Feronia}   \\
2005-07-17          &      11:09:51  &          Kp           &  1.1    &  1.12  &  20  38  01  &  $-$7   21  39  &  1.01  &  Marchis      \\
2002-09-28          &      15:30:05  &          Kp           &  5.0    &  1.55  &  02  08  08  &  14     08  39  &  1.22  &  Merline      \\
\multicolumn{9}{c}  {      121       Hermione}  \\
2003-12-06          &      12:26:43  &          Kp           &  10.0   &  1.02  &  07  38  49  &  26     19  43  &  2.57  &  Marchis      \\
2003-12-06          &      13:55:55  &          Kp           &  15.0   &  1.03  &  07  38  48  &  26     19  46  &  2.57  &  Marchis      \\
2003-12-07          &      12:28:01  &          Kp           &  10.0   &  1.01  &  07  38  20  &  26     23  08  &  2.56  &  Marchis      \\
2005-01-15          &      14:21:45  &          Kp           &  15.0   &  1.01  &  11  35  46  &  12     31  20  &  3.19  &  Marchis      \\
2005-01-15          &      12:42:08  &          Kp           &  15.0   &  1.09  &  11  35  47  &  12     31  35  &  3.19  &  Marchis      \\
2007-08-02          &      07:30:51  &          Kp           &  5.0    &  1.46  &  17  22  47  &  $-$26  43  07  &  2.67  &  Marchis      \\
2008-09-19          &      07:32:15  &          PK50$\_$1.5  &  15.0   &  1.53  &  23  21  12  &  $-$16  25  21  &  2.03  &  Marchis      \\
2008-09-19          &      10:10:15  &          PK50$\_$1.5  &  12.0   &  1.24  &  23  21  07  &  $-$16  25  41  &  2.03  &  Marchis      \\
2009-08-16          &      15:02:37  &          PK50$\_$1.5  &  40.0   &  1.21  &  04  48  44  &  20     11  35  &  3.27  &  Marchis      \\
2002-09-28          &      14:56:32  &          Kp           &  5.0    &  1.45  &  02  04  46  &  03     22  32  &  2.02  &  Merline      \\
2002-09-28          &      15:03:08  &          H            &  5.0    &  1.49  &  02  04  46  &  03     22  32  &  2.02  &  Merline      \\
2002-09-28          &      15:05:54  &          J            &  5.0    &  1.51  &  02  04  46  &  03     22  32  &  2.02  &  Merline      \\
2002-09-28          &      15:12:02  &          Kp           &  5.0    &  1.55  &  02  04  46  &  03     22  30  &  2.02  &  Merline      \\
\multicolumn{9}{c}  {      146       Lucina}    \\
2005-07-17          &      08:10:47  &          Kp           &  6.0    &  1.41  &  16  41  00  &  $-$23  20  33  &  1.74  &  Marchis      \\
2010-06-28          &      10:29:22  &          PK50$\_$1.5  &  30.0   &  2.62  &  21  58  45  &  $-$25  53  44  &  1.88  &  Marchis      \\
\multicolumn{9}{c}  {      250       Bettina}   \\
2008-09-19          &      08:00:49  &          PK50$\_$1.5  &  10.0   &  1.45  &  00  00  47  &  $-$9   25  33  &  2.20  &  Marchis      \\
\multicolumn{9}{c}  {      258       Tyche}     \\
2010-11-30          &      08:38:14  &          PK50$\_$1.5  &  30.0   &  1.03  &  03  01  26  &  05     17  12  &  1.30  &  Marchis      \\
\multicolumn{9}{c}  {      283       Emma}      \\
2003-07-14          &      13:23:23  &          Kp           &  6.0    &  1.26  &  21  24  10  &  $-$14  13  39  &  1.76  &  Merline      \\
2003-07-14          &      13:39:43  &          H            &  6.0    &  1.29  &  21  24  10  &  $-$14  13  39  &  1.76  &  Merline      \\
2003-08-10          &      11:20:19  &          H            &  2.0    &  1.27  &  21  02  58  &  $-$14  18  56  &  1.66  &  Merline      \\
2003-08-14          &      09:56:31  &          H            &  5.0    &  1.21  &  20  59  33  &  $-$14  21  13  &  1.66  &  Margot       \\
2003-08-17          &      11:26:42  &          Kp           &  5.0    &  1.38  &  20  56  57  &  $-$14  22  36  &  1.67  &  Merline      \\
\multicolumn{9}{c}  {      354       Eleonora}  \\
2007-08-02          &      09:08:59  &          PK50$\_$1.5  &  6.0    &  1.20  &  20  08  53  &  $-$12  09  26  &  2.08  &  Marchis      \\
2002-05-07          &      13:57:06  &          H            &  3.0    &  1.10  &  19  02  14  &  $-$3   48  38  &  2.31  &  Margot       \\
2002-05-08          &      13:42:48  &          H            &  2.0    &  1.10  &  19  02  15  &  $-$3   45  46  &  2.30  &  Margot       \\
\multicolumn{9}{c}  {      511       Davida}    \\
2001-12-27          &      05:00:29  &          Kp           &  5.0    &  1.32  &  23  55  36  &  $-$17  26  41  &  2.98  &  Merline      \\
2002-09-22          &      15:28:12  &          H            &  4.0    &  1.07  &  06  34  38  &  14     25  57  &  2.55  &  Dumas        \\
2002-09-22          &      15:37:16  &          Kp           &  4.0    &  1.05  &  06  34  38  &  14     25  57  &  2.55  &  Dumas        \\
2002-12-29          &      12:03:31  &          H            &  2.0    &  1.06  &  06  50  11  &  19     28  10  &  1.62  &  Margot       \\
2007-11-01          &      06:11:50  &          Kp           &  1.8    &  2.06  &  01  59  35  &  $-$13  11  05  &  1.87  &  Engineering  \\
2007-11-01          &      07:24:57  &          Kp           &  1.8    &  1.46  &  01  59  33  &  $-$13  11  08  &  1.87  &  Engineering  \\
2007-11-01          &      07:40:28  &          Kp           &  1.8    &  1.40  &  01  59  32  &  $-$13  11  12  &  1.87  &  Engineering  \\
2007-11-01          &      08:37:46  &          Kp           &  1.8    &  1.24  &  01  59  30  &  $-$13  11  14  &  1.87  &  Engineering  \\
2007-11-01          &      10:06:29  &          Kp           &  1.8    &  1.20  &  01  59  27  &  $-$13  11  17  &  1.87  &  Engineering  \\
2002-12-27          &      11:03:14  &          Kp           &  1.0    &  1.00  &  06  52  02  &  19     12  18  &  1.62  &  Marchis      \\
\multicolumn{9}{c}  {1036  Ganymed}  \\
2011-10-23          &      07:06:19  &          Kp           &  0.181  &  1.42  &  02  03  49  &  21     04  20  &  0.38  &  Merline      \\
2011-10-23          &      08:39:25  &          Kp           &  0.181  &  1.09  &  02  03  51  &  21     00  16  &  0.38  &  Merline      \\
2011-11-13          &      05:26:30  &          Kp           &  0.25   &  1.84  &  02  12  21  &  00     25  33  &  0.51  &  Armandroff   \\
2011-11-13          &      07:57:11  &          H            &  0.25   &  1.11  &  02  12  23  &  00     21  27  &  0.51  &  Armandroff   \\
2011-11-13          &      10:04:47  &          Kp           &  0.25   &  1.10  &  02  12  25  &  00     17  51  &  0.51  &  Armandroff   \\
\hline
\end{longtable}
}
\twocolumn